\documentclass[12pt]{article}
\usepackage{latexsym}
\input{epsf}

\def\hybrid{\topmargin -20pt    \oddsidemargin 0pt
        \headheight 0pt \headsep 0pt
       \textwidth 6.5in        
       \textheight 9in         
       \textwidth 6.25in       
       \textheight 9.5in       
        \textwidth 6.35in
        \textheight 9.65in
        \marginparwidth .875in
        \parskip 5pt plus 1pt   \jot = 1.5ex}

\catcode`\@=11
\newcount\hour
\newcount\minute
\newtoks\amorpm
\hour=\time\divide\hour by60
\minute=\time{\multiply\hour by60 \global\advance\minute by-\hour}
\edef\standardtime{{\ifnum\hour<12 \global\amorpm={am}%
        \else\global\amorpm={pm}\advance\hour by-12 \fi
        \ifnum\hour=0 \hour=12 \fi
        \number\hour:\ifnum\minute<10 0\fi\number\minute\the\amorpm}}
\edef\militarytime{\number\hour:\ifnum\minute<10 0\fi\number\minute}

\def\marginnote#1{}
\def\draftlabel#1{{\@bsphack\if@filesw {\let\thepage\relax
   \xdef\@gtempa{\write\@auxout{\string
      \newlabel{#1}{{\@currentlabel}{\thepage}}}}}\@gtempa
   \if@nobreak \ifvmode\nobreak\fi\fi\fi\@esphack}
        \gdef\@eqnlabel{#1}}
\def\@eqnlabel{}
\def\@vacuum{}
\def\draftmarginnote#1{\marginpar{\raggedright\scriptsize\tt#1}}

\def\draft{\oddsidemargin -.2truein
        \def\@oddfoot{\sl preliminary draft \hfil
        \rm\thepage\hfil\sl\today\quad\militarytime}
        \let\@evenfoot\@oddfoot \overfullrule 3pt
        \let\label=\draftlabel
        \let\marginnote=\draftmarginnote
   \def\@eqnnum{(\theequation)\rlap{\kern\marginparsep\tt\@eqnlabel}%
\global\let\@eqnlabel\@vacuum}  }

\def\preprint{\twocolumn\sloppy\flushbottom\parindent 2em
        \leftmargini 2em\leftmarginv .5em\leftmarginvi .5em
        \oddsidemargin -.5in    \evensidemargin -.5in
        \columnsep .4in \footheight 0pt
        \textwidth 10.in        \topmargin  -.4in
        \headheight 12pt \topskip .4in
        \textheight 6.9in \footskip 0pt
        \def\@oddhead{\thepage\hfil\addtocounter{page}{1}\thepage}
        \let\@evenhead\@oddhead \def\@oddfoot{} \def\@evenfoot{} }
\def\numberbysection{\@addtoreset{equation}{section}
        \def\theequation{\thesection.\arabic{equation}}}

\def\underline#1{\relax\ifmmode\@@underline#1\else
        $\@@underline{\hbox{#1}}$\relax\fi}

\def\titlepage{\@restonecolfalse\if@twocolumn\@restonecoltrue
\onecolumn
     \else \newpage \fi \thispagestyle{empty}\c@page\z@
        \def\thefootnote{\fnsymbol{footnote}} }

\def\endtitlepage{\if@restonecol\twocolumn \else \newpage \fi
        \def\thefootnote{\arabic{footnote}}
        \setcounter{footnote}{0}}  

\catcode`@=12
\relax
\hybrid
\begin{document}

\def\be{\begin{equation}}
\def\ee{\end{equation}}
\def\bea{\begin{eqnarray}}
\def\eea{\end{eqnarray}}
\def\tfrac#1#2{{\textstyle{#1\over #2}}}
\def\r#1{{(\ref{#1})}}
\def\half{\tfrac{1}{2}}
\def\quart{\tfrac{1}{4}}
\def\nn{\nonumber}
\def\na{\nabla}
\def\ms{M_{\rm string}}
\def\gs{g_{\rm string}}
\def\pd{\partial}
\def\a{\alpha}
\def\b{\beta}
\def\g{\gamma}
\def\d{\delta}
\def\m{\mu}
\def\n{\nu}
\def\t{\tau}
\def\l{\lambda}
\def\mh{\hat\m}
\def\nh{\hat\n}
\def\rh{\hat\rho}
\def\th{\vartheta}
\def\s{\sigma}
\def\e{\epsilon}
\def\ap{\alpha'}
\def\al{\alpha}
\font\mybb=msbm10 at 12pt
\def\bb#1{\hbox{\mybb#1}}
\def\Z{\bb{Z}}
\def\R{\bb{R}}
\def\Z{\bb{Z}}
\def\R{\bb{R}}
\def\C{\bb{C}}
\def\square{\hbox{{$\sqcup$}\llap{$\sqcap$}}}
\def\dslash{{\partial\hspace{-7pt}/}}
\hyphenation{re-pa-ra-me-tri-za-tion}
\hyphenation{trans-for-ma-tions}

\begin{titlepage}
\begin{center}
\hfill CERN-TH/97-218\\
\hfill hep-th/9709062\\

\vskip 2cm
{\Large \bf INTRODUCTION  TO  SUPERSTRING THEORY}
\vskip .8in

{\bf Elias Kiritsis}\footnote{e-mail: KIRITSIS@NXTH04.CERN.CH}
\vskip .1in
{\em  Theory Division,  CERN, \\
     CH-1211,  Geneva 23, SWITZERLAND}
\end{center}
\vskip 1in
\begin{abstract}
In these lecture notes, an introduction to superstring theory is
presented. Classical strings, covariant and light-cone
quantization, supersymmetric strings, anomaly cancelation,
compactification, T-duality, supersymmetry breaking,
and threshold corrections to low-energy couplings are discussed.
A brief introduction to non-perturbative duality symmetries is also
included.

\end{abstract}
\vskip 3cm

\centerline{\tt Lectures presented at the Catholic University of Leuven
and}
\centerline{\tt at the
University of Padova during the academic year  1996-97.}
\centerline{\tt To be published
by Leuven University Press.}
\vskip 2cm

\begin{flushleft}
CERN-TH/97-218\\
March 1997\\
\end{flushleft}

\end{titlepage}


\renewcommand{\thepage}{\arabic{page}}
\setcounter{page}{1}
\setcounter{footnote}{1}
\renewcommand{\theequation}{\thesection.\arabic{equation}}
\tableofcontents
\newpage

\section{Introduction}
\setcounter{equation}{0}

String theory has been the leading candidate over the past years for a
theory
that consistently unifies all fundamental forces of nature, including
gravity.
In a sense, the theory predicts gravity and gauge symmetry around flat
space.
Moreover, the theory is UV-finite.
The elementary objects are one-dimensional strings whose vibration
modes
should correspond to the usual elementary particles.

At distances large with respect to the size of the strings, the
low-energy excitations can be described by an effective field theory.
Thus, contact can be established with quantum field theory, which
turned out to be successful in describing the dynamics of the real
world at low energy.

I will try to explain here the basic structure of string theory,
its predictions and problems.

In chapter \ref{history} the evolution of string theory is traced, from
a theory initially built to describe hadrons to a ``theory of
everything".
In chapter \ref{classical} a description of classical bosonic string
theory is given. The oscillation modes of the string are described,
preparing the scene for quantization.
In chapter \ref{quant}, the quantization of the bosonic string is
described.
All three different quantization procedures are presented to varying
depth,
since in each one some specific properties are more transparent than in
others. I thus describe the old covariant quantization, the light-cone
quantization and the modern path-integral quantization.
In chapter \ref{conf} a concise introduction is given, to the central
concepts
of conformal field theory since it is the basic tool in discussing
first quantized string theory.
In chapter \ref{vertex} the calculation of scattering amplitudes is
described.
In chapter \ref{background} the low-energy effective action
for the massless modes is described.

In chapter \ref{super} superstrings are introduced.
They   provide spacetime fermions and realize supersymmetry in
spacetime and on the world-sheet.
I go through quantization again, and describe the different
supersymmetric string theories in ten dimensions.
In chapter \ref{anomal} gauge and gravitational anomalies are
discussed.
In particular it is shown that the superstring theories are
anomaly-free.
In chapter \ref{comp} compactifications of the ten-dimensional
superstring theories are described.
Supersymmetry breaking is also discussed in this context.
In chapter \ref{loop}, I describe how to calculate loop corrections to
effective coupling constants.
This is very important for comparing string theory predictions at low
energy with the real world.
In chapter \ref{nonpert} a brief introduction to non-perturbative
string connections and non-perturbative effects is given.
This is a fast-changing subject and I have just included some basics as
well as tools, so that the reader orients him(her)self in the web of
duality connections.
Finally, in chapter \ref{out} a brief outlook and future problems
are presented.

I have added a number of appendices to make several technical
discussions self-contained.
In Appendix A useful information on the elliptic $\th$-functions is
included.
In Appendix B, I  rederive the various lattice sums that appear in
toroidal compactifications.
In Appendix C the Kaluza-Klein ansatz is described, used to obtain
actions in lower dimensions after toroidal compactification.
In Appendix D some facts are presented about four-dimensional locally
supersymmetric theories with N=1,2,4 supersymmetry.
In Appendix E, BPS states are described along with
their representation theory and
helicity supertrace formulae that can be used to trace their appearance
in a
supersymmetric theory.
In Appendix F  facts about  elliptic modular forms are presented, which
are useful in many contexts, notably in the one-loop computation of
thresholds and counting of BPS multiplicities.
In Appendix G, I present the computation of helicity-generating string
partition functions and the associated  calculation of BPS
multiplicities.
Finally, in Appendix H, I briefly review electric--magnetic duality in
four dimensions.

I have not tried to be complete in my referencing. The focus was to
provide,
in most cases, appropriate reviews for further reading.
Only in the last chapter, which covers very recent topics, I do mostly
refer
to original papers because of the  scarcity of relevant reviews.

\section{Historical perspective\label{history}}
\setcounter{equation}{0}

In the sixties, physicists tried to make sense of a big bulk of
experimental
data relevant to the strong interaction.
There were lots of particles (or ``resonances") and the situation
could best be described as chaotic.
There were some regularities observed, though:

$\bullet$ Almost linear Regge behavior. It was noticed that the large
number of resonances could be nicely put on (almost) straight lines
by plotting their mass versus their spin
\be
m^2={J\over \ap}\;,
\ee
with $\ap\sim 1$ GeV$^{-2}$, and this relation was checked up to
$J=11/2$.

$\bullet$ s-t duality.  If we consider a scattering amplitude of
two$\to$ two hadrons ($1,2\to 3,4$), then it can be described by the
Mandelstam invariants
\be
s=-(p_1+p_2)^2\;\;\;,\;\;\;t=-(p_2+p_3)^2\;\;\;,\;\;\;u=-(p_1+p_3)^2\;,
\ee
with $s+t+u=\sum_i m_i^2$. We are using a metric with signature
$(-+++)$.
Such an amplitude depends on the flavor quantum numbers of hadrons
(for example
SU(3)). Consider the flavor part, which is cyclically symmetric in
flavor
space.
For the full amplitude to be symmetric, it must also be cyclically
symmetric  in the momenta $p_i$.
This symmetry amounts to the interchange $t\leftrightarrow s$.
Thus, the amplitude should satisfy $A(s,t)=A(t,s)$.
Consider a $t$-channel contribution due to the exchange of a spin-$J$
particle of mass $M$. Then, at high energy
\be
A_J(s,t)\sim {(-s)^J\over t-M^2}\,.
\label{3}
\ee
Thus, this partial amplitude increases with $s$ and its behavior
becomes worse
for large values of $J$.
If one sews amplitudes of this form together to make a loop
amplitude, then
there are
uncontrollable UV divergences for $J>1$.
Any finite sum of amplitudes of the form (\ref{3}) has this bad UV
behavior.
However, if one  allows an infinite number of terms then it is
conceivable that
the UV behavior might be different.
Moreover such a finite sum has no $s$-channel poles.

A proposal for such a dual amplitude was made by  Veneziano
\cite{GV}
\be
A(s,t)={\Gamma(-\alpha(s))\Gamma(-\alpha(t))\over
\Gamma(-\alpha(s)-\alpha(t))}\,,
\label{4}\ee
where $\Gamma$ is the standard $\Gamma$-function and
\be
\alpha(s)=\alpha(0)+\ap s\,.
\label{5}\ee
By using the standard properties of the $\Gamma$-function it can be
checked
that the amplitude (\ref{4}) has an infinite number of $s,t$-channel
poles:
\be
A(s,t)=-\sum_{n=0}^{\infty}{(\alpha(s)+1)\dots (\alpha(s)+n)\over
n!}
{1\over \alpha(t)-n}\,.
\label{6}\ee
In this expansion the $s\leftrightarrow t$ interchange symmetry of
(\ref{4})
is not manifest.
The poles in (\ref{6}) correspond to the exchange of an infinite
number of particles of mass $M^2=(n-\alpha(0)/\ap)$ and high spins.
It can also be checked that the high-energy behavior of the Veneziano
amplitude
is softer than any local quantum field theory amplitude, and the
infinite number of poles is crucial for this.

It was subsequently realized by Nambu and Goto that such amplitudes
came out of theories of relativistic strings.
However such theories had several shortcomings in explaining
the dynamics of strong interactions.

$\bullet$ All of them seemed to predict a tachyon.

$\bullet$ Several of them seemed to contain a massless spin-2
particle
that was impossible to get rid of.

$\bullet$ All of them seemed to require a spacetime dimension of 26
in order not to break Lorentz invariance at the quantum level.

$\bullet$ They contained only bosons.

At the same time, experimental data from SLAC showed that at even
higher energies hadrons have a point-like structure; this
opened the way for quantum chromodynamics as the
correct theory that describes strong interactions.

However some work continued in the context of ``dual models" and in
the mid-seventies several interesting breakthroughs were made.

$\bullet$ It was understood by Neveu, Schwarz and Ramond how to
include spacetime fermions in string theory.

$\bullet$ It was also understood by Gliozzi, Scherk and Olive how to
get rid of the omnipresent tachyon. In the process, the
constructed theory had spacetime supersymmetry.

$\bullet$ Scherk and Schwarz, and independently Yoneya, proposed that
closed string theory,
always having
a massless spin-2 particle,  naturally describes gravity and that
the scale
$\ap$ should be identified with the Planck scale.
Moreover, the theory can be defined in four dimensions using the
Kaluza--Klein idea, namely considering the extra dimensions to be
compact and small.

However, the new big impetus for string theory came in 1984.
After a general analysis of gauge and gravitational anomalies
\cite{AW},
it was realized that anomaly-free theories in
higher
dimensions are very restricted.
Green and Schwarz showed in \cite{GS} that open superstrings in 10
dimensions are anomaly-free
if the gauge group is O(32). $\rm E_8\times E_8$ was also
anomaly-free
but could not appear in open string theory.
In \cite{Het} it was shown that another string exists in ten
dimensions,
a hybrid of the superstring and the bosonic string, which can
realize the E$_8\times$E$_8$ or O(32) gauge symmetry.

Since the early eighties, the field of string theory has been
continuously developing and we will see the main points in the rest
of these lectures.
The reader is encouraged to look at  a more detailed discussion in
\cite{GSW}--\cite{Pol}.

\begin{figure}
\begin{center}
\leavevmode
\epsffile[180 0 417 150]{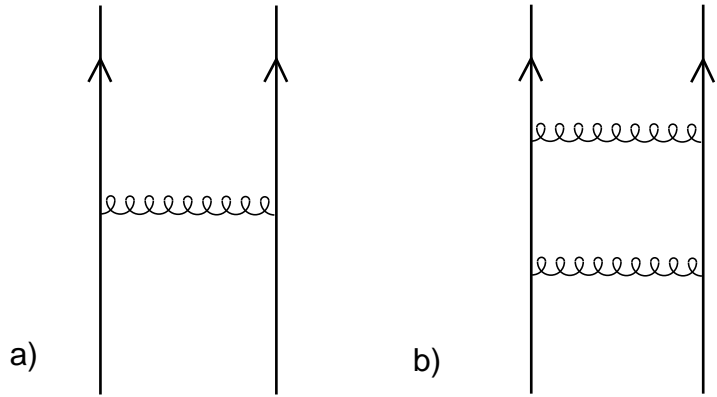}
\end{center}
\caption[]{\it Gravitational interaction between two particles via
graviton exchange.}
\label{f1}\end{figure}

One may wonder what makes string theory so special.
One of its key ingredients is that it provides a finite theory of
quantum gravity, at least in perturbation theory.
To appreciate the difficulties with the quantization of Einstein
gravity, we will look at a single-graviton exchange between two
particles (Fig. \ref{f1}a).
We will set $h=c=1$.
Then the amplitude is proportional to $E^2/M_{\rm Planck}^2$, where $E$
is the
energy of the process and $M_{\rm Planck}$ is the Planck mass, $M_{\rm
Planck}\sim
10^{19}~$GeV.
It is related to the Newton constant $G_N\sim M_{\rm Planck}^2$.
Thus, we see that the gravitational interaction is irrelevant in the
IR ($E<<M_{\rm Planck}$) but strongly relevant in the UV.
In particular it implies that the two-graviton exchange diagram (Fig.
\ref{f1}b) is proportional to
\be
{1\over M_{\rm Planck}^4}\int_{0}^{\Lambda}dE~E^3\sim {\Lambda^4\over
M_{\rm Planck}^4}\,,
\ee
which is strongly UV-divergent.
In fact it is known that Einstein gravity coupled to matter is
non-renormalizable in perturbation theory.
Supersymmetry makes the UV divergence softer but the
non-renormalizability persists.

There are two ways out of this:

$\bullet$ There is a non-trivial UV fixed-point that governs the UV
behavior of quantum gravity. To date, nobody has managed to make sense
out of this possibility.

$\bullet$ There is new physics at $E\sim M_{\rm Planck}$ and Einstein
gravity is
the IR
limit of a more general theory, valid at and beyond the Planck scale.
You could consider the analogous situation with the Fermi theory of
weak interactions. There, one had  a non-renormalizable
current--current interaction
with similar problems, but today we know that this is the IR limit of
the
standard weak interaction mediated by the $W^{\pm}$ and $Z^0$ gauge
bosons.
So far, there is no consistent field theory that can
make sense at
energies beyond
$M_{\rm Planck}$ and contains gravity.
Strings provide precisely a theory that induces new physics at the
Planck scale
due to the infinite tower of string excitations with masses of the
order of the
Planck mass and carefully tuned interactions that become soft at
short distance.

Moreover string theory seems to have all the right properties for
Grand Unification, since it produces and unifies with gravity not only
gauge couplings but also Yukawa couplings.
The shortcomings, to date,  of string theory as an ideal unifying
theory
are its numerous different vacua, the fact that there are
three string theories in 10 dimensions that look different (type-I,
type
II and
heterotic), and most importantly supersymmetry breaking.
There has been some progress recently in these  directions: there is
good evidence that these different-looking string theories might be
non-perturbatively equivalent\footnote{You will find a pedagogical
review
of these developments at the end of these lecture notes as well as in
\cite{P1}.}.

\renewcommand{\theequation}{\thesection.\arabic{equation}}
\section{Classical string theory\label{classical}}

As in field theory there are two approaches to discuss classical and
quantum string theory.
One is the first quantized approach, which discusses the dynamics of
a single string. The dynamical variables are the spacetime
coordinates of the string.
This is an approach that is forced to be on-shell.
The other is the second-quantized or field theory approach.
Here the dynamical variables are functionals of the string
coordinates, or string fields, and
we can have an off-shell formulation.
Unfortunately, although there is an elegant formulation of open
string field theory, the closed string field theory approaches are
complicated and difficult to use. Moreover the open theory is not
complete since we know it also requires the presence of closed
strings.
In these lectures we will follow the first-quantized approach,
although the reader is invited to study the rather elegant formulation
of open string field theory \cite{Op}.

\renewcommand{\theequation}{\thesubsection.\arabic{equation}}
\subsection{The point particle}
\setcounter{equation}{0}

Before discussing strings, it is useful to look first at the
relativistic
point particle. We will use the first-quantized path integral
language.
Point particles classically follow an extremal path when
traveling from one point in spacetime to another.
The natural action is proportional to the length of the world-line
between some initial and final points:
\be
S= m \int^{s_{f}}_{s_{i}} ds =  m \int^{\t_{1}}_{\t_{0}} d\t
\sqrt{-\eta_{\mu \nu} \dot{x}^\mu \dot{x}^\nu }, \label{ngpoint}
\ee
where $\eta_{\mu \nu}=\mbox{diag}(-1,+1,+1,+1)$.
The momentum conjugate to $x^\mu(\t)$ is
\be
p_\mu = -\frac{\delta L}{\delta \dot{x}^\mu} =
\frac{m\dot{x}_\mu}{\sqrt{-\dot{x}^{2}}}, \label{conj}
\ee
and the Lagrange equations coming from varying the action
(\ref{ngpoint})
with respect to $X^{\mu}(\tau)$ read
\be
\pd_{\t}\left( \frac{m\dot{x}_\mu}{\sqrt{-\dot{x}^{2}}} \right) = 0.
\label{eqofmot}
\ee
Equation (\ref{conj}) gives the following mass-shell constraint :
\be
p^{2} + m^{2} = 0.   \label{constraint}
\ee
The canonical Hamiltonian is given by
\be
H_{can} = \frac{\pd L}{\pd \dot{x}^\mu} \dot{x}^\mu - L.
\label{7}\ee
Inserting (\ref{conj}) into (\ref{7}) we can see that $H_{can}$
vanishes identically. Thus, the constraint
(\ref{constraint}) completely governs the dynamics of the system. We
can add it to the Hamiltonian using a Lagrange multiplier. The system
will then be described by
\be
H = \frac{N}{2m} (p^{2} + m^{2}),
\ee
from which it follows that
\be
\dot{x}^\mu = \{ x^\mu , H \} = \frac{N}{m} p^\mu =
\frac{N\dot{x}^\mu}{\sqrt{-\dot{x}^{2}}},
\ee
or
\be
\dot{x}^{2} = - N^{2},
\ee
so we are describing time-like trajectories. The choice N=1
corresponds to a choice of scale for the parameter $\t$, the proper
time.

The square root in (\ref{ngpoint}) is an unwanted feature.
Of course for the free particle it is not a problem, but as we will
see later
it will be a problem for the string case.
Also the action we used above is ill-defined for massless particles.
Classically, there exists an alternative action, which does not
contain the square root and in addition allows the generalization to
the massless case. Consider the following action :
\be
S = -\half \int d\t e(\t) \left( e^{-2}(\t) (\dot{x}^\mu)^{2} - m^{2}
\right). \label{ppoint}
\ee
The auxiliary variable $e(\t)$ can be viewed as an einbein on the
world-line. The associated metric would be $g_{\t\t} = e^{2}$, and
(\ref{ppoint}) could be rewritten as
\be
S = -\half \int d\t \sqrt{ \mbox{det} g_{\t\t} } (
g^{\t\t}\pd_{\t}x\cdot\pd_{\t}x - m^{2} ).
\ee
The action is
invariant under reparametrizations of the world-line. An infinitesimal
reparametrization is given by
\be
\delta x^\mu(\t)=x^{\mu}(\tau+\xi(\tau))-x^{\mu}(\tau) = \xi(\t)
\dot{x}^\mu
+{\cal{O}}(\xi^2).
\ee

Varying $e$ in (\ref{ppoint}) leads to
\be
\delta S = \half \int d\t \left(\frac{1}{e^{2}(\t)}
(\dot{x}^\mu)^{2} + m^{2} \right) \delta e(\t).
\ee
Setting $\delta S = 0$ gives us the equation of motion for $e$ :
\be
e^{-2} x^{2}+m^{2} = 0\;\;\;\to\;\;\;
e = \frac{1}{m} \sqrt{-\dot{x}^{2}}\;.\label{eqe}
\ee
Varying $x$ gives
\be
\delta S = \half \int d\t e(\t) \left(e^{-2}(\t) 2 \dot{x}^\mu
\right) \pd_{\t} \delta x^\mu.
\ee
After partial integration, we find the equation of motion
\be
\pd_{\t} (e^{-1}\dot{x}^\mu) = 0. \label{eqx}
\ee
Substituting (\ref{eqe}) into (\ref{eqx}), we find the same equations
as before (cf. eq. (\ref{eqofmot})). If we substitute (\ref{eqe})
directly into the action (\ref{ppoint}), we find the previous one,
which establishes the classical equivalence of both
actions.

We will derive the propagator for the point particle. By
definition,
\be
\langle x | x' \rangle = N \int_{x(0)=x}^{x(1)=x'} De Dx^\mu
\mbox{exp}\left({1\over 2}\int_{0}^{1}
\left(\frac{1}{e}(\dot{x}^{\mu})^{2}-
e m^{2}\right) d\t \right),
\label{8}\ee
where we have put $\t_{0}=0$, $\t_{1}=1$.

Under reparametrizations of the world-line, the einbein transforms as
a vector. To first order, this means
\be
\delta e = \pd_{\t}(\xi e).
\label{9}\ee
This is the local reparametrization invariance of the path. Since we
are integrating over $e$, this means that (\ref{8}) will give an
infinite
result. Thus, we need to gauge-fix the reparametrization invariance
(\ref{9}).
We can gauge-fix $e$ to be constant. However, (\ref{9}) now indicates
that we cannot fix more. To see what this constant may be, notice
that
 the length
of the path of the particle is
\be
L = \int_{0}^{1} d\t \sqrt{\mbox{det} g_{\t\t}} = \int_{0}^{1} d\t e,
\ee
so the best we can do is $e=L$. This is the simplest example of
leftover
(Teichm\"uller) parameters after gauge fixing.
The $e$ integration contains an integral over the constant mode as
well
as the rest. The rest is the ``gauge volume" and we will throw it
away.
Also, to make the path integral converge, we rotate to Euclidean time
$\tau
\to i\tau$.
Thus, we are left with
\be
\langle x | x' \rangle = N \int_{0}^{\infty} dL
\int_{x(0)=x}^{x(1)=x'} Dx^\mu \mbox{exp} \left(-{1\over
2}\int_{0}^{1}
\left(\frac{1}{L} \dot{x}^{2}+L m^{2} \right) d\t \right).
\ee
Now write
\be
x^\mu (\t) = x^\mu + (x'^\mu - x^\mu) \t + \delta x^\mu (\t),
\ee
where $\delta x^\mu (0) = \delta x^\mu (1) = 0$. The first two terms
in this expansion represent the classical path. The measure for the
fluctuations $\delta x^\mu$ is
\be
\parallel\delta x \parallel^{2} =  \int_{0}^{1} d\t e(\delta
x^\mu)^{2}=  L \int_{0}^{1} d\t (\delta
x^\mu)^{2},
\ee
so that
\be
Dx^\mu \sim \prod_{\t} \sqrt{L} d \delta x^\mu (\t).
\ee
Then
\be
\langle x | x' \rangle = N \int_{0}^{\infty} dL \int \prod_{\t}
\sqrt{L} d \delta x^\mu (\t) e^{ - \frac{(x' - x)^{2}}{2L} - m^{2}L/2
}
e^{-\frac{1}{2L} \int_{0}^{1} (\delta \dot{x^\mu})^{2} }.
\ee
The Gaussian integral involving $\delta \dot{x}^\mu$ can be evaluated
immediately :
\be
\int \prod_{\t} \sqrt{L} d \delta x^\mu (\t) e^{-\frac{1}{L}
\int_{0}^{1} (\delta \dot{x^\mu})^{2} } \sim \left( \mbox{det}
\left(-\frac{1}{L} \pd^{2}_{\t}\right) \right)^{-\frac{D}{2}}.
\label{Gaussian}
\ee
We have to compute the determinant of the operator $-\pd^2_{\tau}/L$.
To do this we will calculate first its eigenvalues. Then the
determinant
will be given as the product of all the eigenvalues.
To find the eigenvalues we consider the eigenvalue problem
\be
-{1\over L}\pd^2_{\tau}\psi(\tau)=\lambda \psi(\tau)
\label{10}
\ee
with the boundary conditions $\psi(0)=\psi(1)=0$.
Note that there is no zero mode problem here because of the
boundary conditions.
The solution is
\be
\psi_n(\tau)=C_n\sin(n\pi\tau)\;\;\;,\;\;\;\lambda_n={n^2\over
L}\;\;\;,\;\;\;
n=1,2,\ldots
\label{11}\ee
and thus
\be
\mbox{det}\left(-\frac{1}{L}
\pd^{2}_{\t}\right)=\prod_{n=1}^{\infty}{n^2\over
L}\,.
\label{12}\ee
Obviously the determinant is infinite and we have to regularize it.
We will use $\zeta$-function regularization in which\footnote{You
will find more details on this in \cite{G}.}
\be
\prod_{n=1}^{\infty}L^{-1}=L^{-\zeta(0)}=L^{1/2}\;\;\;,\;\;\;
\prod_{n=1}^{\infty}n^a=e^{-a\zeta'(0)}=(2\pi)^{a/2}\,.
\label{13}\ee
Adjusting the normalization factor we finally obtain
\be
\langle x | x' \rangle ={1\over 2(2\pi)^{D/2}}  \int_{0}^{\infty} dL
L^{-\frac{D}{2}}
e^{ - \frac{(x' - x)^{2}}{2L} - m^{2} L/2 }=
\ee
$$
={1\over (2\pi)^{D/2}}\left({|x-x'|\over
m}\right)^{(2-D)/2}K_{(D-2)/2}(
m|x-x'|).
$$
This is the free propagator of a scalar particle in D dimensions.
To obtain the more familiar expression, we have to pass to momentum
space
\be
|p\rangle =\int d^D x e^{ip\cdot x}|x\rangle\,,
\label{14}\ee
\bea
\langle p|p'\rangle &=&\int d^D x e^{-ip\cdot x}\int d^D x'~e^{ip'\cdot
x'}
 \langle x | x'\rangle \nn\\
& =&{1\over 2} \int d^D x' e^{i(p'-p)\cdot
x'}\int_0^{\infty}
dL~e^{-{L\over 2}(p^{2} + m^{2})} \\
     & = &(2\pi)^D\delta (p-p')~ \frac{1}{p^2 + m^2}, \nn
\eea
just as  expected.

Here we should make one more comment.
The momentum space amplitude $\langle p|p'\rangle$ can also be
computed
directly if we insert in the path integral $e^{ip\cdot x}$ for the
initial
state and $e^{-ip'\cdot x}$ for the final state.
Thus, amplitudes are given by path-integral averages of the
quantum-mechanical wave-functions of free particles.

\subsection{Relativistic strings}
\setcounter{equation}{0}

We now use the ideas of the previous section to construct actions for
strings. In the case of point particles, the action was proportional
to the length of the world-line between some initial point and final
point.
For strings, it will be related to the surface
 area of the ``world-sheet" swept by the string as it propagates
through spacetime. The Nambu-Goto action is defined as
\be
S_{NG} = -T \int dA.
\ee
The constant factor $T$ makes the action dimensionless; its
dimensions must be $[\mbox{length}]^{-2}$ or $[\mbox{mass}]^{2}$.
Suppose $\xi^{i}$ ($i=0,1$) are coordinates on the world-sheet and
$G_{\mu\nu}$ is the metric of the spacetime in which the string
propagates. Then, $G_{\mu\nu}$ induces a metric on the world-sheet
:
\be
ds^{2}  =  G_{\mu\nu} (X) dX^\mu dX^\nu
       =  G_{\mu\nu} \frac{\pd X^\mu}{\pd \xi^{i}}\frac{\pd
X^\nu}{\pd \xi^{j}}d\xi^{i} d\xi^{j}
       =  G_{ij} d\xi^{i} d\xi^{j}\,,
\ee
where the induced metric is
\be
G_{ij} = G_{\mu\nu} \pd_{i} X^\mu \pd_{j} X^\nu.
\ee
This metric can be used to calculate the surface area. If the
spacetime is
flat Minkowski space then
$G_{\mu \nu} = \eta_{\mu \nu}$ and the
Nambu-Goto action becomes
\be
S_{NG}  = -T \int \sqrt{-\mbox{det} G_{ij} } d^{2} \xi
        =  -T \int \sqrt{(\dot{X}.X')^{2}-(\dot{X}^{2})(X'^{2})}
d^{2}\xi,
\ee
where $\dot{X}^\mu=\frac {\pd X^\mu}{\pd \t}$ and $X'^\mu = \frac
{\pd X^\mu}{\pd \s}$ ($\t = \xi^{0}$, $\sigma = \xi^{1}$ ).
The equations of motion are
\be
\pd_{\t} \left(\frac{\delta L}{\delta \dot{X}^\mu}\right) + \pd_{\s}
\left(\frac{\delta L}{\delta X'^\mu}\right) = 0.
\ee
Depending on the kind of strings, we can impose different boundary
conditions. In the case of closed strings, the world-sheet is a tube.
If we let $\s$ run from $0$ to $\bar{\s} = 2 \pi$, the boundary
condition is periodicity
\be
X^\mu (\s + \bar{\s} ) = X^\mu (\s).
\ee
For open strings, the world-sheet is a strip, and in this case we
will put $\bar{\s} = \pi$. Two kinds of boundary conditions are
frequently used\footnote{One could also impose an arbitrary
linear combination of the two boundary conditions. We will come back to
the interpretation of such boundary conditions in the last chapter.} :
\begin{itemize}
\item Neumann :
\be
\left. \frac{\delta L}{\delta X'^\mu}\right|_{\s=0,\bar{\s}}  = 0;
\ee
\item Dirichlet :
\be
\left. \frac{\delta L}{\delta \dot{X}^\mu}\right|_{\s=0,\bar{\s}} =
0.
\ee
\end{itemize}
As we shall see at the end of this section, Neumann conditions imply
that no momentum flows off the ends of the string. The Dirichlet
condition implies that the end-points of the string are fixed in
spacetime.
We will not discuss them further, but they are relevant
for describing (extended) solitons in string theory also known as
D-branes \cite{P2}.

The momentum conjugate to $X^\mu$ is
\be
\Pi^\mu = \frac{\delta L}{\delta \dot{X}^\mu} = - T \frac{(\dot{X}
\cdot X')X'^\mu - (X')^{2} \dot{X}^\mu}{[(X' \cdot \dot{X})^{2} -
(\dot{X})^{2} (X')^{2}]^{1/2}}.
\ee
The matrix $\frac{\delta^{2}L}{\delta \dot{X}^\mu \delta
\dot{X}^\nu}$ has two zero eigenvalues, with eigenvectors
$\dot{X}^\mu$ and $X'^\mu$. This signals the occurrence of two
constraints that follow directly from the definition of the conjugate
momenta. They are
\be
\Pi \cdot X'  = 0\;\;\;,\;\;\;
\Pi^{2} + T^{2} X'^{2} =0\,.
\ee
The canonical Hamiltonian
\be
H = \int_{0}^{\bar{\s}} d\s (\dot{X} \cdot \Pi - L )
\ee
vanishes identically, just in the case of the point particle.
Again, the dynamics is governed solely by the constraints.

The square root in the Nambu-Goto action makes the treatment of the
quantum
theory quite complicated. Again, we can simplify the action by
introducing an intrinsic fluctuating  metric on the world-sheet.
In this way, we obtain the
Polyakov action for strings moving in flat spacetime \cite{Po}
\be
S_{P} = -{T\over 2} \int d^{2}\xi \sqrt{- \mbox{det} g }~ g^{\al
\beta}
\pd_{\al} X^\mu \pd_{\beta} X^\nu \eta_{\mu \nu}.   \label{Polyakov}
\ee
As is well known from field theory, varying the action with respect
to the metric yields the stress-tensor :
\be
T_{\al \beta} \equiv -\frac{2}{T} \frac {1}{\sqrt{-\mbox{det} g}}
\frac{\delta S_{P}}{\delta g^{\al \beta}} = \pd_{\al} X \cdot
\pd_{\beta} X - \half g_{\al \beta} g^{\gamma \delta} \pd_{\gamma} X
\cdot \pd_{\delta} X.
\ee
Setting this variation to zero and solving for $g_{\al \beta}$, we
obtain, up to a factor,
\be
g_{\al \beta} = \pd_{\al} X \cdot \pd_{\beta} X.
\ee
In other words, the world-sheet metric $g_{\al \beta}$ is classically
equal to the induced metric. If we substitute this back into the
action, we find the Nambu-Goto action. So both actions are
equivalent, at least classically. Whether this is also true
quantum-mechanically is not clear in general.
However, they can be shown to be equivalent in the
critical dimension.
 From now on we will take the Polyakov approach to the quantization
of string theory.

By varying (\ref{Polyakov}) with respect to $X^\mu$, we obtain the
equations of motion:
\be
\frac{1}{\sqrt{-\mbox{det} g}} \pd_{\al} (\sqrt{-\mbox{det}g} g^{\al
\beta} \pd_{\beta} X^\mu ) = 0.
\ee
Thus, the world-sheet action in the Polyakov approach consists of D
two-dimensional  scalar
fields $X^{\mu}$ coupled to the dynamical two-dimensional metric and we
are thus
considering a theory of two-dimensional  quantum gravity coupled to
matter.
One could ask whether there are other terms that can be added to
(\ref{Polyakov}). It turns out that there are only two: the
cosmological term
\be
\lambda_{1} \int \sqrt{-\mbox{det} g} \label{cosmo}
\ee
and the Gauss-Bonnet term
\be
\lambda_{2} \int \sqrt{-\mbox{det} g} R^{(2)}, \label{gaussbonnet}
\ee
where $R^{(2)}$ is the two-dimensional scalar curvature associated with
$g_{\al \beta}$.
This gives the Euler number of the world-sheet,
which is a topological invariant. So this term cannot influence the
local dynamics of the string, but it will give factors that weight
various topologies differently. It is not difficult to prove that
(\ref{cosmo}) has to be zero classically.
In fact the classical equations of motion for $\lambda_1\not=0$ imply
that $g_{\alpha\beta}=0$, which gives trivial dynamics.
We will not consider it further. For the open string, there are
other possible terms, which are defined on the boundary of the
 world-sheet.

We will  discuss the symmetries of the Polyakov action:
\begin{itemize}
\item Poincar\'e invariance :
\be
\delta X^\mu  =  \omega^{\mu}_{\nu} X^\nu +
\al^\mu\;\;\;,\;\;\;\delta g_{\al\beta} =  0\;,
\ee
where $\omega_{\mu\nu} = - \omega_{\nu\mu}$;
\item local two-dimensional reparametrization invariance :
\bea
\delta g_{\al \beta} & = & \xi^{\gamma} \pd_{\gamma} g_{\al\beta} +
\pd_{\al} \xi ^{\gamma} g_{\beta \gamma} + \pd_{\beta} \xi^{\gamma}
g_{\al \gamma}
 =  \nabla_{\al} \xi_{\beta} + \nabla_{\beta} \xi_{\al}, \nn \\
\delta X^\mu & = & \xi^{\al} \pd_{\al} X^\mu, \nn \\
\delta(\sqrt{-\mbox{det} g}) & = & \pd_{\al} (\xi^{\al}
\sqrt{-\mbox{det} g} );
\eea
\item conformal (or Weyl) invariance :
\be
\delta X^\mu  =  0\;\;\;,\;\;\;
\delta g_{\al \beta}  =  2 \Lambda g_{\al \beta}.
\ee
\end{itemize}

Due to the conformal invariance, the stress-tensor will be traceless.
This is in fact true in general.
Consider an action $S(g_{\alpha\beta},\phi^i)$
in arbitrary spacetime dimensions. We assume that it is invariant
under conformal
transformations
\be
\delta g_{\al \beta} = 2 \Lambda(x) g_{\al \beta}\;\;\;,\;\;\;
\delta \phi^i= d_i\Lambda(x)\phi^i\;.
\label{15}\ee
The variation of the action under infinitesimal conformal
transformations is
\be
0 = \delta S = \int d^{2} \xi \left[2 \frac{\delta S}{\delta g^{\al
\beta}} g^{\al \beta} + \sum_{i} d_{i} \frac{\delta
S}{\delta \phi_{i}} \phi_{i}\right]\Lambda.
\ee
Using the equations of motion for the fields $\phi_{i}$, i.e.
$\frac{\delta S}{\delta \phi_{i}} = 0$, we find
\be
T^{\al}_{\al}\sim\frac{\delta S}{\delta g^{\al \beta}} g^{\al
\beta}=0\;,
\ee
which follows without the use of the equations of motion, if and only
if
$d_{i} = 0$. This is the case for the bosonic string, described by
the Polyakov action, but not for fermionic extensions.

Just as we could fix $e(\t)$ for the point particle using
reparametrization invariance, we can reduce $g_{\al \beta}$ to
$\eta_{\al \beta} = \mbox{diag}(-1,+1)$. This is called conformal
gauge. First, we choose a parametrization that makes the metric
conformally flat, i.e.
\be
g_{\al \beta} = e^{2 \Lambda(\xi)} \eta_{\al\beta}.
\ee
It can be proven that in two dimensions, this is always possible for
world-sheets with trivial topology. We will discuss the subtle issues
that appear for non-trivial topologies later on.

Using the conformal symmetry, we can further reduce
the metric to $\eta_{\al \beta}$. We also work with  ``light-cone
coordinates"
\be
\xi_{+}=\t + \s\;\;\;,\;\;\;\xi_{-}=\t - \s.
\ee
The metric becomes
\be
ds^{2} = - d\xi_{+}d\xi_{-}.
\ee
The components of the metric are
\be
g_{++}=g_{--} = 0\;\;\;,\;\;\;g_{+-}=g_{-+} =-{1\over 2}\ee
and
\be
\pd_{\pm}={1\over 2}(\pd_\t\pm\pd_{\s})\;.
\ee
The Polyakov action in conformal gauge is
\be
S_{P}\sim T \int d^{2}\xi ~\pd_{+}X^\mu \pd_{-}X^\nu \eta_{\mu\nu}.
\label{Polconf}
\ee
By going to conformal gauge, we have not completely fixed all
reparametrizations. In particular, the reparametrizations
\be
\xi_{+}  \longrightarrow  f(\xi_{+})\;\;\;,\;\;\;
\xi_{-}  \longrightarrow  g(\xi_{-})
\ee
only put a factor $\pd_{+}f\pd_{-}g$ in front of the metric, so they
can be compensated by the transformation of $d^2\xi$.

Notice that here we have exactly enough symmetry to completely fix
the
metric. A metric on a d-dimensional world-sheet has
d(d+1)/2 independent components. Using reparametrizations,
d of them can be fixed. Conformal invariance fixes one more
component. The number of remaining components is
\be
\frac{d(d+1)}{2} - d - 1.
\ee
This is zero in the case $d=2$ (strings), but not for $d>2$
(membranes). This makes an analogous treatment of higher-dimensional
extended objects problematic.

We will  derive the equations of motion from the Polyakov action in
conformal gauge (eq. (\ref{Polconf})). By varying $X^\mu$, we get
(after partial integration):
\be
\delta S = T \int d^{2}\xi (\delta X^\mu \pd_{+}\pd_{-}X_\mu) - T
\int_{\t_{0}}^{\t_{1}} d\t X'_\mu \delta X^\mu.
\ee
Using periodic boundary conditions for the closed string and
\be
\left. X'^\mu \right|_{\s=0,\bar{\s}} = 0
\ee
for the open string, we find the equations of motion
\be
\pd_{+}\pd_{-} X^\mu = 0.
\ee
Even after gauge fixing, the equations of motion for the metric have to
be imposed. They are
\be
T_{\al\beta} = 0,
\ee
or
\be
T_{10}=  T_{01} = \half \dot{X} \cdot X' = 0\;\;\;,\;\;\;
T_{00} =  T_{11} = \frac{1}{4} (\dot{X}^{2} + X'^{2}) = 0,
\ee
which can also be written as
\be
(\dot{X} \pm X')^{2} = 0.
\label{20}\ee
These are known as the Virasoro constraints.
They are the analog of the Gauss law in the string case.

In light-cone coordinates, the components of the stress-tensor are
\be
T_{++}  =  \half \pd_{+}X \cdot \pd_{+}X\;\;\;,\;\;\;
T_{--} =  \half \pd_{-}X \cdot \pd_{-}X\;\;\;,\;\;\;
T_{+-}  =  T_{-+} = 0. \label{trless}
\ee
This last expression is equivalent to $T^{\al}_{\al} = 0$; it is
trivially satisfied. Energy-momentum conservation, $\nabla^\al
T_{\al\beta} = 0$, becomes
\be
\pd_{-} T_{++} + \pd_{+} T_{-+}  =
\pd_{+} T_{--} + \pd_{-} T_{+-} =  0.
\ee
Using (\ref{trless}), this states
\be
\pd_{-} T_{++}  =
\pd_{+} T_{--} =  0\;
\ee
which leads to conserved charges
\be
Q_{f} = \int_{0}^{\bar{\s}} f(\xi^{+}) T_{++}(\xi^{+}),
\label{charge}
\ee
and likewise for $T_{--}$. To convince ourselves that $Q_f$
is indeed conserved, we need to calculate
\be
0= \int d\s \pd_{-}(f(\xi^{+}) T_{++})=\pd_{\t} Q_{f} +
\left. f(\xi^{+}) T_{++}
\right|^{\bar{\s}}_{0}.
\ee
For closed strings, the boundary term vanishes automatically; for
open strings, we need to use the constraints. Of course, there are
other conserved charges in the theory, namely those associated with
Poincar\'e invariance :
\be
P^{\al}_{\mu}=-T\sqrt{\mbox{det} g} g^{\al \beta} \pd_\beta
X_\mu\,,
\ee
\be
J^{\al}_{\mu \nu} =  -T\sqrt{\mbox{det} g} g^{\al \beta} ( X_\mu
\pd_\beta X_\nu - X_\nu \pd_\beta X_\mu).
\ee
We have $\pd_\al P^{\al}_{\mu} = 0 = \pd_\al J^{\al}_{\mu\nu}$
because of the equation of motion for $X$. The associated charges are
\be
P_\mu =\int_{0}^{\bar{\s}} d\s P^{\t}_{\mu}\;\;\;,\;\;\;
J_{\mu \nu} = \int_{0}^{\bar{\s}} d\s J^{\t}_{\mu \nu}.
\label{17}\ee
These are conserved, e.g.
\bea
\frac{\pd P_\mu}{\pd \t} & = & T \int_{0}^{\bar{\s}} d\s \pd_{\t}^{2}
X_\mu
      =  T \int_{0}^{\bar{\s}} d\s \pd_{\s}^{2} X_\mu \nn \\
      & = & T (\pd_\s X_\mu (\s = \bar{\s}) - \pd_\s X_\mu (\s = 0)
).
\eea
(In the second line we used the equation of motion for $X$.) This
expression automatically vanishes for the closed string. For open
strings, we need Neumann boundary conditions. Here we see that these
conditions imply that there is no momentum flow off the
ends of the string.
The same applies to angular momentum.

\subsection{Oscillator expansions}
\setcounter{equation}{0}

We will  now  solve the equations of motion for the bosonic
string,
\be
\pd_+ \pd_- X^\mu = 0 \;,\label{eom}
\ee
taking into account the proper boundary conditions. To do this we
have to treat the open and closed string cases separately. We will
first
consider the case of the closed string.

\begin{itemize}\item \underline {Closed Strings }

The most general solution to equation \r{eom} that also satisfies the
periodicity condition
\[ X^\mu(\t,\s+2\pi)=X^\mu(\t,\s) \]
can be separated in a left- and a right-moving part:
\be
X^\mu(\t,\s) = X^\mu_L(\t+\s) + X^\mu_R(\t-\s),
\ee
where
\bea
X^\mu_L(\t+\s) = \frac{x^\mu}{2} + \frac{p^\mu}{4\pi T} (\t+\s) +
\frac{i}{\sqrt{4 \pi T}} \sum_{k \neq 0} \frac{\bar \al^\mu_k}{k}
e^{-ik(\t+\s)}, \nn\\
\label{osc}\\
X^\mu_R(\t-\s) = \frac{x^\mu}{2} + \frac{p^\mu}{4\pi T} (\t-\s) +
\frac{i}{\sqrt{4 \pi T}} \sum_{k \neq 0} \frac{ \al^\mu_k}{k}
e^{-ik(\t-\s)}.\nn
\eea
The $\al^\mu_k$ and $\bar\al^\mu_k$ are arbitrary Fourier modes, and
$k$ runs over the integers.
The function $X^\mu(\t,\s)$ must be real, so we know that $x^\mu$ and
$p^\mu$ must also be real and we can derive the following reality
condition for the $\al$'s:
\be  \label{reality}
(\al^\mu_k)^* = \al^\mu_{-k} \quad\quad\mbox{and}\quad\quad
(\bar\al^\mu_k)^* = \bar\al^\mu_{-k}
\ee

If we define $\al^\mu_0=\bar\al^\mu_0 = \frac{1}{\sqrt{4\pi
T}}p^\mu$ we can write
\bea
\pd_- X^\mu_R = \frac{1}{\sqrt{4 \pi T}} \sum_{k \in \Z} \al^\mu_k
e^{-ik(\t-\s)}, \\
\nn\\
\pd_+ X^\mu_L = \frac{1}{\sqrt{4 \pi T}} \sum_{k \in \Z}
\bar\al^\mu_k e^{-ik(\t+\s)}.
\eea

\item \underline{Open Strings}

We will now derive  the oscillator expansion \r{osc} in the case of the
open string. Instead of the periodicity condition, we now have to
impose the Neumann boundary condition
\[ X'^\mu(\t,\s)|_{\s=0,\pi}=0. \]
If we substitute the solutions of the wave equation we obtain the
following condition:
\be
X'^\mu|_{\s=0} = \frac{ p^\mu - \bar p^\mu}{\sqrt{4 \pi T}} +
\frac{1}{\sqrt{ 4 \pi T}} \sum_{k \neq 0} e^{ik\t} ( \bar\al^\mu_k -
\al^\mu_k), \nn
\ee
from which we can draw the following conclusion:
\[ p^\mu = \bar p^\mu \quad\quad \mbox{and} \quad\quad \al^\mu_k =
\bar\al^\mu_k \]
and we see that the left- and right-movers get mixed by the boundary
condition.
The boundary condition at the other end, $\s=\pi$, implies that $k$
is an integer.
Thus, the solution becomes:
\be
X^\mu(\t,\s) = x^\mu + \frac{p^\mu \t}{\pi T} + \frac{i}{\sqrt{\pi
T}} \sum_{k \neq 0} \frac{ \al^\mu_k}{k} e^{-ik\t} \cos (k\s).
\ee

If we again use $\al^\mu_0=\frac{1}{\sqrt{\pi T}} p^\mu$ we can
write:
\be
\pd_\pm X^\mu = \frac{1}{\sqrt{4 \pi T}} \sum_{k \in \Z} \al^\mu_k
e^{-ik(\t\pm\s)}.
\ee

\end{itemize}

For both the closed and open string cases we can calculate the
center-of-mass position of the string:
\be
X^\mu_{CM} \equiv \frac{1}{\bar\s} \int_0^{\bar\s} d\s X^\mu(\t,\s) =
x^\mu +\frac{p^\mu \t}{\pi T},
\ee
Thus, $x^\mu$ is the center-of-mass position at $\t=0$ and is moving
as a free
particle. In the same way
we can calculate the center-of-mass momentum, or just the momentum of
the string. From (\ref{17}) we obtain
\bea
p^\mu_{CM} &=& T \int_0^{\bar\s} d\s \dot X^{\mu} =\frac{T}{\sqrt{4 \pi
T}}
 \int d\s \sum_k (\al^\mu_k +
\bar\al^\mu_k)e^{-ik(\t\pm\s)} \nn\\
&=& \frac{2\pi T}{\sqrt{4 \pi T}} (\al^\mu_0 + \bar\al^\mu_0) = p^\mu\,.
\eea
In the case of the open string there are no $\bar\al$'s.

We observe that the variables that describe the classical motion of the
string
are the center-of-mass position $x^{\mu}$ and momentum $p^{\mu}$
plus an infinite collection of variables $\al_n^{\mu}$ and
$\bar \al_n^{\mu}$. This reflects the fact that the string can move
as a whole,
but it can also vibrate in various modes, and the
oscillator
variables represent precisely the vibrational degrees of freedom.

A similar calculation can be done for the angular momentum of the
string:
\be
J^{\mu\nu} = T \int_0^{\bar\s} d\s ( X^\mu \dot X^\nu - X^\nu \dot
X^\mu)
        = l^{\mu\nu} + E^{\mu\nu} +\bar E^{\mu\nu} \nn\,,
\ee
where
\be
l^{\mu\nu} = x^\mu p^\nu - x^\nu p^\mu\,,
\ee
\bea
E^{\mu\nu} = -i \sum_{n=1}^{\infty} \frac{1}{n} (\al^\mu_{-n}
\al^\nu_n - \al^{\nu}_{-n} \al^\mu_n)\;,\\
\bar E^{\mu\nu} = -i \sum_{n=1}^{\infty} \frac{1}{n}
(\bar\al^\mu_{-n} \bar\al^\nu_n - \bar\al^{\nu}_{-n} \bar\al^\mu_n)
\,.
\eea

In the Hamiltonian picture we have equal-$\t$ Poisson brackets (PB)
for the dynamical variables, the $X^\mu$ fields and
their conjugate momenta:
\be\label{pb}
\{ X^\mu(\s,\t),\dot X^\nu(\s',\t)\}_{PB} = \frac{1}{T}
\delta(\s-\s') \eta^{\mu\nu}\,.
\label{18}\ee
The other brackets $\{X,X\}$ and $\{\dot X,\dot X\}$ vanish.
We can easily derive from (\ref{18}) the PB for the oscillators
and center-of-mass position and momentum:
\bea
\{\al^\mu_m,\al^\nu_n\} &=& \{\bar\al^\mu_m,\bar\al^\nu_n\}
\enspace=\enspace -i m \delta_{m+n,0} ~\eta^{\mu\nu}\,, \nn\\
\{\bar\al^\mu_m,\al^\nu_n\} &=& 0
\;\;\;,\;\;\;\{ x^\mu, p^\nu \} = \eta^{\mu\nu}\,.
\eea
Again for the open string case, the $\bar\al$'s are absent.

The Hamiltonian
\be H = \int d\s (\dot X \Pi - L) = \frac{T}{2} \int d\s (\dot X^2 +
X'^2)
\ee
can also be expressed in terms of oscillators.
In the case of closed strings it is given by
\be \label{hamosc}
H = \half \sum_{n\in Z} (\al_{-n} \al_n + \bar\al_{-n} \bar\al_n),
\ee
while for open strings it is
\be
H = \half \sum_{n\in Z} \al_{-n} \al_n.
\ee

In the previous section we saw that the Virasoro constraints in the
conformal gauge were just $T_{--} = \half (\pd_-X)^2 = 0$ and $T_{++}
= \half(\pd_+ X)^2=0$. We then define the Virasoro operators as the
Fourier modes of the stress-tensor. For the closed string they
become
\be
L_m = 2T \int_0^{2\pi} d\s~ T_{--}  e^{im(\t-\s)}\;\;\;,\;\;\;
\bar L_m = 2T \int_0^{2\pi} d\s~ T_{++}  e^{im(\s+\t)}\,,
\ee
or, expressed in oscillators:
\be
L_m = \half \sum_n \al_{m-n} \al_n\;\;\;,\;\;\;
\bar L_m = \half \sum_n \bar\al_{m-n} \bar\al_n \,.
\ee
They satisfy the reality conditions
\be
L_m^* = L_{-m} \quad\quad {\rm and} \quad\quad \bar L_m^* = \bar
L_{-m}\,.
\ee
If we compare these expressions with \r{hamosc}, we see that we can
write the
Hamiltonian in terms of Virasoro modes as
\be H = L_0 + \bar L_0. \ee
This is one of the classical constraints. The other operator, $\bar
L_0 - L_0$, is the generator of translations in $\s$, as can be shown
with the help of the basic Poisson brackets \r{pb}. There  is no
preferred point on the string, which can be expressed
by the constraint $\bar L_0 -L_0 =0$.

In the case of open strings, there is no difference between the
$\al$'s and $\bar\al$'s and the Virasoro modes are defined as
\be
L_m = 2T \int_0^{\pi} d\s \{ T_{--}  e^{im(\t-\s)} + T_{++}
e^{im(\s+\t)} \}\;.
\ee
Expressed in oscillators, this becomes:
\be
L_m = \half \sum_n \al_{m-n} \al_n.
\ee
The Hamiltonian is then
\[ H = L_0. \]

With the help of the Poisson brackets for the oscillators, we can
derive
the brackets for the Virasoro constraints.
They form an algebra known as the classical
Virasoro algebra:
\bea
\{ L_m,L_n\}_{PB} &=& -i (m-n) L_{m+n} \, ,\nn\\
\{ \bar L_m, \bar L_n \}_{PB} &=& -i (m-n) \bar L_{m+n} \, ,\\
\{ L_m, \bar L_n\}_{PB} &=& 0\, .\nn
\eea
In the open string case, the $\bar L$'s are absent.

\section{Quantization of the bosonic string\label{quant}}
\setcounter{equation}{0}

There are several ways to quantize relativistic strings:

$\bullet$ Covariant Canonical Quantization, in which the classical
variables of the string motion become operators.
Since the string is a constrained system there are two options here.
The first one is to quantize the unconstrained variables and then
impose
the constraints in the quantum theory as conditions on states in the
Hilbert space.
This procedure preserves manifest Lorentz invariance and is known as
the
old covariant approach.

$\bullet$ Light-Cone Quantization. There is another option in the
context of canonical quantization, namely  to solve the constraints
at the level of the classical theory and then quantize.
The solution of the classical constraints is achieved in the
so-called ``light-cone" gauge.
This procedure is also canonical, but manifest Lorentz invariance is
lost,
and its presence has to be checked a posteriori.

$\bullet$ Path Integral Quantization. This can be combined with BRST
techniques
and has manifest Lorentz invariance, but it works in an extended
Hilbert
space that also  contains ghost fields. It is the analogue of the
Faddeev-Popov method of gauge theories.

All three methods of quantization agree whenever all three can be
applied and compared.
Each one has some advantages, depending on the nature of the questions
we
ask in the quantum theory, and all three will be presented.

\subsection{Covariant canonical quantization}
\setcounter{equation}{0}

The usual way to do the canonical quantization is to replace all
fields by operators and replace the Poisson brackets by commutators
\[ \{ \quad,\quad \}_{PB} \quad\longrightarrow\quad -i [ \quad,\quad
]. \]
The Virasoro constraints are then operator constraints
that have to annihilate physical states.

Using the canonical prescription, the commutators for the oscillators
and center-of-mass position and
momentum  become
\bea
[x^\mu,p^\nu] &=& i \eta^{\mu\nu}, \\
{[} \alpha^{\mu}_{m} , \alpha^{\nu}_{n} ] &=& m \delta_{m+n,0}
\eta^{\mu\nu}\;; \label{comm}
\eea
there is  a similar expression for the $\bar\al$'s in the case of
closed
strings, while $\al_n^{\mu}$ and $\bar\al_n^{\mu}$ commute.
The reality condition \r{reality} now becomes a hermiticity
condition on the oscillators. If we absorb the factor $m$ in \r{comm}
in the oscillators, we can write the commutation
 relation as
\be
[a^\mu_m, a^{\nu\dagger}_n] =
\delta_{m,n}\eta^{\mu\nu}\,,
\ee
which is just the harmonic oscillator
commutation relation for an infinite set of oscillators.

The next thing we have to do is to define a Hilbert space on which
the operators act. This is not very difficult since our system is an
infinite collection of harmonic oscillators and we do know how to
construct the Hilbert space.
In this case the negative
frequency modes $\al_m$, $m<0$ are raising operators and the
positive frequency modes are the lowering operators of $L_0$.
We now define
the ground-state of our Hilbert space as the state
that is annihilated by all lowering operators. This does not yet
define the state completely:
we also have to consider the center-of-mass operators $x^{\mu}$ and
$p^{\mu}$.
This however is known from elementary quantum mechanics, and if we
diagonalize $p^{\mu}$ then the states will be also characterized by
the momentum.
If we denote the  state by $|p^\mu \rangle$, we have
\be
\al_m \, |p \rangle  = 0 \quad \forall m>0.
\ee
We can build more states by acting on this ground-state
with the negative frequency modes\footnote{We consider here for
simplicity
the case of the open string.}
\be
|p\rangle \, , \quad \al^\mu_{-1} |p^\mu\rangle\,, \quad \al^\mu_{-1}
\al^\nu_{-1} \al^\nu_{-2} |p^\mu\rangle \;,\,\;\;\;{\rm etc.}
\ee
There seems to be a problem, however: because of the Minkowski metric
in the
commutator for the oscillators we obtain
\be
|\enspace \al^0_{-1} |p\rangle \enspace |^2 = \langle p| \al^0_1
\al^0_{-1} | p\rangle = -1 \,,
\ee
which means that there are negative norm states. But we still have to
impose the classical constraints $L_m=0$. Imposing these
constraints should help us to throw away the states with negative
norm
from the physical spectrum.

Before we go further, however, we have to face a typical ambiguity when
quantizing a classical system.
The classical variables are functions of coordinates and momenta.
In the quantum theory, coordinates and momenta are non-commuting
operators. A specific ordering prescription has to be made in order
to define them as well-defined operators in the quantum theory.
In particular we would like their eigenvalues on physical states
to be
finite; we will therefore have to pick a normal ordering prescription
as in
usual
field theory.
Normal ordering
puts all positive frequency modes to the right of
 the negative frequency modes. The Virasoro operators in the quantum
theory are now defined by their normal-ordered expressions
\be
L_m   = \half \sum_{n \in \Z} :\al_{m-n}\cdot \al_n : \,.
\ee
Only $L_0$  is sensitive to normal ordering,
\be
L_0 = \half \al_0^2 + \sum_{n=1}^\infty \al_{-n}\cdot \al_n \,.
\ee
Since the commutator of two oscillators is a constant, and since we do
not know
in advance what this constant part should be, we include a
normal-ordering constant $a$ in all
expressions containing $L_0$;
thus, we replace $L_0$ by $(L_0-a)$.

We can now calculate the algebra of the $L_m$'s. Because of the
normal ordering this has to be done with great care. The Virasoro
algebra then becomes:
\be
[L_m,L_n] = (m-n) L_{m+n} + \frac{c}{12} m (m^2 -1) \delta_{m+n,0},
\ee
where $c$ is the central charge and in this case $c=d$, the dimension
of the target space or the number of free scalar fields on the
world-sheet.

We can now see that we cannot impose the classical constraints
$L_m=0$ as operator constraints $L_m|\phi\rangle =0$ because
\[ 0=\langle\phi | [L_m,L_{-m}] | \phi\rangle = 2m~\langle\phi  | L_0
|
\phi\rangle + \frac{d}{12} m(m^2-1) \langle\phi | \phi\rangle \neq
0\,. \]
This is analogous to a similar phenomenon that takes place in gauge
theory.
There, one assumes the Gupta-Bleuler approach, which makes sure that
the constraints vanish ``weakly" (their expectation value on physical
states vanishes).
Here the maximal set of constraints we can impose on physical states
is
\be
L_{m>0} | \mbox{phys} \rangle = 0\;\;\;,\;\;\;
(L_0-a) | \mbox{phys} \rangle =0\label{lo-a}
\ee
and, in the case of closed strings, equivalent expressions for the
$\bar L$'s. This is consistent with the classical constraints because
$\langle\mbox{phys}'|L_n|\mbox{phys}\rangle = 0$.

Thus, the physical states in the theory are the states we constructed
so far,
but which also satisfy (\ref{lo-a}).
Apart from physical states, there are the so-called ``spurious states",
$|\mbox{spur}\rangle = L_{-n} |\enspace\rangle$, which are orthogonal
to all physical states. There are even states which are both physical
and spurious, but we would like  them to decouple from the physical
Hilbert space since they can be shown to correspond to
null states.
There is a detailed and complicated analysis of the physical spectrum
of string theory, which culminates with the famous ``no-ghost" theorem;
this states
that if $d=26$, the physical spectrum defined by (\ref{lo-a})
contains only
positive norm states.
We will not pursue this further.

We will  further analyze the $L_0$ condition.
If we substitute the expression for $L_0$ in \r{lo-a} with $p^2=-m^2$
and $\al' = \frac{1}{2 \pi T}$ we obtain the mass-shell
condition
\be
\al'm^2 = 4 ( N -a)\,, \label{massshell}
\ee
where $N$ is the level-number operator:
\be
N = \sum_{m=1}^{\infty} \al_{-m} \cdot \al_m \,.
\ee
We can deduce a similar expression for $(\bar L_0 -a)$, from which it
follows that $\bar N = N$.

\subsection{Light-cone quantization}
\setcounter{equation}{0}

In this approach we first  solve  the classical
constraints. This will leave us with a smaller number of classical
variables.
Then we quantize them.

There is a gauge in which the solution of the Virasoro constraints is
simple.
This is the light-cone gauge.
Remember that we still had some
invariance leftover after going to the conformal gauge:
\[ \xi_+' = f(\xi_+)\,, \quad\quad \xi_-' = g(\xi_-)\,. \]
This invariance can be used to set
\be X^+ = x^{+}+\al' p^+ \t \,.\ee
This gauge can indeed be reached because, according to the gauge
transformations, the transformed coordinates $\s'$ and $\t'$ have to
satisfy the wave equation in terms of the old coordinates and $X^+$
clearly does so.
The light-cone coordinates are defined as
\[ X^\pm = X^0 \pm X^1\,. \]
Imposing now the classical Virasoro constraints (\ref{20}) we can
solve
for $X^-$ in
terms of the transverse coordinates $X^i$, which means that we can
eliminate both $X^+$ and $X^-$ and only work with the transverse
directions. Thus, after solving the constraints we are left with all
positions and momenta of the string, but only the transverse
oscillators.

The light-cone oscillators can then be expressed in the following way
(closed strings):
\bea \al^+_n &=& \bar \al^+_n \enspace = \enspace \sqrt{
\frac{\al'}{2}} p^+ \delta_{n,0}\,, \nn \\
\al^-_n &=& \frac{1}{\sqrt{2\al'} p^+} \left\{ \sum_{m\in \Z} :
\al^i_{n-m} \al^i_m   : - 2 a \delta_{n,0} \right\} \,,
\eea
and a similar expression for $\bar \al^-$.

We have now explicitly solved the Virasoro constraints and we can now
quantize,
that is replace $x^{\mu}$, $p^{\mu}$ , $\al_n^{i}$ and
$\bar\al^{i}_n$ by operators. The index $i$ takes values in the
transverse directions.
However, we have given up the manifest Lorentz covariance of the
theory. Since this theory in the light-cone gauge
originated from a manifest Lorentz-invariant theory in d dimensions,
one
would expect that after fixing the gauge this invariance is still
present. However, it turns out that in the quantum theory this is
only true in 26 dimensions, i.e. the Poincar\'e algebra
only closes if $d=26$.

\subsection{Spectrum of the bosonic string}
\setcounter{equation}{0}

So we will assume $d=26$ and analyze the spectrum of the theory.
In the light-cone gauge we have solved almost all of the Virasoro
constraints.
However we still have to impose $(L_0-a)|{\rm phys}\rangle=0$ and a
similar
one
$(\bar L_0-\bar a)|{\rm phys}\rangle$ for the closed string.
It is left to the reader  as an exercise to show that only $a=\bar a$
gives a
non-trivial spectrum consistent with Lorentz invariance.
In particular this implies that $L_0=\bar L_0$ on physical states.
The states are constructed in a fashion similar to that of the previous
section.
One starts from the state $|p^{\mu}\rangle$, which is the
vacuum for the transverse oscillators, and then creates more
states by acting with the negative frequency modes
of the transverse oscillators.

We will start from the closed string.
The ground-state is $| p^{\mu} \rangle$, for which we have the
mass-shell
condition
$ \al' m^2 = -4 a$
and, as we will see later, $a=1$ for a consistent theory; this
state is the infamous tachyon.

The first excited level will be (imposing $L_0=\bar L_0$)
\be
\al^i_{-1} \bar\al^j_{-1} \,|p\rangle\,.
\ee
We can decompose this into irreducible representations of the
transverse rotation group SO(24) in the following manner
\bea
\al^i_{-1} \bar\al^j_{-1} \,|p \rangle = \al^{ [i }_{-1} \bar\al^{ j]
}_{-1} \,|p \rangle + \left[ \al^{ \{ i }_{-1} \bar\al^{j\}}_{-1} -
\frac{1}{d-2} \delta^{ij} \al^k_{-1} \bar\al^k_{-1} \right] \,|p
\rangle + \nn\\
\hskip 5cm + \frac{1}{d-2} \delta^{ij} \al^k_{-1} \bar\al^k_{-1} |p
\rangle \,.
\eea
These states can be interpreted as a spin-2 particle $G_{\mu\nu}$
(graviton), an antisymmetric tensor $B_{\mu\nu}$ and a scalar $\Phi$.

Lorentz invariance requires physical states to be representations of
the little group of the Lorentz group SO(d-1,1), which is SO(d-1)
for massive states and SO(d-2) for massless states. Thus, we
conclude that states at this first excited level must
be massless, since the representation content is such that they
cannot
be assembled into SO(25) representations.
Their mass-shell condition is
\[ \al' m^2 = 4( 1- a)\,, \]
from which we can derive the value of the normal-ordering constant,
$a=1$, as we claimed before. This constant can also be expressed in
terms of the target space dimension $d$ via
$\zeta$-function regularization: one then finds that $a=
\frac{d-2}{24}$. We conclude that Lorentz invariance requires that
$a=1$ and $d=26$.

What about the next level? It turns out that higher excitations, which
are naturally tensors of SO(24), can be uniquely combined in
representations
of SO(25). This is consistent with Lorentz invariance for  massive
states and can be shown to hold for
all higher-mass excitations \cite{GSW}.

Now consider the open string: again the ground-state is tachyonic.
The
first excited level is
\[ \al^i_{-1} | p\rangle \,,\]
which is again massless and is the vector representation of SO(24),
as it should be for a massless vector in 26 dimensions.
The second-level excitations are given by
\[ \al^i_{-2} |p\rangle\,, \quad \al^i_{-1} \al^j_{-1} |p\rangle
\,,\]
which are tensors of SO(24); however , the last one can be decomposed
into a symmetric part and a trace part and, together with the SO(24)
vector, these three parts uniquely combine into a symmetric SO(25)
massive tensor.

In the case of the open string we see that at level $n$ with
mass-shell condition $\al' m^2 = (n-1)$ we always have a state
described by a symmetric tensor of rank $n$ and we can conclude that
the maximal spin at level $n$ can be expressed in terms of the
 mass
\[ j^{{\rm max}} = \al' m^2 +1 \,.\]

Open strings are allowed to carry charges at the end-points.
These are known as Chan-Paton factors and give rise to non-abelian
gauge groups of the type  Sp(N) or O(N) in the unoriented case and U(N)
in the oriented case.
To see how this comes about, we will attach charges labeled by an index
$i=1,2,\cdots,$N at the two end-points of the open string.
Then, the ground-state is labeled, apart from the momentum, by the
end-point charges: $|p,i,j\rangle$, where $i$ is on one end and $j$ on
the other.
In the case of oriented strings, the massless states are
$a^{\mu}_{-1}|p,i,j\rangle$ and they give a collection of $N^2$
vectors.
It can be shown that the gauge group is U(N) by studying the scattering
amplitude of three vectors.

In the unoriented case, we will have to project by the transformation
that interchanges the two string end-points $\Omega$ and also reverses
the orientation of the string itself:
\be
\Omega |p,i,j\rangle=\epsilon |p,j,i\rangle
\ee
where $\epsilon^2=1$ since $\Omega^2=1$.
Thus, from the N$^2$ massless vectors, only N(N+1)/2 survive
when $\epsilon=1$
forming the adjoint of Sp(N), while when $\epsilon=-1$, N(N-1)/2
survive
forming the adjoint of O(N).

We have seen that a consistent quantization of the bosonic string
requires 26 spacetime dimensions. This dimension is called the
critical dimension. String theories can also be defined in less then
26 dimensions and are
therefore called non-critical. They are not Lorentz-invariant.
For more details see \cite{Pol}.

\subsection{Path integral quantization}
\setcounter{equation}{0}

In this section we will use the path integral approach to quantize
the
string, starting from the Polyakov action. Consider the bosonic
string partition function
\be
Z = \int {{\cal D}g {\cal D} X^\mu\over V_{\rm gauge}}
 e^{ i S_p(g,X^\mu)}\,,
\ee
The measures are defined from the norms:
\bea
|| \delta g || &=& \int d^2\s \sqrt{g} g^{\al\beta} g^{\delta\gamma}
\delta g_{\al\gamma} \delta g_{\beta\delta} \,,\nn\\
|| \delta X^\mu || &=& \int d^\s \sqrt{g} \delta X^\mu \delta X^\nu
\eta_{\mu\nu} \,.\nn
\eea
The action is Weyl-invariant, but the measures are not. This implies
that
generically in the quantum theory the Weyl factor will couple to the
rest of the fields. We can use
conformal reparametrizations to rescale our metric
\[ g_{\al\beta} = e^{2\phi} h_{\al\beta}\,. \]
The variation of the metric under
reparametrizations and Weyl rescalings can be decomposed into
\be
\delta g_{\al\beta} =  \nabla_{\al} \xi_{\beta} + \nabla_{\beta}
\xi_{\al} + 2 \Lambda g_{\al\beta}
 =  (\hat{P}\xi)_{\al\beta} + 2 \tilde{\Lambda} g_{\al\beta},
\ee
where $(\hat{P}\xi)_{\al\beta} = \nabla_{\al} \xi_{\beta} +
\nabla_{\beta} \xi_{\al} - (\nabla_{\gamma}
\xi^{\gamma})g_{\al\beta}$ and $\tilde{\Lambda} = \Lambda + \half
\nabla_{\gamma} \xi^{\gamma}$.
The integration measure can be written as
\be
{\cal D} g = {\cal D}(\hat P \xi){\cal D}(\tilde \Lambda) =
{\cal D} \xi {\cal D} \Lambda \left| \frac{ \pd (P\xi ,\tilde \Lambda
) }{ \pd ( \xi , \Lambda ) } \right| \,,
\ee
where the Jacobian is
\be
 \left| \frac{\pd(P\xi,\tilde \Lambda)}{\pd(\xi,\Lambda)} \right| =
\left| \,\det \left( \begin{array}{cc} \hat P & 0 \\ {*} & 1
\end{array} \right) \right| = \left| \det P \right| = \sqrt{ \det
\hat P \hat P^\dagger } \,.
\ee
The $*$ here means some operator that is not important for the
determinant.

There are two sources of Weyl non-invariance in the path integral:
the
Faddeev-Popov determinant and the $X^{\mu}$ measure.
As shown by Polyakov \cite{Po}, the Weyl factor of the
metric decouples also in the quantum theory only if $d=26$. This is the
way that the
critical dimension is singled out in the path integral approach.
If $d\not=26$, then the Weyl factor has to be kept;  we are dealing
with the
so-called non-critical string theory, which we will not discuss here
(but those who are
interested are referred to \cite{Pol}).
In our discussion here, we will always assume that we are in the
critical dimension.
We can factor out the integration over the reparametrizations and the
Weyl group, in which case the partition function becomes:
\be
Z = \int {\cal D} X^\mu \sqrt{ \det PP^\dagger}
e^{iS_p(\hat h_{\al\beta},X^\mu)}\,,
\ee
where $\hat h_{\alpha\beta}$ is some fixed reference metric that can
be chosen at will.
We can now use the so-called Faddeev-Popov trick: we can
exponentiate the determinant using anticommuting ghost variables
$c^\al$ and $b_{\al\beta}$, where $b_{\al\beta}$ (the antighost) is
a symmetric and traceless tensor:
\be
\sqrt{ \det PP^\dagger} = \int {\cal D}c{\cal D}b e^{i\int d^2 \s
\sqrt{g} g^{\al\beta} b_{\al\gamma} \nabla_\beta c^\al }\,.
\ee
If we now choose $h_{\al\beta} = \eta_{\al\beta}$ the partition
function becomes:
\be
Z = \int {\cal D}X {\cal D}c{\cal D}b e^{i( S_p[X] + S_{gh}[c,b])}\,,
\ee
where
\bea
S_p[X] &=& T \int d^2\s \pd_+ X^\mu \pd_- X_\mu \,,\\
S_{gh}[b,c] &=& \int b_{++} \pd_- c^+ + b_{--} \pd_+ c^- \,.
\eea

\subsection{Topologically non-trivial world-sheets}
\setcounter{equation}{0}

We have seen above that gauge fixing the diffeomorphisms and Weyl
rescalings gives rise to a Faddeev-Popov determinant.
Subtleties arise when this determinant is zero, and we will discuss
the appropriate treatment here.

As already mentioned, under the combined effect of reparametrizations
and
Weyl rescalings
the metric transforms as
\be
\delta g_{\al\beta}  =  \nabla_{\al} \xi_{\beta} + \nabla_{\beta}
\xi_{\al} + 2 \Lambda g_{\al\beta}
 =  (\hat{P}\xi)_{\al\beta} + 2 \tilde{\Lambda} g_{\al\beta},
\label{decomp}
\ee
where $(\hat{P}\xi)_{\al\beta} = \nabla_{\al} \xi_{\beta} +
\nabla_{\beta} \xi_{\al} - (\nabla_{\gamma}
\xi^{\gamma})g_{\al\beta}$ and $\tilde{\Lambda} = \Lambda + \half
\nabla_{\gamma} \xi^{\gamma}$.
The operator $\hat P$ maps vectors to traceless symmetric tensors.
Those reparametrizations satisfying
\be
\hat{P} \xi^{*} = 0  \label{CKE}
\ee
do not affect the metric. Equation (\ref{CKE}) is called the
conformal Killing equation, and its solutions are the conformal
Killing vectors. These are the zero modes of $\hat P$.
When a surface admits conformal Killing vectors then there are
reparametrizations that cannot be fixed by fixing the metric but
have to be fixed separately.

Now define the natural  inner product for vectors and tensors:
\be
(V_{\al},W_{\al}) = \int d^{2}\xi \sqrt{\mbox{det} g} g^{\al\beta}
V_{\al}W_{\beta}
\ee
and
\be
(T_{\al\beta},S_{\al\beta}) = \int d^{2}\xi \sqrt{\mbox{det} g}
g^{\al\gamma} g^{\beta \delta} T_{\al\beta}S_{\gamma\delta}.
\ee
The decomposition (\ref{decomp}) separating the traceless part
from the trace is orthogonal. The Hermitian
conjugate with respect to this product maps traceless symmetric
tensors $T_{\al\beta}$ to vectors:
\be
(\hat{P}^{\dagger} t)_{\al} = -2 \nabla^{\beta} t_{\al\beta}.
\ee
The zero modes of $\hat P^{\dagger}$  are the solutions of
\be
\hat{P}^{\dagger} t^{*} = 0  \label{0Pdagg}
\ee
and correspond to symmetric traceless tensors, which cannot be written
as
$(\hat{P}\xi)_{\al\beta}$ for any vector field $\xi$. Indeed, if
(\ref{0Pdagg}) is satisfied, then for all $\xi^{\al}$, $0 =
(\xi,\hat{P}^{\dagger} t^{*}) = - (\hat{P} \xi , t^{*} )$. Thus,
zero modes of $\hat{P}^{\dagger}$ correspond to deformations of the
metric that cannot be compensated for by reparametrizations and Weyl
rescalings. Such deformations cannot be fixed by fixing the gauge and
are called Teichm\"uller deformations.
We have already seen an example of this in the point-particle case.
The length of the path was a Teichm\"uller parameter, since it
could not
 be changed by diffeomorphisms.

The following table gives the number of conformal Killing vectors and
zero modes of $\hat{P}^{\dagger}$, depending on the topology of the
closed string world-sheet. The genus is essentially the number of
handles of the closed surface.

\begin{center}
\begin{tabular}{c|cc}
Genus & $\#$ zeros of $\hat{P}$& $\#$ zeros of $\hat{P}^{\dagger}$
\\
\hline
0 & 3 & 0 \\
1 & 1 & 1 \\
$\geq 2$ & 0 & $3g-3$
\end{tabular}
\end{center}

The results described above will be important for the calculation
of loop corrections to scattering amplitudes.

\subsection{BRST primer}
\setcounter{equation}{0}

We will  take a brief look at the BRST formalism in general. Consider a
theory
with fields $\phi_i$, which has a certain gauge symmetry. The gauge
transformations will satisfy an algebra\footnote{This is not the most
general algebra possible, but it is sufficient for our purposes.}
\be [\delta_\al , \delta_\beta] = f_{\al\beta}{}^\gamma
\delta_\gamma\,.
\label{22}\ee
We can now fix the gauge by imposing some appropriate gauge
conditions
\be
F^A(\phi_i) = 0\,.
\ee
Using again the Faddeev--Popov trick, we can write the path integral
as
\bea
\int \frac{{\cal D}\phi}{V_{{\rm gauge}}} e^{-S_0} &\sim& \int {\cal
D}\phi \delta(F^A(\phi)=0) {\cal D} b_A {\cal D}c^\al e^{-S_0 - \int
b_A (\delta_\al F^A )c^\al } \quad\quad\quad\nn\\
&\sim& \int {\cal D}\phi{\cal D}B_A{\cal D}b_A{\cal D}c^\al e^{ -S_0
- i\int B_A F^A(\phi) - \int b_A (\delta_\al F^A) c^\al  }\nn\\
&=&\int {\cal D}\phi{\cal D}B_A{\cal D}b_A{\cal D}c^\al e^{ -S}\,,
\label{brstaction}
\eea
where
\be
S=S_0+S_1+S_2\;\;\;,\;\;\;S_1=i\int B_A F^A(\phi)\;\;\;,\;\;\;S_2=
\int b_A (\delta_\al F^A) c^\al \,.
\label{21}\ee
Note that the index $\alpha$ associated with the ghost $c_{\alpha}$
is in one-to-one correspondence with the parameters of
the gauge transformations in (\ref{22}).
The index $A$ associated with the ghost $b_A$ and the antighost $B_A$
are in one-to-one correspondence with the gauge-fixing conditions.

The full gauge-fixed action $S$  is invariant under the {\it
Becchi-Rouet-Stora-Tyupin
(BRST) transformation},
\bea
\delta_{BRST}~ \phi_i &=& -i \epsilon c^\al \delta_\al \phi_i
\,,\nn\\
\delta_{BRST}~ b_A &=& -\epsilon B_A \,,\\
\delta_{BRST}~ c^\al &=& -\half \epsilon c^\beta c^\gamma
f_{\beta\gamma}{}^\al \,,\nn\\
\delta_{BRST}~B_A &=& 0\,.\nn
\eea
In these transformations, $\epsilon$ has to be anticommuting. The
first transformation is just the original gauge transformation on
$\phi_i$, but with the gauge parameter replaced by the ghost
$c_{\alpha}$.

The extra terms in the action due to the ghosts and gauge fixing in
\r{brstaction} can be written in terms of a BRST transformation:
\be
\delta_{BRST} (b_A F^A) =  \epsilon [ B_A F^A (\phi) + b_A c^\al
\delta_\al F^A(\phi) ] \,.
\ee
The concept of the BRST symmetry is important for the following
reason.
When we introduce the ghosts during gauge-fixing the theory
is no longer invariant under the original symmetry.
The BRST symmetry is an extension of the original symmetry, which
remains intact.

Consider now a small change in the gauge-fixing condition $\delta F$,
and look at the change induced in a physical amplitude
\be
\epsilon \delta_F \langle \psi | \psi'\rangle = -i \langle \psi |
\delta_{BRST} (b_A \delta F^A) |\psi'\rangle = \langle \psi| \{
Q_{B}, b_A \delta F^A \} |\psi' \rangle\,,
\ee
where $Q_{B}$ is the conserved charge corresponding to the BRST
variation. The amplitude should not change under variation of the
gauge condition and we conclude that ($Q^\dagger_{B} = Q_{B}$)
\be
Q_{B} | \mbox{phys} \rangle = 0 \,.
\ee
Thus, {\it all physical states must be BRST-invariant}.

Next, we have to check whether this BRST charge is conserved, or
equivalently
whether it commutes with the change in the Hamiltonian under
variation of the gauge condition.
The conservation of the BRST charge is equivalent to the statement
that our original gauge symmetry is intact and we do not want
to compromise its conservation in the quantum theory
just by changing our gauge-fixing condition:
\bea
0 &=& [Q_{B},\delta H] =[ Q_{B} , \delta_{B} (b_A
\delta F^A) ] \nn\\ &=& [ Q_{B} , \{ Q_{B} , b_A \delta F^A \} ]
                = [ Q_{B}^2 , b_A \delta F^A ] \,.
\eea
This should be  true for an arbitrary change in the gauge condition
and we
conclude
\be
Q_B^2 =0 \,,
\ee
that is, the BRST charge has to be  nilpotent for our description of
the quantum theory to be consistent.
If for example there is an anomaly in the gauge symmetry at the
quantum
level
this will show up as a failure of the nilpotency of the BRST charge
in the quantum theory.
This implies that the quantum theory as it stands is inconsistent: we
have
fixed a classical symmetry that is not a symmetry at the quantum
level.

The nilpotency of the BRST charge has strong
consequences. Consider the state $Q_{B} | \chi \rangle$.
This state will be annihilated by $Q_B$ whatever  $|\chi\rangle$ is,
so it is
physical. However, this state is orthogonal to all physical states
including itself and therefore it is a {\it null state}.
Thus, it should be ignored when we discuss quantum dynamics.
Two states related by
\[ | \psi'\rangle = |\psi\rangle + Q_B |\chi\rangle \]
have the same inner products and are indistinguishable.
This is the remnant in the gauge-fixed version of the original
gauge symmetry. The Hilbert
space of physical states is then the cohomology of $Q_B$, i.e.
physical states are the BRST closed states modulo the BRST exact
states:
\bea
Q_B | \mbox{phys} \rangle &=&  0 \,,\nn\\
\mbox{and} \quad \quad | \mbox{phys} \rangle &\neq& Q_B |
\mbox{something} \rangle \,.
\eea

\subsection{BRST in string theory and the physical spectrum}
\setcounter{equation}{0}

We are now ready to apply this formalism to the  bosonic string.
We can also get rid of the antighost $B$ by explicitly solving the
gauge-fixing condition as we did before, by setting the two-dimensional
 metric to be
equal to some fixed reference metric.
Expressed in the world-sheet light-cone coordinates, we obtain the
following BRST
transformations:
\bea
\delta_B X^\mu &=& i \epsilon ( c^+ \pd_+ + c^- \pd_- ) X^\mu \,,
\nn\\
\delta_B c^\pm &=& \pm i\epsilon ( c^+ \pd_+ + c^- \pd_- ) c^\pm \,,
\\
\delta_B b_\pm &=& \pm i \epsilon ( T_\pm^X + T_\pm^{gh} ) \,. \nn
\eea
We used the short-hand notation $T^X_\pm = T_{\pm\pm}(X)$, etc. The
action containing the ghost terms is
\be
S_{gh}  =  \int d^2\s \left( b_{++} \pd_- c^+ + b_{--} \pd_+ c^-
\right)\,.
\ee
The stress-tensor for the ghosts has the non-vanishing
terms
\bea
T^{gh}_{++} &=& i ( 2 b_{++}\pd_+ c^+ + \pd_+ b_{++} c^+ ) \,,\nn\\
T^{gh}_{--} &=& i ( 2 b_{--}\pd_- c^- + \pd_- b_{--} c^- ) \,,
\eea
and its conservation becomes
\be
\pd_- T^{gh}_{++} = \pd_+ T^{gh}_{--} = 0  \,.
\ee
The equations of motion for the ghosts are
\be
\pd_- b_{++} = \pd_+ b_{--} =
\pd_- c^+ = \pd_+ c^- = 0\,.
\ee

We have to impose again the appropriate periodicity (closed strings)
or
boundary (open strings) conditions on the ghosts, and then we can
expand the fields in Fourier modes again:
\bea
c^+ = \sum \bar c_n e^{-in(\t+\s)}\,, & c^- = \sum  c_n
e^{-in(\t-\s)}\,,
\nn\\
b_{++} = \sum \bar b_n e^{-in(\t+\s)}\,, & b_{--} =\sum c_n
e^{-in(\t-\s)}\,. \nn
\eea
The Fourier modes can be shown to satisfy the following
anticommutation relations
\be
\{ b_m,c_n \} = \delta_{m+n,0}\;\;\;,\;\;\;
\{ b_m,b_n\} = \{ c_m,c_n\} =0 \,.
\label{24}\ee

We can define the Virasoro operators for the ghost system as the
expansion modes of the stress-tensor. We then find
\be
L^{gh}_m = \sum_n (m-n) : b_{m+n} c_{-n} : \;\;\;,\;\;\;
\bar L^{gh}_m = \sum_n (m-n) : \bar b_{m+n} \bar c_{-n} : \,.
\ee
 From this we can compute the algebra of Virasoro operators:
\be
[ L^{gh}_m,L^{gh}_n] = (m-n) L^{gh}_{m+n} + \tfrac{1}{6}(m-13m^3)
\delta_{m+n,0} \,.
\ee
The total Virasoro operators for the combined system of $X^\mu$
fields and ghost then become
\be
L_m = L_m^X + L_m^{gh} - a \delta_{m} \,,
\ee
where the constant term is due to normal ordering of $L_0$. The
algebra of the combined system can then be written as
\be
[L_m,L_n] = (m-n) L_{m+n} + A(m) \delta_{m+n} \,,
\ee
with
\be
A(m) = \frac{d}{12} m(m^2 -1) + \frac{1}{6} (m-13m^3) + 2 a m \,.
\ee
This anomaly vanishes, if and only if $d=26$ and $a=1$, which is
exactly the same result we obtained from requiring Lorentz invariance
after quantization in the light-cone gauge.

This can also be shown using the BRST formalism. Invariance under
BRST transformation induces, via Noether's theorem, a BRST current:
\be
j_B = cT^X + \half : cT^{gh}: = cT^X + : bc\pd c : \,,
\ee
and the BRST charge becomes
\[ Q_B = \int d\s j_B \,.\]
The anomaly now shows up in $Q_B^2$: the BRST charge is
nilpotent if and only if  $d=26$.

We can express the BRST charge in terms of the $X^{\mu}$ Virasoro
operators
and the ghost oscillators as
\be
Q_B = \sum_n c_n L^X_{-n} + \sum_{m,n} \frac{m-n}{2} : c_m c_n
b_{-m-n} : -c_0 \,,\\
\label{27}\ee
where the $c_0$ term comes from the normal ordering of $L^X_0$. In
the
case of closed strings there is of course also a $\bar Q_B$, and the
BRST
charge is $Q_B+\bar Q_B$.

We will  find the physical spectrum in the BRST context.
According to our previous discussion, the physical states will have
to be annihilated by the BRST charge, and not be of the form
$Q_B|\phantom{\chi}\rangle$.
It turns out that we have to impose one more condition, namely
\be
b_0|{\rm phys}\rangle =0\,.
\label{25}\ee
This is known as the ``Siegel gauge" and although it seems mysterious
to impose it at this level, it is needed for the following
reason\footnote{Look also
at the discussion in \cite{GSW}.}:
when computing scattering amplitudes of physical states the
propagators
always come with factors of $b_0$, which effectively projects the
physical states to those satisfying (\ref{25}) since $b_0^2=0$.
Another way to see this from the path integral is that, when inserting
vertex operators to compute scattering amplitudes, the position of
the vertex operator
is a Teichm\"uller modulus and there is always a $b$ insertion
associated to every such modulus.

First we have to describe our extended Hilbert space that includes
the ghosts.
As far as the $X^{\mu}$ oscillators are concerned the situation is
the same as in the previous sections, so we need only be concerned
with the ghost Hilbert space.
The full Hilbert space will be a tensor product of the two.

First we must describe the ghost vacuum state.
This should be annihilated by the positive ghost oscillator modes
\be
b_{n>0}|{\rm ghost~vacuum}\rangle=c_{n>0}|{\rm ghost~vacuum}\rangle=0
\;.\label{23}\ee
However, there is a subtlety because of the presence of the zero
modes
$b_0$ and $c_0$ which, according to (\ref{24}),
satisfy
$b_0^2 = c_0^2 = 0$ and $\{b_0,c_0\}=1$.

These anticommutation relations are the same as those of the
 $\gamma$-matrix algebra in two spacetime dimensions
in light-cone coordinates.
The simplest representation of this algebra is two-dimensional
and is realized by $b_0=(\sigma^1+i\sigma^2)/\sqrt{2}$ and
$c_0=(\sigma^1-i\sigma^2)/\sqrt{2}$.
Thus, in this representation, there should be two states: a ``spin up"
and a ``spin down" state, satisfying
\bea
b_0 | \downarrow \rangle = 0\,, \;\;\;&\;\;\; b_0 |\uparrow\rangle =
|
\downarrow
\rangle\,, \nn\\
c_0 | \uparrow \rangle
= 0\,, \;\;\;&\;\;\; c_0 | \downarrow\rangle = |
\uparrow
\rangle \,.\nn
\eea
Imposing also (\ref{25}) implies that the correct ghost vacuum is
$|\downarrow\rangle$.
We can now create states from this vacuum by acting with the negative
modes
of the ghosts $b_m,c_n$.
We cannot act with $c_0$ since the new state does not satisfy the
Siegel condition (\ref{25}).
Now, we are ready to describe the physical states in the open string.
Note that since $Q_B$ in (\ref{27}) has ``level" zero\footnote{By level
here we mean total mode number. Thus, $L_0$ and $L_{-n}L_n$ both have
level zero.}
, we can impose
BRST invariance on physical states level by level.

At level zero there is only one state, the total vacuum
$|\downarrow,p^{\mu}\rangle$
\be
0 = Q_B | \downarrow , p \rangle = ( L^X_0 - 1 ) c_0 | \downarrow,
p\rangle\,.
\ee
BRST invariance
gives the same mass-shell condition, namely $L_0^X-1=0$ that we
obtained
in the previous  quantization
scheme. This state cannot be a BRST exact state; it is therefore
 physical: it is the tachyon.

At the first level, the possible operators are $\al^\mu_{-1}$,
$b_{-1}$ and $c_{-1}$. The most general state of this form is then
\be
|\psi\rangle = ( \zeta \cdot \al_{-1} + \xi_1 c_{-1} + \xi_2 b_{-1} )
| \downarrow,p\rangle \,, \label{n=1}
\ee
which has 28 parameters: a 26-vector $\zeta_\mu$ and two more
constants
$\xi_1$,$\xi_2$. The BRST condition demands
\be
 0 = Q_B |\psi\rangle = 2 ( p^2 c_0 + (p\cdot \zeta) c_{-1} + \xi_1
p\cdot \al_{-1} ) | \downarrow, p \rangle \,.
\ee
This only holds if $p^2 =0$ (massless) and $p\cdot\zeta=0$ and
$\xi_1=0$. So there are only 26 parameters left. Next we have to make
sure that this state is not Q-exact: a general state $|\chi\rangle$
is of the same form as \r{n=1}, but with parameters
$\zeta'^{\mu}$, $\xi_{1,2}'$. So the most general Q-exact state at
this level
with $p^2=0$ will be
\[ Q_B |\chi \rangle = 2 ( p\cdot \zeta' c_{-1} + \xi_1'
p\cdot\al_{-1} ) | \downarrow, p\rangle \,. \]
This means that the $c_{-1}$ part in \r{n=1} is BRST-exact and that
the polarization has the equivalence relation $\zeta_\mu \sim
\zeta_\mu + 2 \xi_1' p_\mu $. This leaves us with the 24 physical
degrees of freedom we expect for a massless vector particle
 in 26 dimensions.

The same procedure can be followed for the higher  levels. In the
case
of the closed string we have to include the barred operators, and of
course we have to use $Q_B+\bar Q_B$.

\section{Interactions and loop amplitudes}

\begin{figure}
\begin{center}
\leavevmode
\epsfxsize=12cm
\epsffile{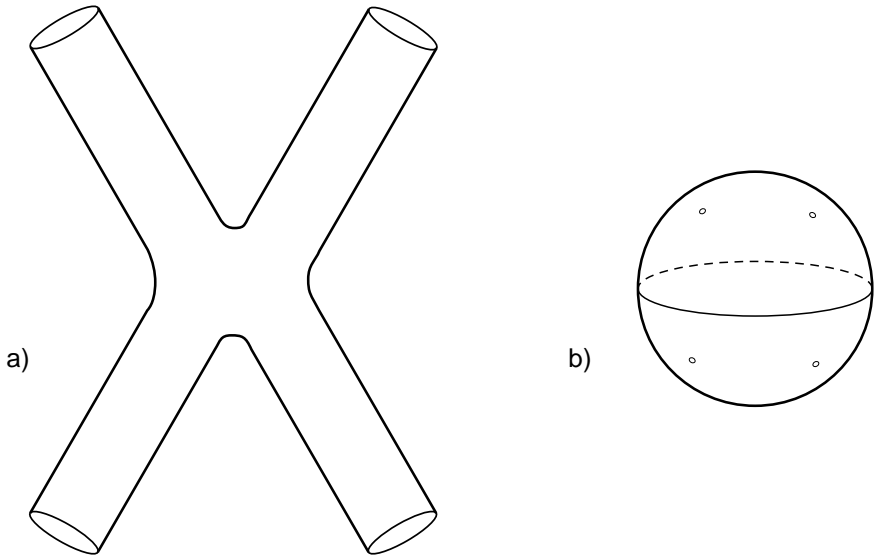}
\end{center}
\caption[]{\it a) Tree closed string diagram describing four-point
scattering.
b) Its conformal equivalent, the four-punctured sphere}
\label{f2}\end{figure}

The obvious next question is how to compute scattering amplitudes of
physical states.
Consider two closed strings, which enter, interact and leave at
tree level (Fig. \ref{f2}a).

By a conformal transformation we can map the diagram to a
sphere
with four infinitesimal holes (punctures) (Fig. \ref{f2}b).
At each puncture we have to put appropriate boundary conditions that
will specify which is the external physical state that participates
in the interaction. In the language of the path integral we will have
to insert
a ``vertex operator", namely the appropriate wavefunction as we have
done in the case of the point particle.
Then, we will have to take the path-integral average of these
vertex operators weighted with the Polyakov action on the sphere.
In the operator language, this amplitude (S-matrix element) will be
given
by a correlation function of these vertex operators in the
two-dimensional
world-sheet
quantum theory. We will also have to integrate over the positions
of these vertex operators. On the sphere there are three conformal
Killing vectors,
which implies that there are three reparametrizations that have not
been
fixed.
We can fix them by fixing the positions of three vertex operators.
The positions of the rest are Teichm\"uller moduli and should be
integrated over.

What is the vertex operator associated to a given physical state?
This can be found directly from the two-dimensional world-sheet
theory.
The correct vertex operator will produce the appropriate physical
state
as it comes close to the out vacuum, but more on this will follow
in the next section.

\begin{figure}
\begin{center}
\leavevmode
\epsfxsize=12cm
\epsffile{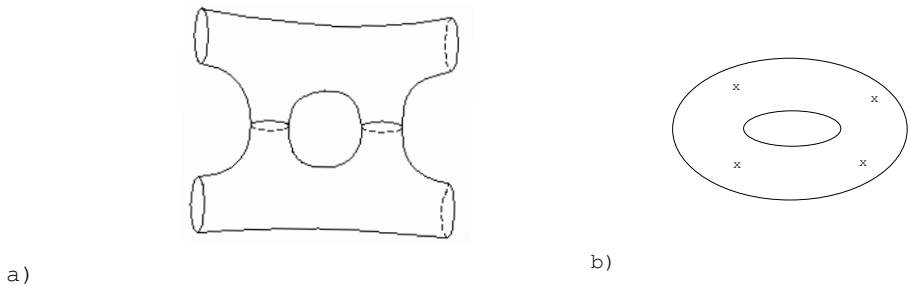}
\end{center}
\caption[]{\it a) World-sheet relevant for the one-loop contribution to
four-point scattering;
b) Its conformal transform where the holes become punctures on a
torus.}
\label{f3}\end{figure}

One more word about loop amplitudes.
Consider the string diagram in Fig. \ref{f3}a. This is the string
generalization of a one-loop amplitude contribution to the scattering
of four particles in Fig. \ref{f2}a.
Again by a conformal transformation it can be deformed into a torus
with
four punctures (Fig. \ref{f3}b).
The generalization is straightforward.
An N-point amplitude (S matrix element) at g-loop order is given by
the average of the N appropriate vertex operators, the average taken
with the Polyakov action on a two-dimensional
surface with g handles (genus g
Riemann surface).
For more details, we refer the reader to \cite{GSW}.

{}From this discussion, we have seen that the zero-, one- and two-point
amplitudes
on the sphere are not defined.
This is consistent with the fact that such amplitudes
do not exist on-shell.
The zero-point amplitude at one loop is not defined either.
When we will be talking about the one-loop vacuum amplitude below, we
will
implicitly consider the one-point dilaton amplitude at zero momentum.

\section{Conformal field theory\label{conf}}
\setcounter{equation}{0}

We have seen so far that the world-sheet quantum theory that
describes
the bosonic string is a conformally invariant quantum field theory in
two dimensions.
In order to describe more general ground-states of the string, we will
need
to study this concept in more detail.
In this chapter we will give a basic introduction to conformal field
theory and its application in string theory.
We will assume Euclidean signature in two dimensions.
A more complete discussion can be found in \cite{G}.

\subsection{Conformal transformations}
\setcounter{equation}{0}

Under general coordinate transformations, $x \to x'$,
the metric transforms as
\[
g_{\mu\nu} \to g'_{\mu\nu}(x') = \frac{\pd x^\al}{\pd
x'^\mu}\frac{\pd x^\beta}{\pd x'^\nu} g_{\al\beta}(x)       \,.
\]
The group of conformal transformations, in any dimension, is then
defined as the
subgroup of these coordinate transformations that leave the metric
invariant up to a scale change:
\be
g_{\mu\nu}(x) \to g'_{\mu\nu}(x') = \Omega(x) g_{\mu\nu}(x) \,.
\ee
These are precisely the coordinate transformations that preserve the
angle between two vectors, hence the name conformal transformations.
Note that the Poincar{\' e} group is a subgroup of the conformal
group (with $\Omega =1$).

We will  examine the generators of these transformations. Under
infinitesimal coordinate transformations, $x^\mu \to x'^\mu = x^\mu +
\epsilon^\mu$, we obtain
\[
ds'^2 = ds^2 - ( \pd_\mu \e_\nu + \pd_\nu \e_\mu ) dx^\mu dx^\nu \,.
\]
For it to be a conformal transformation, the second term on the
right-hand side has to be proportional to $\eta_{\mu\nu}$, or
\be \label{epsilon}
\pd_\mu \e_\nu + \pd_\nu \e_\mu = \frac{2}{d}
(\pd\cdot\e)\eta_{\mu\nu} \,,
\ee
where the proportionality factor can be found by contracting both
sides with $\eta^{\mu\nu}$. If we act on both sides of this equation
with $\pd^\mu$ we obtain
\[ \Box \e_\nu + \left( 1- \frac{2}{d} \right) \pd_\nu ( \pd\cdot\e
)=0
\,, \]
or if we act on both sides of \r{epsilon} with $\Box = \pd_\mu
\pd^\mu$ we obtain
\[ \pd_\mu \Box \e_\nu + \pd_\nu \Box \e_\mu = \frac{2}{d}
\eta_{\mu\nu} \Box (\pd\cdot\e) \,.\]
With these two equations, we can write the constraints on the
parameter as follows
\be
\left[ \eta_{\mu\nu} \Box + (d-2) \pd_{\mu} \pd_{\nu} \right]
\pd\cdot\e =
0 \,.
\label{see}\ee
We can already see in (\ref{see}) that $d=2$ will be a special case.
Indeed for $d>2$, (\ref{see}) implies that the parameter $\e$ can be at
most
quadratic in $x$.
We can then identify the following possibilities for $\e$:
\bea
\e^\mu = a^\mu &\quad& \mbox{translations}\,, \nn\\
\e^\mu = \omega^\mu_\nu x^\nu && \mbox{rotations} \quad
(\omega_{\mu\nu} = - \omega_{\nu\mu}) \,,\\
\e^\mu = \l x^\mu && \mbox{scale transformations} \nn
\eea
and
\bea
\e^\mu = b^\mu x^2 - 2 x^\mu (b\cdot x)\,,
\eea
which are the special conformal transformations. Thus,
we have a total of
\[ d + \half d(d-1) + 1 + d = \half ( d+2 )(d+1) \]
parameters.
In a space of signature $(p,q)$ with $d=p+q$, the Lorentz group
is O(p,q).

\vskip .4cm
\noindent\hrulefill
\vskip .2cm
{\large\bf Exercise}: Show that the algebra of conformal
transformations
is isomorphic to the Lie algebra of O(p+1,q+1).
\nopagebreak\vskip .2cm
\noindent\hrulefill
\vskip .4cm

We will now investigate  the special case $d=2$. The restriction that
$\epsilon$ can be at most of second order does not apply anymore,
but \r{epsilon} in Euclidean space ($g_{\mu\nu} = \d_{\mu\nu}$)
reduces to
\be
\pd_1 \e_1 = \pd_2 \e_2\,, \quad \quad \pd_1 \e_2 = - \pd_2 \e_1 \,.
\ee
This can be further simplified by going to complex coordinates,
$z,\bar z=x^1\pm ix^2$.
If we  define  the complex parameters $\e,\bar \e = \e_1 \pm i \e_2
$, the equations for the parameters become
\be
\pd \bar \e =0\,, \quad \quad \bar \pd \e =0 \,,
\ee
where we used the short-hand notation $\bar\pd = \pd_{\bar z}$. This
means that $\e$ can be an arbitrary function of $z$, but it is
independent of $\bar z$ and vice versa for $\bar \e$. Globally, this
means that conformal transformations in two dimensions
consist of the analytic coordinate transformations
\be
z \to f(z) \quad\mbox{and}\quad \bar z \to \bar f (\bar z)\,.
\ee

We can expand the infinitesimal transformation parameter
\[ \e(z) = - \sum a_n z^{n+1}\,. \]
The generators corresponding to these transformations are then
\be
{\ell}_n = - z^{n+1} \pd_z \,,
\ee
i.e. ${\ell}_n$ generates the transformation with $\e = - z^{n+1}$.
The
generators satisfy the following algebra
\be
[ {\ell}_m ,{\ell}_n ] = (m-n) {\ell}_{m+n}\,, \quad\quad
[\bar{\ell}_m,\bar{\ell}_n ] = (m-n) \bar{\ell}_{m+n}\,,
\ee
and $[\bar{\ell}_m,{\ell}_n] = 0$.
Thus, the conformal group in two dimensions is infinite-dimensional.

An interesting subalgebra of this algebra is spanned by the
generators ${\ell}_{0,\pm 1}$ and $\bar {\ell}_{0,\pm 1}$. These are
the only generators that are
globally well-defined on the Riemann sphere $S^2 = \C \cup \infty$.
They form the algebra of O(2,2)$\sim$SL(2,$\C)$.
They generate the following transformations:
\begin{center}
\begin{tabular}{cccl} & Infinitesimal & Finite
\\Generator&transformation&transformation
\\
            $ {\ell}_{-1}$ & $z \to z - \e$ & $z \to z + \al$
&Translations\\
            $ {\ell}_0$ & $z \to z - \e z$ &  $z \to \l z$ & Scaling \\
            $ {\ell}_1$ & $z \to z - \e z^2$ & $z \to
\frac{z}{1-\beta z}$ & Special conformal
\end{tabular}
\end{center}
with equivalent expressions for the barred generators. From this, it
is immediately clear that the generator $i( \ell_0 - \bar\ell_0 )$
generates a rescaling of the phase or, in other words, it generates
rotations in the $z$-plane.
Dilatations are generated by $\ell_0 + \bar\ell_0$. These
transformations generated by ${\ell}_{0,\pm 1}$ can be summarized by
the expression
\be\label{sl2c}
z \to \frac{ a z +b }{c z +d} \,,
\ee
where $a,b,c,d \in \C$ and $ad-bc =1$. This is the group
SL(2,$\C)/\Z_2$, where the $\Z_2$ fixes the freedom to replace all
parameters $a,b,c,d$ by minus themselves, leaving the transformation
\r{sl2c} unchanged.
We will call this finite-dimensional subgroup of the conformal group
the {\it restricted conformal group}.

\subsection{Conformally invariant field theory}
\setcounter{equation}{0}

A two-dimensional theory will be called conformally invariant if the
trace of its stress-tensor vanishes in the quantum theory in flat
space.
Such a theory has the following properties:

1) There is an (infinite)  set of fields $\{ A_i \}$. In
particular, this set will contain all the derivatives of the fields.

2) There exists a subset $\{ \phi_j \} \subset \{ A_i \} $, called
quasi-primary fields, that transforms under $restricted$ conformal
transformations
\be
z\to f(z)={az+b\over cz+d}\;\;\;,\;\;\;\bar z\to \bar f(\bar z)={\bar
a\bar z+\bar b\over \bar c\bar z+\bar d}
\ee
in the following way,
\be\label{prrest}
\Phi(z,\bar z) \to \left(\frac{\pd f}{\pd z} \right)^h
\left(\frac{\pd \bar f}{\pd \bar z} \right)^{\bar h}   \Phi\left(
f(z), \bar f (\bar z) \right) \,.
\ee
As we shall see later all fields that are not derivatives of other
fields
are quasi-primary.

3) Finally there are the so-called primary fields, which transform as
in (\ref{prrest}) for all conformal transformations;
$h,\bar h$ are real-valued ($\bar h$ is not  the complex
conjugate of $h$). Note that this transformation property is very
similar to the transformation property of tensors. As for tensors,
the expression
\[ \Phi(z,\bar z) dz^h d \bar z^{\bar h} \]
is invariant under conformal transformations;
$(h,\bar h)$ are the conformal weights of the primary field.

The theory is covariant under conformal transformations. Consequently,
the correlation functions satisfy
\be
\left\langle\,\prod_{i=1}^N \Phi_i(z_i,\bar z_i) \,\right\rangle =
\prod_{i=1}^N \left( \frac{\pd f}{\pd z} \right)^{h_i}_{z\to z_i}
\left( \frac{
\pd \bar f}{\pd \bar z} \right)^{\bar h_i}_{\bar z\to\bar z_i}
\left\langle\,\prod_{j=1}^N\Phi_j\left(
f(z_j),\bar f(\bar z_j)\right)\right\rangle\;.
\label{28}\ee
As we shall see later on, the conformal anomaly spontaneously breaks
the invariance
of the full conformal group.
On the sphere, the unbroken subgroup is the restricted conformal
group
and (\ref{28}) is, thus, valid only for SL(2,$\C)$.
However, there will be Ward identities that will encode the full
conformal
covariance of the theory.

Infinitesimally, under $z\to z + \e(z)$ and $\bar z \to \bar z + \bar
\e(\bar z)$, a primary field transforms as
\be\label{deephi1}
\d_{\e ,\bar\e} \Phi(z,\bar z) = \left[ ( h\pd\e + \e\pd ) + (\bar
h\bar\pd\bar\e + \bar\e\bar\pd ) \right] \Phi(z,\bar z) \,,
\ee
and the two-point function $G^{(2)}(z_i,\bar z_i) = \langle\,
\Phi(z_1,\bar z_1) \Phi(z_2,\bar z_2) \,\rangle$ transforms as
\[
\d_{\e,\bar \e} G^{(2)}(z_i,\bar z_i) = \langle\, \d_{\e,\bar
\e}\Phi_1 ,\Phi_2\,\rangle +\langle\, \Phi_1 , \d_{\e,\bar \e}\Phi_2
\,\rangle = 0 \,.
\]
If we put these two together, it results in the following
differential equation for the two-point function
\be\label{difg2}
\left[ \left( \e(z_1)\pd_{z_1} + h_1 \pd \e(z_1) + \e(z_2)\pd_{z_2} +
h_2 \pd \e(z_2) \right) + \left( \mbox{barred terms} \right) \right]
G^{(2)}(z_i,\bar z_i) =0 \,.
\ee
We can now use the series expansion  of $\e(z)$ to analyze this
equation. If we first take $\e(z)=1$ and $\bar\e(\bar z) =1$
(remember, this corresponded to translations), then \r{difg2} tells
us that $G^{(2)}(z_i,\bar z_i)$ only depends on $z_{12} = z_1 -
 z_2$, $\bar z_{12} = \bar z_1 -\bar z_2$. This is not very
surprising because in a translationally invariant theory we would
expect the correlation functions to only depend on the relative
distance.
If we next use $\e(z) = z$, $\bar\e(\bar z) = \bar z$ (rotational
invariance), we find $G^{(2)} \sim 1/(z_{12}^{h_1+h_2} \bar
z_{12}^{\bar h_1 + \bar h_2} )$ and if we finally use $\e(z) = z^2$
(special conformal transformation) we find the restriction $h_1=h_2=h$
and $
\bar h_1 = \bar h_2 =\bar h$. The conclusion is that the
two-point function is completely fixed up to a constant:
\be
G^{(2)}(z_i,\bar z_i) = \frac{C_{12}}{z_{12}^{2h} \bar z_{12}^{2\bar
h} } \,.
\ee
This constant can be set to 1, by normalizing the operators.

A similar analysis can be done for the three-point function and it
turns out to be  also completely determined up to a constant:

\vskip .4cm
\noindent\hrulefill
\nopagebreak\vskip .2cm
{\large\bf Exercise}: Solve the Ward identities and show that the most
general
form allowed for the three-point function is
\be
G^{(3)}(z_i,\bar z_i) = \frac{ C_{123} }{z_{12}^{\Delta_{12}}
z_{23}^{\Delta_{23}} z_{31}^{\Delta_{31}} \bar
z_{12}^{\bar\Delta_{12}} \bar z_{12}^{\bar \Delta_{12}} \bar
z_{12}^{\bar\Delta_{12}}} \,,
\ee
where $\Delta_{12} = h_1 + h_2 - h_3$, $\bar\Delta_{12} = \bar h_1 +
\bar h_2 - \bar h_3$, etc.
\nopagebreak
\nopagebreak\vskip .2cm
\noindent\hrulefill
\vskip .4cm

The next correlation function, however, the four-point function, is not
fully determined. Conformal invariance restricts it, using the
procedure outlined above,  to have the following
form
\be
G^{(4)}(z_i,\bar z_i) = f(x,\bar x) \prod_{i<j} z_{ij}^{-(h_i+h_j) +
h/3 }\prod_{i<j} \bar z_{ij}^{-(\bar h_i+\bar h_j) + \bar h/3 } \,,
\ee
where $h=\sum h_i$, $\bar h = \sum \bar h_i$. The function
$f$ is arbitrary, but only depends on the cross-ratio $x = z_{12}
z_{23}/z_{13} z_{24}$ and $\bar x$.

The general N-point function of quasiprimary fields on the sphere
\be
G^N(z_1,\bar z_1,\dots z_N,\bar z_N)=\left\langle \prod_{i=1}^N \Phi_i
(z_i,\bar z_i)\right\rangle
\ee
satisfies the following constraints coming from SL(2,$\C)$ covariance
\be
\sum_{i=1}^N\pd_i~G^N=0\,,
\label{0}\ee
\be
\sum_{i=1}^N(z_i\pd_i+h_i)~G^N=0\,,
\label{00}
\ee
\be
 \sum_{i=1}^N(z_i^2\pd_i+2z_i h_i)~G^N=0
\label{000}\ee
and similar ones with $z_i\to \bar z_i$, $h_i\to \bar h_i$.
These are the Ward identities reflecting SL(2,$\C)$ invariance of the
correlation functions on the sphere.

\subsection{Radial quantization}
\setcounter{equation}{0}

\begin{figure}
\begin{center}
\leavevmode
\epsfxsize=13cm
\epsffile{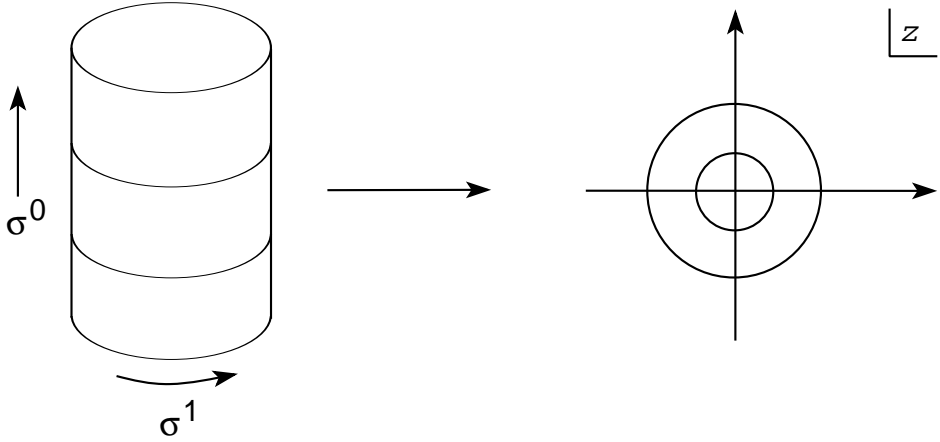}
\end{center}
\vspace{-.5cm}
\caption[]{\it The map from the cylinder to the compactified complex
plane}
\label{f4}\end{figure}

We will now  study the Hilbert space of a conformally invariant theory.
We start from a
two-dimensional Euclidean space with coordinates $\t$ and $\s$. (Note
that we can go from a two-dimensional Euclidean space to
Minkowski space by means of a Wick rotation, $\t \to i\t$.) To avoid
IR problems we will
compactify the space direction, $\s = \s + 2 \pi$, and the
two-dimensional space becomes a cylinder. Next, we make the conformal
transformation
\[ z = e^{\t + i\s}\,, \quad\quad \bar z = e^{\t - i \s}\,, \]
which maps the cylinder onto the complex plane (topologically a
sphere) as shown in Fig. \ref{f4}.

Surfaces of equal time on the cylinder will become circles of equal
radius on the complex plane. This means that the infinite past ($\t =
-\infty$) gets mapped onto the origin of the plane ($z=0$) and
the infinite future becomes $z=\infty$. Time reversal becomes
$z\to 1/z^*$ on the complex plane, and parity $z\to z^{*}$.

We already saw that $\ell_0$ was the generator of dilatations on
the cylinder, $z \to \l z$ so $\ell_o + \bar \ell_0$ will move us
in the radial direction
on the plane, which corresponds to the time direction on the
cylinder. This means that the dilatation operator is the
Hamiltonian of our system\footnote{Since we are
in Euclidean space the term Hamiltonian may appear bizarre.
The proper name should be transfer operator, which upon Wick rotation
becomes the Hamiltonian. Similarly the exponential of the transfer
operator
gives the transfer matrix, which would become the time evolution
operator upon Wick rotation.}
\[ H = \ell_0 + \bar \ell_0 \,. \]

An integral over the space direction $\s$ will become a contour
integral on the complex plane. This enables us to use all the
powerful techniques developed in complex analysis.

Infinitesimal coordinate transformations are generated by the
stress-tensor, which is traceless in the case of a Conformal Field
Theory (CFT)\footnote{This is only true in flat space.
In general  ${T_{\mu}}^{\mu}\sim c R^{(2)}$ where $c$ is a number
known as the conformal anomaly
and will appear also in the Virasoro algebra; $R^{(2)}$ is
the two-dimensional curvature scalar.},
\be
T_\mu{}^\mu =0.
\ee
In complex coordinates this means that the stress-tensor has
non-vanishing components $T_{zz}$ and $T_{\bar z\bar z}$, while
$T_{z\bar z}=0$ since $T_{z\bar z}$ is the trace of the
stress-tensor. This can be shown by expressing them back in Euclidean
coordinates, $z = x + iy$,
\[ T_{z\bar z} = T_{\bar zz} = \quart ( T_{00} + T_{11} ) = \quart
T_\mu{}^\mu \,. \]
The conservation law $\pd^\mu T_{\mu\nu} =0 $ gives us, together with
the traceless condition,
\be
\pd_z T_{\bar z\bar z} =0 \quad\mbox{and}\quad \pd_{\bar z} T_{zz} =0
\,,
\ee
which implies that the two non-vanishing components of the
stress-tensor are holomorphic and antiholomorphic
respectively:
\be
T(z) \equiv T_{zz} \quad\mbox{and}\quad \bar T(\bar z) \equiv T_{\bar
z\bar z}\,.
\ee
Thus, we can construct an  infinite number of conserved currents,
because
if $T(z)$ is conserved, then $\e(z) T(z)$ is also conserved, for
every holomorphic function $\e(z)$.

These currents produce  the following conserved charges
\be
Q_\e = \frac{1}{2\pi i} \oint dz \e(z) T(z)\;\;\;,\;\;\;
Q_{\bar\e} = \frac{1}{2\pi i} \oint d\bar z \bar \e(\bar z) \bar
T(\bar z)\,.
\ee
These charges are the generators of the infinitesimal conformal
transformations
\[ z \to z + \e(z)\,, \quad\quad\quad \bar z \to \bar z + \bar
\e(\bar
z)\,. \]
The variation of fields under these transformations is given, as
usual, by the
commutator of the fields with the generators:
\be\label{deephi}
\d_{\e,\bar\e} \Phi(z,\bar z) = \left[ Q_\e + Q_{\bar \e} ,
\Phi(z,\bar z) \right]\,.
\ee

\begin{figure}
\begin{center}
\leavevmode
\epsfxsize=13cm
\epsffile{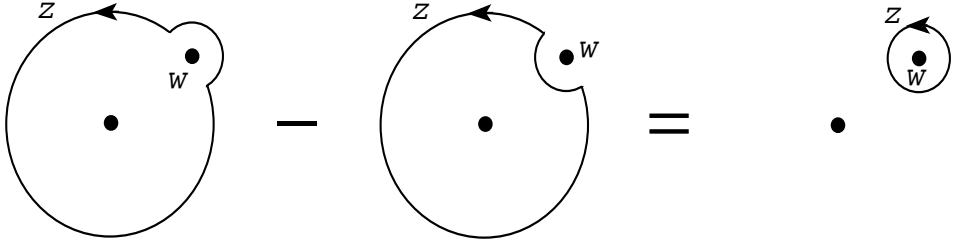}
\end{center}
\caption[]{\it Rearrangement of contours relevant for commutators.}
\label{f5}\end{figure}

We know that products of operators are only well-defined in a quantum
theory if they are time-ordered. The analog of this in radial
quantization on the complex plane is radial ordering. The
radial-ordering operator $R$ is defined as:
\be
R( A(z) B(w) ) = \left\{ \begin{array}{cc}
A(z)B(w)&|z|>|w|\\(-1)^FB(w)A(z)&|z|<|w| \end{array} \;. \right.
\ee
In the case of fermionic operators, there appears of course a minus
sign if we interchange them.
With the help of this ordering we can write an equal-time commutator
of an operator
with a spatial integral over another operator as a contour integral
over the radially-ordered product of the two operators:
\[ \left[ \int d\s B , A \right] = \oint dz R( B(z)A(w) ) \]\,
as shown in Fig. \ref{f5}.
This means that we can rewrite \r{deephi} as
\bea
\d_{\e,\bar\e} \Phi(z,\bar z) &=& \frac{1}{2\pi i} \oint\left( dz
\e(z) R( T(z) \Phi(w,\bar w)) + d\bar z \bar \e(\bar z) R( \bar
T(\bar z) \Phi(w,\bar w))\right) \nn\\
&=& \left[ ( h\pd\e(w) + \e(w)\pd ) + (\bar h\bar\pd\bar\e(\bar w) +
\bar\e(\bar w)\bar\pd ) \right] \Phi(w,\bar w) \,, \nn
\eea
where the last line is the desired result copied from \r{deephi1}.
This equality will only hold if $T$ and $\bar T$ have the following
short-distance singularities with $\Phi$:
\bea
R( T(z) \Phi(w,\bar w) ) = \frac{h}{(z-w)^2} \Phi(w,\bar w) +
\frac{1}{z-w} \pd_w \Phi(w,\bar w) + \dots\,, \\
R( \bar T(\bar z) \Phi(w,\bar w) ) = \frac{\bar h}{(\bar z-\bar w)^2}
\Phi(w,\bar w) + \frac{1}{\bar z-\bar w} \pd_{\bar w} \Phi(w,\bar w)
+ \dots\,,
\label{36}\eea
where the dots (which we write at first, but end up  being implicit)
denote regular terms. From now on we shall drop the
$R$ symbol and assume that the operator product expansion (OPE) is
always radially ordered. The OPE  with the
stress-tensor can be used as a definition of a conformal field of
weight $(h,\bar h)$ instead of \r{prrest}.

We will describe here the general Ward identities for insertions
of the stress-tensor.
Consider the correlation function
\be
F^N(z,z_i,\bar z_i)=\left\langle T(z)\prod_{i=1}^N\Phi_i(z_i,\bar z_i)
\right\rangle\label{012}\,,\ee
where $\Phi_i$ are primary fields.
Viewed as a function of $z$, $F^N$ is meromorphic with poles
when $z\to z_i$. The residues of these poles can be calculated with
the help of
(\ref{36}). A meromorphic function on the sphere is uniquely
specified by its poles and residues. Thus, we obtain
\be
F^N(z,z_i,\bar z_i)=\sum_{i=1}^N\left({h_i\over
(z-z_i)^2}+{\pd_{z_i}\over z-z_i}\right) \left\langle
\prod_{i=1}^N\Phi_i(z_i,\bar z_i)
\right\rangle\,.
\label{011}\ee
This Ward identity expresses correlation functions of primary fields
with an insertion of the stress-tensor in terms of the correlator of
the primary fields themselves.
Multiple insertions can also be handled using in addition (\ref{TT}).

In general, the product of two operators can be expanded in terms of
a complete set of orthonormal local operators
\be
\Phi_i(z,\bar z) \Phi_j(w,\bar w) = \sum_k C_{ijk} (z-w)^{h_k -h_i -
h_j} (\bar z -\bar w)^{\bar h_k - \bar h_i - \bar h_j} \Phi_k(w,\bar
w) \,,
\label{90}\ee
where the numerical constants $C_{ijk}$ can be shown to coincide with
the constants in the three-point function $\langle\, \Phi_i \Phi_j
\Phi_k
\,\rangle$.
This is true in any quantum field theory; here, however, because of
conformal invariance there is no mass scale that appears in the OPE.
This type of expansion can be thought of as a way to encode the
correlation
functions, since knowledge of (\ref{90}) determines them completely
in a unitary theory and vice versa.

\subsection{Example: the free boson\label{boson}}
\setcounter{equation}{0}

The action for a non-compact free boson in two dimensions as we
encountered it
in string theory is
\be\label{actionboson} S = \frac{1}{4\pi} \int d^2z ~\pd X \bar \pd X
\,. \ee
The field $X(z,\bar z)$ has the propagator
\be
\langle\, X(z,\bar z)  X(w,\bar w) \,\rangle = - \log (| z-w|^2
\mu^2) \,.
\label{31}\ee
This is obtained by taking the massless limit of the massive scalar
propagator in two dimensions; $\mu$ is an IR cutoff.
Equation (\ref{31}) can be obtained by starting with the massive
propagator,
with mass $\mu$, and taking the limit $\mu\to 0$, keeping terms that do
not vanish in the limit.
The dependence on $\mu$ should disappear from correlation functions.
Note that $X$ itself is not a conformal field since its correlation
functions
are IR-divergent. Its
derivative $\pd_z X$,  however, is well behaved. The OPE of the
derivative with itself is
\bea
\pd_z X(z) \pd_w X(w) &=& \pd_z\pd_w \langle XX\rangle + : \pd_z
X\pd_wX: \nn\\
                      &=& -\frac{1}{(z-w)^2} + : \pd_z X\pd_wX:
\eea
and $\pd_zX$ is a  conformal field of weight $(1,0)$.
Note that $\mu$ has disappeared.
We will now calculate its OPE with the stress-tensor.

According to the action \r{actionboson} the stress-tensor for the
free boson is given by
\bea
T(z) = -\half : \pd X\pd X : = - \half \lim_{z\to w} \left[ \pd_z X
\pd_w X + \frac{1}{(z-w)^2} \right]\,, \\
\bar T(\bar z) = -\half : \bar\pd X \bar\pd X : = - \half \lim_{\bar
z\to
\bar w} \left[ \pd_{\bar z} X \pd_{\bar w} X + \frac{1}{(\bar z-\bar
w)^2} \right]\,.
\eea
Using Wick's theorem,  we can calculate
\bea
T(z) \pd X(w) &=& -\half : \pd X(z) \pd X(z): \pd X(w)\nn\\
              &=& -\pd X(z) \langle \pd X(z) \pd
X(w) \rangle + \dots \nn\\
              &=& \pd X(z) \frac{1}{(z-w)^2} + \dots \nn\\
              &=& \frac{\pd X(w)}{(z-w)^2} + \frac{1}{z-w} \pd^2 X(w)
+ \dots \,,
\eea
where the dots indicate terms that are not singular as $z\to w$.
Similarly we find $\bar T \pd X = \mbox{regular}$.
Thus, $\pd X$ is a $(1,0)$ primary field.
In the same way we find that $\bar\pd X$ is a $(0,1)$ primary field.

Are there any other primary fields? The answer is yes. There are
certainly
several, constructed  out of products of derivatives of $X$.
We will consider another interesting class, the ``vertex"
operators $V_a(z)$=$:e^{ ia X(z)}:$. The OPE with the stress-tensor
is
\be
T(z) V_a(w,\bar w) = - \half : \pd X(z) \pd X(z) : \sum_{n=0}^\infty
\frac{i^n a^n}{n!} :X^n(w,\bar w): \,.
\ee
For all terms in the expansion there can be either one or two
contractions. We obtain
\bea
T(z) V_a(w) &=& -\half \left[ ia\pd \langle XX \rangle \right]^2
e^{iaX(w)} - \half 2 i a :\pd X(z) \pd \langle XX \rangle e^{iaX(w)}: +
\dots\nn\\
        &=& \frac{a^2/2}{(z-w)^2} e^{ia X(w)} + \frac{ia\pd X(z)}{z-w}
e^{iaX(w)} + \dots \nn\\
        &=& \frac{a^2/2}{(z-w)^2} V_a(w) + \frac{1}{z-w} \pd V_a(w) +
\dots\,.
\eea
Thus, the vertex operator $V_a$ is a conformal field of weight
$(a^2/2,0)$.

Consider now a correlation function of vertex operators
\be
G^N=\left\langle \prod_{i=1}^N V_{a_i}(z_i,\bar z_i)
\right\rangle=\exp\left[{1\over 2}
\sum_{i,j=1;i\ne j}^Na_i a_j \langle X(z_i,\bar z_i)X(z_j,\bar z_j)\rangle\right]
\label{30}\,,\ee
 where the second step in the above formula is due to the fact that we
have a free
(Gaussian)  field theory.
Using the propagator (\ref{31}) we can see that the IR divergences
cancel only if
\be
\sum_i a_i=0.
\label{91}\ee
This a charge-conservation condition.

For the two-point function we obtain
\bea
\langle V_{a}(z) V_{-a}(w) \rangle &=& \langle : e^{ia X(z)}:
\,:e^{-ia X(w)}: \rangle \nn\\
&=& e^{-a^2 \log|z-w|^2  }= \frac{1}{|z-w|^{2a^2}} \,,
\eea
which confirms that  $a^2 = 2h=2\bar h$.

In this theory the operator $i\pd X$ is a U(1) current, which is
chirally conserved.
It is associated to the symmetry of the action under $X\to
X+\epsilon$.
The zero mode of the current is the charge operator.
{}From
\be
i\pd_z X~V_a(w,\bar w)=a ~{V_a(w,\bar w)\over (z-w)}+{\rm finite}
\label{92}\ee
we can tell that the operator $V_a$ carries charge $a$.
The charge-conservation condition (\ref{91}) is precisely due to the
U(1)
invariance of the theory.
In the case of string theory, this type of U(1) invariance is
essentially
momentum conservation.

\subsection{The central charge}
\setcounter{equation}{0}

The stress-tensor $T_{\mu\nu}$ is conserved so it has scaling
dimension two.
In particular $T(z)$ has conformal weight (2,0) and $\bar T(\bar z)$
(0,2).
They are obviously quasiprimary fields.
 From these properties we can write the most general OPE between two
stress-tensors compatible with conservation (holomorphicity) and
conformal invariance.
\be\label{TT}
T(z)T(w) = \frac{c/2}{(z-w)^4} + 2 \frac{T(w)}{(z-w)^2} + \frac{\pd
T(w)}{z-w} + \dots \,.
\ee
The fourth-order pole can only be a constant. This constant has to be
positive in a unitary theory since
$\langle T(z)T(w)\rangle =c/2(z-w)^4$.
There can be no third-order pole since the OPE has to be symmetric
under $z\leftrightarrow w$.
Finally the rest of the singular terms are fixed by the fact that $T$
has conformal weight (2,0).
We have a similar OPE for $\bar T$ with $z\to \bar z$ and $c\to \bar
c$
and
\be
T(z)\bar T(\bar w)={\rm regular}\,.
\ee

Comparing (\ref{TT}) with (\ref{36}) we can conclude that  $T(z)$
itself is not a primary field due to the presence of the most
singular term. The constant $c$ is called
the
(left) central charge and $\bar c$ the right central charge.
Modular invariance implies that for a left-right asymmetric
theory $c-\bar c = 0 (\,\,{\rm mod}\,\, 24)$ and two-dimensional
Lorentz invariance
requires $c = \bar c$.

We will calculate the value of $c,\bar c$ for the free boson theory.
With
the stress-tensor
$T(z) = -\half : \pd X \pd X : $ we can calculate the OPE
\bea
T(z)T(w) &=& \quart \left\{ 2 \left( \pd\pd\langle XX \rangle
\right)^2 + 4 :\pd X(z) \pd X(w): \pd\pd \langle XX \rangle + \dots
\right\} \quad\quad\nn\\
&=& \frac{1/2}{(z-w)^4} + \frac{2}{(z-w)^2} T(w) + \frac{1}{z-w} \pd
T(w) + \dots \,,
\eea
and we see that a single free boson has central charge $c=\bar c=1$.
In
the bosonic string theory we have $d$ free bosons, consequently the
central charge
is $c=\bar c=d$.

\vskip .4cm
\noindent\hrulefill
\nopagebreak\vskip .2cm
{\large\bf Exercise}: Consider another (2,0) operator
\[ \tilde T = -\half: \pd X \pd X : + i Q \pd^2 X \,,\]
where the second term is a total derivative.
This is the stress-tensor of a modified theory for the free boson
where there is some background charge $Q$. Follow the same
procedure as above and show that the OPE of the two stress-tensors is
again of the same form as \r{TT}, but with central charge:
\be c = 1 - 12 Q^2\,.
\ee
Verify that  the conformal weight of the vertex operator $V_{\a}$ is
now
$\Delta=\a(\a-2Q)/2$. In particular, $V_{\a}$ and $V_{-\a+2Q}$ have the
same
conformal weight.
The charge neutrality condition (\ref{91}) now becomes $\sum_i
\a_i=2Q$.
\nopagebreak\vskip .2cm
\noindent\hrulefill
\vskip .4cm

\subsection{The free fermion}
\setcounter{equation}{0}

We will now analyze the conformal field theory, which describes a free
massless fermion. In two dimensions,
it is possible to have spinors that are both Majorana and Weyl, and
these will have only one component. The gamma matrices can be
represented by the Pauli matrices, i.e. $\gamma^1 =
\s^1$, $\gamma^2 = \s^2$, so that the chirality projectors are
$\half(1 \pm \s^3)$. The Dirac operator becomes
\be
\dslash = \s^1 \pd_1 + \s^2 \pd_2 = \left( \begin{array}{cc} 0 &
\pd_1 -
i\pd_2 \\ \pd_1 + i\pd_2 & 0 \end{array} \right) \sim \left(
\begin{array}{cc} 0 & \pd \\ \bar{\pd} & 0 \end{array} \right).
\ee
The action for a Majorana spinor $\left( \begin{array}{c} \psi \\
\bar{\psi} \end{array} \right)$ is
\be
S = -\frac{1}{8 \pi} \int d^2 z (\psi \bar{\pd} \psi + \bar{\psi} \pd
\bar{\psi} ).
\ee
The equations of motion are
\be
\bar{\pd} \psi = \pd \bar{\psi}=0,
\ee
which means that the left and right chiralities are represented by a
holomorphic and an anti-holomorphic spinor, respectively.

The operator product expansion of $\psi$ and $\bar{\psi}$ with
themselves can be found either by transforming the action into
momentum space or by explicitly writing down the most general power
expression with the correct conformal dimension. They are given by
\be
\psi(z)\psi(w) = \frac{1}{z-w}\;\;\;,\;\;\;
\bar{\psi}(\bar{z}) \bar{\psi}(\bar{w}) = \frac{1}{\bar{z} -
\bar{w}}
\label{32}\,.\ee
Up to a constant factor, the only expressions with conformal
dimension $(2,0)$ and $(0,2)$ respectively are
\be
T(z)  =  -\half : \psi(z) \pd \psi(z) :\;\;\;,\;\;\;
\bar{T}(\bar{z}) =  -\half : \bar{\psi}(\bar{z}) \bar{\pd}
\bar{\psi}(\bar{z}) :\,.
\ee
This stress-tensor has the correct operator product expansion
\be
T(z)T(w) = \frac{1/4}{(z-w)^4} + \frac{2}{(z-w)^2} T(w) +
\frac{1}{z-w} \pd T(w),
\ee
and a similar expression for $\bar{T}(\bar{z})$, so that
$c=\bar{c}=\half$.

\vskip .4cm
\noindent\hrulefill
\nopagebreak\vskip .2cm
{\large\bf Exercise}: By calculating the expansions of $T(z)\psi(w)$
and
$\bar{T}(\bar{z})\bar{\psi}(\bar{w})$, show that $\psi$ and
$\bar{\psi}$ are primary fields of conformal weight $(\half,0)$ and
$(0,\half)$, respectively.
\nopagebreak\vskip .2cm
\noindent\hrulefill
\vskip .4cm

\subsection{Mode expansions}
\setcounter{equation}{0}

We will write the mode expansion for the stress-tensor as
\be
T(z)  =  \sum_{n \in \Z} z^{-n-2} L_n\;\;\;,\;\;\;
\bar{T}(\bar{z})  =  \sum_{n \in \Z} \bar{z}^{-n-2} \bar{L}_n.
\label{stress}
\ee
The exponent $-n-2$ is chosen such that for the scale change $z
\rightarrow \frac{z}{\lambda}$, under which $T(z) \rightarrow
\lambda^2 T\left(\frac{z}{\lambda}\right)$, we have $L_{-n}
\rightarrow \lambda^n L_{-n}$. $L_{-n}$ and $\bar{L}_{-n}$, then have
scaling dimension $n$. If we consider a theory on a closed string
world-sheet, the transformation from the Euclidean space cylinder to
the complex plane is given by
\be
w = \t +i\s \rightarrow z = e^w. \label{defw}
\ee
For a holomorphic field $\Phi$ with conformal weight $h$, we would
write
\be
\Phi_{\mbox{\tiny cyl}}(w) = \sum_{n \in \Z} \phi_n e^{-nw} = \sum_{n
\in \Z} \phi_n e^{in(i\t - \s)} = \sum_{n \in  \Z}\phi_n z^{-n}.
\ee
When going to the plane and using (\ref{prrest}) this becomes,
for primary fields:
\be
\Phi(z) = \sum_{n \in \Z} \phi_n z^{-n-h}.
\label{mod}\ee
Non-primary fields also have an inhomogeneous piece in
(\ref{prrest}).
In particular the correct transformation of the stress-tensor is
\cite{BPZ}
\be
T(z)\to (f')^2 T(f(z))+{c\over 12}\left[{f'''\over f'}-{3\over
2}\left(
{f''\over f'}\right)^2\right]
\label{93}\,.\ee
This justifies the expansion of the stress-tensor
(\ref{stress}).

The mode expansion can be inverted by
\be
L_n  =  \oint \frac{dz}{2\pi i} z^{n+1} T(z)\;\;\;,\;\;\;
\bar{L}_n  =  \oint \frac{d\bar{z}}{2\pi i} \bar{z}^{n+1}
\bar{T}(\bar{z}).
\label{s2}\ee
The operator product expansions of $T(z)T(w)$ and
$\bar{T}(\bar{z})\bar{T}(\bar{w})$ can now be written in terms of the
modes. We have
\bea
[L_n,L_m] & = & \left(\oint \frac{dz}{2\pi i} \oint \frac{dw}{2\pi i}
- \oint \frac{dw}{2\pi i} \oint \frac{dz}{2\pi i} \right) z^{n+1}
T(z) w^{m+1} T(w) \nn \\
& = & \oint \frac{dw}{2\pi i} \oint_{C_w} \frac{dz}{2\pi i} z^{n+1}
w^{m+1} \left(\frac{c/2}{(z-w)^4} + \frac{2T(w)}{(z-w)^2} + \frac{\pd
T(w)}{z-w}  +\ldots \right)\nn\\
 &=&   \oint \frac{dw}{2\pi i}
\left(\frac{c}{12}(n+1)n(n-1)w^{n-2}w^{m+1}+ \right.\nn\\
&&\left. \quad\quad + 2~(n+1) w^n w^{m+1} T(w)
+ w^{n+1}w^{m+1} \pd T(w)
\right)\,.
\eea
The residue of the first term comes from $\left. \frac{1}{3!}
\pd^{3}_{z} z^{n+1} \right|_{z=w} = \frac{1}{6} (n+1)n(n-1) w^{n-2}$.
We integrate the last term by parts and combine it with the second
term. This gives $(n-m)w^{n+m+1} T(w)$. Performing the $
w$ integration leads to the Virasoro algebra
\be
[L_n,L_m] = (n-m) L_{n+m} + \frac{c}{12} (n^3 - n) \delta_{n+m,0}.
\ee
The analogous calculation for $\bar{T}(\bar{z})$ yields
\be
[\bar{L}_n,\bar{L}_m] = (n-m) \bar{L}_{n+m} + \frac{\bar{c}}{12} (n^3
- n) \delta_{n+m,0}.
\ee
Since $T\bar{T}$ has no singularities in its OPE,
\be
[L_n,\bar{L}_m] = 0.
\ee
Every conformally invariant theory realizes the conformal  algebra, and
its
spectrum forms representations of it.
For $c=\bar{c}=0$, it reduces to the classical algebra.
A consequence of the conformal anomaly is that
\be
T^{\al}_{\al} ={c\over 96\pi^3 }\sqrt{g}~ R^{(2)},
\label{33}\ee
where $R^{(2)}$ is the two-dimensional scalar curvature.
In a generic non-con\-formally invariant theory the trace can be a
generic
function of its various fields.
In a CFT it is proportional only to the scalar curvature.
This implies that in a CFT with $c\not= 0$ the theory depends on the
conformal factor of the metric, but in a very specific form
implied by (\ref{33}).
If we remember that the stress-tensor is the variation of the action
with respect to the metric, we can integrate (\ref{33}) to obtain the
dependence
of the quantum theory on the conformal factor. Let $\hat
g_{\alpha\beta}
=e^{\phi}g_{\alpha\beta}$. Then
\be
\int [DX]_{\hat g} e^{-S[\hat
g_{\alpha\beta},X]}=e^{-cS_{L}[g_{\alpha\beta},\phi]}
\int [DX]_{g} e^{-S[g_{\alpha\beta},X]}\,,
\ee
where $X$ is a generic set of fields and
\be
S_{L}[g_{\alpha\beta},\phi]={1\over 96\pi} \int \sqrt{\det
g}g^{\alpha\beta}
\pd_{\alpha}\phi\pd_{\beta}\phi+{1\over 48\pi}\int \sqrt{\det
g}R^{(2)}~\phi\,.
\label{34}\ee
This is the Liouville action.
In critical string theory, the ghost system cancels the central charge
of the string coordinates and the full theory is independent of
the scale factor.

\subsection{The Hilbert space}
\setcounter{equation}{0}

To describe the Hilbert space we will use the standard formalism of
in and out states of quantum field theory adapted to our coordinate
system.
For quasi-primary fields $A(z,\bar{z})$, the in-states are
defined as
\be
|A_{\mbox{\tiny in}}\rangle =\lim_{\t \rightarrow -\infty}
A(\t,\s) |0\rangle
= \lim_{z \rightarrow 0} A(z,\bar{z}) |0\rangle.
\ee
For the out-states, we need a description in the neighborhood of $z
\rightarrow \infty$. If we define $z=\frac{1}{w}$, then this is the
point $w=0$. The map $f:w \rightarrow z=\frac{1}{w}$ is a conformal
transformation, under which $A(z,\bar{z})$ transforms as
\be
\tilde{A}(w,\bar{w}) = A(f(w),\bar{f}(\bar{w}))(\pd f(w))^h
(\bar{\pd}\bar{f}(\bar{w}))^{\bar{h}}.
\ee
Substituting $f(w)=\frac{1}{w}$, we find
\be
\tilde{A}(w,\tilde{w}) =
A\left(\frac{1}{w},\frac{1}{\bar{w}}\right)(-w^{-2})^h
(-\bar{w}^{-2})^{\bar{h}}.
\ee
It is natural to define
\be
\langle A_{\mbox{\tiny out}} | = \lim_{w,\bar{w}\rightarrow 0}
\langle 0 | \tilde{A}(w,\bar{w}).
\ee
We would like $\langle A_{\mbox{\tiny out}}|$ to be the Hermitian
conjugate of $|A_{\mbox{\tiny in}} \rangle$. Hermitian conjugation of
operators of weight $(h,\bar{h})$ is defined by
\be
\left[ A(z,\bar{z}) \right]^{\dagger} = A\left(
\frac{1}{\bar{z}},\frac{1}{z}\right) \bar{z}^{-2h} z^{-2\bar{h}}.
\label{hermconj}
\ee
This definition finds its justification in the continuation from
Euclidean space back to Minkowski space. The
missing factor of $i$ in Euclidean time evolution $A(\s,\t) = e^{\t
H}A(\s,0)e^{-\t H}$ must be compensated for in the definition of
the adjoint by a Euclidean time reversal, which is
implemented on the plane by $z \rightarrow 1/z^\ast$. With
the definition (\ref{hermconj}), we find
\bea
\langle A_{\mbox{\tiny out}} |  &=&  \lim_{w \rightarrow 0}
\langle 0 | \tilde{A}(w,\bar{w})
=   \lim_{z\rightarrow0} \langle 0 |
A\left(\frac{1}{z},\frac{1}{\bar{z}}\right)\bar{z}^{-2h}z^{-2\bar{h}}\nn\\
 &=&  \lim_{z\rightarrow 0} \langle 0 | \left[ A(z,\bar{z})
\right]^{\dagger}
=  | A_{\mbox{\tiny in}} \rangle^{\dagger}.
\eea
The fact that the stress-tensor is a Hermitian operator can be
expressed using (\ref{hermconj}) in the following way :
\be
T^\dagger (z) = \sum_{m} \frac{L^{\dagger}_{m}}{\bar{z}^{m+2}} \equiv
\sum_{m} \frac{L_{m}}{\bar{z}^{-m-2}} \frac{1}{\bar{z}^4},
\ee
or in terms of the oscillator modes :
\be
L_{m}^{\dagger} = L_{-m},  \label{hermosc}
\ee
and analogously $\bar{L}_{m}^{\dagger} = \bar{L}_{-m}$.

These conditions can also be derived from the hermiticity of $T$ in
Minkowski space.

Conditions on the vacuum follow from the regularity of
\be
T(z) | 0 \rangle  = \sum_{m \in \Z} L_m z^{-m-2} | 0 \rangle
\ee
at $z=0$. Only positive powers of $z$ are allowed, so we must demand
\be
L_m | 0 \rangle = 0, \quad m \geq -1. \label{vc}
\ee
The same condition for $\lim_{ w \rightarrow 0}\langle 0 |
\tilde{T}(w)$ gives
\be
\langle 0 | L_m = 0, \quad m \leq 1.
\ee
Equation  (\ref{vc}) states that the in-vacuum is SL(2,$\C)$-invariant,
along with extra conditions for $m > 1$. The rest of the
Virasoro operators create non-trivial states out of the vacuum.
The only operators that annihilate both $\langle 0
|$ and $| 0 \rangle$ are generated by $L_{\pm 1,0}$ and
$\bar{L}_{\pm 1,0}$ and constitute the SL(2,$\C)$
subgroup of the conformal group.

If we consider holomorphic fields with mode expansion (\ref{mod}),
conformal invariance and the SL(2,$\C)$ invariance of the vacuum
imply
\be
\Phi_{n>-h}|0\rangle=0\,.
\label{mod1}\ee

\subsection{Representations of the conformal algebra}
\setcounter{equation}{0}

In CFT, the spectrum decomposes into representations of the
generic symmetry algebra, namely two copies of the  Virasoro algebra.
We will describe here only the left algebra with operators $L_m$
to avoid repetition.

The Cartan subalgebra of the Virasoro algebra is generated by $L_0$.
The positive modes are raising operators and the negative
ones are lowering operators.
Highest-weight (HW) representations are constructed by starting from
a state that is annihilated by all raising operators.
The representation is then generated by acting on the HW state by the
lowering operators.

Suppose $\Phi$ is a primary field (operator) of left weight $h$. From
the operator product expansion with the stress-tensor (\ref{36}), we
find
\be
[L_n,\Phi(w)]= \oint \frac{dz}{2\pi i} z^{n+1} T(z) \Phi(w)
= h(n+1) w^n \Phi(w) + w^{n+1} \pd \Phi(w).
\label{37}\ee
The state associated with this operator is
\be
|h\rangle \equiv \Phi(0)|0\rangle. \label{defhw}
\ee
First of all, $[L_n,\Phi(0)] = 0$, $n > 0$, so
\be
L_{m>0} | h \rangle  =  L_{m>0} \Phi(0) | 0 \rangle
= [L_m,\Phi(0)] | 0 \rangle + \Phi(0) L_{m>0} | 0 \rangle
=  0\,.
\ee
Thus, primary fields are in one-to-one correspondence with HW states.
Each primary field then generates a representation of the Virasoro
algebra.
Also, $L_0 | h \rangle = h | h \rangle$. More generally, in-states $|
h , \bar{h} \rangle$, defined by (\ref{defhw}) with $\Phi$ of
conformal dimension $(h,\bar{h})$, also satisfy $\bar{L}_0 |h,\bar{h}
\rangle = \bar{h} | h,\bar{h} \rangle$ and $\bar{L}_{n>
0} |h,\bar{h} \rangle = 0$.

The rest of the states in the representation generated by $|h\rangle
$
are of the form
\be
|\chi\rangle=L_{-n_{1}} L_{-n_{2}} \ldots L_{-n_{k}} | h \rangle,
\label{s3}\ee
where all $n_{i}>0$, and are called descendants. They are $L_0$
eigenstates with eigenvalues $h + \sum_{k} n_{k}$.
This type of representation is called a Verma module.

We have seen that we can have a one-to-one correspondence with HW
states $|h\rangle$ and primary fields $\Phi_h(z)$ given by
(\ref{defhw}).
A similar statement can be made for descendants.
Consider the state $L_{-1}|h\rangle$. It is not difficult to show
that
the operator that creates this state out of the vacuum is
\be
(L_{-1}\Phi)(z)\equiv \oint_{C_z}{dw\over 2\pi i}~T(w)\Phi_h(z)\;,
\label{s1}\ee
using (\ref{s2}).
For the general state (\ref{s3}) we have to use nested contours
\be
\Phi_{\chi}(z)=\prod_{i=1}^k\oint~{dw_i\over 2\pi
i}~(w_i-z)^{-n_i+1}T(w_i)
\Phi_h(z)\;.
\label{s4}\ee
Thus, a general correlation function of descendant operators can be
written in terms of multiple contour integrals of a correlation
function of the associated primary fields and several insertions of
the stress-tensor.
However, in a previous section we have seen that conformal Ward
identities
express such a correlation function in terms of the one with primary
fields
only.
Thus, knowledge of the correlators of primary fields determines all
correlators
of the CFT.

We will also discuss quasiprimary fields.
We have seen that on the sphere $L_{-1}$ is the translation operator
\be
[L_{-1},O(z,\bar z)]=\pd_z~O(z,\bar z)\;.
\ee
The quasiprimary states are the HW states of the global conformal
group.
Consider the part generated by $L_{\pm 1},L_0$.
The raising operator is $L_1$, while $L_{-1}$ is the lowering
operator.
The HW states are annihilated by $L_1$.
The rest of the representation is generated by acting several times
with $L_{-1}$.
Thus, the descendant (non-quasiprimary) states are derivatives of
quasiprimary ones.

An interesting function of a conformal representation generated by a
primary
of dimension $h$ is the character
\be
\chi_{h}(q)\equiv {\rm Tr}[q^{L_0-{c\over 24}}]\,,
\label{44}\ee
where the trace is taken over the whole representation.
There is an  extra shift of $L_0$ in (\ref{44}) proportional to the
central charge. The reason is that
characters will appear when discussing the partition function on the
torus which can be thought of as the cylinder with the two end-points
identified (with a twist).
Going from the sphere to the cylinder there is precisely this shift
of $L_0$
and is due to the fact that the stress-tensor is not a primary field
but transforms as in (\ref{93}).

\vskip .4cm
\noindent\hrulefill
\nopagebreak\vskip .2cm
{\large\bf Exercise}: For a generic representation without null
vectors, calculate  the character and show that it is given by
\be
\chi_h(q)={q^{h-c/24}\over \prod_{n=1}^{\infty}(1-q^n)}\,.
\label{45}\ee
{}From this expression we can read off the multiplicities of states
at any given level.
\nopagebreak\vskip .2cm
\noindent\hrulefill
\vskip .4cm

There is a special representation, which is called the vacuum
representation. If one starts with the unit operator, the state
associated with it via (\ref{defhw}) is the vacuum state.
The rest of the representation is generated by the negative Virasoro
modes.
Note, however, that from (\ref{37}) $L_{-1}$ acts as a $z$ derivative.
However the $z$ derivative of the unit operator is zero.
This is equivalent to the statement that $L_{-1}$ annihilates the
vacuum state.
For $c\geq 1$ the vacuum character is given by
\be
\chi_0(q)={q^{-c/24}\over \prod_{n=2}^{\infty}(1-q^n)}\,.
\label{46}\ee
The term with $n=1$ is missing here since $L_{-1}$ does not
generate any states out of the vacuum.

In a positive (unitary) theory the norms of states have to be
positive.
The norm of the state $L_{-n} | 0
\rangle$, $n>0$, is
\bea
\parallel L_{-n} | 0 \rangle \parallel^2 &=&  \langle 0 |
L_{-n}^{\dagger} L_{-n} | 0 \rangle
 =  \langle 0 | \left[ \frac{c}{12}(n^3 - n) + 2n L_0 \right] | 0
\rangle \nn\\&=&\frac{c}{12} (n^3 - n),
\eea
where we have used the commutation relations of the Virasoro algebra
and the SL(2,$\C)$-invariance of the vacuum. Unitarity demands this
to be positive. For large enough $n$, this means $c \geq 0$ (if
$c=0$, the Hilbert space is one-dimensional and spanned by $|0
\rangle$). A more detailed investigation shows that for $c\geq 1$ we
cannot obtain direct constraints from unitarity.
However, when $0<c<1$, unitarity implies that $c$ must
be of the form
\be
c = 1 - \frac{6}{m(m+1)}.
\label{010}\ee
An example for $m=3$ is the Ising model, for $m=4$ the tricritical
Ising model, and for $m=5$ the 3-state Potts model;
$m=2$ is the trivial theory with $c=0$.

Generically the Verma modules described above correspond to
irreducible representations of the Virasoro algebra.
However, in special cases, it may happen that the Verma module
contains
``null" states (states of zero norm that are orthogonal to any other
state).
Then, the irreducible representation is obtained by factoring out null
states.
Such representations are called degenerate.
We will give here an example of a null state.
Consider the Ising model, $m=3$ above, with $c=1/2$. This is
essentially the conformal field theory of a Majorana fermion that we
discussed earlier.
Consider the primary state with $h=1/2$ corresponding to the fermion
$|1/2\rangle$
and the following descendant state
\be
|\chi\rangle =\left(L_{-2}-{3\over 4}L_{-1}^2\right)|1/2\rangle\,.
\label{nnul}\ee

\vskip .4cm
\noindent\hrulefill
\nopagebreak\vskip .2cm
{\large\bf Exercise}: Show that $|\chi\rangle$ in (\ref{nnul}),
although being
a descendant, is also primary and that its norm is zero.
\nopagebreak\vskip .2cm
\noindent\hrulefill
\vskip .4cm

\subsection{Affine algebras\label{affine}}
\setcounter{equation}{0}

So far we have seen that in any CFT there is a holomorphic
stress-tensor
of weight (2,0).
However a CFT can also have chiral symmetries whose conserved
currents
have
weight (1,0).
Chiral conservation implies $\bar \pd J=0$. Thus, such currents are
holomorphic.
Consider the whole  set of such holomorphic currents, $J^a (z)$,
present in the theory.
We can write the most general OPE of these currents compatible
with chiral conservation and conformal invariance:
\be
J^a (z) J^b (w) = \frac{G^{ab}}{(z-w)^2} + \frac{if^{ab}{}_c J^c
(w)}{z-w} + \mbox{finite},   \label{KM}
\ee
where $f^{ab}{}_c$ is antisymmetric in the upper indices and $G^{ab}$
is symmetric. Using
associativity of the operator products, it can be shown that the
$f^{ab}{}_c$ also satisfy a Jacobi identity and $f^{abc}={f^{ab}}_d
G^{dc}$
is totally antisymmetric. Therefore they must be
the structure constants of a Lie group with invariant metric
$G^{ab}$.

Expanding $J^a(z)=\sum_n J^a_n z^{-n-1}$ we can translate (\ref{KM})
into commutation relations for the modes of the currents
\be
[J^{a}_{m},J^{b}_{n}] = m~G^{ab}~\delta_{m+n,0} +
if^{ab}{}_cJ^{c}_{m+n}.
\label{38}\ee
This algebra is an infinite-dimensional generalization of Lie
algebras and
is known as an affine algebra.
Clearly, the subalgebra of the zero modes $J^{a}_{0}$ constitutes a
Lie algebra with structure constants $f^{ab}{}_c$.

\vskip .4cm
\noindent\hrulefill
\nopagebreak\vskip .2cm
{\large\bf Exercise}: Show that a conformal field of weight (1,0) is
necessarily primary in a positive theory.
\nopagebreak\vskip .2cm
\noindent\hrulefill
\vskip .4cm

Thus, the OPE with the stress-tensor should be
\be
T(z)J^{a} (w) = \frac{J^{a} (w)}{(z-w)^2} + \frac{\pd J^{a}
(w)}{z-w} \label{TJ}
\ee
and $\bar T(\bar z)J^a(w)=$ regular.

This type of algebra is realized as we will see in many CFTs.
The prototype is the non-linear $\s$-model with a Wess-Zumino term
\cite{W}.
This is a theory in two dimensions, where the basic field $g(x)$ is
in a matrix representation of a group $G$.
The action is
\be
S = \frac{1}{4 \lambda^2} \int_{M_2} d^2 \xi ~\mbox{Tr} ( \pd_\mu g
\pd^\mu
g^{-1} ) + \frac{ik}{8\pi} \int_{B;\pd B = M_2} d^3 \xi ~\mbox{Tr}
(\epsilon_{\al \beta \gamma} U^\al U^\beta U^\gamma ),
\label{wzw}\ee
where $U_\mu = g^{-1} \pd_\mu g$.
The second term in the action is an integral over a
three-dimensional manifold B whose boundary is the two-dimensional
 space $M_2$ we
define the theory on.
This is the WZ term and it has the special property that its
variation
gives two-dimensional instead of three-dimensional equations of motion.
There is a consistency condition that has to be imposed, however.
Consider another three-manifold with the same boundary.
We would like the theory to be the same. This gives a quantization
condition
on the coupling $k$\footnote{There is another way to show the
quantization
of $k$. If we demand positivity of the quantum theory then we obtain
the same
quantization condition.}.
The above theory has two different couplings, $\lambda$ and $k$.
It can be shown that when $\lambda^2 = 4\pi/k$ then the theory is
conformally invariant (this is called the WZW model).
In this case it can be verified that the matrix currents
$J = g^{-1}\pd g$
and $\bar{J}=\bar{\pd} g g^{-1}$ are chirally conserved
$\bar{\pd} J = \pd \bar{J} =0 $. This
is a reflection of the symmetry of the action (\ref{wzw}) under $g\to
h_1~g~h_2$,
where $h_{1,2}$ are arbitrary $G$ elements.
Thus, the currents $J$ generate a $G_L$ affine algebra while the
currents
$\bar J$ generate a $G_R$ current algebra.

An interesting phenomenon in this theory (which turns out to be
generic) is that the stress-tensor can be written as a bilinear in
terms of the currents.
This is known as the affine-Sugawara construction.
Consider the group $G$ to be simple. Then by a change of basis
in (\ref{KM}) we can set $G^{ab}=k\delta^{ab}$. Choose a
basis
where the long roots have square equal to 2. Then the (2,0)
operator
\be
T_G(z) = {1\over 2(k+\tilde h)} :J^{a} (z)  J^{a} (z):
\ee
satisfies the Virasoro algebra with central charge
\be
c_G = \frac{k D_G}{k + \tilde{h}}\;.
\label{ccc}\ee
$\tilde h$ is the dual Coxeter number of the group $G$.
In the case of SU(N), we have $\tilde{h} =$N; for SO(N),
$\tilde{h} = $N-2 etc.
With this normalization, $k$ should be a positive integer
in order to have a positive theory.
It is called the level of the affine algebra.

In this type of theories, the affine symmetry is ``larger" than the
Virasoro
symmetry since we can construct the Virasoro operators out of the
current operators.
In particular the spectrum will form representations of the affine
algebra.
To describe such representations we will use a procedure similar to
the case of a Virasoro algebra.
The representation is generated by a set of states $|R_i\rangle$ that
transform
in the representation $R$ of the zero-mode subalgebra and are
annihilated by the positive modes of the currents
\be
J^a_{m>0}|R_i\rangle =0\;\;\;,\;\;\;J_0^a|R_i\rangle
=i(T^a_R)_{ij}|R_j\rangle\,.
\label{40}\ee
The rest of the affine representation is generated from the states
$|R_i\rangle$ by the action of the negative modes of the currents.
The states $|R_i\rangle$ are generated as usual, out of the vacuum, by
local operators $R_i(z,\bar z)$.
Then conditions (\ref{40}) translate into the following OPE
\be
J^a(z)R_i(w,\bar w)=i~{(T^a_R)_{ij}\over (z-w)}R_j(w,\bar w)+\dots\,.
\label{41}\ee
This is the definition of {\it affine primary} fields that play the
same role
as the primary fields in the case of the conformal algebra \cite{KZ}.

The conformal weight of affine primaries can be calculated from the
affine-Sugawara form of the stress-tensor and is given by
\be
h_R={C_R\over k+\tilde h}\,,
\label{42}\ee
where $C_R$ is the quadratic Casimir for the representation $R$.
For example the spin $j$ representation of SU(2) has
$h_j=j(j+1)/(k+2)$.

We have seen so far that the irreducible representations of the
affine
algebra $\hat g$ are in one-to-one correspondence with those of the
finite Lie algebra $g$.
This is not the end of the story, however.
It turns out that not all representations of the finite algebra can
appear,
but only the ``integrable" ones.
In the case of SU(2) this implies $j\leq k/2$.
For $\rm SU(N)_k$ the integrable representations are those with at most
k columns in their Young tableau.

The non-integrable representations are not unitary, and they can be
shown to decouple from the correlation functions.

\vskip .4cm
\noindent\hrulefill
\nopagebreak\vskip .2cm
{\large\bf Exercise}: The Coset Construction:
Consider the affine-Sugawara stress-tensor $T_G$ associated to the
group
G. Pick a subgroup $\rm H\subset G$ with regular embedding
and consider its associated affine-Sugawara
stress-tensor $T_{\rm H}$ constructed out of the H currents,
which are a subset of the G currents.
Consider also $T_{\rm G/H}=T_{\rm G}-T_{\rm H}$.
Show that
\be
T_{\rm G/H}(z)J^{\rm H}(w)={\rm regular}\;\;\;,\;\;\;T_{\rm
G/H}(z)T_{\rm H}(w)=
{\rm regular}\,.
\label{09}\ee
Show also that $T_{\rm G/H}$ satisfies the Virasoro algebra with
central charge $c_{\rm G/H}=c_{\rm G}-c_{\rm H}$.
The interpretation of the above construction is that, roughly
speaking, the  G-WZW theory can be decomposed into the
H-theory and the G/H theory described by the
stress-tensor $T_{\rm G/H}$.
As an application, show that if you choose
$\rm G=SU(2)_m\times SU(2)_1$ and H to be the diagonal
subgroup $\rm SU(2)_{m+1}$ then the G/H theory is that of the
minimal models with central charge (\ref{010}).
For a generalization of this construction, see \cite{K}.
\nopagebreak\vskip .2cm
\noindent\hrulefill
\vskip .4cm

The interested reader can find more details on affine algebras and
related theories in
\cite{go,Rep}.

\subsection{Free fermions and O(N) affine symmetry\label{Nfermions}}
\setcounter{equation}{0}

Free fermions and bosons can be used to realize
particular representations of current algebras. It will be useful for
our later purposes to consider the
CFT of  $N$ free Majorana-Weyl fermions $\psi^i$:
\be
S = -{1\over 8\pi}\int d^2 z ~\psi^i \bar{\pd} \psi^i\,. \label{S}
\ee
Clearly, this model exhibits a global O(N) symmetry, $\psi^i\to
\Omega_{ij}\psi_j$, $\Omega^T\Omega=1$, which leads
to the chirally conserved
Hermitian ($J^{ij\dagger}_m=J^{ij}_{-m}$) currents
\be
J^{ij} (z) = i~:\psi^i (z) \psi^j (z):\,\;\;,\;\;\;i<j\,.
\label{47}\ee
Using the OPE
\be
\psi^i (z) \psi^j (w) =\frac{\delta^{ij}}{z-w} \label{psipsi}
\ee
and Wick's theorem, we can calculate
\be
J^{ij}(z)J^{kl}(w)={G^{ij,kl}\over
(z-w)^2}+i~{f^{ij,kl}}_{mn}{J^{mn}(w)\over (z-w)}+\dots \,,
\label{70}\ee
where $G^{ij,kl}=(\delta^{ik}\delta^{jl}-\delta^{il}\delta^{jk})$
is the invariant O(N) metric and
\be
2~{f^{ij,kl}}_{mn}=(\delta^{ik}\delta^{ln}-
\delta^{il}\delta^{kn})\delta^{jm}
+(\delta^{jl}\delta^{kn}-\delta^{jk}\delta^{ln})\delta^{im}-
(m\leftrightarrow n)
\label{71}\ee
are the structure constants of O(N) in a basis where the long roots
have square equal to 2.
Thus, $N$ free fermions realize the O(N) current algebra at level
$k=1$.

We can construct the
affine-Sugawara stress-tensor
\be
T(z) = \frac{1}{2(N-1)}\sum_{i<j}^N:J^{ij}(z) J^{ij}(z):\,.
\label{48}\ee
As discussed previously, $T(z)$ will satisfy an operator product
expansion
\be
T(z)T(w) = \frac{c_G / 2}{(z-w)^4} + \frac{2T(w)}{(z-w)^2} +
\frac{\pd T(w)}{z-w}\,,
\ee
where
\be
c_G = \frac{k D}{k + \tilde{h}}\,.
\ee
For SO(N), one has $\tilde{h} = N-2$ and $D = \half N(N-1)$. With
$k=1$, this gives
\be
c_G =\frac{N(N-1)/2}{1+N-2} = \frac{N}{2},
\ee
i.e. each fermion contributes $\half$ to the central charge.
This is expected since the central charge of the tensor product of
two theories is the sum of the two central charges.
Moreover if we use the explicit form of the currents in terms of the
fermions we can directly evaluate the normal-ordered product in
(\ref{48})
with the result
\be
T(z)=-{1\over 2}\sum_{i=1}^N ~:\psi^i\pd \psi^i:\,,
\ee
which is the stress-tensor we would compute directly from
the free fermion action.

Since $N$ free fermions realize the $\rm O(N)_1$ affine symmetry, we
should be able
to classify the spectrum into irreducible representations
of the $\rm O(N)_1$ current algebra.
{}From the representation theory of current algebra we learn that at
level
1 there exist the following integrable (unitary) representations.
The unit (vacuum) representation constructed by acting on the vacuum
with the negative current modes, the vector $V$ representation
and the spinor representation. If N is odd there is a single spinor
representation of dimension $2^{(N-1)/2}$.
When N is even, there are two inequivalent spinor representations of
dimension
$2^{N/2-1}$: the spinor $S$ and the conjugate spinor $C$.
{}From now on we will assume N to be even because this is the case of
interest in what follows.
Applying (\ref{42}) to our case we find that the conformal weight of
the vector
is
\be
h_V={(N-1)/2\over 1+N-2}={1\over 2}\,.
\label{53}\ee
The candidate affine primary fields for the vectors are the
fermions themselves, whose conformal weight is $1/2$ and
which transform as a vector under the global O(N) symmetry.
This can be verified by computing
\be
J^{ij}(z)\psi^{k}(w)=i~{T^{ij}_{kl}\over z-w}\psi^l(w)+\dots\,,
\label{54}\ee
where
$T^{ij}_{kl}=(\delta^{il}\delta^{jk}-\delta^{ik}\delta^{jl})$
are the representation matrices of the vector.
Comparing (\ref{54}) with (\ref{41}) we indeed see that
$\psi^i$ are the affine primaries of the vector representation.

The conformal weights of the spinor and conjugate spinor are equal
and, from (\ref{42}), we obtain $h_S=h_C=N/16$.
Operators with such a conformal weight do not exist in the free
fermion theory in the way it has been described so far.

Things get better if we notice that the action (\ref{S}) has
a $\Z_2$ symmetry
\be
\psi^i \rightarrow - \psi^i.
\ee
Because of this symmetry, we can choose two different boundary
conditions on the cylinder:
\begin{itemize}
\item{Neveu-Schwarz :} $\psi^i(\s+2\pi) = -\psi^i(\s)$,
\item{Ramond :} $\psi^i(\s+2\pi) = \psi^i(\s)$.
\end{itemize}
We will impose the same boundary condition on all fermions, otherwise
we will break the O(N) symmetry.

The mode expansion of a periodic holomorphic field on the cylinder
is
\be
\psi(\t+i\s) = \sum_{n} \psi_{n} e^{-n(\t+i\s)},
\label{50}\ee
where $n$ is an integer. Thus, in the Ramond ($R$) sector $\psi$ is
integer modded.
In the Neveu-Schwarz ($NS$) sector, $\psi$ is antiperiodic so its
Fourier
expansion is like (\ref{50}) but now $n$ is half-integer.
As discussed before, when we go from the cylinder to the sphere,
$z=e^{\t+i\s}$ the mode expansion becomes
\be
\psi^i(z)=\sum_{n}\psi^i_n z^{-n-h}=\sum_{n}\psi^i_n z^{-n-1/2}\,.
\label{73}\ee
We observe that in the $NS$ sector (half-integer n) the field
$\psi^i(z)$ is single-valued (invariant under $z\to z e^{2\pi i}$)
while in the $R$ sector it has a $Z_2$ branch cut.
To summarize
\begin{itemize}
\item{$n \in \Z$ (Ramond),}
\item{$n \in \Z + \half$ (Neveu-Schwarz).}
\end{itemize}

The OPE (\ref{psipsi}) implies the following
anticommutation
relations for the fermionic modes
\be
\{\psi^i_m,\psi^j_n\}=\delta^{ij}\delta_{m+n,0}\,,
\label{51}\ee
in both the $NS$ and the $R$ sector.

We will first look at the $NS$ sector.
Here the fermionic oscillators are half-integrally modded
and (\ref{51}) shows that $\psi^i_{-n-\frac{1}{2}}$, $n \leq 0$, are
creation
operators, while $\psi^i_{n+\frac{1}{2}}$ are annihilation operators.
Consequently, the vacuum satisfies
\be
\psi^i_{n>0}|0\rangle =0
\label{52}\ee
and the full spectrum is generated by acting on the vacuum with
the negative modded oscillators.
We would like to decompose the spectrum into affine representations.
We have argued above that we expect to obtain here the vacuum
and the vector representation.
The primary states of the vector
are
\be
|i\rangle =\psi^i_{-{1\over 2}}|0\rangle
\label{55}\ee
and the rest of the representation is constructed from the above states
by acting with the negative current modes.

At this point it is useful to introduce the fermion number operator
$F$ and
$(-1)^F$, which essentially counts the number of fermionic modes
modulo 2.
The precise way to say this is
\be
\{(-1)^F,\psi^i_n\}=0
\label{56}\ee
and that the vacuum has eigenvalue 1: $(-1)^F|0\rangle =|0\rangle$.
Using (\ref{56}) we can calculate that the vector primary states
(\ref{55})
have $(-1)^F=-1$.
Since the currents contain an even number of fermion modes we can
state the following:
\begin{itemize}
\item{All states of the vacuum (unit) representation have $(-1)^F=1$}.
The first non-trivial states correspond to the currents themselves:
\be
J^{ij}_{-1}|0\rangle=i\psi^i_{-{1\over
2}}\psi^j_{-{1\over 2}}|0\rangle\,.
\label{57}\ee

\item All states of the vector representation have $(-1)^F=-1$.
The first non-trivial states below the primaries are
\be
J^{ij}_{-1}|k\rangle=i\left[\delta^{jk}\psi^i_{-{3\over
2}}-\delta^{ik}\psi^j_{-{3\over 2}}+\psi^i_{-{1\over 2}}
\psi^j_{-{1\over 2}}\psi^k_{-{1\over 2}}\right]|0\rangle\,.
\label{58}\ee
\end{itemize}

We will now calculate the characters (multiplicities) in the $NS$
sector.
We will first calculate the trace of $q^{L_0-c/24}$ in the full $NS$
sector.
This is not difficult to do since every negative modded fermionic
oscillator
$\psi^i_{-n-{1\over 2}}$ contributes $1+q^{n+1/2}$. The first term
corresponds to the oscillator being absent, while the second
corresponds to it being present.
Since the oscillators are fermionic, their square is zero and
therefore no more terms can appear.
Putting everything together, we obtain
\be
{\rm Tr}_{NS}[q^{L_0-c/24}]=q^{-{N\over
48}}\prod_{n=1}^{\infty}(1+q^{n-{1\over 2}})^N\,.
\label{59}\ee
Using (\ref{t8}) and (\ref{t10}) from Appendix A, we can write this as
\be
{\rm Tr}_{NS}[q^{L_0-c/24}]=\left[{\th_3\over \eta}\right]^{N/2}\,,
\label{60}\ee
where $\th_i=\th_i(0|\tau)$.
In order to separate the contributions of the unit and vector
representations, we also need to calculate the same trace but with
$(-1)^F$
inserted. Then $\psi^i_{-n-{1\over 2}}$ contributes $1-q^{n+1/2}$
and
\be
{\rm Tr}_{NS}[(-1)^F~q^{L_0-c/24}]=q^{-{N\over
48}}\prod_{n=1}^{\infty}(1-q^{n-{1\over 2}})^N
=\left[{\th_4\over \eta}\right]^{N/2}\,.
\label{61}\ee
Now we can project onto the vector or the unit representation:
\be
\chi_{0}={\rm Tr}_{NS}\left[{(1+(-1)^F)\over
2}~q^{L_0-c/24}\right]={1\over
2}
\left(\left[{\th_3\over \eta}\right]^{N/2}+
\left[{\th_4\over \eta}\right]^{N/2}\right)\,,
\label{62}\ee
\be
\chi_{V}={\rm Tr}_{NS}\left[{(1-(-1)^F)\over
2}~q^{L_0-c/24}\right]={1\over
2}
\left(\left[{\th_3\over \eta}\right]^{N/2}-
\left[{\th_4\over \eta}\right]^{N/2}\right)\,.
\label{63}\ee

It turns out that, sometimes, inequivalent current algebra
representations
have the same conformal weight and same multiplicities, and therefore
the same
characters. This will happen for the spinors.
To distinguish them we will define a refined character (the $affine$
character), where we insert an arbitrary affine group element in the
trace.
By an adjoint action (that leaves the trace invariant) we can bring
this element into the Cartan torus. In this case the group element
can be written as an exponential of the Cartan generators $g=e^{2\pi
i\sum_i v_iJ^i_0}$. We will consider
\be
\chi_R(v_i)={\rm Tr}_R\left[q^{L_0-c/24}~e^{2\pi i\sum_i
v_iJ^i_0}\right]\,,
\label{64}\ee
where $i$ runs over the Cartan subalgebra and $J^i_0$ are the zero
modes of the Cartan currents.
The Cartan subalgebra of O(N) for N even is generated by $J^{12}_0$,
$J^{34}_0
,\dots$ ,$J^{N/2-1,N/2}_0$ and has dimension $N/2$.
We will  calculate the  affine characters of the unit and vector
representations.
Consider the contribution of the fermions $\psi^1$ and $\psi^2$.
By going to the basis $\psi^{\pm}=\psi^1\pm i\psi^2$ we can see that
the $J_0^{12}$ eigenvalues of $\psi^{\pm}_{n}$ are $\pm 1$.
Putting everything together and using the $\vartheta$-function product
formulae from Appendix A we obtain
\be
\chi_0(v_i)={1\over 2}\left[\prod_{i=1}^{N/2}{\th_3(v_i)\over \eta}+
\prod_{i=1}^{N/2}{\th_4(v_i)\over \eta}\right]\,,
\label{65}\ee
\be
\chi_V(v_i)={1\over 2}\left[\prod_{i=1}^{N/2}{\th_3(v_i)\over \eta}-
\prod_{i=1}^{N/2}{\th_4(v_i)\over \eta}\right]\,.
\label{66}\ee

We will  now move to the Ramond sector and construct the Hilbert space.
Here the fermions are integrally modded. For $\psi^i_n$ with
$n\not=0$
the same discussion as before applies. We separate creation and
annihilation operators, and the vacuum should be annihilated by the
annihilation operators.
However, an important difference here is the presence of
anticommuting
zero modes
\be
\{\psi^{i}_{0},\psi^{j}_{0}\} =\delta^{ij}\;.
\label{67}\ee
This situation occurred when discussing the ghost system.
Equation (\ref{67}) is the O(N) Clifford algebra and it is realized by
the Hermitian O(N)
$\gamma$-matrices. Consequently, the ``vacuum" must be a (Dirac) spinor
$\hat
S$ of O(N) with $2^{N/2}$ components.
We label the $R$ vacuum by
$|\hat S_{\alpha}\rangle$
and we have
\be
\psi^i_{m>0}| \hat S_{\alpha}\rangle=0\;\;\;,\;\;\;\psi^i_0
| \hat S_{\alpha}\rangle=\gamma^i_{{\alpha}\b}| S_{\b}\rangle\,.
\label{68}\ee
Consider also
\be
\gamma^{N+1}=\prod_{i=1}^N~(\psi^i_0/\sqrt{2})\;\;\;,\;\;\;
\{\gamma^{N+1},\psi^i_0\}=0\;\;\;,\;\;\;[\gamma^{N+1}]^2=1\,.
\label{g5}\ee
This matrix plays the role of $\gamma^5$ in order to define Weyl
spinors.
Thus, we obtain the spinor $S=(1+\gamma^{N+1})/2~\hat S$ and the
conjugate spinor $C=(1-\gamma^{N+1})/2~\hat S$.
In fact, in the Ramond sector
\be
(-1)^F=\gamma^{N+1}~(-1)^{\sum_{n=1}^{\infty}\psi^i_{-n}\psi^i_n}
\label{1f}\ee
and with this definition
\be
(-1)^F|S\rangle =|S\rangle \;\;\;,\;\;\;(-1)^F|C\rangle=-|C\rangle\,.
\label{chir}\ee
By now acting with the negative modded
fermionic oscillators we construct the full
spectrum of the Ramond sector.

Does the $R$ vacuum, the way we constructed it, have the correct
conformal weight?
We can verify this as follows.
Consider the two-point function of fermions in the Ramond vacuum
\be
G^{ij}_R(z,w)=\langle \hat S|\psi^i(z)\psi^j(w)|\hat S\rangle\,.
\label{72}\ee
This can be evaluated directly using the mode expansion (\ref{73})
and
the commutation relations (\ref{51})
and (\ref{68}) to be
\be
G^{ij}_R(z,w)=\delta^{ij}\frac{z+w}{2\sqrt{z w}}{1\over z-w}\,.
\label{74}\ee
Note also that for any state $|X\rangle$ in CFT corresponding to an
operator
with conformal weight $h$ we have
\be
\langle X|T(z)|X\rangle ={h\over z^2}\,.
\label{75}
\ee
Finally, remember the definition of the stress-tensor
\be
T(w)=\lim_{z\to w}\left[-{1\over 2}\sum_{i=1}^N\psi^i(z)\pd_w
\psi^i(w)
+{N\over 2(z-w)^2}\right]\,,
\label{76}\ee
where we subtract the singular part of the OPE.
Putting all these ingredients together we can calculate
\be
\langle \hat S|T(z)|\hat S\rangle ={N\over 16~z^2}\,,
\ee
which gives the correct conformal weight for the spinor.
We will now compute the multiplicities in the Ramond sector.
We will first evaluate the direct trace. Every fermionic oscillator
$\psi^i_{-n}$ with $n>0$ will give a contribution $1+q^n$.
There will also be the multiplicity $2^{N/2}$ from the S and C
ground-states.
Thus,
\be
{\rm Tr}_R[q^{L_0-c/24}]=2^{N/2}~q^{{N\over 16}-{N\over
48}}\prod_{n=1}^{\infty}(1+q^n)^N=\left[{\th_2\over
\eta}\right]^{N/2}\,.
\label{77}\ee
If we consider the trace with $(-1)^F$ inserted, we will obtain 0
since, for any state, there is another one of opposite $(-1)^F$
eigenvalue
related by the zero modes.
The fact that ${\rm Tr}[(-1)^F]=0$ translates into the statement that
the $R$
spectrum is non-chiral (both C and S appear).
So
\be
\chi_S=\chi_C={1\over 2}\left[{\th_2\over \eta}\right]^{N/2}\,.
\label{79}\ee
The affine character does distinguish between the C  and S
representations:
\be
\chi_S(v_i)={1\over 2}\left[\prod_{i=1}^{N/2}{\th_2(v_i)\over \eta}+
\prod_{i=1}^{N/2}{\th_1(v_i)\over \eta}\right]\,,
\label{80}\ee
\be
\chi_C(v_i)={1\over 2}\left[\prod_{i=1}^{N/2}{\th_2(v_i)\over \eta}-
\prod_{i=1}^{N/2}{\th_1(v_i)\over \eta}\right]\,.
\label{81}\ee
For $v_i=0$ they reduce to (\ref{79}).

Finally, the $R$ vacua corresponding to the C and S representations
are created out of the $NS$ vacuum $|0\rangle$ by
affine primary fields $\hat S_{\alpha}(z)$:
\be
|\hat S_{\alpha}\rangle=\lim_{z\to 0}\hat S_{\alpha}(z)|0\rangle\,.
\label{82}\ee
We will raise and lower spinor indices with the O(N)
antisymmetric charge conjugation matrix $C^{\a\b}$.
We have then the following OPEs:
\be
\psi^i(z)\hat S_{\a}(w)=\g^i_{\a\b}{\hat S_{\b}(w)\over
\sqrt{z-w}}+\dots
\label{8880}\ee
\be
J^{ij}(z)\hat S_{\alpha}(w)={i\over 2}[\g^i,\g^j]_{\alpha\beta}
{\hat S_{\beta}(w)\over(z-w)}+\dots\label{85}\ee
\be
\hat S_{\alpha}(z)\hat S_{\beta}(w)={\delta_{\alpha\beta}\over
(z-w)^{N/8}}+\g^{i}_{\a\b}{\psi^i(w)\over (z-w)^{N/8-1/2}}
+{i\over 2}[\gamma^i,\gamma^j]_{\alpha\beta}{J^{ij}(w)\over
(z-w)^{N/8-1}}+\dots
\label{86}\ee

\subsection{N=1 superconformal symmetry\label{superconformal}}
\setcounter{equation}{0}

We have seen that the conformal symmetry of a CFT is encoded in
the
OPE of the stress-tensor $T$ which is a chiral (2,0) operator.
Other chiral operators encountered which generate symmetries
include chiral fermions (1/2,0) and currents (1,0).
Here we will study symmetries whose conserved chiral currents have
spin
$3/2$. They are associated with fermionic symmetries known as
supersymmetries.

Consider the theory of a free scalar and Majorana fermion with action
\be
S={1\over 2\pi}\int d^2 z \pd X\bar\pd X+{1\over 2\pi}\int d^2
z(\psi\bar\pd \psi+\bar\psi\pd \bar\psi)\,.
\label{198}\ee
The action is invariant under a left-moving supersymmetry
\be
\delta X=\epsilon(z)\psi\;\;\;,\;\;\;\delta\psi=-\epsilon(z)\pd
X\;\;\;,\;\;\;
\delta\bar\psi=0
\label{199}\ee
and a right-moving  one
\be
\delta X=\bar\epsilon(\bar
z)\bar\psi\;\;\;,\;\;\;\delta\bar\psi=-\bar\epsilon(\bar z)
\bar\pd X\;\;\;,\;\;\;
\delta\psi=0\;,
\label{200}\ee
where are $\epsilon$ and $\bar\epsilon$ are anticommuting.

The associated conservation laws can be written as $\pd\bar G=\bar\pd
G=0$ and the conserved chiral currents are
\be
G(z)=i\psi\pd X\;\;\;,\;\;\;\bar G(\bar z)=i\bar\psi\bar\pd X\,.
\label{201}\ee
We can easily obtain the OPE
\bea
G(z)G(w)={1\over (z-w)^3}+2{T(w)\over z-w}+\dots\;,\nn\\
T(z)G(w)={3\over 2}{G(w)\over (z-w)^2}+{\pd G(w)\over z-w}+\dots\,,
\label{202}\eea
where $T(z)$ is the total stress-tensor of the theory
satisfying (\ref{TT}) with $c=3/2$,
\be
T(z) = - \half :\pd X \pd X: - \half :\psi \pd \psi: \,.
\ee
(\ref{202}) implies that $G(z)$ is a primary field of dimension 3/2.
The algebra generated by $T$ and $G$ is known as the N=1
superconformal
algebra since it encodes the presence of conformal invariance and one
supersymmetry.
The most general such algebra can be written down using conformal
invariance and associativity. Define $\hat c=2c/3$. Then, the algebra,
apart from (\ref{TT}), contains the following OPEs
\bea
G(z)G(w)={\hat c\over (z-w)^3}+2{T(w)\over z-w}+\dots\;,\nn\\
T(z)G(w)={3\over 2}{G(w)\over (z-w)^2}+{\pd G(w)\over z-w}+\dots \,.
\label{203}
\eea
Introducing the modes of the supercurrent $G(z)=\sum G_r/z^{r+3/2}$
we obtain the following (anti)commutation relations
\bea
 \{G_r,G_s\}&=&{\hat c\over 2}\left(r^2-{1\over 4}\right)
\delta_{r+s,0}+2L_{r+s} \,,\nn\\
{[} L_{m} , G_{r} {]} &=& \left({m\over 2}-r\right)G_{m+r}\,,
\label{204}
\eea
along with the Virasoro algebra.

This algebra has the symmetry (external automorphism) $G\to -G$ and
$T\to T$.
Consequently, $NS$ or $R$ boundary conditions are possible for the
supercurrent.
In the explicit realization (\ref{201}), they correspond to the
respective boundary conditions for the fermion.

In the $NS$ sector, the supercurrent modes are half-integral
and $G_r|0\rangle=0$ for $r>0$.
Primary states are annihilated by the positive modes of $G$ and $T$
and the superconformal representation is generated by the action of
the
negative modes of $G$ and $T$.
The generic character is
\be
\chi^{NS}_{N=1}={\rm Tr}[q^{L_0-c/24}]=q^{h-c/24}\prod_{n=1}^{\infty}{
1+q^{n-{1\over 2}}\over 1-q^n}\,.
\label{205}\ee

In the Ramond sector, $G$ is integrally modded and has in particular a
zero
mode, $G_0$, which satisfies according to (\ref{204})
\be
\{G_0,G_0\}=2L_0-{\hat c\over 8}\,.
\label{206}\ee
Primary states are again annihilated by the positive modes.
In a unitary theory, (\ref{206}) indicates that for any state $h\geq
\hat c/16$.
When the right-hand side of (\ref{206}) is non-zero the state is
doubly degenerate and $G_0$ moves between the two degenerate states.
There is no degeneracy when $h=\hat c/16$ since from (\ref{206})
$G_0^2=0$, which implies that $G_0=0$ on such a
state.
As in the case of free fermions, we can introduce the operator
$(-1)^F$,
which anticommutes with $G$ and counts fermion number modulo 2.

The pairing of states in the $R$ sector due to $G_0$ can be stated as
follows:
the trace of $(-1)^F$ in the $R$ sector has contributions only from the
ground-states with $\Delta= \hat c/16$.
This trace is known as the $elliptic$ $genus$ of the N=1 superconformal
field theory and is the CFT generalization of the Dirac index\footnote{
For a further discussion, see \cite{lnw}.}.

The generic $R$ character is given by
\be
\chi^R_{N=1}={\rm Tr}[q^{L_0-c/24}]=q^{h-c/24}\prod_{n=1}^{\infty}{
1+q^{n}\over 1-q^n}\,.
\label{207}\ee

The N=1 superconformal theories have an elegant formulation in N=1
superspace
where, along with the coordinates $z,\bar z$, we introduce two
anticommuting
variables $\theta,\bar\theta$ and the covariant derivatives
\be
D_{\theta}={\partial\over \partial\theta}+\theta\pd_z\;\;\;,\;\;\;
\bar D_{\bar\theta}={\partial\over
\partial\bar\theta}+\bar\theta\pd_{\bar z}\,.
\label{208}\ee
The fields $X$ and $\psi,\bar\psi$ can now be described by a
function in superspace, the  scalar superfield
\be
\hat X(z,\bar
z,\theta,\bar\theta)=X+\theta\psi+\bar\theta\bar\psi+\theta\bar\theta
F\,,
\label{209}\ee
where $F$ is an auxiliary field with no dynamics.
The action (\ref{198}) becomes
\be
S={1\over 2\pi}\int d^2 z\int d\theta d\bar\theta ~D_\theta \hat
X~\bar D_{\bar\theta}\hat X \,.
\label{210}\ee

\vskip .4cm
\noindent\hrulefill
\nopagebreak\vskip .2cm
{\large\bf Exercise}. Write  the action (\ref{210}) in components by
doing the integral over the anticommuting coordinates, and show that
it is equivalent to (\ref{198}).
\nopagebreak\vskip .2cm
\noindent\hrulefill
\vskip .4cm

\subsection{N=2 superconformal symmetry\label{N=2superconformal}}
\setcounter{equation}{0}

There are further generalizations of superconformal symmetry.
The next simplest case is the N=2 superconformal algebra that contains,
apart from the stress-tensor, two supercurrents $G^{\pm}$ and a U(1)
current $J$.
Its OPEs, apart from the Virasoro one, are
\be
G^{+}(z)G^-(w)={2c\over 3}{1 \over (z-w)^3}
+\left({2J(w)\over (z-w)^2}+{\pd J(w)\over z-w}
\right)
+\frac{2}{z-w}T(w)+\dots
\label{211}\,,\ee
\be
G^{+}(z)G^{+}(w)={\rm regular}\;\;\;,\;\;\; G^{-}(z)G^{-}(w)={\rm
regular}\,,
\ee
\be
T(z)G^{\pm}(w)=\frac{3}{2}{G^{\pm}(w)\over (z-w)^2}+\frac{\pd
G^{\pm}(w)}{z-w}
+\dots\,,
\label{212}\ee
\be
J(z)G^{\pm}(w)=\pm{G^{\pm}(w)\over z-w}+\dots\,,
\label{213}\ee
\be
T(z)J(w)=\frac{J(w)}{(z-w)^2}+\frac{\pd J(w)}{z-w}+\dots\,,
\label{214}\ee
\be
J(z)J(w)={c/3 \over (z-w)^2}+\dots\,.
\label{215}\ee

The symmetry of the N=2 superconformal algebra is a continuous O(2)
symmetry,
which rotates the two supercharges, in a real basis $G^1=G^+ +G^-$,
$G^2=i(G^+-G^-)$.
The SO(2) part is an internal automorphism.
The extra $Z_2$ transformation $G^1\to G^1$, $G^2\to -G^2$ is an
external
automorphism.
We can use the symmetry to impose various boundary conditions.
Twisting with the external automorphism provides an inequivalent
algebra, the twisted N=2 algebra, where $G^{1,2}$ have opposite
boundary conditions.
More interesting for our purposes is to
use the SO(2)$\sim$U(1) symmetry in order to impose
\be
G^{\pm}(e^{2\pi i}z)=e^{\mp 2\pi i\a}G^{\pm}(z)\;,
\label{boun}\ee
while $T,J$ are single-valued. The parameter $\a$ takes values in
[0,1].
For $\a=0$ we have the $NS$ sector, where both supercharges are
half-integrally
modded.
For $\a=\pm 1/2$ we obtain the Ramond sector, where both supercharges
are
integrally modded.
Since the U(1) symmetry we used is an internal automorphism, the
algebras
obtained for the various boundary conditions, labeled by $\a$, are
isomorphic.
We can write this isomorphism (known as ``spectral flow") explicitly:
\bea
J^{\a}_{n}=J_n-\a{c\over 3}\delta_{n,0} &,& L^{\a}_{n}=L_n-\a
J_n+
\a^2{c\over 6}\delta_{n,0}\;\;, \\
\label{sf1}
G^{\a,+}_{r+\a}=G^+_{r} &,& G^{\a,-}_{r-\a}=G^-_{r}\;\;,
\label{sf2}\eea
where $n\in Z$ and $r\in Z+{1\over 2}$.
The spectral flow provides a continuous map between the $NS$ and $R$
sectors.

In the $NS$ sector HW irreducible representations are generated by a HW
state
$|h,q\rangle$, annihilated by the positive modes of $T,J,G^{\pm}$ and
characterized by the eigenvalues $h$ of $L_0$ and $q$ of $J_0$.
The SL(2,$\C$)-invariant vacuum has $h=q=0$.
The rest of the states of the representations are generated by the
action
of the negative modes of the superconformal generators.

In the $R$ sector, HW states are again annihilated by the positive
modes.
Here, however, we have also the zero modes of the supercurrents
$G^{\pm}_0$ satisfying
\be
G^{\pm}_0G^{\pm}_0=0\;\;\;,\;\;\;\{G^+_0,G^-_0\}=2\left(L_0-{c\over
24}\right)
\;.
\label{n2r}\ee
Unitarity again implies that $\Delta\geq c/24$ in the $R$ sector.
When $\Delta > c/24$ both $G^{\pm}_0$ act non-trivially and the HW
state is
a collection of four states.
When $\Delta=c/24$, then $G^{\pm}_0$ are null and the HW vector is a
singlet.
Here again we can introduce the $(-1)^F$ operator in a way analogous to
N=1.
The trace of $(-1)^F$ in the $R$ sector (elliptic genus) obtains
contributions
only from states with $\Delta=c/24$.

Using (\ref{sf1}) we deduce that
\be
J_0^{R^{\pm}}=J^{NS}_0\mp{c\over 6}\;\;\;,\;\;\;L_0^{R^{\pm}}-{c\over
24}=L_0^{NS}-{1\over 2}J^{NS}_0\;\;.
\label{sf3}\ee
Thus, the positivity condition $L_0^R-c/24\geq 0$ translates in the
$NS$ sector
to $2h-|q|\geq 0$.
The Ramond ground-states correspond to $NS$ states with $2h=|q|$ known
as chiral
states.
They are generated from the vacuum by the chiral field operators.
Because of charge conservation, their OPE at short distance is regular
and can be written as a ring, the {\em chiral ring}
\be
O_{q_1}(z)O_{q_{2}}(z)=O_{q_1+q_2}(z)\;\;.
\ee
This chiral ring contains most of the important information about the
N=2
superconformal theory.

{}From (\ref{sf3}) we can deduce that the unit operator (h=q=0) in the
$NS$
sector is mapped, under spectral flow, to an operator with
(h=c/24,q=$\pm $c/6)
in the $R$ sector.
This is the maximal charge ground-state in the $R$ sector, and applying
the spectral flow once more we learn that there must be a chiral
operator
with (h=c/6,q=$\pm $c/3) in the $NS$ sector. As we will see later on,
this operator
is very important for spacetime supersymmetry in string theory.

N=2 superconformal theories can be realized as $\sigma$-models on
manifolds with SU(N) holonomy. The six-dimensional case corresponds
to Calabi-Yau (CY) manifolds, that are Ricci-flat. The central charge
of the N=2 algebra in the CY case is c=9.
The (h=3/2,q=$\pm$3) state, mentioned above, corresponds to the unique
(3,0) form of the CY manifold.

This symmetry will be relevant for superstring ground-states with N=1
spacetime supersymmetry in four dimensions.
An extended description of the superspace geometry, representation
theory,
and dynamics of CFTs with N=2 superconformal symmetry can be found in
\cite{n2,warner,ketov}.

\subsection{N=4 superconformal symmetry\label{N=4superconformal}}
\setcounter{equation}{0}

Finally another extended superconformal algebra that is useful in
string theory
is the ``short" N=4 superconformal algebra that contains,
apart from the stress-tensor, four supercurrents and three currents
that form the current
algebra of $\rm SU(2)_k$.
The four supercurrents transform as two conjugate spinors under the
$\rm SU(2)_k$.
The Virasoro central charge $c$ is related to the level $k$ of the
SU(2)
current algebra as $c=6k$.
The algebra is defined in terms of the usual Virasoro OPE, the
statement
that $J^{a}$, $G^{\a},\bar G^{\a}$ are primary with the appropriate
conformal weight
and the following OPEs
\be
J^{a}(z)J^{b}(w)={k\over 2}{\delta^{ab}\over
(z-w)^2}+i\epsilon^{abc}{J^{c}(w)\over (z-w)}+\dots\,,
\label{400}\ee
\be
J^{a}(z)G^{\a}(w)={1\over 2}\s^{a}_{\b\a}{G^{\b}(w)\over
(z-w)}+\dots\;\;,\;\;
J^{a}(z)\bar G^{\a}(w)=-{1\over 2}\s^{a}_{\a\b}{\bar G^{\b}(w)\over
(z-w)}+\dots\,,
\label{401}\ee
\be
G^{\a}(z)\bar G^{\b}(w)={4k\delta^{\a\b}\over
(z-w)^3}+2\s^{a}_{\b\a}\left[
{2J^{a}(w)\over (z-w)^2}+{\d J^{a}(w)\over
(z-w)}\right]+2\delta^{\a\b}{T(w)\over (z-w)}+\dots\,,
\label{402}
\ee
\be
G^{\a}(z)G^{\b}(w)={\rm regular}\;\;\;,\;\;\;\bar G^{\a}(z)\bar
G^{\b}(w)=\rm regular\;\;.
\ee

As in the N=2 case there are various conditions we can impose, but we
are eventually interested in $NS$ and $R$ boundary conditions.
There is again a spectral flow, similar to the N=2 one, that
interpolates
between $NS$ and $R$ boundary conditions.

In the $NS$ sector, primary states are annihilated by the positive
modes
and are characterized by their conformal weight $h$ and $\rm SU(2)_k$
spin $j$.
As usual, for unitarity we have $j\leq k/2$.

\vskip .4cm
\noindent\hrulefill
\nopagebreak\vskip .2cm
{\large\bf Exercise}: Use the same procedure as that used in the N=2
superconformal case to show that in the $NS$ sector $h-j\geq 0$,
while in the $R$
sector, $h\geq k/4$.
\nopagebreak\vskip .2cm
\noindent\hrulefill
\vskip .4cm

The representations saturating the above bounds are  called
``massless", since they would correspond to massless states in the
appropriate string context.
In the particular case of k=1, relevant for string compactification,
the N=4 superconformal algebra can be realized in terms of a $\s$-model
on a four-dimensional Ricci-flat, K\"ahler manifold with SU(2)
holonomy.
In the compact case this is the K3 class of manifolds.
In the $NS$ sector, the two massless representations have (h,j)=(0,0)
and (1/2,1/2), while in the $R$ sector, (h,j)=(1/4,0) and (1/4,1/2).

Again the trace in the $R$ sector of $(-1)^F$ obtains contributions
from ground-states only and provides the elliptic genus of the N=4
superconformal theory.

More information on the N=4 representation theory can be found in
\cite{et}

\subsection{The CFT of ghosts\label{ghosts}}
\setcounter{equation}{0}

We have seen that in the covariant quantization of the
string
we had to introduce an anticommuting ghost system containing
the $b$ ghost with conformal weight 2 and the $c$ ghost with
conformal
weight $-1$.
Here, anticipating further applications, we will describe in general
the CFT of such ghost systems.
The field $b$ has conformal weight $h=\lambda$ while $c$
has $h=1-\lambda$.
We will also consider them to be anticommuting ($\epsilon=1$)
or commuting $\e=-1$.
They are governed by the free action
\be
S_{\l}={1\over \pi}\int d^2 z~b\bar\pd c\,,
\label{100}\ee
from which we obtain the OPEs
\be
c(z)b(w)={1\over z-w}\;\;\;,\;\;\;b(z)c(w)={\e\over z-w}\,.
\label{101}\ee
The equations of motion $\bar\pd b=\bar \pd c=0$ imply that the
fields are
holomorphic.
Their conformal weights determine their mode expansions on the sphere
and hermiticity properties
\be
c(z)=\sum_{n\in
Z}~z^{-n-(1-\l)}~c_n\;\;\;,\;\;\;c^{\dagger}_n=c_{-n}\,,
\label{102}\ee
\be
b(z)=\sum_{n\in Z}~z^{-n-\l}~b_n\;\;\;,\;\;\;b^{\dagger}_n=\e
b_{-n}\,.
\label{103}\ee
Thus, their (anti)commutation relations are
\be
c_mb_n+\e b_nc_m=\delta_{m+n,0}\;\;\;,\;\;\;c_mc_n+\e c_nc_m=
b_mb_n+\e b_nb_m=0\,.
\label{104}\ee
Here also, due to the $Z_2$ symmetry $b\to -b,c\to -c$,
we can introduce the analog of $NS$ and $R$ sectors
(corresponding to antiperiodic and periodic boundary conditions
on the cylinder):
\be
{\rm NS}\;:\;\;b_n,\;\;\;n\in \Z-\l\;\;\;,\;\;\;c_n,\;\;\;n\in \Z+\l\,,
\label{105}\ee
\be
{\rm R}\;:\;\;b_n,\;\;\;n\in {1\over
2}+\Z-\l\;\;\;,\;\;\;c_n,\;\;\;n\in
{1\over 2}
+\Z+\l \,.\label{106}\ee
The stress-tensor is fixed by the conformal properties
of the
$bc$ system to be
\be
T=-\l b\pd c+(1-\l)(\pd b)c\,.
\label{107}\ee
Under this stress-tensor, $b$ and $c$ transform as primary fields with
 conformal weight ($\l$,0) and (0,1-$\l$).

\vskip .4cm
\noindent\hrulefill
\nopagebreak\vskip .2cm
{\large\bf Exercise}. Show that $T$ satisfies the Virasoro algebra
with central charge
\be
c=-2\e (6\l^2-6\l +1)=\e(1-3 Q^2)\;\;\;,\;\;\;Q=\e (1-2\l)\,.
\label{108}\ee
\nopagebreak\vskip .2cm
\noindent\hrulefill
\vskip .4cm

There are two special cases of this system that we have encountered
so far. The first is $\l=2$ and $\e=1$, which corresponds to the
reparametrization ghosts with $c=-26$.
The second is $\l=1/2$ and $\e=1$, which corresponds to a complex
(Dirac) fermion or equivalently to two Majorana fermions with $c=1$.

There is a classical U(1) symmetry in (\ref{100}): $b\to
e^{i\theta}b$,
$c\to e^{-i\theta}c$.
The associated U(1) current is
\be
J(z)=-:b(z)c(z):=\sum_{n\in \Z} z^{-n-1}J_n\,,
\label{109}\ee
where the normal ordering is chosen with respect to the standard
SL(2,$\C)$-invariant vacuum $|0\rangle$, in which $\langle
c(z)b(w)\rangle
=1/(z-w)$.
It generates a U(1) current algebra
\be
J(z)J(w)={\e\over (z-w)^2}+\dots\,,
\label{110}\ee
under which $b,c$ are affine primary
\be
J(z)b(w)=-{b(w)\over z-w}+\dots\;\;\;,\;\;\;J(z)c(w)={c(w)\over
z-w}\dots\,.
\label{111}\ee
A direct computation of the $TJ$ OPE gives
\be
T(z)J(w)={Q\over (z-w)^3}+{J(w)\over (z-w)^2}+{\pd_wJ(w)\over
z-w}+\dots \,.
\label{112}\ee
Note the appearance of the central term in (\ref{112}), which makes it
different from (\ref{TJ}).
Translating into commutation relations, we obtain
\be
[L_m,J_n]=-nJ_{m+n}+{Q\over 2}m(m+1)\delta_{m+n,0}\,.
\label{113}\ee
The central term implies an ``anomaly" in the current algebra:
{}from (\ref{113}) we obtain
\be
[L_1,J_{-1}]=J_0+Q\;\;\;,\;\;\;Q+J_0^{\dagger}=
[L_1,J_{-1}]^{\dagger}=-J_0\,,
\label{114}\ee
so that $J_0^\dagger=-(J_0+Q)$ and the U(1) charge conservation
condition is modified to $\sum_i q_i=Q$
($Q$ is a background charge for the system).
This is a reflection of the zero-mode structure of the $bc$ system
and translates into
\be
\# \mbox{ zero modes of } c-\# \mbox{ zero modes of } b =-{\e\over
2}Q\chi\,,
\label{115}\ee
where $\chi=2(1-g)$ is the Euler number of a genus $g$ surface.

According to (\ref{mod1}) we obtain ($NS$ sector)
\be
b_{n>-\l}|0\rangle =c_{n>\l-1}|0\rangle=0\,.
\label{116}\ee
Consequently, for the standard reparametrization ghosts ($\l=2$) the
lowest
state
is not the vacuum but $c_1|0\rangle$ with $L_0$ eigenvalue equal to
$-1$.

We will also describe here the rebosonization of the bosonic ghost
systems since it will be needed in the superstring case.
{}From now on we assume $\e=-1$.
We first bosonize the U(1) current:
\be
J(z)=-\pd \phi\;\;\;,\;\;\;\langle \phi(z)\phi(w)\rangle =-\log(z-w)\,.
\label{403}\ee
The stress-tensor that gives the OPE  (\ref{112}) is
\be
\hat T={1\over 2}:J^2:+{1\over 2}Q\pd J={1\over 2}(\pd\phi)^2-{Q\over
2}\pd^2\phi\,.
\label{404}\ee
The boson $\phi$ has ``background charge'' because of the derivative
term in its stress-tensor. It is described by the following action
\be
S_Q={1\over 2\pi}\int d^2 z\left[\pd \phi\bar \pd \phi-{Q\over
4}\sqrt{g}R^{(2)}~\phi
\right]\,,
\label{409}\ee
where $R^{(2)}$ is the two-dimensional scalar curvature.
Using (\ref{256}) we see that there is a background charge of
$-Q\chi/2$, where
$\chi=2(1-g)$ is the Euler number of the surface.

However, a direct computation shows that $\hat T$ has central charge
$\hat c=
1+3Q^2$.
The original central charge of the theory was $c=\hat c-2$, as
can be seen from (\ref{108}).
Thus, we must also add an auxiliary Fermi system with $\l=1$,
composed of a dimension-one field $\eta(z)$ and a dimension-zero
field $\xi(z)$.
This system has central charge $-$2. The stress-tensor of the
original system can be written as
\be
T=\hat T+T_{\eta\xi}\,.
\label{405}\ee
Exponentials of the scalar $\phi$ have the following OPEs with
the stress-tensor and the U(1) current.
\be
T(z):e^{q\phi(w)}:=\left[-{q(q+Q)\over (z-w)^2}+{1\over
z-w}\pd_w\right]:e^{q\phi(w)}:+\dots \,,
\label{406}\ee
\be
J(z):e^{q\phi(w)}:={q\over
z-w}:e^{q\phi(w)}:\dots\;\;\to\;\;[J_0,:e^{q\phi(w)}:]
=q~:e^{q\phi(w)}: \,.
\label{407}\ee
In terms of the new variables we can express the original $b,c$
ghosts as
\be
c(z)=e^{\phi(z)}\eta(z)\;\;\;,\;\;\;b(z)=e^{-\phi(z)}\pd\xi(z)\,.
\label{408}\ee

\vskip .4cm
\noindent\hrulefill
\nopagebreak\vskip .2cm
{\large\bf Exercise}. Use the expressions of (\ref{408}) to verify by
direct computation (\ref{101}), (\ref{107}) and (\ref{109}).
\nopagebreak\vskip .2cm
\noindent\hrulefill
\vskip .4cm

Finally, the spin fields of $b,c$ that interpolate between $NS$ and $R$
sectors are given by $e^{\pm\phi/2}$ with conformal weight $-(1\pm
2Q)/8$.
Note that the zero mode of the field $\xi$ does not enter in the
definition
of $b,c$. Thus, the bosonized Hilbert space provides two copies of the
original Hilbert space since any state $|\rho\rangle$ has a
degenerate partner
$\xi_0|\rho\rangle$.

We will not delve any further into the structure of the CFT of the
$bc$ system, but we will refer the interested reader to \cite{FMS},

\renewcommand{\theequation}{\thesection.\arabic{equation}}
\section{CFT on the torus}
\setcounter{equation}{0}

\begin{figure}
\begin{center}
\leavevmode
\epsffile{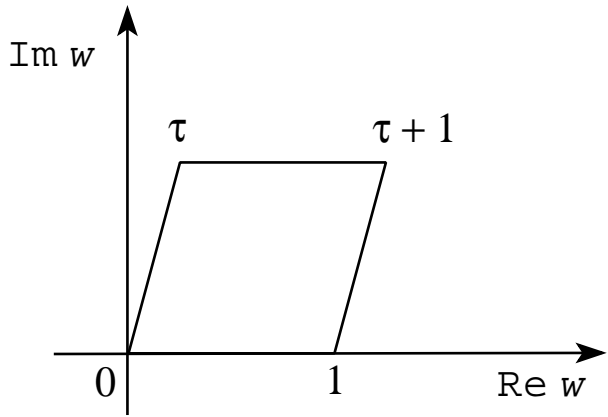}
\vspace{-.5cm}
\end{center}
\caption[]{\it The torus as a quotient of the complex plane.}
\label{f6}\end{figure}

Consider the next simplest closed Riemann surface after the sphere.
It has genus $g=1$ and Euler number $\chi=0$.
By using conformal symmetry we can pick a constant metric so
that the volume is normalized to 1.
Pick coordinates $\s_1,\s_2\in [0,1]$.
Then the volume is 1 if the determinant of the metric  is 1.
We can parametrize the metric, which is also  a symmetric and
positive-definite matrix, by a single complex number
$\tau=\tau_1+i~\tau_2$,
with positive
imaginary part $ \tau_2\geq 0$ as follows:
\be
g_{ij}={1\over \tau_2}\left(\matrix{1&\tau_1\cr
\tau_1&|\tau|^2\cr}\right)\,.
\label{117}\ee
The line element is
\be
ds^2=g_{ij}d\s_i d\s_j={1\over \tau_2}|d\s_1+\tau~d\s_2|^2={dw~d\bar
w\over \tau_2}\,,
\label{118}\ee
where
\be
w=\s_1+\tau\s_2\;\;\;,\;\;\; \bar w=\s_1+\bar\tau\s_2
\label{119}\ee
are the complex coordinates of the torus.
This is the reason why the parameter $\tau$ is known as the complex
structure (or modulus) of the torus.
It cannot be changed by infinitesimal diffeomorphisms or Weyl
rescalings
and is thus the complex Teichm\"uller parameter of the torus.
The periodicity properties of $\sigma_1,\sigma_2$ translate to
\be
w\to w+1\;\;\;\;,\;\;\;\;w\to w+\tau\,.
\label{120}\ee
The torus can be thought of as the points of the complex plane
$w$
identified under two translation vectors corresponding to the complex
numbers
1 and $\tau$, as suggested in Fig. \ref{f6}.

\begin{figure}
\begin{center}
\leavevmode
\epsfxsize=11cm
\epsffile{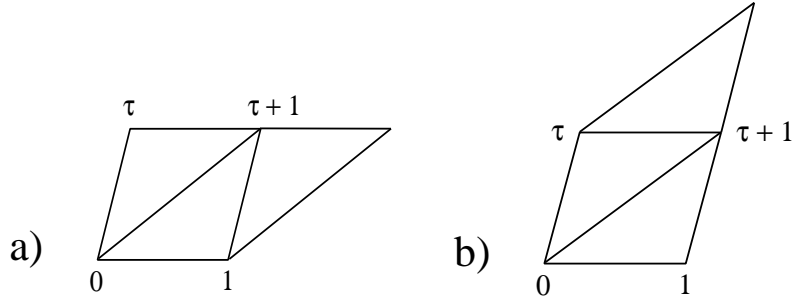}
\end{center}
\caption[]{\it a) The modular transformation $\tau\to\tau+1$. b)
The modular transformation $\tau\to \tau/(\tau+1)$.}
\label{f7}\end{figure}

Although $\tau$ is invariant under infinitesimal diffeomorphisms, it
does transform under some ``large" transformations.
Consider instead of the parallelogram in Fig. \ref{f6} defining the
torus, the one in
Fig. \ref{f7}a. Obviously, they are equivalent, due to the periodicity
conditions
(\ref{120}). However, the second one corresponds to a modulus
$\tau+1$.
We conclude that two tori with moduli differing by 1 are
equivalent.
Thus, the transformation
\be
T\;\;:\;\;\tau\to\tau +1
\label{121}\ee
leaves the torus invariant.
Consider now another equivalent choice of a parallelogram, that
depicted
in Fig. \ref{f7}b, characterized by complex numbers $\tau$ and
$\tau+1$.
To bring it to the original form (one side on the real axis) and
preserve its orientation, we have to scale both sides down by a factor
of $\tau+1$.
It will then correspond to an equivalent  torus with modulus
$\tau/(\tau+1)$.
We have obtained a  second modular transformation
\be
TST\;\;:\;\;\tau\to {\tau\over \tau+1}\,.
\label{122}
\ee
It can be shown that taking products of these transformations
generates the full modular group of the torus. A convenient set of
generators is given
also by $T$ in (\ref{121}) and
\be
S\;\;:\;\;\tau\to -{1\over
\tau}\;\;\;,\;\;\;S^2=1\;\;\;,\;\;\;(ST)^3=1\,.
\label{123}
\ee
The most general transformation is of the form
\be
\tau'={a\tau+b\over
c\tau+d}\;\;\;\leftrightarrow\;\;\;A=\left(\matrix{a&b\cr
c&d\cr}\right)\,,
\label{124}\ee
where the matrix $A$ has integer entries and determinant 1.
Such matrices form the group SL(2,$\Z)$.
Since changing the sign of the matrix does not affect the modular
transformation in (\ref{124}) the modular group is
PSL(2,$\rm \Z)=SL(2,\Z)/\Z_2$.

\begin{figure}[t]
\begin{center}
\leavevmode
\epsffile[173 421 452 663]{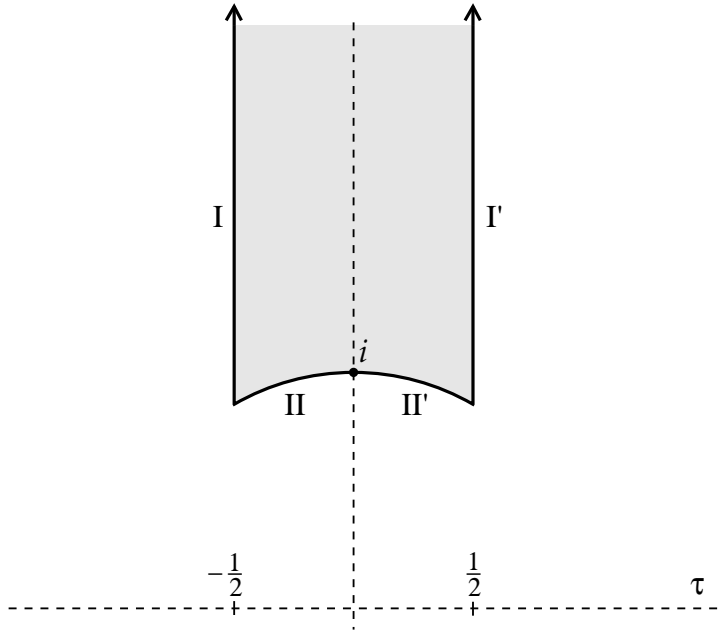}
\end{center}
\vspace{-.5cm}
\caption[]{\it The moduli space of the torus.}
\label{f8}\end{figure}

As mentioned above, the modulus takes values in the upper-half plane
${\cal H}$ ($\tau_2\geq 0$), which is the Teichm\"uller space of the
torus.
However, to find the moduli space of  truly inequivalent tori
we have to quotient this with the modular group.
It can be shown that the fundamental domain ${\cal F}={\cal
H}$/PSL(2,$\Z)$
of the modular group is the area contained in between the lines
$\tau_1=\pm 1/2$
and above the unit circle with center at the origin.
It is shown in Fig. \ref{f8}.

There is an interesting construction of the torus starting from the
cylinder.
Consider a cylinder of length $2\pi\tau_2$ and circumference  1.
Take one end, rotate it by an angle $2\pi\tau_1$ and glue it to the
other
end. This produces a torus with modulus $\tau=\tau_1+i\tau_2$.
This construction gives a very useful relation between the path
integral
of a CFT on the torus and a trace over the Hilbert space.
First, the propagation along the cylinder is governed by the
``Hamiltonian" (transfer matrix) $H=L_0^{cyl}+\bar L_0^{cyl}$.
The rotation around the cylinder is implemented by the ``momentum"
operator
$P=L_0^{cyl}-\bar L_0^{cyl}$.
Gluing together the two ends gives a trace in the Hilbert space.
{}From (\ref{93})
\be
L_0^{cyl}=L_0-{c\over 24}\;\;\;,\;\;\;\bar L_0^{cyl}=\bar L_0-{\bar
c\over 24}\,,
\label{125}\ee
where $L_0,\bar L_0$ are the operators on the sphere.
Putting everything together, we obtain
\bea
\int~ e^{-S}&=&{\rm Tr}\left[e^{-2\pi\tau_2~ H}~e^{2\pi
i\tau_1~P}\right]={\rm Tr}\left[e^{2\pi i\tau
L_0^{cyl}}~e^{-2\pi i\bar \tau\bar L_0^{cyl}}\right]=
\label{126}\\
&=&{\rm Tr}\left[q^{L_0-c/24}~\bar q^{\bar L_0-\bar c/24}\right]\,,
\nn\eea
where $q=\exp[2\pi i\tau]$.
The trace includes also possible continuous parts of the spectrum.
This is a very useful relation and also provides the correct
normalization of the path integral.

\renewcommand{\theequation}{\thesubsection.\arabic{equation}}
\subsection{Compact scalars\label{cscalars}}
\setcounter{equation}{0}

In section \ref{boson} we have described the CFT of a non-compact
real scalar field.
Here we will consider a compact scalar field $X$ taking values on a
circle of radius $R$.
Consequently, the values $X$ and $X+2\pi m R$, $m\in Z$ will be
considered
equivalent.

We will first evaluate the path integral of the theory on the torus.
The action is
\bea
S &=& \frac{1}{4\pi} \int d^2\s \sqrt{g}~g^{ij} \pd_i X {\pd}_j
X \nn\\&=& \frac{1}{4\pi} \int_{0}^1 d\s_1\int_{0}^1 d\s_2{1\over \tau_2}|\tau
\pd_1X-\pd_2 X|^2 \nn\\&=&-{1\over 4\pi}\int d^2\s ~X\square ~X\,,
\label{127}
\eea
where the Laplacian is given by
\be
\square ={1\over \tau_2}|\tau\pd_1-\pd_2|^2\,.
\label{128}
\ee
We wish to evaluate the path integral
\be
Z(R)=\int~DX~e^{-S}
\label{129}\ee
on the torus.
As usual, we will have to find the classical solutions of finite
action
(instantons) and calculate the fluctuations around them.
The field $X$ should be periodic on the torus and is a map from the
torus
(topologically $S^1\times S^1$) to the circle $S^1$. Such maps are
classified
by two integers that specify how many times $X$ winds around the two
cycles of the torus.
The equation of motion $\square X=0$ has the following instanton
solutions
\be
X_{class}=2\pi R(n\s_1+m\s_2)\;\;\;,\;\;\;m,n\in \Z\;\;\;.
\label{130}\ee
They have the correct periodicity properties
\be
X_{class}(\s_1+1,\s_2)=X(\s_1,\s_2)+2\pi
nR\;\;,\;\;X_{class}(\s_1,\s_2+1)=X(\s_1,\s_2)+2\pi mR\;,
\label{131}\ee
and the following classical action
\be
S_{m,n}={\pi R^2\over\tau_2}|m-n\tau|^2\,.
\label{132}\ee
Thus, we can separate $X=X_{class}+\chi$,
and the path integral can be written as
\bea
Z(R)&=& \sum_{m,n\in \Z}\int~D\chi~e^{-S_{m,n}-S(\chi)}\nn\\&=&\sum_{m,n\in
\Z}e^{-S_{m,n}}~\int~D\chi~e^{-S(\chi)}\,.
\label{133}\eea
What remains to be done is the path integral over $\chi$. There is
always the
constant
zero mode that we can separate, $\chi(\s_1,\s_2)=\chi_0+
\delta\chi(\s_1,\s_2)$, with $0\leq \chi_0\leq 2\pi R$.
The field $\delta\chi$ can be expanded in the eigenfunctions of the
Laplacian
\be
\square \psi_i =-\lambda_i~\psi_i\,.
\label{134}\ee
It is not difficult to see that these eigenfunctions are
\be
\psi_{m_1,m_2}=e^{2\pi
i(m_1\s_1+m_2\s_2)}\;\;\;,\;\;\;\lambda_{m_1,m_2}={4\pi^2\over
\tau_2}
|m_1\tau-m_2|^2\,.
\label{135}\ee
The eigenfunctions satisfy
\be
\int
d^2\s~\psi_{m_1,m_2}\psi_{n_1,n_2}=\delta_{m_1+n_1,0}
\delta_{m_2+n_2,0}\,,\label{136}\ee
so we expand
\be
\delta\chi={\sum_{m_1,m_2\in {\Z}}}' A_{m_1,m_2}\psi_{m_1,m_2}\,,
\label{137}\ee
where the prime implies omission of the constant mode
$(m_1,m_2)=(0,0)$.
Reality implies $A^{*}_{m_1,m_2}=A_{-m_1,-m_2}$.
The action becomes
\be
S(\chi)={1\over
4\pi}{\sum_{m_1,m_2}}^{'}~\lambda_{m_1,m_2}~|A_{m_1,m_2}|^2\,.
\label{138}
\ee
We can specify the measure
from
\be
\| \d X \| = \int d^2 \s \sqrt{\mbox{det} G} (d
\chi)^2={\sum_{m_1,m_2}}^{'}
|dA_{m_1,m_2}|^2
\ee
to be
\be
\int ~D\chi=\int_{0}^{2\pi R}d\chi_0~{\prod_{m_1,m_2}}^{'}
{dA_{m_1,m_2}\over 2\pi}\,.
\label{139}\ee
Putting everything together we obtain
\be
\int D\chi e^{-S(\chi)}={2\pi R\over
{\prod_{m_1,m_2}}^{'}\lambda_{m_1,m_2}^{1/2}}
={2\pi R\over \sqrt{{\rm det}'\square }}\,.
\label{140}\ee
Using the explicit form of the eigenvalues, the determinant of the
Laplacian can be calculated using a $\zeta$-function regularization
\cite{G},
\be
{\rm det}'\square =4\pi^2\tau_2 \eta^2(\tau)\bar\eta^2(\bar \tau)\,,
\label{141}\ee
where $\eta$ is the Dedekind function defined in (\ref{t10}).
Collecting all terms in (\ref{133}) we obtain
\be
Z(R)={R\over \sqrt{\tau_2}|\eta|^2}\sum_{m,n\in Z}e^{-{\pi R^2\over
\tau_2}|m-n\tau|^2}\,.
\label{142}\ee
This is the Lagrangian form of the partition function.
We have mentioned in the previous chapter that the partition function
on the torus can also be written in Hamiltonian form as in
(\ref{126}).
To do this we have to perform a Poisson resummation (see appendix A)
on the
integer $m$. We obtain
\be
Z(R)=\sum_{m,n\in \Z}~{q^{{P^2_L\over 2}}~\bar q^{P^2_R\over 2}
\over \eta\bar \eta}
\label{143}\ee
with
\be
P_L={1\over \sqrt{2}}\left({m\over R}+nR\right)\;\;\;,\;\;\;
P_R={1\over \sqrt{2}}\left({m\over R}-nR\right)\,\,.
\label{144}\ee
This is in the form (\ref{126}) and from it we can read off the
spectrum
of conformal weights and multiplicities of the theory.
Before we do this, however, let us return to the sphere and discuss the
current algebra structure of the theory.
As in section \ref{boson} there is a holomorphic and
anti-holomorphic U(1) current:
\be
J(z)=i\pd X\;\;\;,\;\;\;\bar J(\bar z)=i\bar \pd X
\label{145}\ee
satisfying the U(1) current algebra
\be
J(z)J(w)={1\over (z-w)^2}+{\rm finite}\;\;,\;\; \bar J(\bar z)\bar
J(\bar w)={1\over (\bar z-\bar w)^2}+{\rm finite}\,.
\label{146}\ee
We can write the stress-tensor in the affine-Sugawara form
\be
T(z)=-{1\over 2}(\pd X)^2={1\over 2}:J^2:\;\;\;,\;\;\;
\bar T(\bar z)=-{1\over 2}(\bar \pd X)^2={1\over 2}:\bar J^2:\,.
\label{147}\ee
The spectrum can be decomposed into affine HW representations as
discussed
in section \ref{affine}.
An affine primary field is specified by its charges $Q_L$ and $Q_R$
under the left- and right-moving current algebras.
{}From (\ref{147}) we obtain that its conformal weights are given by
\be
\Delta={1\over 2}Q_L^2\;\;\;,\;\;\;\bar \Delta={1\over 2}Q_R^2\,.
\label{148}\ee
The rest of the representation is constructed by acting on the affine
primary state with the negative current modes $J_{-n}$ and $\bar
J_{-n}$.
We can easily compute the character of such a representation ($c=\bar
c=1$):
\be
\chi_{Q_L,Q_R}(q,\bar q)={\rm Tr}[q^{L_0-1/24}\bar q^{\bar
L_0-1/24}]={q^{Q_L^2/2}~\bar q^{Q_R^2/2}\over \eta\bar \eta}\,.
\label{149}\ee
A comparison with (\ref{143}) shows that the spectrum contains an
infinite
number of affine U(1) representations labeled by $m,n$ with $Q_L=P_L$
and $Q_R=P_R$.
For $m=n=0$ we have the vacuum representation whose HW state is the
standard
vacuum.
The other HW states, labeled by $m,n$ satisfy
\be
J_0|m,n\rangle =P_L~ |m,n\rangle \;\;\;,\;\;\;\bar J_0 |m,n\rangle
=P_R~|m,n\rangle\,\,.
\label{150}\ee
In the operator picture, they are created out of the vacuum by the
vertex operators (we split $X(z,\bar z)\sim X(z)+\bar X(\bar z)$ as
usual):
\be
V_{m,n}=:\exp[i~p_L ~X+i~p_R~\bar X]:\,,
\label{151}\ee
\bea
J(z)V_{m,n}(w,\bar w)&=&p_L{V_{m,n}(w,\bar w)\over
z-w}+\dots\;, \nn\\
\bar J(\bar z)V_{m,n}(w,\bar w)&=&p_R{V_{m,n}(w,\bar w)\over \bar
z-\bar w}
+\dots\,.
\label{152}\eea
Their correlators are again given by the Gaussian formula
\be
\left\langle ~\prod_{i=1}^N~V_{m_i,n_i}(z_i,\bar z_i)~\right\rangle
=\prod_{i<j}^N~
z_{ij}^{p^i_Lp^j_L}~\bar z_{ij}^{p^i_Rp^j_R}\,,
\label{153}\ee
where as usual $z_{ij}=z_i-z_j$.
Using the Gaussian formula
\bea
:e^{ia\phi(z)}::e^{ib\phi(w)}:&=&(z-w)^{ab}:e^{ia\phi(z)+ib\phi(w)}
:\nn\\ &=&(z-w)^{ab}
\left[:e^{i(a+b)\phi(w)}:+{\cal O}(z-w)\right]\;,
\eea
we obtain the following OPE rule for U(1)
representations
\be
[V_{m_1,n_1}]\cdot [V_{m_2,n_2}]\sim [V_{m_1+m_2,n_1+n_2}]\,,
\label{154}\ee
compatible with U(1) charge conservation.
Under the $\rm U(1)_L\times U(1)_R$ transformation
$e^{i\theta_L+i\theta_R}$
the oscillators are invariant but the states $|m,n\rangle$ pick up a
phase
$e^{i(m+n)\theta_L+i(m-n)\theta_R}$.

In the canonical representation, the momentum operator is taking
values
$m/R$ as required by the usual (point-particle) quantum mechanical
quantization
condition on a circle of radius $R$.
The existence of the extra spatial dimension of the string allows for
the
possibility
of $X$ winding around the circle $n$ times. This is precisely the
interpretation
of the integer $n$ in (\ref{144}).
It has no point-particle (one-dimensional) analogue.

In a CFT, there is a special class of operators, known as marginal
operators,
with $(\Delta,\bar \Delta)=(1,1)$.
For such an operator $\phi_{1,1}$, the density $\phi_{1,1}dzd\bar z$
is conformally invariant.
If we perturb our action by $g\int \phi_{1,1}$ we would expect
that the theory remains conformally invariant.
There are subtleties in the quantum theory, however (short distance
singularities) which sometimes spoil conformal invariance.
When conformal invariance persists, we call $\phi_{1,1}$ exactly
marginal.
In this way, perturbing by $\phi_{1,1}$ we obtain a continuous family
of CFTs parametrized by the coupling $g$. The central charge cannot
change during a marginal perturbation.

In our present example we have an occurrence of this phenomenon.
There is a (1,1) operator namely $\phi=\pd X\bar \pd X=J\bar J$.
By adding this to the action (\ref{127}), it is easy to see that the
effect
of the perturbation is to change the effective radius $R$.
The theory, being again a free field theory, remains conformally
invariant.
In this case the operation seems trivial, however, marginal operators
exist in more complicated CFTs.

Finally let us go back to the torus and take another look at the
partition function.
For string theory purposes we would like it to be invariant under the
full
diffeomorphism group. In particular it should be invariant under the
large transformations, namely the modular transformations.
This is important in string theory since modular invariance is at the
very
heart
of finiteness of string theory and is essential for the
cancelation of
spacetime anomalies.

It suffices to prove invariance under the two generating
transformations
$T$ and $S$, since these generate the modular group.
We will use the Lagrangian representation of the partition function
(\ref{142}).
It is not difficult to verify, using the formulae of appendix A, that
$\sqrt{\tau_2}\eta\bar \eta$ is separately modular-invariant.
Thus, we only need to consider the instanton sum.
Under $\tau\to \tau+1$ we can change the summation $(m,n)\to
(m+n,n)$,
and the full sum is invariant.
Under $\tau\to -1/\tau$ we can again change the summation  $(m,n)\to
(-n,m)$
and again the sum is invariant. We conclude that the torus
partition
function of the compact boson is modular-invariant.
It is interesting to note that invariance under the $T$
transformation
in the Hamiltonian representation (\ref{126}) implies
\be
\Delta-\bar\Delta -{c-\bar c\over 24}={\rm integer}
\label{155}\ee
for the whole spectrum. In particular, for the vacuum state
$\Delta=\bar \Delta
=0$, it implies that $c-\bar c=0~$mod$~(24)$.
In our case $c=\bar c=1$ and from (\ref{144}), (\ref{148})
$P^2_L/2-P^2_R/2=mn\in \Z$.

Another comment concerns the partition function on the torus of a
non-compact boson. This can be obtained by taking the limit $R\to
\infty$.
We expect that the partition function in this limit diverges like
the volume of our space, so we have to divide first by the volume. The
free energy per unit volume is finite.
{}From (\ref{126}) we note that as $R$ gets large, the only term that
is not exponentially suppressed is the one with $m=n=0$, so
\be
\lim_{R\to\infty}{Z(R)\over R}={1\over \sqrt{\tau_2}\eta\bar \eta}\,.
\label{156}
\ee

Before we proceed, we will derive the torus propagator
for the boson.
It can be directly written in terms of the non-zero eigenvalues and the
associated eigenfunctions of the Laplacian:
\be
\Delta(\s_1,\s_2)\equiv\langle
\delta\chi(\s_1,\s_2)\delta\chi(0,0)\rangle
=-{\sum_{m,n}}'{1\over |m\tau-n|^2}e^{2\pi i(m\s_1+n\s_2)}\,.
\label{pr1}\ee
The sum is conditionally convergent and has to be regularized  using
$\zeta$-function regularization.
We obtain
\be
\square~\Delta(\s_1,\s_2)={4\pi^2\over
\t_2}\left[\d(\s_1)\d(\s_2)-1\right]\,,
\label{pr2}\ee
so that the integral over the torus gives zero, in accordance with the
fact that
we have omitted the zero mode.
It can also be expressed in complex coordinates in terms of
$\th$-functions as
\be
\Delta(\s_1,\s_2)=-\log G(z,\bar z)\;\;\;,\;\;\;G=e^{-2\pi{Imz^2\over
\t_2}}
{}~\left|{\th_1(z)\over \th_1'(0)}\right|^2\,\,.
\label{pr3}\ee

The above discussion can easily be generalized to the case of $N$ free
compact
scalar fields $X^i$, $i=1,2,\dots ,N$. We will take them to have
values in $[0,2\pi]$.
They parametrize an $N$-dimensional torus.
The most general quadratic action is
\be
S = {1\over 4\pi}\int d^2\s\sqrt{\mbox{det}~g}g^{ab}G_{ij}\pd_a X^i
\pd_b
X^j
+{1\over 4\pi}\int d^2\s \epsilon^{ab}B_{ij}\pd_a X^i \pd_b X^j\;,
\label{157}\ee
where $g_{ab}$ is the torus metric (\ref{117}), $\epsilon^{ab}$ is the
usual
$\epsilon$-symbol, $\epsilon^{12}=1$;
$G_{ij}$ is a constant symmetric positive-definite matrix that plays
the role
 of metric in the space of the $X^i$ (target-space torus).
The constant matrix $B_{ij}$ is antisymmetric.
It is the analogue of the $\theta$-term in four-dimensional gauge
theories.

An analogous calculation of the path integral produces
\be
\Z_{\mbox{\tiny d,d}}(G,B) = \frac{\sqrt{\mbox{det}~G}}{(\sqrt{\t_{2}}
\eta \bar{\eta})^N}\sum_{\vec{m},\vec{n}} e^{-\frac{\pi(G_{ij} +
B_{ij})}{\t_2} (m_i + n_i\t) (m_j + n_j\bar{\t})}\,.
\label{158}\ee
This partition function reduces for N=1, $G=R^2$, $B=0$ to
(\ref{126}).
Using a multiple Poisson resummation on the $m_i$, it can be
transformed
in the Hamiltonian representation:
\be
 \Z_{\mbox{\tiny d,d}}(G,B)={\Gamma_{d,d}(G,B)\over
\eta^d\bar\eta^d}=\sum_{\vec m,\vec n\in \Z^N}{q^{{1\over 2}
P_L^2}~\bar q^{{1\over 2}
P_R^2}\over \eta^N\bar \eta^N}\,,
\label{159}
\ee
where
\be
P_{L,R}^2\equiv P^i_{L,R}~G_{ij}P^j_{L,R}\,,
\label{160}\ee
\be
P^i_L={G^{ij}\over\sqrt{2}}(m_j+(B_{jk}+G_{jk})n_k)\;\;\;,
\;\;\;P^i_R={G^{ij}\over \sqrt{2}}(m_j+(B_{jk}-G_{jk})n_k)\,.
\label{161}\ee
The theory has  a left-moving and a right-moving U(1)$^N$ current
algebra
generated by the currents
\be
J^i(z)=i\pd X^i\;\;\;,\;\;\;\bar J^i=i\bar \pd X^i\,\,,
\label{162}\ee
\be
J^i(z)J^j(w)={G^{ij}\over (z-w)^2}+\dots
\label{163}\ee
and similarly for $\bar J^i$.
The stress-tensor is again of the affine-Sugawara form
\be
T(z)=-{1\over 2}G_{ij}\pd X^i\pd X^j={1\over 2}G_{ij}:J^iJ^j:\,.
\label{164}\ee
Affine primaries are characterized by charges $Q^i_{L,R}$ and
\be
\Delta={1\over 2}G_{ij}Q^i_LQ^j_L\;\;\;,\;\;\;\bar \Delta={1\over
2}G_{ij}Q^i_RQ^j_R\,.
\label{165}\ee
Comparing this with (\ref{159}) we obtain $Q^i_{L,R}=P^i_{L,R}$.

It can be shown that the partition function (\ref{159}) is
modular-invariant.

\subsection{Enhanced symmetry and the string Higgs effect\label{su2}}
\setcounter{equation}{0}

Something special happens to the CFT of the single compact boson
when the radius is $R=1$.
The conformal weights of the primaries are now given by
\be
\Delta={1\over 4}(m+n)^2\;\;\;,\;\;\;\bar \Delta={1\over 4}(m-n)^2\,.
\label{169}\ee
Notice that the two states with $m=n=\pm 1$ are (1,0) operators.
For generic $R$ the only (1,0) (chiral) operator is the U(1) current
$J(z)=i\pd X$. Now we have two more. We expect that the current
algebra
becomes larger if we also include these operators.
Similarly, the states with $m=-n=\pm 1$ are (0,1) operators and the
right-moving current algebra is also enhanced.
We will discuss only the left-moving part, since the right-moving part
behaves in a similar way.
The two operators that become (1,0) can be written as vertex
operators
(\ref{151})
\be
J^{\pm}(z)={1\over \sqrt{2}}:e^{\pm i\sqrt{2}X(z)}:\,.
\label{173}\ee
Define also
\be
J^3(z)={1\over \sqrt{2}}J(z)={i\over \sqrt{2}}\pd X(z)
\,.\label{171}\ee
They satisfy the following OPEs, which can be computed
directly using $\langle X(z)X(0)\rangle =-\log z$,
\bea
J^3(z)J^{\pm}(w)=\pm {J^{\pm}(w)\over
z-w}+\dots\;\;\;,\;\;\;J^+(z)J^+(w)=\dots\;\;,\nn\\
J^-(z)J^-(w)=\dots
\,,\label{170}\eea
\bea
J^+(z)J^-(w)={1/2\over (z-w)^2}+{J^3(w)\over z-w}+\dots\;\;,\nn\\
J^3(z)J^3(w)={1/2\over (z-w)^2}+\dots
\,.\label{172}\eea
It is not difficult to realize that this is the SU(2) current
algebra with level $k=1$.
This is not too surprising, since the central charge of $\rm SU(2)_k$
is
given
by (\ref{ccc}) to be $c=3k/(k+2)$. It indeed becomes $c=1$ when
$k=1$.
This realization of current algebra at level 1 in terms of free bosons
is known as the Halpern-Frenkel-Kac-Segal construction.

We have seen before that $\rm SU(2)_1$ has two integrable affine
representations,
the vacuum representation with $j=0$ and the $j=1/2$ representation
with
conformal
weight $\Delta=1/4$ (from (\ref{42})).
The primary state of the $j=0$ representation is the vacuum.
The primary operators of the $j=1/2$ representation transform as a
two-component
spinor of $\rm SU(2)_L$ and a two-component spinor of $\rm SU(2)_R$
with
conformal weights (1/4,1/4).
They are represented by the four vertex operators $V_{m,n}$ with
$(m,n)=(0,\pm 1)$ and $(\pm 1,0)$. They have the correct conformal
weight and OPEs with the currents (\ref{173}).

This phenomenon generalizes to the N-dimensional toroidal models.
The U(1) charges $p^i_{L,R}$ take values on an N-dimensional lattice
that depends on $G_{ij},B_{ij}$. For special values of G,B this lattice
coincides with the root lattice of a Lie group $G$ with rank N. Then
some vertex operators become
extra chiral currents and along with the $N$ abelian currents $J^i$
form an affine $G$ algebra at level $k=1$.

When the toroidal CFT acquires enhanced current algebra symmetry then
the associated string theory acquires enhanced gauge symmetry.
Consider the bosonic string with one of the 26 dimensions (say
$X^{25}$)
compactified on a circle of radius $R$.
Then the massless states are again similar, but with a slightly
different interpretation.
There are now 25  non-compact dimensions, so we have 25-dimensional
Lorentz invariance.
The massless states are
\be
a^{\m}_{-1}\bar a^{\nu}_{-1}|0\rangle\;\;\;,\;\;\;a^{\m}_{-1}\bar
a^{25}_{-1}
|0\rangle\;\;\;,\;\;\;a^{25}_{-1}\bar a^{\m}_{-1}|0\rangle\;\;\;,\;\;\;
a^{25}_{-1}\bar a^{25}_{-1}|0\rangle
\,,\label{1000}\ee
which are the graviton, antisymmetric tensor, dilaton, two U(1) gauge
fields and a scalar.
Note that the scalar state is generated by $\pd X^{25}\bar \pd X^{25}$,
which is the perturbation that changes the radius.
Thus, the expectation value of the scalar is the radius $R$.
There are other massive states, among them
\be
|A^{\pm}_{\m}\rangle=\bar a^{\m}|m=\pm 1,n=\pm 1\rangle
\,,\label{1001}\ee
which are massive vectors with mass $m^2=(R-1/R)^2/4$
and
\be
|\bar A^{\pm}_{\m}\rangle=a^{\m}|m=\pm 1,n=\mp 1\rangle
\label{1002}\ee
with the same mass as above.
As we vary $R$ the mass changes, and at $R=1$ they become massless.
At that point, the string theory acquires an SU(2)$\times$SU(2)
gauge symmetry.
Moving away from $R=1$, SU(2)$\times$SU(2) gauge symmetry is
spontaneously broken to U(1)$\times$U(1).
This is the usual Higgs effect and the scalar whose expectation value
is the
radius plays the role of the Higgs scalar (although there is no
potential here).

\subsection{T-duality\label{tduality}}
\setcounter{equation}{0}

We now return to the example of a single scalar, compactified on a
circle of radius $R$, discussed in the previous section.
As we have seen the primaries have
\be
H=L_0+\bar L_0={1\over 2}\left({m^2\over R^2}+n^2
R^2\right)\;\;\;,\;\;\;
P=L_0-\bar L_0=mn
\,.\label{166}\ee
It is obvious that the above spectrum is invariant under
\be
R\to {1\over R}\;\;\;,\;\;\;m\leftrightarrow n
\,.\label{167}\ee
This corresponds to the following transformation of the U(1) charges
\be
P_L\to P_L\;\;\;,\;\;\;P_R\to -P_R
\,.\label{168}\ee
Only the right charge  changes  sign.
The action on the respective currents is analogous
\be
J(z)\to J(z)\;\;\;,\;\;\;\bar J(\bar z)\to - \bar J(\bar z)
\,.\label{dual}\ee
It can be easily checked that not only the spectrum but also
the interactions respect this property.
This is a peculiar property since it implies that a CFT cannot
distinguish
a circle of radius $R$ from another of radius $1/R$.
This is, strictly speaking, not a symmetry of the two-dimensional
theory.
It states that two a priori different theories are in fact
equivalent.
However, in the context of string theory it will become a true
symmetry
and is known under the name $T$-duality.
Notice that the presence of winding modes is essential for the
presence
of $T$-duality. Therefore it can appear in string theory but not in
point-particle field theory.

There is an interesting interpretation of $T$-duality in string theory.
We start from the CFT with $R=1$. We have seen that at this point there
is an enhanced symmetry $\rm SU(2)_L\times SU(2)_R$.
Then, at this point, the duality transformation (\ref{dual}) is an
$\rm SU(2)_R$ Weyl transformation, which is an obvious symmetry of the
CFT.
This explains the self-duality at $R=1$.
Move now infinitesimally away from $R=1$ by perturbing the CFT with the
marginal operator $\e\int J^3\bar J^3=\e\int \pd X\bar \pd X$.
Because of the self-duality of the unperturbed theory, the $\e$
perturbation
and $-\e$ perturbation give identical theories. This is the
infinitesimal version of $R\to 1/R$ duality around $R=1$.
We can further extend this duality on the whole line.
In this sense the duality is a consequence of SU(2) symmetry at $R=1$
and the duality transformation is an $\rm SU(2)_R$ transformation.
Consider again the bosonic string with one dimension compactified.
At $R=1$ the $\rm SU(2)_L$ transformation is a gauge transformation.
Away from $R=1$ the gauge symmetry is broken and the duality symmetry
is
a discrete remnant of the original gauge symmetry.

We can generalize the $T$-duality symmetry to the
N-dimensional toroidal models;
here the duality transformations form an infinite discrete group,
unlike the one-dimensional case where the group was $Z_2$.

First observe that the partition function (\ref{158}) is invariant
under
shifts of $B_{ij}$ by any antisymmetric matrix with integer entries.
Also by construction the theory is invariant under $GL(N)$ rotations
of the scalars $G_{ij}$ and $B_{ij}$.
However, since the rotations also act on $m_i,n_i$ they have to
rotate them
back to integers. The  $GL(N)$ matrix must have integer entries and
such matrices form the discrete group $GL(N,\Z)$.
Finally there are transformations such as the radius inversion, which
leave the spectrum invariant.
Together, all of these transformations combine into an infinite
discrete group
O(N,N,$\Z)$.
It is described by $2N\times 2N$ integer-valued matrices of the form
\be
\Omega = \left(\begin{array}{cc} A & B \\ C & D \end{array}\right)
\,,\label{174}\ee
where $A,B,C,D$ are $N\times N$ matrices.
Define also the O(N,N)-invariant metric
\be
L=\left(\matrix{0&{\bf 1}_N\cr {\bf 1}_N&0\cr}\right)
\,,\label{175}\ee
where ${\bf 1}_N$ is the $N$-dimensional unit matrix.
$\Omega$ belongs to O(N,N,$\Z)$ if it has integer entries and
satisfies
\be
\Omega^T~L~\Omega=L
\,.\label{176}\ee
Define $E_{ij} = G_{ij} + B_{ij}$.
Then the duality transformations are
\be
E \rightarrow (AE + B)(CE + D)^{-1}\;\;\;,\;\;\;
\left(\begin{array}{c} \vec{m} \\ \vec{n} \end{array}\right)
\rightarrow \Omega \left(\begin{array}{c} \vec{m} \\ \vec{n}
\end{array}\right)
\,.\label{177}\ee
In the special (but useful) case N=2 we can parametrize
\be
G_{ij}={T_2\over U_2}\left(\matrix{1&U_1\cr
U_1&U_1^2+U_2^2\cr}\right)\;\;\;,
\;\;\;B_{ij}=\left(\matrix{0&T_1\cr -T_1&0\cr}\right)\;,
\label{178}\ee
with $T_2,U_2\geq 0$.
Defining the complex parameters $T=T_1+iT_2$, $U=U_1+iU_2$,
the lattice sum (\ref{159}) becomes
\bea
\Gamma_{2,2}(T,U)=\sum_{\vec m,\vec n}\exp\left[-{\pi\t_2\over T_2U_2}
|-m_1U+m_2+T(n_1+Un_2)|^2 +\right.\nn\\\left.\rule[-1.5ex]{0pt}{4ex} +2\pi i\t(m_1n_1+m_2n_2)\right]
\,.\label{l22}\eea
The duality group O(2,2,$\Z)$ acts on $T$ and $U$ with independent
PSL(2,$\Z)$ transformations (\ref{124}) as well as with the exchange
$T\leftrightarrow U$.

$T$-duality can be generalized to $s$-models that have a curved
target space. For a more detailed discussion, see \cite{dual}.

\subsection{Free fermions on the torus\label{tfermion}}
\setcounter{equation}{0}

In section \ref{Nfermions} we have analyzed the CFT of $N$ free
Majorana-Weyl fermions. We will now consider the partition function
of this theory   on the torus.
The action was given in (\ref{S}).
To do the path integral, we have to choose boundary conditions for the
fermions
around the two cycles of the torus.
For each cycle  we have the choice between periodic and antiperiodic
boundary conditions. In total we have four possible sectors.
The fermionic path integral will give a power of the fermionic
determinant
defined with the appropriate boundary conditions (also known as
spin-structures).
\be
\int e^{-S}=({\rm det} ~\pd)^{N/2}
\,.\label{179}\ee
This can be computed by finding the appropriate eigenvalues and
taking the
$\zeta$-regularized product.
We will first consider antiperiodic boundary conditions on both cycles
(A,A).
Then the eigenvalues are
\be
\lambda_{AA}\sim \left(\left(m_1+{1\over
2}\right)\tau+\left(m_2+{1\over 2}\right)
\right)\;\;\;,\;\;\;m_{1,2}\in\Z
\,.\label{180}\ee
A calculation of the regularized product gives
\be
({\rm det} ~\pd)_{AA}={\th_{3}(\tau)\over \eta(\tau)}
\,.\label{181}\ee
For  (A,P) boundary conditions we obtain
\be
\lambda_{AP}\sim \left(\left(m_1+{1\over
2}\right)\tau+m_2
\right)\;\;\;,\;\;\;m_{1,2}\in\Z
\,,\label{182}\ee
\be
({\rm det} ~\pd)_{AP}={\th_{4}(\tau)\over \eta(\tau)}
\,.\label{183}\ee
For (P,A) boundary conditions we have
\be
\lambda_{PA}\sim \left(m_1\tau+\left(m_2+{1\over
2}\right)
\right)\;\;\;,\;\;\;m_{1,2}\in\Z
\,,\label{184}\ee
\be
({\rm det} ~\pd)_{PA}={\th_{2}(\tau)\over \eta(\tau)}
\,.\label{185}\ee
Finally for (P,P) boundary conditions the determinant vanishes, since
 these boundary conditions now allow  zero modes.
By coupling to constant gauge fields (which act as sources for the zero
modes) it can be seen that the
determinant
here is proportional to $\th_1(\tau)$, which indeed is identically
zero.

We can summarize the above results as follows.
Let $a=0,1$ indicate A,P boundary conditions respectively around
the first cycle
and $b=0,1$ indicate A,P around the second.
Then
\be
(\det \pd)[^a_b]={\th[^a_b](\tau)\over \eta(\tau)}
\,.\label{186}\ee
The (P,P) spin-structure is known as the odd spin-structure,
the rest as even spin-structures.
{}From appendix A we can see that modular transformations permute the
various
boundary conditions since they permute the various cycles.
To construct something that is modular-invariant, we will have to sum
over all boundary conditions.
Including also the right-moving fermions
we can write the  full partition function as
\be
Z_{N}^{\rm  fermionic}=\half \sum_{a,b=0}^1\left|{\th[^a_b]\over
\eta}\right|^N
\,.\label{187}\ee
It can be checked directly that it is modular-invariant.
To expose the spectrum, we can express the partition function in
terms
of the characters (\ref{65}), (\ref{66}), (\ref{80}), (\ref{81})
as
\be
Z_{N}^{\rm fermionic}=|\chi_0|^2+|\chi_V|^2+|\chi_S|^2+|\chi_C|^2
\,,\label{188}\ee
from which we see that all $\rm O(N)_1$ integrable representations
participate.

The two-point functions of the fermions in the even spin-structures can
be
fixed in terms of their pole structure and transformation properties
under modular transformations.
They are given by the Szeg\"o kernel
\be
\langle\psi^i(z)\psi^j(0)\rangle=\d^{ij}~S[^a_b](z)\;\;\;,
\;\;\;S[^a_b](z)=
{\th[^a_b](z)\th_1'(0)\over \th_1(z)\th[^a_b](0)}
\,.\label{fprop}\ee

We will  also discuss the zero modes in the odd spin-structure further.
Each real fermion has a zero mode, and the path integral vanishes.
The first non-zero correlation function must contain $N$ fermions so
that they soak up all the zero modes.
The integral over the zero modes gives a completely antisymmetric
tensor, which we normalize to the invariant $\e$-tensor.
The rest of the contribution is given by the partition function in the
absence
of zero modes. Since the oscillators are integrally moded and since
there is a $(-1)^F$ insertion, the non-zero mode contribution is
\be
q^{-N/24}\prod_{n=1}^{\infty}(1-q^n)^N=\eta^N=\left[{1\over
2\pi}{\pd_v\th_1(v)|_{v=0}\over \eta}\right]^{N/2}
\,.\label{1888}\ee
Thus,
\be
\left\langle \prod_{k=1}^N\psi^{i_{k}}(z_k)\right\rangle_{\rm odd}
=\e^{i_1,\dots,i_N}~\eta^N
\,.\label{1889}\ee

\subsection{Bosonization\label{bosonization}}
\setcounter{equation}{0}

Consider two Majorana-Weyl fermions $\psi^i(z)$
with
\be
\psi^i(z)\psi^j(w)={\delta^{ij}\over z-w}+\dots
\,.\label{189}\ee
We can change basis to
\be
\psi={1\over \sqrt{2}}(\psi^1+i\psi^2)\;\;\;,\;\;\;\bar\psi={1\over
\sqrt{2}}(\psi^1-i\psi^2)
\,.\label{190}\ee
The theory contains a U(1) current algebra generated by the (1,0)
current
\be
J(z)=:\psi\bar \psi:\;\;\;,\;\;\;J(z)J(w)={1\over (z-w)^2}+\dots
\,,\label{191}\ee
\be
J(z)\psi(w)={\psi(w)\over z-w}+\dots\;\;\;,\;\;\;J(z)\bar
\psi(w)=-{\bar \psi(w)\over z-w}+\dots
\,.\label{192}\ee
Equation (\ref{192}) states that $\psi,\bar\psi$ are affine primaries
with
charges 1 and $-1$.
The stress-tensor is
\be
T(z)=-{1\over 2}:\psi^i\pd \psi^i:={1\over 2}:J^2:
\,.\label{193}\ee
It has central charge $c=1$.

We can represent the same operator algebra using a single chiral
boson $X(z)$.
Namely
\be
J(z)=i\pd X\;\;\;,\;\;\;\psi=:e^{i X}:\;\;\;,\;\;\;\bar\psi=:e^{-iX}:
\,.\label{194}\ee

\vskip .4cm
\noindent\hrulefill
\nopagebreak\vskip .2cm
{\large\bf Exercise}: Verify that the above definitions reproduce the
same OPEs
as in the fermionic theory.
\nopagebreak\vskip .2cm
\noindent\hrulefill
\vskip .4cm

Moreover, applying these definitions to (\ref{193}) they produce the
correct stress-tensor
of the scalar, namely $T=-\half :\pd X^2:$.
This chiral operator construction suggests that two Majorana-Weyl
fermions
and a chiral boson might give equivalent theories.
However, the full theories contain also right-moving parts.
When included, we are considering on the one hand a Dirac fermion and
on the other a scalar.
For the scalar theory, however, we have to specify the radius $R$.
To do this we start from the partition function of the torus
for a Dirac fermion (\ref{187}) for N=2.

Applying a Poisson resummation to the $\th$-functions we can show
that
\be
|\th[^a_b]|^2={1\over \sqrt{2\tau_2}}\sum_{m,n\in Z}
\exp\left[-{\pi\over 2\tau_2}|n-b+\tau(m-a)|^2+i\pi mn\right]
\label{195}\ee
$$={1\over \sqrt{2\tau_2}}\sum_{m,n\in Z}
\exp\left[-{\pi\over 2\tau_2}|n+\tau m|^2+i\pi(m+a)(n+b)\right]\;.
$$
The second equation is valid when $a,b\in Z$.
Then,
\bea
Z&=&{1\over 2}\sum_{a,b=0}^1\left|{\th[^a_b]\over
\eta}\right|^2 \nn\\ &=&
{1\over 2\sqrt{2\tau_2}}\sum_{a,b=0}^1\sum_{m,n\in \Z}
\exp\left[-{\pi\over 2\tau_2}|n+\tau m|^2+i\pi(m+a)(n+b)\right]
\,.\label{196}\quad\quad\eea
Summation over $b$ gives a factor of $2$ and sets $m+a$ to be even.
Thus, $m=2\tilde m+a$.
Summing over $a$ resets $m$ to be an arbitrary integer.
Thus,
\be
Z_{\rm Dirac}=
{1\over \sqrt{2\tau_2}}\sum_{m,n\in \Z}
\exp\left[-{\pi\over 2\tau_2}|n+\tau m|^2\right]
\label{197}\ee
and comparing with (\ref{142}) we see that it is the same as that of
a boson
with radius $R=1/\sqrt{2}$.

To summarize, a Dirac fermion is equivalent to a compact boson with
radius $R=1/\sqrt{2}$.

\subsection{Orbifolds}
\setcounter{equation}{0}

The notion of orbifold arises when we consider a manifold $M$ that
has a discrete symmetry group $G$.
We may  consider a new manifold $\tilde M\equiv M/G$, which is
obtained from the old one by modding out the symmetry group $G$.
If $G$ is freely acting ($M$ has no fixed-points under the $G$
action)
then $M/G$ is a smooth manifold.
On the other hand, if $G$ has fixed-points, then $M/G$ is no longer a
smooth manifold but has conical singularities at the fixed-points known
as
orbifold singularities.
We will now  provide examples of the above.

Consider the real line $\R$. It has a $Z_2$ symmetry $x\to -x$.
This symmetry has one fixed-point, namely $x=0$.
The orbifold $\R/Z_2$ is the half-line with an orbifold point
(singularity)
at the boundary $x=0$.
On the other hand the real line $\R$ has another discrete infinite
symmetry group, namely translations $x\to x+2\pi \lambda$.
This symmetry is freely acting, and the resulting orbifold is a
smooth
manifold, namely a circle of radius $\lambda$.

Orbifolds are interesting in the context of CFT and string theory,
since they provide spaces for string compactification that are
richer than tori,
but admit an exact CFT description.
Moreover, although their classical geometry can be singular, strings
propagate smoothly on them. In other words, the correlation functions
of the associated CFT are finite.

We will describe here some simple examples of orbifolds in order to
indicate the important issues.
They will be useful later on, in order to break supersymmetry in
string theory.
More can be found in the original papers \cite{orb1}.

\begin{figure}
\begin{center}
\leavevmode
\epsffile{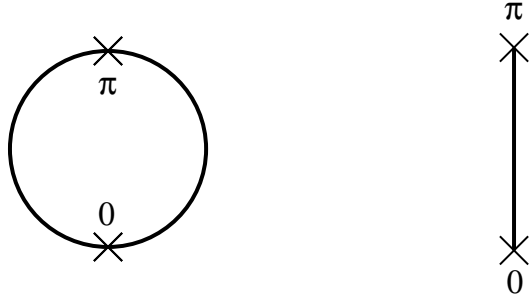}
\end{center}
\caption[]{\it The orbifold $S^1/Z_2$.}
\label{f10}\end{figure}

Consider first a simple example of a non-freely acting
orbifold.
Consider a circle of radius $R$, parametrized by $x\in[0,2\pi]$,
and mod out the symmetry $x\to -x$.
There are two fixed-points under the symmetry action, $x=0$ and
$x=\pi$.
The resulting orbifold is a line segment with the fixed-points at the
boundaries, (Fig. \ref{f10}).

It is not very difficult to construct the CFT of the orbifold.
Every operator in the original Hilbert space has a well defined
behaviour
under the $Z_2$ orbifold transformation, $X\to -X$, and for the vertex
operators, $V_{m,n}\to V_{-m,-n}$.

The orbifold construction indicates that we should keep only the
operators invariant under the orbifold transformation.
Thus, the orbifold theory contains the $Z_2$ invariant operators and
their
correlators are the same as in the original theory.
In particular the invariant vertex operators are $V^+_{m,n}=\half
(V_{m,n}+
V_{-m,-n})$.

However, this is not the end of the story. What we have constructed so
far is
the ``untwisted sector".
An indication that we must have more can be seen from the torus
partition
function.
We will start from (\ref{143}) in the Hamiltonian representation.
In order to keep only the invariant states we will have to insert a
projector
in the trace.
This projector is $(1+g)/2$, where $g$ is the non-trivial orbifold
group element
acting on states as
\be
g\left[~\prod_{i=1}^N a_{-n_i}\prod_{j=1}^{\bar N}\bar a_{\bar
n_j}|m,n\rangle \right]=(-1)^{N+\bar N}
\prod_{i=1}^N a_{-n_i}\prod_{j=1}^{\bar N}\bar a_{\bar
n_j}|-m,-n\rangle\,.
\label{217}\ee
Thus,
\be
Z(R)^{\rm invariant}={1\over 2}Z(R)+{1\over 2}{\rm
Tr}[g~q^{L_0-1/24}~\bar
q^{\bar L_0-1/24}]
\,.\label{218}\ee
To evaluate the second trace we note that $\langle
m_1,n_1|m_2,n_2\rangle\sim
\delta_{m_1+m_2}\delta_{n_1+n_2}$, which implies that only states with
$m=n=0$ contribute to that trace.
These are pure oscillator states (the vacuum module) and every
oscillator is weighted with a factor of $-1$ due to the action
of $g$.
We obtain
\be
{1\over 2}{\rm Tr}[g~q^{L_0-1/24}~\bar q^{\bar L_0-1/24}]={1\over
2}(q\bar
q)^{-1/24}
\prod_{n=1}^{\infty}{1\over (1+q^n)(1+\bar q^n)}=
\left|{\eta\over \th_2}\right|
\label{219}\ee
and
\be
Z(R)^{\rm invariant}={1\over 2}Z(R)+\left|{\eta\over \th_2}\right|
\,.\label{220}\ee
A simple look at the modular properties of the $\th$-functions
indicates that
this partition function is not modular-invariant. Something is
missing.
This is precisely the set of $twisted$ states.
There is now another boundary condition possible for the field $X$,
namely
$X(\sigma+2\pi)=-X(\sigma)$. This is again a periodicity condition,
since now
$X$ and $-X$ are identified.
In this sector (which is similar to the Ramond sector for the
fermions)
the momentum and winding are forced to be zero by the boundary
condition
and the oscillators are half-integrally modded.
Imposing the boundary condition above to the solution of the Laplace
equation,
we obtain the following mode expansion in the twisted sector
\be
X(\s,\tau)=x_0+{i\over \sqrt{4\pi T}}\sum_{n\in Z}
\left({a_{n+1/2}\over n+1/2}e^{i(n+1/2)(\s+\tau)}+{\bar a_{n+1/2}\over
n+1/2}e^{-i(n+1/2)(\s-\tau)}\right)\;.
\ee
The zero mode $x_0$ is forced to lie at the two fixed-points:
$x_0=0,\pi R$.
This  indicates the presence of two ground-states
$|H^{0,\pi}\rangle$
in this sector, which are primaries under the Virasoro algebra and
invariant
under the orbifold transformation.
They satisfy
\be
a_{n+1/2}|H^{0,\pi}\rangle=\bar
a_{n+1/2}|H^{0,\pi}\rangle=0\;\;\;\;\;n\geq 0
\;\;.
\ee
Their conformal weight can be computed in the same way as we did for
the spin
fields in the fermionic case. It is $h=\bar h=1/16$.
The rest of the states are generated by the action of the negative
modded
oscillators on the ground-states. However, not all of the states are
invariant.
To pick the invariant states we will have to do a trace with our
projector
in the twisted sector:
\bea
Z^{\rm twisted}&=&{1\over 2}{\rm Tr}[(1+g)q^{L_0-1/24}~\bar q^{\bar
L_0-1/24}] \nn\\&=&
{1\over 2}\frac{1}{(q\bar q)^{48}}\left[\prod_{n=1}^{\infty}{1\over
(1-q^{n-{1\over 2}})(1-\bar q^{n-{1\over 2}})}
\label{221}
+\prod_{n=1}^{\infty}{1\over (1+q^{n-{1\over 2}})(1+\bar
q^{n-{1\over 2}})}\right]\nn\\
&=&\left|{\eta\over \th_4}\right|+\left|{\eta\over \th_3}\right|\,.
\eea
The full partition function
\be
Z^{\rm orb}(R)=Z^{\rm untwisted}+Z^{\rm twisted}={1\over
2}Z(R)+\left|{\eta\over \th_2}\right|+\left|{\eta\over
\th_4}\right|+\left|{\eta\over \th_3}\right|
\label{222}\ee
is modular-invariant.
In fact, the four different parts in (\ref{222}) can be interpreted as
the result of performing the path integral on the torus, with the four
different boundary conditions around the two cycles, as in the case
of the fermions.
We will introduce the notation $Z[^h_g]$ where $h,g$ take values
$0,1$;
$h=0$ labels the untwisted sector, $h=1$ the twisted sector;
$g=0$ implies no projection, while $g=1$ implies a projection.
In this notation the orbifold partition function can be written as
\be
Z^{\rm orb}={1\over 2}\sum_{h,g=0}^1~Z[^h_g]\;\;,
\label{231}\ee
with $Z[^0_0]=Z(R)$ and
\be
Z[^h_g]=2\left|{\eta\over \th[^{1-h}_{1-g}]}\right|\;\;\;,\;\;\;
(h,g)\not=(0,0)
\,.\label{232}\ee
They transform as follows under modular transformations
\be
\tau\to\tau+1\;\;:\;\;\;Z[^h_g]\to Z[^h_{h+g}]
\,,\label{235}\ee
\be
\tau\to-{1\over \tau}\;\;:\;\;\;Z[^h_g]\to Z[^g_{h}]
\label{236}\ee
and we conclude that (\ref{231}) is modular-invariant.

Notice also that the whole twisted sector does not depend on the
radius.
This is a general characteristic of non-freely acting orbifolds.
As we will see later, the situation is different for freely acting
orbifolds.

The twisted ground-states are generated from the SL(2,$\C)$-invariant
vacuum by the twist operators $H^{0,\pi}(z,\bar z)$.
Correlation functions of twist operators are more difficult to
compute, but this calculation can be done (see \cite{orb,DVV} for more
details).
The following schematic OPEs can be established \cite{orb,DVV}
\be
[H^0]\cdot[H^0]\sim
\sum_{n,m}~C^{2m,2n}[V^{+}_{2m,2n}]+C^{2m,2n+1}[V^{+}_{2m,2n+1}]
\,,\label{223}\ee
 \be
[H^{\pi}]\cdot[H^{\pi}]\sim
\sum_{n,m}~C^{2m,2n}[V^{+}_{2m,2n}]-C^{2m,2n+1}[V^{+}_{2m,2n+1}]
\,,\label{224}\ee
\be
[H^0]\cdot[H^{\pi}]\sim \sum_{n,m}~C^{2m+1,2n}[V^{+}_{2m+1,2n}]
\,.\label{225}\ee
Here $[V^+_{m,n}]$ stands for the whole U(1) representation generated
from the primary vertex operator
$V^+_{m,n}=(V_{m,n}+V_{-m,-n})/\sqrt{2}$
by the action of the U(1) current modes.
The OPE coefficients are given by
\be
C^{m,n}=\sqrt{2}2^{-2(h_{m,n}+\bar h_{m,n})}\;\;\;,\;\;\;C_{0,0}=1
\ee
and
\be
h_{m,n}=(m/R+nR)^2/4\;\;\;,\;\;\;\bar h_{m,n}=(m/R-nR)^2/4\;.
\ee

Notice that the two U(1) currents $\pd X$ and $\bar \pd X$ of the
original theory have been projected out.
Consequently, in the orbifold theory we do not expect to have the
continuous
$\rm U(1)_L\times U(1)_R$ invariance any longer.
This is already obvious in the twisted OPEs, which show that the
charges m,n
are no longer conserved.
There remains however a residual $Z_2\times Z_2$ symmetry
\be
(H^0,H^{\pi},V^+_{m,n})\to (-H^0,H^{\pi},(-1)^m~V^+_{m,n})
\,,\label{226}\ee
 \be
(H^0,H^{\pi},V^+_{m,n})\to (H^{\pi},H^{0},(-1)^n~V^+_{m,n})
\,.\label{227}\ee
When these transformations are combined with the extra symmetry that
changes the sign of the
twist fields
\be
(H^0,H^{\pi},V^+_{m,n})\to (-H^0,-H^{\pi},V^+_{m,n})
\,,\label{228}\ee
they generate the group $D_4$, which is the invariance group of the
orbifold.

The orbifold theory depends also on a continuous parameter, the radius
$R$.
Moreover, we also have here the duality symmetry $R\to 1/R$, since
from
(\ref{222}):
\be
Z^{\rm orb}(R)=Z^{\rm orb}(1/R)
\,.\label{229}\ee

\vskip .4cm
\noindent\hrulefill
\nopagebreak\vskip .2cm
{\large\bf Exercise}: Use the OPEs in (\ref{223})-(\ref{225}) to deduce
the following transformation rule for the twist fields under $R\to 1/R$
duality,
\be
\left(\matrix{H^0\cr H^{\pi}\cr}\right)\to {1\over
\sqrt{2}}\left(\matrix{1&1\cr 1&-1\cr}\right)\left(\matrix{H^0\cr
H^{\pi}\cr}\right)\;.
\ee
\vskip .4cm

{\large\bf Exercise}. Consider further ``orbifolding" the orbifold
theory by the $Z_2$
transformation in (\ref{228}). Show that the resulting theory is the
original toroidal theory. In this respect the toroidal theory is
not any more fundamental than the orbifold one.

\vskip .4cm

{\large\bf Exercise}. Show that when $R=\sqrt{2}$ the orbifold
partition function
becomes the square of the Ising partition function
\be
Z^{\rm Ising}={1\over 2}\left[\left|{\th_2\over \eta}\right|+
\left|{\th_3\over \eta}\right|+\left|{\th_4\over \eta}\right|
\right]
\,,\label{230}\ee
which was computed in (\ref{187}). Here N=1.
You will also need (\ref{t13}).
\nopagebreak\vskip .2cm
\noindent\hrulefill
\vskip .4cm

The orbifold from above can be easily generalized in various
directions.
First we can consider other starting CFTs, such as higher-dimensional
tori or interacting CFTs. Moreover the symmetry we mod out can
be a bigger abelian
or non-abelian discrete group.
We will not delve further in this direction for the moment.

We will now discuss a simple example of a freely acting orbifold
group.
We will start again from the theory of a scalar on a circle of radius
$R$.
However, here we will use a $Z_2$ subgroup of the U(1) symmetry
that
acts as $|m,n\rangle\to (-1)^m~|m,n\rangle$ and leaves the
oscillators invariant.
The geometrical action is a half-lattice shift: $X\to X+\pi~R$.
We will calculate the partition function using the same method as
above.
It will be written again in the form (\ref{231}) with $Z[^0_0]=Z(R)$.
$Z[^0_1]$ must contain the group element:
\be
Z[^0_1]=\sum_{m,n\in \Z}~(-1)^m~{\exp\left[{i\pi\tau\over
2}\left({m\over R}+nR\right)^2-{i\pi\bar \tau\over 2}\left({m\over
R}-nR\right)^2\right]
\over \eta\bar\eta}
\,.\label{233}\ee
The computation of $Z[^1_0]$ can be made by noting that the twisted
boundary
condition
is similar to that of a circle of half the radius, so that $n\to
n+1/2$,
or by performing a $\tau\to-1/\tau$ transformation on $Z[^0_1]$.
Both methods give
\be
Z[^1_0]=\sum_{m,n\in \Z}~{\exp\left[{i\pi\tau\over 2}\left({m\over
R}+\left(n+{1\over 2}\right)R\right)^2-{i\pi\bar \tau\over
2}\left({m\over R}-\left(n+{1\over 2}\right)R\right)^2\right]\over
\eta\bar\eta}
\,.\label{234}\ee
Finally $Z[^1_1]$ can be obtained from $Z[^1_0]$ by a $\tau\to\tau+1$
transformation or by inserting the group element:
\be
Z[^1_1]=\sum_{m,n\in \Z}~(-1)^m~{\exp\left[{i\pi\tau\over
2}\left({m\over R}+\left(n+{1\over 2}\right)R\right)^2-{i\pi\bar
\tau\over 2}\left({m\over R}-\left(n+{1\over
2}\right)R\right)^2\right]\over \eta\bar\eta}
\,.\label{237}\ee
We can summarize the above by
\be
Z[^h_g]=\sum_{m,n\in \Z}~(-1)^{gm}~{\exp\left[{i\pi\tau\over
2}\left({m\over R}+\left(n+{h\over 2}\right)R\right)^2-{i\pi\bar
\tau\over 2}\left({m\over R}-\left(n+{h\over
2}\right)R\right)^2\right]\over \eta\bar\eta}
\label{238}\ee
or, in the Lagrangian representation, by
\be
Z[^h_g]={R\over \sqrt{\tau_2}\eta\bar\eta}\sum_{m,n,\in\Z}
{}~\exp\left[-{\pi R^2\over \tau_2}\left|m+{g\over 2}+\left(n+{h\over
2}\right)
\tau\right|^2\right]
\,.\label{239}\ee

Summing up the contributions as in (\ref{231}) we obtain, not to our
surprise,
the partition function for a boson compactified on a circle of radius
$R/2$.
This is what we would have expected from the geometrical action of
the orbifold element.
Note also that here the twisted sectors have a non-trivial dependence
on the radius.  This is a generic feature of freely acting orbifolds.

Although this orbifold example looks trivial, it can be combined with
other projections to make non-trivial orbifold CFTs.

\vskip .4cm
\noindent\hrulefill
\nopagebreak\vskip .2cm
{\large\bf Exercise}. Consider the CFT of a two-dimensional torus,
which is a direct product of two circles of radii $R_{1,2}$ and
coordinates $X_{1,2}$.
This theory has, among others, the $Z_2$ symmetry, which acts
simultaneously as $X_1\to -X_1$ and $X_2\to X_2+\pi R_2$.
It is a freely acting symmetry.
Construct the orbifold partition function.
\nopagebreak\vskip .2cm
\noindent\hrulefill
\vskip .4cm

We will comment here on the most general orbifold group of a toroidal
model.
The generic symmetry of a d-dimensional toroidal CFT contains the
$\rm U(1)^d_L\times U(1)_R^d$ chiral symmetry. The transformations
associated with it are arbitrary lattice translations.
They act on a state with  momenta $m_i$ and windings $n_i$ as
\be
g_{\rm translation}= \exp\left[2\pi
i\sum_{i=1}^d(m_i\theta_{i}+n_i\phi_{i})\right]
\,,\label{240}\ee
where $\theta_i,\phi_i$ are rational in order to obtain a discrete
group.
There are also symmetries that are subgroups of the O(d,d) group
not
broken by the moduli $G_{ij}$ and $B_{ij}$. These depend on the point
of the moduli space.
Consequently, the generic element is a combination of a translation and
a
rotation
acting on the left part of the theory and an a priori different
rotation and
translation acting on the right part of the theory.

\vskip .4cm
\noindent\hrulefill
\nopagebreak\vskip .2cm
{\large\bf Exercise}. Consider the CFT of the product of two circles
with equal
radii. It is invariant under the interchange of the two circles.
This transformation forms a $Z_2$ subgroup of the rotation group
O(2).
Orbifold by this symmetry and construct the orbifold blocks of the
partition function.
Is the partition function modular-invariant? You will need
(\ref{tu14}).
\nopagebreak\vskip .2cm
\noindent\hrulefill
\vskip .4cm

There are constraints imposed by modular invariance that restrict
the choice of orbifold groups.
The orbifolding procedure can be viewed as a gauging of a discrete
symmetry.
It can happen that the discrete symmetry is anomalous.
Then, the theory will not be modular-invariant.

\vskip .4cm
\noindent\hrulefill
\nopagebreak\vskip .2cm
{\large\bf Exercise}. Redo the freely acting orbifold of a free
scalar, but now use
the following group element: $g=(-1)^{m+n}$. It corresponds to a
non-geometric translation.
Show that it is impossible to construct a modular-invariant partition
function.
Thus, this is an anomalous symmetry, something to be expected since
it corresponds to a gauging of a $Z_2$ subgroup of the chiral
$\rm U(1)_L$.
\nopagebreak\vskip .2cm
\noindent\hrulefill
\vskip .4cm

\begin{figure}
\begin{center}
\leavevmode
\epsfxsize=13cm
\epsffile{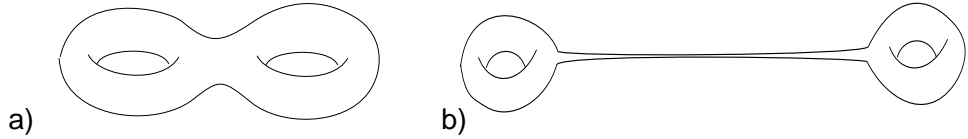}
\end{center}
\caption[]{\it a) The double torus. b) The degeneration limit into two
tori.}
\label{f9}\end{figure}

\subsection{CFT on higher-genus Riemann surfaces}
\setcounter{equation}{0}

So far we have analyzed CFT on surfaces of low genus, namely the
Riemann
sphere $(g=0)$ and the torus ($g=1$).
Similarly, we can define and analyze various CFTs on surfaces with more
handles.
A general N-point function on a genus $g\geq 2$ Riemann surface
depends on the N (complex) positions of the operators and on $3(g-1)$
complex numbers that are the moduli of the surface.
They are the generalizations of the modulus $\tau$ of the torus.
There is also the notion of a modular group that acts on the moduli.
The partition function must be invariant under the modular
group of a genus $g$ surface with N punctures.

\begin{figure}
\begin{center}
\leavevmode
\epsffile{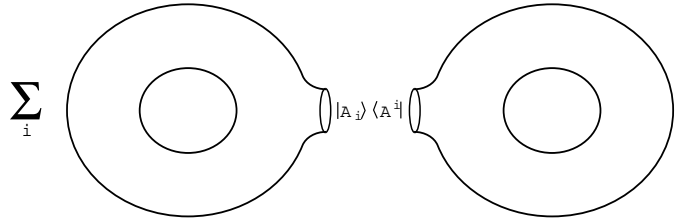}
\end{center}
\caption[]{\it The Hamiltonian description of its
degeneration into a pair of tori.}
\label{f99}\end{figure}

There is a set of relations, however, between correlation functions
of the same CFT defined on various Riemann surfaces.
This is known as factorization.
Consider as an example the partition function of a CFT on a genus-2
surface depicted in Fig. \ref{f9}a.
It depends on three complex moduli.
In particular there is a modulus, which we will denote by $q$, such
that
as $q\to 0$ the surface develops a long cylinder in between and, at
$q=0$, degenerates into two tori with one puncture each (Fig.
\ref{f9}b).
This implies a Hamiltonian degeneration formula for the partition
function
\be
\langle~1~\rangle_{g=2}=\sum_{i}q^{h_i-c/24}~\bar q^{\bar h_i-\bar
c/24}~\langle \phi_i\rangle
_{g=1}~\langle \phi_i\rangle_{g=1}
\,,\label{216}\ee
where the sum is over all states of the theory and the one-point
functions
are evaluated on the once-punctured tori.
This happens because as the intermediate cylinder becomes long we can
use the cylinder Hamiltonian to describe this part of the theory.
Equation (\ref{216}) is schematically represented in Fig. \ref{f99}.
This can be generalized to arbitrary correlation functions and
arbitrary
degenerations.

Factorization is important since it will imply perturbative unitarity
in the underlying string theory.
For example a $g=2$ amplitude is a two-loop correction to a
scattering amplitude and we should be able to construct it, in
perturbation theory,
by sewing one-loop amplitudes.

\renewcommand{\theequation}{\thesection.\arabic{equation}}
\section{Scattering amplitudes and vertex operators of bosonic
strings\label{vertex}}
\setcounter{equation}{0}

We have seen in the previous chapter that to each state in the CFT
there corresponds a local operator that creates the respective state
out of the SL(2,$\C)$-invariant vacuum.
So to all states in string theory there correspond local operators
on the world-sheet.
However, we need only consider physical states. How do the physical
state
conditions translate to local operators?

We will work in the old covariant approach and consider the case
of closed strings.
We have seen that physical states had to satisfy $L_0=\bar L_0=1$
and they should be annihilated by the positive modes of the Virasoro
operators. In CFT language they have to be primary of conformal
weight (1,1).
Moreover, from their definition, spurious states
correspond to Virasoro descendants.

The Polyakov action in the conformal gauge is
\be
S_P={1\over 4\pi\alpha'}\int d^2\sigma~\pd X^{\mu}\bar \pd
X^{\nu}\eta_{\mu\nu}
\,,\label{241}\ee
from which we obtain the two-point function
\be
\langle X^{\mu}(z,\bar z)X^{\nu}(w,\bar w)\rangle=-{\alpha'\over
4}\eta^{\mu\nu}\log|z-w|^2
\label{242}\ee
and the stress-tensor
\be
T=-{2\over \alpha'}\eta_{\mu\nu}\pd X^{\mu}\pd X^{\nu}
\label{243}\ee
and similarly for $\bar T$.

The states at level zero $|p\rangle $ correspond to the
operators
$V_{p}=:e^{ip^{\mu}X^{\mu}}:$.
Under $T,\bar T$ they are primary
of  conformal weights $\Delta=\bar\Delta=\alpha' p^2/4$.
In order for them to be (1,1), and thus physical, we need
$p^2=-m^2=4/\alpha'$,
which is the tachyon mass-shell condition.

The next set of states is $a^{\mu}_{-1}\bar a^{\nu}_{-1}|p\rangle$
and corresponds to the operators
\linebreak[4]
$:\pd X^{\mu}\bar\pd X^{\nu} V_p:$.
We
consider the linear combination
\mbox{$O(\epsilon)=\epsilon_{\mu\nu} :\pd X^{\mu}\bar\pd X^{\nu} V_p:$}
of these operators and compute their OPE with $T,\bar T$:
\be
T(z)O(w,\bar w)=
-ip^{\mu}\epsilon_{\mu\nu}{\alpha'\over  4}{\bar \pd x^{\nu}V_p\over
(z-w)^3}
+\left(1+{\alpha'p^2\over 4}\right){O(w, \bar w)\over (z-w)^2}
+{\pd_w O(w,\bar w)\over z-w}+\dots
\label{244}\ee
and a similar one for $\bar T$.
In order to have a primary (1,1) operator we must have the
third-order pole vanish
\be
p^{\mu}\epsilon_{\mu\nu}=p^{\nu}\epsilon_{\mu\nu}=0
\label{245}\ee
and $p^2=0$, which are the mass-shell and transversality conditions
for the graviton ($\epsilon $ symmetric and traceless), the
antisymmetric tensor
($\epsilon $ antisymmetric) and the dilaton
($\epsilon_{\mu\nu}\sim\eta_{\mu\nu}-p_{\mu}\bar p_{\nu}-\bar
p_{\mu}p_{\nu}$
with $\bar p^2=0$, $p\cdot\bar p=1$).
Higher levels work in a similar fashion.

In modern (BRST) covariant quantization the physical state condition
translates into $[Q_{BRST},V_{phys}(z,\bar z)]=0$, which reduces to
the usual condition on physical states.
In this case the physical vertex operators are the ones we found in
the old covariant case multiplied by $c\bar c$.

$N$-point scattering amplitudes ($S$-matrix elements) on the sphere
are constructed by calculating the appropriate $N$-point correlator of
the associated vertex operators
and integrating it over the positions of the insertions.
As we mentioned before, there is a residual SL(2,$\C)$ invariance on
the sphere
that was not fixed by going to the conformal gauge.
This can be used to set the positions of three vertex operators to
three
fixed points taken conventionally to be $0,1,\infty$.
Then we integrate over the remaining N-3 positions.
See \cite{GSW} for explicit calculations of tree (sphere) scattering
amplitudes.

Moving to one-loop diagrams we have a similar prescription.
For an $N$-point one-loop amplitude we first have to calculate the
$N$-point function of the appropriate vertex operators on the torus.
Due to translational symmetry of the torus ($c,\bar c$ zero modes)
the correlator depends on N-1 positions as well as on the modulus of
the
torus $\tau$.
These are moduli, and they should be integrated over.
Diffeomorphism invariance implies that the correlator integrated over
the N-1 positions should
be invariant under modular transformation.
Finally we have to integrate over $\tau$ in the fundamental domain
(Fig. \ref{f8}).

We will calculate the one-loop vacuum energy of the closed bosonic
string.
This is the one-loop bubble diagram, and corresponds to calculating the
torus
partition function of the underlying CFT and integrating over the
torus moduli space.
We will do this computation in the covariant approach.
We have seen that the torus partition function for a single
non-compact boson
is given by  $1/(\sqrt{\tau_2}\eta\bar\eta)$, and we have 26 of these.
The $b,c$ ghosts contribute $\eta^2$ to the partition function and
cancel the contribution of two left-moving oscillators. Similarly the
$\bar b,\bar c$
ghosts contribute $\bar \eta^2$.
Finally the integration measure contains an integral over $\tau_1$,
which imposes $L_0=\bar L_0$ and the usual Schwinger measure
$d\tau_2/\tau_2$.
Putting everything together, we obtain
\be
\Lambda^4=\int_{\cal F}{d^2\tau\over \tau_2^2}~Z_{\rm
bosonic}(\tau,\bar\tau)=
 \int_{\cal F}{d^2\tau\over \tau_2^2}~{1\over
(\sqrt{\tau_2}\eta\bar\eta)^{24}}
\,,\label{246}\ee
where $\cal F$ is the fundamental domain.
Note that $Z_{\rm bosonic}$ is just the contribution of the 24
transverse
non-compact coordinates.
In the light-cone gauge a similar calculation would include
$Z_{\rm bosonic}$ for the transverse coordinates, an extra $1/\tau_2$
factor
from the light-cone zero modes and the Schwinger measure giving again
the same result as in (\ref{246}).

In field theory, the $\tau_2$ integration extends down to $\tau_2=0$
and this is the region where UV divergences come from.
We see that in string theory this region is absent, due to modular
invariance.
This provides a (technical)  explanation for the absence of UV
divergences in string theory.
Of course in the case of bosonic strings the vacuum energy is IR
divergent
due to the presence of the tachyon.

In a similar fashion, g-loop diagrams can be calculated by integrating
correlators on the CFT on a genus-g Riemann surface.
One final ingredient is the string coupling $g_{\rm string}$.
A g-loop contribution has to be additionally weighted by a factor
$g_{\rm string}^{-\chi}$, where $\chi=2(1-g)$ is the Euler number of
the
Riemann surface.
The perturbative expansion is a topological expansion.
Notice also that the insertion of a vertex operator creates an
infinitesimal
hole in the Riemann surface and increases its Euler number by 1.
It is thus accompanied by a factor of $1/\gs$, as it should.

\begin{figure}
\begin{center}
\leavevmode
\epsfxsize=13cm
\epsffile{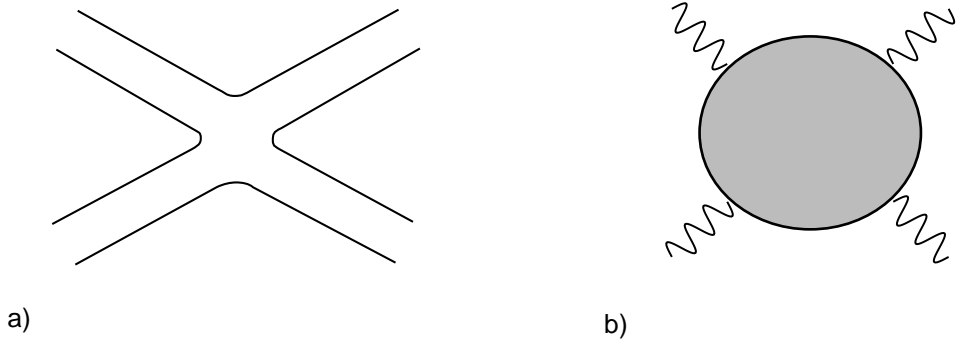}
\end{center}
\caption[]{\it a) The four-point open string tree amplitude. b)
Its conformal transform to a disk with four points marked
on the boundary.}
\label{f11}\end{figure}

We will briefly describe here the topological expansion for the case
of
open strings.
A tree-level four-point diagram in this case is shown in Fig.
\ref{f11}a.
By a conformal transformation it can be mapped to a disk with four
points marked on the boundary (Fig. \ref{f11}b). These are the
positions
of insertion of the  appropriate vertex operators.
Thus, the open string vertex operators are inserted at the boundary of
the surface.
Open strings can also emit closed strings. In such amplitudes, closed
string emission is represented by the insertion of closed string vertex
operators
in the interior of the surface.

\begin{figure}
\begin{center}
\leavevmode
\epsfxsize=13cm
\epsffile{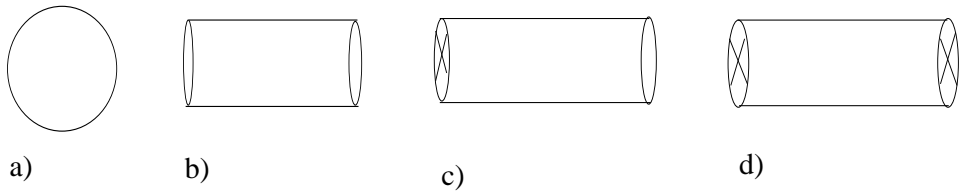}
\end{center}
\caption[]{\it a) The disk. b) The annulus. c) The M\"obius strip.
d) the Klein bottle.}
\label{f12}\end{figure}

Here the topological expansion also includes Riemann surfaces with
boundary.
Moreover, we can consider orientable strings (where the string is
oriented) as
well as non-orientable strings (where the orientation of the string is
immaterial).
In the second case we will have to include non-orientable Riemann
surfaces
in the topological expansion.
Such a surface is characterized by the number of handles $g$
the number of boundaries $B$, and the number of cross-caps $C$ that
introduce the non-orientability of the surface.
A cross-cap is a boundary that  instead of being $S^1$, is
$S^1/Z_2=RP^1$.
The Euler number is given by
\be
\chi=2(1-g)-B-C
\,.\label{247}\ee
In Fig. \ref{f12} we show the four simplest surfaces with
boundaries:
the disk with $\chi=1$, the annulus with $\chi=0$, as well as two
non-orientable surfaces, the M\"obius strip
with $\chi=0$ and the Klein bottle with $\chi=0$.

As we will see later on, consistent theories of open unoriented strings
necessarily include couplings to closed unoriented strings. An easy way
to see this is to consider the annulus diagram (Fig. \ref{f12}b).
If we take time to run upwards, then it describes a one-loop diagram
of an open string.
If, however, we take time to run sideways then it describes the
tree-level propagation of a closed string.

\renewcommand{\theequation}{\thesection.\arabic{equation}}
\section{Strings in background fields and low-energy effective actions\label{background}}
\setcounter{equation}{0}

So far, we have described the propagation of strings in flat
26-dimensional
Minkowski space. We would like, however, to be able to describe string
physics when
the massless fields $G_{\mu\nu}$, $B_{\mu\nu}$ and $\Phi$ have
non-trivial
VEVs.
This can be done by a finite perturbation of the flat CFT, using the
vertex
operators for the massless fields. The  correct prescription for
coupling the dilaton was given by  Fradkin and Tseytlin.
The Polyakov action becomes\footnote{We are discussing here closed
oriented
strings.}
\bea
S_P={1\over 4\pi\alpha'}\int d^2\xi\left[\sqrt{g}g^{ab}G_{\mu\nu}(X)
+\epsilon^{ab}B_{\mu\nu}(X)\right]\pd_a X^{\mu}\pd_b X_{\nu}
+\nn\\ +{1\over 8\pi}\int d^2\xi\sqrt{g}R^{(2)}\Phi(X)
\,,\label{248}\eea
where $R^{(2)}$ is the scalar curvature of the intrinsic word-sheet
metric
$g_{ab}$.
An interesting observation is the following:
consider the constant part of the dilaton field (VEV) $\Phi_0$.
Since the Euler character of the world-sheet is given by
\be
\chi={1\over 4\pi}\int \sqrt{g}~R^{(2)}
\,,\label{256}\ee
we observe that  we have a factor
$e^{-\chi\Phi_0/2}$ in front of $e^{-S_{P}}$.
Thus, the string coupling is essentially given by the dilaton VEV
\be
g_{\rm string}=e^{\Phi_0/2}
\,.\label{257}\ee

For general backgrounds, this $\sigma$-model is not conformally
invariant.
Rather
\be
{{T_{a}}^{a}\over \sqrt{g}}={\beta^{\Phi}\over
96\pi^3}R^{(2)}+{1\over 2\pi}(\beta_{\mu\nu}^G
{}~g^{ab}+\beta_{\mu\nu}^B\epsilon^{ab})
\pd_a X^{\mu}\pd_b X_{\nu}
\,,\label{249}\ee
where the $\beta$-functions can be obtained perturbatively in the weak
coupling expansion of the $\s$-model, $\alpha'\to 0$.
To leading non-trivial order,
\be
{\beta_{\mu\nu}^{G}\over \alpha'}= R_{\mu\nu}-{1\over
4}H_{\mu\rho\s}{H_{\nu}}^{\rho\s}+\nabla_{\mu}\nabla_{\nu}\Phi
+{\cal O}(\alpha')
\,,\label{250}\ee
\be
{\beta_{\mu\nu}^B\over \alpha'}=
\nabla^{\mu}\left[e^{-\Phi}~H_{\mu\nu\rho}\right]
+{\cal O}(\alpha')
\label{251}\ee
and
\be
\beta^{\Phi}=D-26+3\alpha'\left[(\nabla\Phi)^2-2\square
\Phi-R+{1\over 12}H^2\right]+{\cal O}(\alpha'^2)
\,,\label{252}
\ee
where $H_{\mu\nu\rho}$ is the totally antisymmetric field strength of
$B_{\mu\nu}$,
\be
H_{\mu\nu\rho}=\pd_{\mu}B_{\nu\rho}+\pd_{\nu}B_{\rho\mu}+\pd_{\rho}
B_{\mu\nu}
\label{254}\ee
and $D$ is the spacetime dimension.

When $G,B,\Phi$ are such that $\beta^G_{\mu\nu}=\beta^B_{\mu\nu}=0$,
then the $\s$-model describes a CFT with central charge
$c=\beta^{\Phi}$.
It can be shown in particular that $\beta^{\Phi}$ is a constant when
the other $\beta$-functions vanish.

The conditions for conformal invariance and thus consistent string
propagation are given by the equations
\be
\beta^{\Phi}=\beta^G_{\mu\nu}=\beta^B_{\mu\nu}=0
\,.\label{253}\ee

These conditions are second-order equations for the background fields
and can be obtained from an action
\be
\alpha'^{D-2}S^{tree}\sim \int d^D x\sqrt{-{\rm det}~G}e^{-\Phi}
\left[R+(\nabla\Phi)^2-{1\over 12}H^2+{D-26\over 3}\right] +{\cal
O}(\alpha')
\,.\label{255}\ee

At this point it is useful to pause and ask the question: Why do we
need conformal invariance?
After all, since we are integrating over two-dimensional metrics
we expect conformal invariance to be enforced by the integration.
In fact we will consider two distinct cases:

$\bullet$ The background fields and spacetime dimension are such that
all $\beta$-functions vanish.
Then we have conformal invariance at the quantum level, the conformal
factor decouples, and we have to factor out its volume from the
definition
of the path integral.
This is the case we considered so far.
Moreover, the vanishing of the $\beta$-functions implies, to leading
order in $\alpha'$, second-order equations for the background fields.

$\bullet$ The $\beta$-functions are not zero. Then the conformal
factor
does not decouple. Since ${T_{a}}^a\sim \delta_{\phi} \log Z$,
where $\phi$ is the conformal factor, we can solve this equation to
derive the dependence of the effective action on $\phi$.
Apart from the Liouville action discussed in (\ref{34}), we will also
have
couplings of $\phi$ to the scalars $X^{\mu}$ if $\beta^{B,G}$ do not
vanish.
Effectively we have a new $\s$-model in $D+1$ dimensions ($\phi$
provides the extra coordinate), which is by construction
Weyl-invariant.\footnote{There are subtleties having to do with the
signature of the extended spacetime.
We refer the interested reader to \cite{Pol}.}
Thus, we are back to the first case.

We will now consider the notion of the effective action for the light
fields.
We have seen that (ignoring the tachyon for the moment)
the massless fields are $G,B,\Phi$.
All other particles have masses of the order of the string scale.
We can imagine integrating out the heavy fields that will induce
corrections to the action of the light fields.
This is the definition of the  low-energy effective action.
This effective action contains only the light fields and is valid
up to energies of the order of the mass of the heavy fields.
At tree level, this procedure can be implemented by considering the
full
(on-shell) scattering amplitudes of the light fields from string
calculations, expanding them in $\alpha'$ and finding the extra
interactions
induced on the light fields.
A comprehensive exposition of this procedure can be found in
\cite{gs}.
Since the amplitudes used are on-shell, the effective action can be
calculated up to terms that vanish by using the equations of motion.

It can be shown that, up to terms that vanish on-shell, the $\s$-model
conformal invariance conditions and the string amplitude calculations
produce the same low-energy effective action (\ref{255}).

In (\ref{255}) the fields that appear are those that couple to the
string
$\s$-model. This is known as the ``string frame".
In this frame, the kinetic terms of the metric $G$ and the dilaton are
not diagonal. They become diagonal in the ``Einstein frame", related
to the string frame by a conformal rescaling of the metric.
Separating the expectation value
of the dilaton $\Phi\to \Phi_0+\Phi$ and defining the Einstein metric
as
\be
G^{E}_{\mu\nu}=e^{-{2\Phi\over D-2}}G_{\mu\nu}
\,,\label{258}\ee
we obtain the action in the Einstein frame:
\bea
S_E^{\rm tree}\sim {1\over \kappa^2}\int d^D x\sqrt{G^E}
\left[R-{1\over D-2}(\nabla\Phi)^2-{e^{-4\Phi/(D-2)}\over
12}H^2 \right. \nn\\ \left. +e^{2\Phi/(D-2)}{D-26\over 3}\right] +{\cal O}(\alpha')
\,,\label{259}\eea
where the gravitational constant is given by
\be
\kappa= g_{\rm string}\alpha'^{(D-2)/2}
\,.\label{260}\ee

\renewcommand{\theequation}{\thesection.\arabic{equation}}
\section{Superstrings and supersymmetry\label{super}}
\setcounter{equation}{0}

We have seen so far that bosonic strings suffer from two major
problems:

$\bullet$ Their spectrum always contains a tachyon. In that respect
their
vacuum is unstable.

$\bullet$ They do not contain spacetime fermions.

On the other hand, we have already seen, during our study of free
fermion CFTs that
they contain states that transform as spinors under the associated
orthogonal symmetry.
Therefore we should be willing to add free fermions on the world-sheet
of the
string in order to obtain states that transform as spinors.
These fermions should carry a spacetime index, i.e. $\psi^\mu$, $\bar
\psi^{\mu}$,
in order for the spinor to be a spacetime spinor.
However, in such a case there will be additional negative norm states
associated with the modes of $\psi^0$.
In order for these to be removed from the physical spectrum, we need
more
constraints than the Virasoro constraints alone.
The appropriate result comes from considering an N=1 superconformal
algebra of
constraints.
In the bosonic case, we started with two-dimensional gravity coupled to
$D$ scalars
$X^{\mu}$
on the world-sheet, which eventually boiled down to a set of Virasoro
constraints on the Hilbert space.
Here, we would like to start from the two-dimensional N=1 supergravity
coupled to $D$
N=1 superfields, each containing a bosonic coordinate $X^{\mu}$ and
two fermionic coordinates, one left-moving $\psi^{\mu}$ and one
right-moving $\bar \psi^{\mu}$.
The two-dimensional  N=1 supergravity multiplet contains the  metric
and a
gravitino $\chi_a$.

The analog of the bosonic Polyakov action is
\bea
S^{II}_{P}={1\over
4\pi\alpha'}\int\sqrt{g}\left[g^{ab}\pd_{a}X^{\mu}\pd_{b}X^{\mu}
+{i\over 2}\psi^{\mu}\dslash \psi^{\mu}+\quad\quad\quad\quad \right.
\nn\\ \left. {i\over 2}(\chi_a\gamma^b\gamma^a\psi^{\mu})\left(\pd_{b}X^{\mu}-{i\over
4}\chi_b\psi^{\mu}\right)
   \right]
\,.\label{261}\eea
It is invariant under local N=1 left-moving (1,0) supersymmetry
\be
\delta g_{ab}=i\epsilon (\gamma_{a}\chi_{b}+\gamma_{b}\chi_{a})
\;\;\;,\;\;\;\delta \chi_a=2\nabla_a \epsilon
\,,\label{262}\ee
\be
\delta X^{\mu}=i\epsilon \psi^{\mu}\;\;\;,\;\;\;\delta\psi^{\mu}=
\gamma^a\left(\pd_a X^{\mu}-{i\over
2}\chi_a\psi^{\mu}\right)\epsilon\;\;\;,\;\;\;\delta\bar\psi^{\mu}=0
\,,\label{263}\ee
where $\epsilon$ is a left-moving Majorana-Weyl spinor.
There is a similar right-moving (0,1) supersymmetry involving
a right-moving Majorana-Weyl spinor $\bar \epsilon$ and the fermions
$\bar \psi^{\mu}$.
In our notation we have (1,1) supersymmetry.

The analog of the conformal gauge is the superconformal gauge
\be
g_{ab}=e^{\phi}\delta_{ab}\;\;\;,\;\;\;\chi_a=\gamma_a \zeta
\,,\label{264}\ee
where $\zeta$ is a constant Majorana spinor;
$\phi$ and $\zeta$ decouple from the classical action (\ref{261}).
Apart from the Virasoro operators we also have the supercurrents
\be
G_{\rm matter}=i\psi^{\mu}\pd X^{\mu}\;\;\;,\;\;\;\bar G_{\rm
matter}=i\bar
\psi^{\mu}
\bar \pd X^{\mu}\,.\label{274}\ee
We  also have to introduce the appropriate ghosts. We will still have
the usual $b,c$ system with $\lambda=2$ associated with
diffeomorphisms,
but now we also need  a commuting set of ghosts $\beta,\gamma$
with $\epsilon=-1$, $\lambda=\tfrac{3}{2}$ associated with the
supersymmetry.
Superconformal invariance will be present at the quantum level,
provided
the ghost central charge cancels the matter central charge.
Each bosonic and fermionic coordinate contributes $3/2$ to the
central charge.
Since we have $D$ of them, the matter central charge is $c_{\rm
matter}=\tfrac{3}{2} D$.
The $b,c$ system contributes $-26$ to the central charge while the
$\beta,\gamma$
system contributes $+11$.
The total central charge vanishes, provided that $D=10$.

The classical constraints imply the vanishing of $T,G,\bar T,\bar G$.
Consequently, we have enough constraints to remove the negative norm
states.

The BRST current is \cite{FMS}
\be
j_{BRST}=\gamma G_{\rm matter}+cT_{\rm matter}+{1\over 2}\left(
cT_{\rm ghost}+\gamma G_{\rm ghost} \right)
\,,\label{265}\ee
where
\be
G_{\rm matter}=i\psi^{\mu}\pd X^{\mu}\;\;\;,\;\;\;T_{\rm matter
}=-{1\over 2}\pd X^{\mu}\pd X^{\mu}-{1\over 2}\psi^{\mu}\pd
\psi^{\mu}
\,,\label{266}\ee
\be
G_{\rm ghost}=-i(c~\pd \beta-{1\over 2}\gamma ~b+{3\over 2}\pd c
{}~\beta)
\;\;\;,\;\;\;T_{\rm ghost}=T_{bc}-{1\over 2}\gamma~\pd \beta-{3\over
2}\pd \gamma~\beta
\,.\label{267}\ee

\vskip .4cm
\noindent\hrulefill
\nopagebreak\vskip .2cm
{\large\bf Exercise}. Verify that $G_{\rm ghost}$ and $T_{\rm ghost}$
satisfy the OPEs of the N=1 superconformal algebra
(\ref{TT}),(\ref{203}) with the correct central charge.
\nopagebreak\vskip .2cm
\noindent\hrulefill
\vskip .4cm

The BRST charge is
\be
Q={1\over 2\pi i}\left[\oint dz~j_{BRST}+\oint d\bar z~\bar
j_{BRST}\right]
\,.\label{268}\ee
It is nilpotent for $D=10$ and can be used in the standard way to
define physical states.

\renewcommand{\theequation}{\thesubsection.\arabic{equation}}
\subsection{Closed (type-II) superstrings\label{typeII}}
\setcounter{equation}{0}

We will consider first the closed (type-II) superstring case.
We will work in a physical gauge and derive the spectrum.
The  analogue of the light-cone gauge in the supersymmetric case
is\footnote{
There is a subtlety here concerning the super-light-cone gauge.
If $\psi^+$ for example has $NS$ boundary conditions, then it can be
set
to zero.
If it has $R$ boundary conditions, then it can be set to zero except
for its
zero mode. A similar remark applies to $\bar\psi^+$.}:
\be
X^+=x^++p^+\tau\;\;\;,\;\;\;\psi^+=\bar\psi^+=0
\,.\label{269}\ee
As in the bosonic case, we can explicitly solve the constraints by
expressing
$X^-,\psi^-,\bar\psi^-$ in terms of the transverse modes.
Then, the physical states can be constructed out of the transverse
bosonic
and fermionic oscillators.
However, all zero modes are present.

As we mentioned before, we have two left-moving sectors corresponding
to $NS$ and $R$ boundary conditions for $\psi^{\mu}$ and $G$ and
another two sectors
corresponding to $\overline{NS}$ and $\bar R$ boundary conditions for
$\bar\psi^{\mu}$ and $\bar G$.

For the moment we will discuss only the left sector to avoid
repetition.
We will introduce as usual the modes $L_n$ and $G_r$ of the
superconformal generators\footnote{Remember that in the light-cone
gauge there are no
ghosts and only transverse (bosonic and fermionic) oscillators.}.

In the light-cone gauge we have solved the constraints, apart from
those associated with the zero modes.
In the $NS$ sector, $G$ is half-integrally modded and the only zero
mode
is $L_0$.
There is also a normal-ordering constant, which can be calculated
either
by demanding Lorentz invariance of the physical spectrum, as we have
done
for the bosonic string, or by realizing that in the covariant
formulation
the lowest ``energy'' state is not the usual vacuum $|0\rangle$ but
$c_{1}\gamma_{-1/2}|0\rangle$. Both approaches result in a
normal-ordering constant equal to $a = \half$,
and  $L_0-\half$ should be zero on physical states.
The state $|p\rangle$ is a physical state with $p^2=-m^2=2/\alpha'$,
and it is a tachyon.
The next states are of the form $\psi^i_{-1/2}|p\rangle$ and satisfy
$L_0=\half$ if $p^2=0$. These states are  massless.
However, we would prefer not to have a tachyonic state.
Since the tachyon has $(-1)^{F_L}=1$ we would like to impose the
extra constraint (GSO projection):
physical states in the $NS$ sector should have odd fermion
number.

In the $R$ sector we have two zero modes: $L_0$ and $G_0$.
The $L_0$ constraint is the same as in the $NS$ sector.
A quick look at the expression for $G_0$ in (\ref{274}) is enough to
convince us that there
can be no normal-ordering constant, and that  $G_0$ should be zero
on physical states.
On the other hand we know from the superconformal algebra
\be
0=\{G_0,G_0\}=2\left(L_0-{D-2\over 16}\right)
\,.\label{270}\ee
Compatibility with the $L_0$ constraint implies again that $D=10$.
The $R$ ground-states are spinors of O(10).
Consequently, these states satisfy the $L_0$ constraint.
Also remember that $G_0=\psi^{\mu}_0
a_{0}^{\mu}+2\sum_{n\not=0}^{\infty}\psi^i_n a^i_{-n}$.
As shown in section \ref{Nfermions}, the operator $\psi^{\mu}_0$ is
represented by $\gamma^{\mu}$ and $a_0^{\mu}$ by
$p_{\mu}$.
The other terms in $G_0$ do not contribute  to the ground-states.
$G_0=0$ implies the Dirac equation $p\!\!\!
/\equiv\gamma^{\mu}~p_{\mu}=0$.
Thus, the potentially massless states in the $R$ sector are a spinor
$S$ and a conjugate spinor $C$ of O(10) satisfying the massless Dirac
equation.
Under $(-1)^{F_L}$, $S$ has eigenvalue 1 and $C$ has $-1$.
All other states are built on these ground-states and are massive.
So far, there is no a priori reason to impose also a GSO projection in
the $R$
sector. As we  will see later on, one-loop modular invariance
will force us to do so. Anticipating this fact, we will also fix the
fermion parity in the $R$
sector. Since $(-1)^F=$ $plus$ or $minus$ is a matter of convention in
the $R$
sector,
we will allow both possibilities. We will only keep the $S$  or $C$
spinor ground-states, but not both.

A similar discussion applies to the right-moving sector.
Combining the two we have overall four sectors:

$\bullet$ ($NS$-$\overline{NS}$): These are bosons since they transform
in tensor representations of the rotation group. The projection here
is $(-1)^{F_L}
=(-1)^{F_R}=-1$.
The lowest states allowed by the constraint and the GSO projection
are of the
form $\psi^i_{-1/2}\bar \psi^j_{-1/2}|p\rangle$, they are massless
and correspond to a symmetric traceless tensor (the graviton), an
antisymmetric tensor and a scalar. The tachyon is gone!

$\bullet$ ($NS$-$\overline R$): These are fermions.
The GSO projection here is
$(-1)^{F_L}=-1$
and by convention we keep the $S$ representation in the $\bar R$
sector.
The lowest-lying states, $\psi^i_{-1/2}|p,\bar S\rangle$ are massless
spacetime fermions
and contain a $C$ Majorana-Weyl gravitino and an $S$ fermion.

$\bullet$ ($R$-$\overline{NS}$): Here the GSO projection in
$\overline{NS}$
is $(-1)^{F_R}=-1$, but in the $R$ sector we have two physically
distinct
options:
keep the S spinor (type-IIB) or the $C$ spinor (type-IIA).
Again the lowest-lying states $\bar \psi^i_{-1/2}|p, S~{\rm
or}~C\rangle$
are massless spacetime fermions.

$\bullet$ ($R$-$\bar R$): In the IIA case the massless states are
$|S,\bar C\rangle$, which decomposes into a vector and a three-index
antisymmetric tensor, as will be shown below.
In type-IIB they are $|S,\bar S\rangle$,
which decomposes into a scalar, a two-index antisymmetric tensor, and a
self-dual four-index antisymmetric tensor.

There are also the bosonic oscillators for us to use but, as we have
seen, they are not
involved in the massless states. They do, however, contribute to
the massive spectrum.

Both type-IIA and IIB theories have two gravitini and are thus expected
to have N=2 local supersymmetry in ten dimensions.
In type-IIB  the gravitini have the same spacetime chirality, while
the two
spin $\half$ fermions have opposite chirality. Thus, the theory
is chiral.
The type-IIA theory is non-chiral since the gravitini and $\half$
fermions have opposite chiralities.

In the light-cone gauge, the left-over constraints are essentially the
linearized equations of motion.
In the $NS$-$\overline{NS}$ sector the constraints are
\be
L_0=\bar L_0\;\;\;,\;\;\;L_0-{1\over 2}=0\;;
\label{273}\ee
and, as we  have seen already in the bosonic case, it gives the
mass-shell condition.
This
corresponds to
the free Klein-Gordon equation.
The  $R$-$\overline{NS}$ sector contains spacetime fermions and the
constraints
are as in (\ref{273}), plus the $G_0=0$ constraint, which provides as
we showed
above, the Dirac equation both for massless and massive states.
Its square, from (\ref{270}), gives the Klein-Gordon equation; the
independent equations are thus $G_0=0$ and $L_0=\bar L_0$.
Similar remarks apply for the $NS$-$\overline{R}$ sector.
Finally in the $R$-$\overline{R}$ sector the states are bispinors and
they satisfy two Dirac equations: $G_0=\bar G_0=0$.
Ramond-Ramond massless states are special for two reasons. First,
they are always forms and always coupled to other states via
derivatives.
No perturbative states are charged under them.
As we will see later on, they are in the heart of non-perturbative
duality conjectures. A more detailed discussion of their properties can
be found in the next section.

We will  examine more closely the spacetime meaning of the
operators
$(-1)^{F_{L,R}}$.
In the $NS$ sector
\be
(-1)^F=\exp\left[i\pi \sum_{r\in Z+1/2}\psi^i_{r}\psi^i_{-r}\right]
\,.\label{271}\ee
In the $R$ sector
\be
(-1)^F=\prod_{\mu=0}^9\psi^{\mu}_0~\exp\left[i\pi \sum_{n=1}^{\infty}
\psi^i_{n}\psi^{i}_{-n}\right]=\Gamma^{11}~\exp\left[i\pi
\sum_{n=1}^{\infty} \psi^i_{n}\psi^{i}_{-n}\right]
\,,\label{272}\ee
where $\Gamma^{11}$ is the analog of $\gamma^5$ in ten dimensions.
We can deduce from the form of the supercurrents (\ref{274})
that the zero modes satisfy
\be
\{(-1)^{F_L},G_0\}=0\;\;\;,\;\;\;\{(-1)^{F_R},\bar G_0\}=0
\,.\label{275}\ee
These generalize the field theoretic relation
\be
\{\Gamma^{11},\dslash\}=0
\,.\label{276}\ee
Note that in string theory this equation holds also for the massive
Dirac
operator $G_0$.

\vskip .4cm
\noindent\hrulefill
\nopagebreak\vskip .2cm
{\large\bf Exercise}. Show that at the massless level there is an
equal number of on-shell fermionic and bosonic degrees of freedom.
This would be necessary if the theory has (at least one) spacetime
supersymmetry. Find also the physical states at the next (massive)
level
both in type-IIA and type-IIB theory.
Show that they combine into SO(9) representations, as they should,
and that there is again an equal number of fermionic and bosonic
degrees of freedom.
\nopagebreak\vskip .2cm
\noindent\hrulefill
\vskip .4cm

We will now study the one-loop vacuum amplitude (or partition
function).
For the bosonic part of the action we have eight transverse oscillators
and, in analogy
with  the case of the bosonic string, we will get a contribution of
$(\sqrt{\tau_2}\eta\bar\eta)^{-8}$.
Here, however, we also have the contribution of the world-sheet
fermions.
We will consider first the IIB case.
In the $NS$-$\overline{NS}$ sector the two GSO projections imply that
we
have the
vector on both sides.
So the contribution to the partition function is
$\chi_V\bar\chi_V$.
{}From the $R$-$\overline R$ sector we have projected out the $C$
representation
so the contribution is
$\chi_S\bar\chi_S$.
In the $R$-$\overline{NS}$ and $NS$-$\overline{R}$ sectors we obtain
$-\chi_{S}\bar\chi_V$ and $-\chi_{V}\bar\chi_S$ respectively.
The minus sign is there since spacetime fermions contribute with a
minus sign relative to spacetime bosons.
So
\be
Z^{IIB}={(\chi_V-\chi_S)(\bar\chi_V-\bar\chi_S)\over
(\sqrt{\tau_2}\eta~\bar\eta)^{8}}
\,.\label{277}\ee
Using the formulae (\ref{63}) and (\ref{80}) for the SO(8) characters,
we
can write the partition function as
\be
Z^{IIB}={1\over(\sqrt{\tau_2}\eta\bar\eta)^{8}}~{1\over 2}\sum
_{a,b=0}^1~(-1)^{a+b+ab}~{1\over 2}\sum_{\bar a,\bar b=0}^1~(-1)^
{\bar a+\bar b+\bar a\bar b}~{\th^4[^a_b]~\bar\th^4[^{\bar a}_{\bar
b}]\over \eta^4~\bar\eta^4}
\,,\label{278}\ee
$a=0$ labels the $NS$ sector, $a=1$ the $R$ sector and similarly for
the right-movers.

In the type-IIA case, the only difference is that $\chi_S$ should be
substituted by $\chi_C$.
The partition function becomes
 \be
Z^{IIA}={1\over(\sqrt{\tau_2}\eta\bar\eta)^{8}}~{1\over 2}\sum
_{a,b=0}^1~(-1)^{a+b}~{1\over 2}\sum_{\bar a,\bar b=0}^1~(-1)^
{\bar a+\bar b+\bar a\bar b}~{\th^4[^a_b]~\bar\th^4[^{\bar a}_{\bar
b}]\over \eta^4~\bar\eta^4}
\,.\label{279}\ee

\vskip .4cm
\noindent\hrulefill
\nopagebreak\vskip .2cm
{\large\bf Exercise}. Show that $Z^{IIB},Z^{IIA}$ are
modular-invariant.
Using (\ref{tt14}), show that they  also are  identically zero.
This implies that there is at each mass level an equal number of
bosonic and fermionic degrees of freedom, consistent with the
presence of spacetime supersymmetry.
\nopagebreak\vskip .2cm
\noindent\hrulefill
\vskip .4cm

\subsection{Massless $R$-$R$ states\label{RRR}}
\setcounter{equation}{0}

We will now consider  in more detail the massless $R$-$R$ states of
type-IIA,B string theory, since they have unusual properties and play a
central role
in non-perturbative duality symmetries. Further reading is to be found
in  \cite{bac}.

I will first start by describing in detail the $\Gamma$-matrix
conventions
in flat ten-dimensional Minkowski space \cite{GSW}.

The ($32\times 32$)-dimensional $\Gamma$-matrices satisfy
\be
\{\Gamma^{\m},\Gamma^{\n}\}=-2\eta^{\mu\nu}\;\;\;,
\;\;\;\eta^{\m\n}=(-++\ldots +)
\,.\label{GG1}\ee
The $\Gamma$-matrix indices are raised and lowered with the flat
Minkowski
metric $\eta^{\m\n}$:
\be
\Gamma_{\m}=\eta_{\m\n}\Gamma^{\n}\;\;\;\,\;\;\;
\Gamma^{\m}=\eta^{\m\n}\Gamma_{\n}\;.
\ee
We will be in the Majorana representation where the $\Gamma$-matrices
are
purely imaginary, $\Gamma^{0}$ is antisymmetric, the rest symmetric.
Also
\be
\Gamma^{0}\Gamma_{\m}^{\dagger}\Gamma^0=\Gamma_{\m}\;\;\;,\;\;\;
\Gamma^{0}\Gamma_{\m}\Gamma^0=-\Gamma_{\m}^T
\,.\label{GG2}\ee
Majorana spinors $S_{\a}$ are real: $S^{*}_{\a}=S_{\a}$;
\be
\Gamma_{11}=\Gamma_{0}\ldots\Gamma_9\;\;\;,\;\;\;
(\Gamma_{11})^2=1\;\;\;,\;\;\;
\{\Gamma_{11},\Gamma^{\m}\}=0\;;
\,.\label{GG3}\ee
$\Gamma_{11}$ is symmetric and real.
This is the reason why, in ten dimensions, the Weyl condition
$\Gamma_{11}S=\pm S$
is compatible with the Majorana condition.\footnote{In a space with
signature
(p,q) the Majorana and Weyl conditions are compatible, provided $|p-q|$
is a multiple of 8.}
We use the convention that for the Levi-Civita tensor, $\e^{01\ldots
9}=1$.
We will define the antisymmetrized products of $\Gamma$-matrices
\be
\Gamma^{\m_1\ldots\m_k}={1\over k!}\Gamma^{[\m_1}\ldots\Gamma^{\m_k]}=
{1\over k!}\left(\Gamma^{\m_1}\ldots\Gamma^{\m_k}\pm{\rm
permutations}\right)
\,.\label{GG4}\ee

We can derive by straightforward computation the following
identities among $\Gamma$-matri-\\ces:
\bea
\Gamma_{11} \Gamma^{\mu_1...\mu_k} &=&
{(-1)^{[{k\over 2}]} \over (10-k)!}
 \epsilon^{\mu_1...\mu_{10}} \Gamma_{\mu_{k+1}...\mu_{10}}\;,
\\
\Gamma^{\mu_1...\mu_k} \Gamma_{11} &=&
 {(-1)^{[{k+1\over 2}]}\over (10-k)!}
 \epsilon^{\mu_1...\mu_{10}} \Gamma_{\mu_{k+1}...\mu_{10}}
\,,\label{GG5}\eea
with $[x]$ denoting the integer part of $x$.
Then
\be
\Gamma^\mu \Gamma^{\nu_1...\nu_k} =  \Gamma^{\mu\nu_1...\nu_k}
-{1\over (k-1)!} \eta^{\mu [\nu_1} \Gamma^{\nu_2...\nu_k]}
\,,\label{GG6}\ee
\be
 \Gamma^{\nu_1...\nu_k} \Gamma^\mu =  \Gamma^{ \nu_1...\nu_k\mu}
-{1\over (k-1)!} \eta^{\mu [\nu_k} \Gamma^{\nu_1...\nu_{k-1}]}
\,,\label{GG7}\ee
with square brackets denoting the  alternating sum
over all permutations of the enclosed indices.
The invariant Lorentz scalar product of two spinors $\chi,\phi$ is
$\chi^{*}
_{\a}(\Gamma^{0})_{\a\b}\phi_{\b}$.

Now consider the ground-states of the $R$-$R$ sector.
On the left, we have a Majorana spinor $S_{\a}$ satisfying $\Gamma_{11}
S=S$ by convention.
On the right, we have another Majorana spinor $\tilde S_{\a}$
satisfying $
\Gamma_{11}\tilde S=\xi \tilde S$, where $\xi=1$ for the type-IIB
string
and $\xi=-1$ for the type-IIA string.
The total ground-state is the product of the two.
To represent it, it  is convenient to define the following bispinor
field
\be
F_{\a\b}=S_{\a}(i\Gamma^{0})_{\b\g}\tilde S_{\g}
\,.\label{GG8}\ee
With this definition, $F_{\a\b}$ is real and the trace
$F_{\a\b}\delta^{\a\b}$ is Lorentz-invariant.
The chirality conditions on the spinor translate into
\be
\Gamma_{11}F=F\;\;\;,\;\;\;F\Gamma_{11}=-\xi F
\,,\label{gg9}\ee
where we have used the fact that $\Gamma_{11}$ is symmetric and
anticommutes
with $\Gamma^0$.

We can now expand the bispinor $F$ into the complete set of
antisymmetrized $\Gamma$'s:
\be
F_{\a\b}=\sum_{k=0}^{10}{i^k\over k!}F_{\m_1\ldots\m_k}
(\Gamma^{\m_1\ldots\m_k})_{\a\b}
\,,\label{gg10}\ee
where the $k=0$ term is proportional to the unit matrix and the tensors
$F_{\m_1\ldots \m_k}$ are real.

We can now translate the first of the chirality conditions in
(\ref{gg9})
using (\ref{GG5}) to obtain the following equation:
\be
F^{\m_1\ldots\m_k}={(-1)^{\left[{k+1\over 2}\right]}\over (10-k)!}\e
^{\m_1\ldots\m_{10}}F_{\m_{k+1}\ldots\m_{10}}
\,.\label{gg11}\ee
The second chirality condition implies
\be
F^{\m_1\ldots\m_k}=\xi{(-1)^{\left[{k\over 2}\right]+1}\over (10-k)!}\e
^{\m_1\ldots\m_{10}}F_{\m_{k+1}\ldots\m_{10}}
\,.\label{gg12}\ee
Compatibility between (\ref{gg11}) and (\ref{gg12}) implies that
type-IIB
theory ($\xi=1$) contains tensors of odd rank (the independent ones
being
k=1,3 and k=5 satisfying a selfduality condition) and type-IIA theory
($\xi=-1)$
contains tensors of even rank (the independent ones having k=0,2,4).
The number of independent tensor components adds up in both cases to
$16\times 16=256$.

As mentioned in section \ref{typeII},
the mass-shell
conditions imply that the bispinor field (~\ref{GG1})
obeys two
 massless Dirac equations coming from $G_0$ and $\bar G_0$:
\be
 (p_\mu \Gamma^\mu) F = F (p_\mu \Gamma^\mu) = 0 \
\,.\label{G6}\ee
To convert these to equations for the tensors, we use the gamma
identities (\ref{GG6}) and (\ref{GG7}).
After some
straightforward algebra one finds
\be
 p^{[\mu} F^{\nu_1...\nu_k]} = p_\mu F^{\mu \nu_2...\nu_k} =0
\,,\label{G9}\ee
which are the Bianchi identity and the free massless equation for
an antisymmetric tensor field strength. We may write these in
economic form as
\be
 d F = d \ ^*F = 0
\,.\label{G10}\ee
Solving the Bianchi identity locally
allows us to express the  $k$-index field strength as the
exterior derivative of a $(k-1)$-form potential
\be
F_{\mu_1...\mu_k} =
{1\over (k-1)!} \partial_{[\mu_1} C_{\mu_2...\mu_k]}
\,,\label{G11}\ee
or in short-hand notation
\be
 F_{(k)} = d C_{(k-1)}
\,.\label{G12}\ee
Consequently, the type-IIA theory has a vector ($C^\mu$) and a
three-index
tensor potential ($C^{\mu\nu\rho}$) , in addition to a constant
non-propagating zero-form field strength ($F$), while the
type-IIB theory has a zero-form ($C$), a two-form ($C^{\mu\nu}$)
and a four-form potential ($C^{\mu\nu\rho\sigma}$), the latter
 with self-dual field strength. The number of physical transverse
degrees of freedom adds up in both cases to $64=8\times 8$.

It is not difficult to see that in the perturbative string
spectrum there are no states charged under the $R$-$R$ forms.
First, couplings of the form $\langle s|R\overline{R}|s\rangle$ are not
allowed
by the separately conserved left and right fermion numbers.
Second, the $R$-$R$ vertex operators contain the field strengths rather
than the potentials and equations of motion and Bianchi identities
enter on an equal footing. If there were electric states in
perturbation theory we would also
have magnetic states.

$R$-$R$ forms have another peculiarity.
There are various ways to deduce that their couplings to the dilaton
are exotic.
The dilaton dependence of an $F^{2m}$ term at the k-th order of
perturbation theory is $e^{(k-1)\Phi}e^{m\Phi}$ instead of the usual
$e^{(k-1)\Phi}$ term for $NS$-$NS$ fields.
For example, at tree-level, the quadratic terms are
dilaton-independent.

\renewcommand{\theequation}{\thesubsection.\arabic{equation}}
\subsection{Type-I superstrings}
\setcounter{equation}{0}

We will now consider open
superstrings. There are two possibilities: oriented and unoriented open
superstrings.
Unoriented open strings are obtained by identifying open strings
by the operation that exchanges the two end-points.
The bosonic case was discussed earlier.
We had seen that we can add Chan-Paton factors at the end-points.
In the oriented case we obtained a U(N) gauge group, while in the
unoriented case
we obtained O(N) or Sp(N) gauge groups.
As usual, in the superstring case, the GSO projection will remove the
tachyonic ground-state
and the lowest bosonic states will be a collection of massless vectors.

{}From the Ramond sector we will obtain a Majorana-Weyl spinor in the
adjoint
of the gauge group.
Thus, the massless spectrum in the open sector would consist of a
ten-dimensional Yang-Mills
supermultiplet.
Anticipating the discussion on anomalies in the next chapter we will
point out that Yang-Mills theory in ten dimensions has gravitational
and gauge anomalies for any gauge group.
It is necessary to couple the open strings with appropriate closed
strings
for  the resulting theory
to be anomaly-free.
The only anomaly-free possibility turns out to be the unoriented O(32)
open string theory.

The modern way to make this construction uses the concept of the
orientifold.
An orientifold is a generalization of the orbifold concept:
along with the projection acting on the target space, there is a
projection
acting also on the world-sheet.
This projection is an orientation-reversal (parity) operation on the
world-sheet $\Omega$: $z\leftrightarrow \bar z$ or in terms of the
cylinder
coordinates $\t,\s$: $\s\to -\s$.
For this to be a symmetry we must start with a left-right-symmetric
closed superstring theory.
{}From the type-IIA,B theories that we have considered so far only
IIB is left-right-symmetric (Ramond ground-states of same chirality on
left and right).
We will thus  consider the IIB string and construct its orientifold
using the world-sheet parity operation as the projection operator.
As in the case of standard orbifolds, we will have an untwisted sector
that  contains the invariant states of the original theory.
We will first find what the untwisted sector is, in our case.

In the $NS$-$\overline{NS}$ sector, the states that survive are the
graviton
and a scalar (the dilaton). The antisymmetric tensor, being an
antisymmetrized product of left and right oscillators,  is projected
out.
In the $R$-$\overline{R}$ sector the two-index antisymmetric tensor
survives, but the scalar and the four-index self-dual antisymmetric
tensor are projected out.
Finally from the fermionic sectors that contain two Majorana-Weyl
gravitini
and two Majorana-Weyl fermions we obtain just half of them.
Thus, in total, we have the graviton, a scalar and antisymmetric
tensor,
as well as a Majorana-Weyl gravitino and a Majorana-Weyl fermion.
This is the content of the (chiral) N=1 supergravity multiplet
in ten dimensions.
To summarize, the untwisted sector of the orientifold contains
unoriented
closed strings.

What is the twisted sector?
We usually define it by imposing a periodicity condition together with
the orbifold transformation, which we will also do here:
\be
X(\s+2\pi)=X(2\pi-\s)
\ee
and similarly for the fermions.
Using also the fact that they satisfy the two-dimensional Laplace
equation $\pd^2_{\t}-
\pd_{\s}^2=0$, we can show that the solutions with these boundary
conditions can be written as a sum of open string coordinates
satisfying Neumann boundary conditions at both end-points and open
string coordinates satisfying Dirichlet-Neumann boundary  conditions.
It turns out that consistency (tadpole cancelation) demands that the
second kind of oscillators to be absent.
The upshot of all this is  that the twisted sector is the open
superstring.
In this context, we can interpret the Chan-Paton factors as labels of
the twisted sector ground-states.
Thus, together with the untwisted (closed string)
sector, we obtain that the
massless sector is ten-dimensional N=1 supergravity coupled to N=1
super-Yang-Mills.
This is the (unoriented) type-I superstring theory.
As we will see later on, anomaly cancelation restricts the gauge
group to be O(32).

\subsection{Heterotic superstrings}
\setcounter{equation}{0}

So far, we have seen that we could use either the Virasoro algebra
(bosonic strings) or the N=1 superconformal algebra (superstrings) to
remove ghosts
from string theories.
Moreover, the closed theories were left-right symmetric, in the sense
that
a similar algebra is acting on both the left and right.
We might however envisage the possibility of using a Virasoro algebra
on the
left and the superconformal algebra on the right.

Consider a string theory where we have on the left side a
number
of bosonic coordinates and an equal number of left-moving word-sheet
fermions.
The left constraint algebra  will be that of the superstring and, the
absence
of Weyl anomaly will imply that the number of left-moving coordinates
must
be 10.
In the right-moving sector, we will include just a number of bosonic
coordinates. The constraint algebra will be the Virasoro algebra and
the Weyl
anomaly cancelation implies that the number of right-moving
coordinates is 26.
Together, we have ten left+right bosonic coordinates $X^{\mu}(z,\bar
z)$, ten left-moving fermions $\psi^{\mu}(z)$ and an extra sixteen
right-moving coordinates $\phi^{I}(\bar z)$, $I=1,2,\dots,16$.
The $X^{\mu}$ are non-compact, but the $\phi^I$ are necessarily compact
(for reasons of modular invariance)
and must take values in some sixteen-dimensional lattice $L_{16}$.
To remove the tachyon, we will also impose the usual GSO projection on
the
left, namely $(-1)^F=-1$.
Here, we will have two sectors, generated by the left-moving fermions,
the $NS$ sector (spacetime bosons) and the $R$ sector (spacetime
bosons).
Also the non-compact spacetime dimension is ten, the $\phi^I$ being
compact (``internal") coordinates.

We will try to compute the one-loop partition function in this case
(light-cone gauge).
The eight transverse non-compact bosons contribute as usual
$(\sqrt{\tau_2}\eta\bar\eta)^{-8}$.
The left-moving fermions contribute (due to the GSO projection)
$\chi_V-\chi_S$.
Finally the contribution of the right-moving compact bosons
$\phi^I$
can be obtained by taking the right-moving part of the toroidal CFT
(\ref{159}):
\be
Z_{\rm compact}(\bar q)=\sum_{L_{16}}~{\bar q^{\vec p_R^2\over
2}\over \bar\eta^{16}}={\bar\Gamma_{16}(\bar q)\over \bar\eta^{16}}
\,,\label{280}\ee
where $\vec p_R$ is a lattice vector.
Putting everything together we obtain
\be
Z^{\rm heterotic}={1\over
(\sqrt{\tau_2}\eta\bar\eta)^{8}}~{\bar\Gamma_{16}\over
\bar\eta^{16}}~{1\over 2}\sum_{a,b=0}^1~(-1)^{a+b+ab}{\th[^a_b]^4\over
\eta^4}
\,.\label{281}\ee

In order for $Z^{\rm heterotic}$ to be modular-invariant, the lattice
sum
$\bar\Gamma_{16}$ must be invariant under $\tau\to \tau+1$, which
implies that the lattice must be even (${\vec p_R}^2=$ even integer).
It must also  transform as
\be
\tau\to-{1\over \tau}\;\;:\;\;\bar \Gamma_{16}\to
\bar\tau^8~\bar\Gamma_{16}
\,,\label{282}\ee
which implies that the lattice is self-dual (the dual of the lattice
coincides with the lattice itself).
There are two sixteen-dimensional lattices that satisfy the above
requirements:

$\bullet$ $\rm E_8\times E_8$ lattice. This is the root lattice of the
group
$\rm E_8\times E_8$.
The roots of E$_8$ are composed of the roots of O(16), $\vec\e_{ij}$,
which are
eight-dimensional vectors with a $\pm 1$ in position i, a $\pm 1$ in
position
j and zero elsewhere, as well as the spinor weights of O(16),
$\vec\e^s_{\a}=
(\zeta_1,\zeta_2,\cdots,\zeta_8)/2$, $\a=1,2,\cdots,128$, with
$\zeta_i=\pm 1$, and
$\sum_i\zeta_i=0~{\rm mod}~(4)$.
The roots have squared length equal to 2.
A general lattice vector can be written as
$\sum_{i<j}n_{ij}\vec\e_{ij}+\sum_{\a}m_{\a}
\vec\e^s_{\a}$, with $n_{ij},m_{\a}\in Z$.
The lattice sum can be written in terms of $\th$-functions as
\be
\bar\Gamma_{\rm E_8\times E_8}=(\bar\Gamma_8)^2=\left[{1\over
2}\sum_{a,b=0,1}~\bar\th[^a_b]^8\right]^2=1+2\cdot240~\bar q+{\cal O}
(\bar q^2)
\,.\label{283}\ee
Combining it with the oscillators, we observe that there are $2\cdot
240+16=2\cdot 248$ states with $\bar L_0=1$, which make the adjoint
representation
of $\rm E_8\times E_8$.
In fact this left-moving theory realizes the current algebra of
$\rm E_8\times E_8$ both at level 1. The only integrable representation
is
the vacuum representation, and the first non-trivial states above the
vacuum
are generated by the current modes $\bar J_{-1}^a$.

$\bullet$ $\rm O(32)/Z_2$ lattice. This is the root lattice of O(32)
augmented
by one of the two spinor weights.
The roots of O(32) are $\vec\e_{ij}$, which are
sixteen-dimensional vectors with a $\pm 1$ in position i, a $\pm 1$ in
position
j and zero elsewhere.
The spinor weights are $\vec\e^s_{\a}=
(\zeta_1,\zeta_2,\cdots,\zeta_{16})/2$, $\a=1,2,3,\cdots,2^{16}$, with
$\zeta_i=\pm 1$, and
$\sum_i\zeta_i=0~{\rm mod}~(4)$.
The roots have squared length equal to 2.
The generic lattice vector is
$\sum_{i<j}n_{ij}\vec\e_{ij}+\sum_{\a}m_{\a}
\vec\e^s_{\a}$, with $n_{ij},m_{\a}\in Z$.
The lattice sum can be written as
\be
\bar\Gamma_{\rm O(32)/Z_2}={1\over
2}\sum_{a,b=0,1}~\bar\th[^a_b]^{16}=1+480~\bar q+{\cal O}
(\bar q^2)
\,.\label{284}\ee
This theory has a O(32) right-moving current algebra at level 1.
The integrable representations that participate are the vacuum and the
spinor
and again the states at $\bar L_0=1$ come from the current modes
$\bar J^a_{-1}$.
The spinor ground-states  have $\bar L_0=2$.

Both right-moving current algebra theories can also be constructed
from 32
free  right-moving fermions $\bar\psi^i$, $i=1,2,\dots,32$.
We will start from the O(32)/Z$_2$ theory. The currents
\be
\bar J^{ij}=i\bar \psi^{i}\bar\psi^{j}
\label{287}\ee
form the level-one O(32) current algebra.
In the Ramond sector, all fermions are periodic, in which case O(32)
invariance is not broken and we obtain the two spinors $S,C$ of
O(32).
Finally imposing a GSO-like projection $(-1)^F=1$  keeps the vacuum
representation
in the $NS$ sector and one of the spinors in the $R$ sector.

For the $\rm E_8\times E_8$ theory we will consider separate periodic
or
antiperiodic
conditions for the two groups of sixteen fermions.
In this case the O(32) invariance is broken to $\rm O(16)\times O(16)$.
In the Ramond sector, however, we obtain one of the spinors of
O(16) with
$\bar L_0=1$. This spinor combines with the adjoint of O(16) to
make
the adjoint of $\rm E_8$.

We can now describe the massless spectrum of the heterotic string
theory
(light-cone gauge).
In the $NS$ sector the constraints are
\be
L_0={1\over 2}\;\;\;,\;\;\;\bar L_0=1
\,.\label{285}\ee
Taking also into account the GSO projection, we find that there is
no tachyon and the massless states are $\psi^{i}_{-1/2}\bar
a^{j}_{-1}|p\rangle$,   which gives the graviton, antisymmetric tensor
and dilaton, and $\psi^{i}_{-1/2}\bar J^{a}_{-1}|p\rangle$, which
gives
vectors in the adjoint of $\rm G=E_8\times E_8$ or O(32).

In the $R$ sector the independent constraints are
\be
G_0=0\;\;\;,\;\;\;\bar L_0=1
 \,,\label{286}\ee
which, together with the GSO condition, give a Majorana-Weyl gravitino,
a
Majorana-Weyl
fermion,
and a set of Majorana-Weyl fermions in the adjoint of the gauge group
G.
The theory has N=1 supersymmetry in ten dimensions
and contains at the massless level the supergravity multiplet, and a
vector
supermultiplet in the adjoint of G.
Moreover, the theory is chiral.

There is another interesting heterotic theory we can construct in ten
dimensions.
This can be obtained as a $Z_2$ orbifold of the $\rm E_8\times E_8$
theory.
The first symmetry we will use is $(-1)^F$.
In each of the  two $\rm E_8$'s there is also  a symmetry that leaves
the vector of the O(16) subgroup invariant and changes the sign of
the O(16) spinor. We will call this symmetry generator  ${\cal
S}_i$, $i=1,2$ acting on
the first, respectively second $\rm E_8$'s.
The $Z_2$ element by which we will orbifold is $(-1)^{F+1}~{\cal
S}_1~{\cal S}_2$.
We will construct the orbifold blocks.
In the sector of the left-moving world-sheet fermions only $(-1)^{F+1}$
acts non-trivially.
Using (\ref{63}) and (\ref{80}) we can see that $(-1)^{F+1}$ acts as
unity on
the vector
and as $-1$ on the spinor.
The twisted blocks are
\be
Z_{\rm fermions}[^h_g]={1\over
2}\sum_{a,b=0}^1~(-1)^{a+b+ab+ag+bh+gh}~{\th^4[^a_b]
\over \eta^4}
\,.\label{288}\ee
On each of the $\rm E_8$'s the non-trivial projection is ${\cal S}_i$,
which gives the following orbifold blocks
\be
\bar Z_{E_8}[^h_g]={1\over 2}\sum_{\gamma,\delta=0}^1~(-1)^{\gamma
g+\delta h}
{}~{\bar\th^8[^{\gamma}_{\delta}]\over \bar\eta^8}
\,.\label{289}\ee
The total partition function is
\be
Z^{\rm heterotic}_{\rm O(16)\times O(16)}={1\over
2}\sum_{h,g=0}^1~{\bar
Z_{E_8}[^h_g]^2\over (\sqrt{\tau_2}\eta\bar\eta)^{8}}~{1\over 2}
\sum_{a,b=0}^1~(-1)^{a+b+ab+ag+bh+gh}~{\th^4[^a_b]
\over \eta^4}
\,.\label{290}\ee

\vskip .3cm
\noindent\hrulefill
\nopagebreak\vskip .2cm
{\large\bf Exercise}. Show that (\ref{290}) is modular-invariant.
Show also that it describes a ten-dimensional theory with gauge group
$\rm O(16)\times
O(16)$ and find the massless spectrum.
Is this theory supersymmetric ?
\vskip .6cm
{\large\bf Exercise}. To construct the partition functions of
ten-dimensional heterotic theories we need in general the characters
of O(8)
for the left-moving fermions and the characters of a rank 16, level
one current algebra for the internal right-moving part (the bosonic
contribution is always
the same).
Consider first the case G=O(32). Write the most general partition
function
as linear combinations of the characters, and then impose the
following constraints:

$\bullet$ Normalization of the vacuum contribution to 1.

$\bullet$ Modular invariance.

$\bullet$ Correct spin-statistics relation.

$\bullet$ Absence of tachyons.

How many theories do you find? How many are supersymmetric?

Repeat the procedure above for $\rm G=E_8\times O(16)$ and $\rm
O(16)\times
O(16)$.
\nopagebreak\vskip .2cm
\noindent\hrulefill
\vskip .4cm

\subsection{Superstring vertex operators\label{svertex}}
\setcounter{equation}{0}

In analogy with the bosonic string, the vertex operators must be
primary states
of the superconformal algebra.
Using chiral superfield language (see (\ref{208}),(\ref{209}))
where
\be
\hat X^{\m}(z,\theta)=X^{\mu}(z)+\theta \psi^{\mu}(z)\,.
\label{410}\ee
The left-moving vertex operators can be written in the form:
\be
\int dz~\int d\theta ~V(z,\theta)=\int dz~\int d\theta
{}~(V_0(z)+\theta V_{-1}(z))=
\int dz~V_{-1}
\,.\label{411}\ee
The conformal weight of $V_0$ is $\half$ while that of $V_{-1}$ is 1.
The integral of $V_{-1}$ has conformal weight zero.
For the massless spacetime bosons the vertex operator is
\be
V^{\rm boson}(\e,p,z,\theta)=\e_{\m}:D\hat X^{\m}~e^{ip\cdot \hat
X}\,,\label{412}\ee
\be
V_0^{\rm boson}=\e_{\m}\psi^{\m}e^{ip\cdot X}\;\;\;,\;\;\;V^{\rm
boson}_{-1}(\e,p,z)
=\e_{\m}:(\pd X^{\mu}+ip\cdot\psi~\psi^{\mu})e^{ip\cdot X}:
\,,\label{4122}\ee
where $\e\cdot p=0$.
In the covariant picture this vertex operator becomes
\be
V^{\rm boson}_{-1}(\e,p,z)=[Q_{\rm BRST},\xi(z)e^{-\phi(z)}\e\cdot
\psi~e^{ip\cdot X}]
\,.\label{414}\ee

The spacetime fermion vertex operators can only be constructed in the
covariant
formalism.
For the massless states ($p^2=0$) they are of the form
\be
V^{\rm
fermion}_{-1/2}(u,p,z)=u^{\a}(p):e^{-{\phi(z)/2}}~
S_{\a}(z)~e^{ip\cdot X}:
\,,\label{413}\ee
$\phi$ is the boson coming from the bosonization of the $\b,\g$
superconformal ghosts, $e^{-\phi/2}$ is the spin field of the
$\beta,\g$ system
of conformal weight 3/8 (see section  \ref{ghosts}) and $S_{\a}$ is
the spin field
of the fermions $\psi^{\m}$ forming an O(10)$_1$ current algebra,
with weight
5/8 (see section \ref{Nfermions}).
The subscript $-1/2$ indicates the $\phi$-charge.
The total conformal weight of $V_{-1/2}$ is 1.
Finally, $u^{\a}$ is a spinor satisfying the massless Dirac equation
$p\hspace{-6pt}/u=0$.

There is a subtlety in the case of fermionic strings having to do with
the $\b,\g$
system.
As we have seen, in the bosonized form, the presence of the
background
charge alters the charge neutrality condition\footnote{The charge
neutrality
condition was given in (\ref{91}). It states that the sum of the
charges
of vertex operators in a non-zero correlation function has to vanish.}.
This is related to the existence of supermoduli or superkilling
spinors.
Thus, depending on the correlation function and surface we must have
different representatives for the vertex operators of a given
physical state with different $\phi$-charges.
This can be done in the following way. Consider a physical vertex
operator
with $\phi$ charge $q$, $V_q$.
It is BRST invariant, $[Q_{\rm BRST},V_q]=0$.
We can construct another physical vertex operator representing the
same physical state but with charge $q+1$ as $V_{q+1}=[Q_{\rm BRST},\xi
V_q]$
since $Q_{\rm BRST}$ carries charge 1.
Since it is a BRST commutator, $V_{q+1}$ is also BRST-invariant.
However, we have seen that states that are BRST commutators of
physical states
are spurious. In this case this is avoided since the
$\xi$ field appears  in the commutator and its zero mode lies outside
the ghost Hilbert space.
The different $\phi$ charges are usually called pictures in the
literature.
The $\half$   picture for the fermion vertex can be computed to be
\bea
V_{1/2}^{\rm fermion}(u,p)&=&[Q_{\rm BRST},\xi(z)V_{-1/2}^{\rm
fermion}(u,p,z)]\nn\\&=& u^{\a}(p)e^{\phi/2}S_{\a}e^{ip\cdot X}+\cdots
\,,\label{418}\eea
where the ellipsis involves terms that do not contribute to four-point
amplitudes.
The ten-dimensional spacetime supersymmetry charges can be
constructed
from the fermion vertex at zero momentum,
\be
Q_{\a}={1\over 2\pi i}\oint dz ~:e^{-{\phi(z)/2}}~S_{\a}(z)
\,.\label{415}\ee
It transforms fermions into bosons and vice versa
\be
[Q_{\a},V^{\rm fermion}_{-1/2}(u,p,z)]=V^{\rm
boson}_{-1}(\e^{\m}=u^{\b}\gamma^{\m}_{\b\a},p,z)
\,,\label{416}\ee
\be
[Q_{\a},V^{\rm boson}_{0}(\e,p,z)]=V^{\rm
fermion}_{-1/2}(u^{\b}=ip^{\m}\e^{\n}(\gamma_{\m\n})^{\b}_{\a},p,z)
\,.\label{417}\ee
There are various pictures for the supersymmetry charges also.

\subsection{Supersymmetric effective actions\label{SEA}}
\setcounter{equation}{0}

So far we have seen that, in ten dimensions, there are the following
spacetime supersymmetric string theories.

$\bullet$ Type-I theory (chiral) with N=1 supersymmetry and gauge
group O(32).

$\bullet$ Heterotic theories (chiral) with N=1 supersymmetry and
gauge groups O(32) and E$_8\times $E$_8$.

$\bullet$ Type-IIA theory (non-chiral) with N=2 supersymmetry.

$\bullet$ Type-IIB theory (chiral) with N=2 supersymmetry.

We would like to find the effective field theories that describe the
dynamics
of the massless fields. A straightforward approach would be the one we
used in the bosonic case, namely either extracting them from
scattering amplitudes or
requiring Weyl invariance of the associated $\s$-model in general
background fields.
In the presence of supersymmetry, however, these effective actions are
uniquely fixed.
They have been constructed during the late seventies, early eighties,
as supergravity theories.

First we would like to obtain the low-energy effective action at the
leading order approximation.
When only bosonic fields are present, we just have to keep terms of
up to two derivatives.
In the presence of fermions, however, we would like to modify our
counting rules a bit so that the kinetic terms $\phi\square \phi$ for
bosons and $\bar\psi
\dslash \psi$ for fermions are equally important at low energy.
We will give weight 0 to bosons, weight $\half$ to fermions and 1 to a
derivative.
Then, both kinetic terms have the same weight, namely 2.
These weights are also respected by supersymmetry (SUSY) as can be
directly
verified from the generic SUSY transformations
\be
\delta_{\epsilon}\phi\sim
\phi^m~\psi\epsilon\;\;\;,\;\;\;\delta_{\epsilon}\psi\sim \pd
\phi^m~\epsilon
+\phi^m~\psi^2\epsilon
\,.\label{291}\ee
The effective actions in the leading order must have weight 2,
and this is true for all supergravity actions.

In ten dimensions, in order to have massless fields with spin not
greater than 2 we have to restrict ourselves to N$\leq 2$ SUSY.

We will  first consider N=1 supersymmetry.
There are two massless supersymmetry representations
(supermultiplets).
The vector multiplet contains a vector ($A_{\mu}$) and a Majorana-Weyl
fermion
($\chi_{\alpha}$).
The supergravity multiplet contains the graviton ($g_{\mu\nu}$),
antisymmetric tensor ($B_{\mu\nu}$) and a scalar $\phi$ (dilaton) as
well as a Majorana-Weyl gravitino ($\psi^{\mu}_{\alpha}$) and a
Majorana-Weyl fermion
($\lambda_{\alpha}$).
The effective action of an N=1 supergravity coupled to super
Yang-Mills
is fixed by supersymmetry, the only  choice that remains being  that of
the
gauge group.
The super Yang-Mills action is (in the absence of gravity)
\be
L_{\rm YM}=-{1\over
4}F^a_{\mu\nu}F^{a,\mu\nu}-\bar\chi^a\Gamma^{\mu}D_{\mu}\chi^a
\,,\label{292}\ee
where
\be
F^a_{\mu\nu}=\pd_{\mu}A^a_{\nu}-\pd_{\nu}A^a_{\mu}+
g{f^a}_{bc}A^b_{\mu}A^c_{\nu}
\,,\label{294}\ee
\be
D_{\mu}\chi^a=\pd_{\mu}\chi^a+g{f^{a}}_{bc}A^b_{\mu}\chi^c
\label{295}\ee
and $g$ is the Yang-Mills coupling constant.
The pure N=1 supergravity action is
\bea
L_{\rm SUGRA}^{N=1}&=&-{1\over 2\kappa^2}R-{3\over
4}\phi^{-3/2}H_{\mu\nu\rho}
H^{\mu\nu\rho}-{9\over 16\kappa^2}{\pd_{\mu}\phi\pd^{\mu}\phi\over
\phi^2}
-{1\over 2}\bar \psi^{\mu}\Gamma^{\mu\nu\rho}\nabla_{\nu}\psi_{\rho}-\nn\\
&&-{1\over 2}\bar\lambda\Gamma^{\mu}\nabla_{\mu}\lambda
-{3\sqrt{2}\over 8}{\pd_{\nu}\over
\phi}\bar\psi^{\mu}\Gamma^{\nu}\Gamma^{\mu}\lambda
+{\sqrt{2}\kappa\over 16}\phi^{-3/4}H_{\nu\rho\s}\left[
\bar
\psi_{\mu}\Gamma^{\mu\nu\rho\s\tau}\psi_{\tau}+
\right.\nn\\
&&\left. +6\bar\psi^{\nu}\Gamma
^{\rho}\psi^{\s}-\sqrt{2}\bar\psi_{\mu}\Gamma^{\nu\rho\sigma}
\Gamma^{\mu}\lambda\right]+({\rm Fermi})^4\,,
\eea
where $\kappa$ is Newton's constant, $\Gamma^{\mu_1\dots\mu_n}$
stands for the completely antisymmetrized
product of $\Gamma$ matrices, $H_{\mu\nu\rho}$ is given in
(\ref{254})
and we did not write explicitly terms involving four fermions.

The two actions can be coupled together
\be
L^{N=1}_{\rm SUGRA+YM}={L^{N=1}_{\rm SUGRA}}'+\phi^{-3/4}~L'_{\rm YM}
\,.\label{296}\ee
The prime in the Yang-Mills action implies that covariant derivatives
now contain the spin connection.
The prime in the supergravity action implies that we have to modify
the definition of the field strength of $B$:
\be
\hat H_{\mu\nu\rho}=H_{\mu\nu\rho}-{\kappa\over
\sqrt{2}}\omega^{CS}_{\mu\nu\rho}
\,,\label{297}\ee
where the Chern-Simons form is
\be
\omega^{CS}_{\mu\nu\rho}=A^a_{\mu}F^a_{\nu\rho}-{g\over
3}f_{abc}A^a_{\mu}A^b_{\nu}A^c_{\rho}+{\rm cyclic}
\,.\label{298}\ee
This modification implies that in order for the full theory to be
gauge-invariant the antisymmetric tensor must transform under gauge
transformations
$\delta A\to d\Lambda+[A,\Lambda]$ as
\be
 \delta B={\kappa\over \sqrt{2}}{\rm Tr}[\Lambda d A]
\,,\label{299}\ee
so that the modified field strength $\hat H$ is invariant.

The theory contains a single parameter, since the combination
$g^4/\kappa^3$ is dimensionless and can be scaled to 1 by a rescaling
of the
field $\phi$.

When we have two supersymmetries, there are only two possibilities:

$\bullet$ \underline{Type-IIA supergravity}. This is the low energy
limit of the type-IIA superstring in ten dimensions.
It contains a single supermultiplet of N=2 supersymmetry containing
the graviton ($g_{\mu\nu}$), an antisymmetric tensor ($B_{\mu\nu}$), a
scalar $\phi$ (dilaton), a vector $A_{\mu}$ and a three-index
antisymmetric tensor $C_{\mu\nu\rho}$ as well as a Majorana
gravitino ($\psi^{\mu}_{\alpha}$) and a Majorana fermion
($\lambda_{\alpha}$).
The supergravity action is completely fixed and can be obtained by
dimensional reduction
of the eleven-dimensional N=1 supergravity \cite{cj} containing the
eleven-dimensional metric
$G_{\mu\nu}$ and a three-index antisymmetric tensor $\hat
C_{\mu\nu\rho}$.
The action is
\bea
L^{D=11}&=&{1\over 2\kappa^2}\left[R-{1\over 2\cdot
4!}G_4^2\right]-i\bar\psi_{\mu}\Gamma^{\mu\nu\rho}
\tilde\nabla_{\nu}\psi_{\rho}+{1\over 2\kappa^2(144)^2}
G_4\wedge G_4\wedge \hat C+
\nn\\
&&+{1\over 192}\left[\bar
\psi_{\mu}\Gamma^{\mu\nu\rho\s\tau\upsilon}\psi_{\upsilon}
+12\bar\psi^{\nu}\Gamma^{\rho\s}\psi^{\tau}\right]
(G+\hat G)_{\nu\rho\s\tau}\,,
\eea
where $\tilde \nabla$ is defined with respect to the connection
$(\omega+\tilde\omega)/2$,  and $\omega$ is the spin connection while
\be
\tilde \omega_{\mu,ab}=\omega_{\mu,ab}+{i\kappa^2\over
4}\left[-\bar\psi^{\nu}\Gamma_{\nu\mu
ab\rho}\psi^{\rho}+2(\bar\psi_{\mu}
\Gamma_b\psi_{a}-\bar\psi_{\mu}
\Gamma_a\psi_{b}+\bar\psi_{b}
\Gamma_{\mu}\psi_{a})\right]
\label{301}\ee
is its supercovariantization.
Finally, $G_4$ is the field strength of $\hat C$,
\be
G_{\m\n\rho\s}=\pd_{\mu}\hat C_{\n\rho\s}-\pd_{\nu}\hat C_{\rho\s\mu}
+\pd_{\rho}\hat C_{\s\m\n}-\pd_{\s}\hat C_{\m\n\rho}
\ee
and $\tilde G_4$ is its supercovariantization
\be
\tilde G_{\m\n\rho\s}=G_{\m\n\rho\s}-6\kappa^2\bar
\psi_{[\m}\Gamma_{\n\rho}
\psi_{\s]}\;.
\ee
Upon dimensional reduction, the eleven-dimensional metric gives rise
to a ten-dimensional metric, a gauge field and a scalar as follows
(see Appendix C):
\be
G_{\mu\nu}=\left(\matrix{g_{\mu\nu}+
e^{2\s}A_{\mu}A_{\nu}&e^{2\s}A_{\mu}\cr
e^{2\s}A_{\mu}&e^{2\s}\cr}\right)
\,.\label{302}\ee
The three-form $\hat C$ gives rise to a three-form and a two-form in
ten dimensions
\be
C_{\mu\nu\rho}=\hat C_{\mu\nu\rho}-\left(\hat
C_{\nu\rho,11}A_{\mu}+{\rm cyclic}\right)\;\;,\;\;
B_{\mu\nu}=\hat C_{\mu\nu,11}\;.
\ee

The ten-dimensional action can be directly obtained from the
eleven-dimensional one using the formulae of Appendix C.
For the bosonic part we obtain,
\bea
S^{IIA}&=&{1\over 2\kappa^2}\int d^{10}x\sqrt{g}e^{\s}\left[R-{1\over
2\cdot 4!}\hat G^2-{1\over 2\cdot 3!}e^{-2\s}H^2 -{1\over
4}e^{2\s}F^2\right]+\nn\\
&&+{1\over 2\kappa^2 (48)^2}\int B\wedge G\wedge G\;,
\eea
where
\be
F_{\mu\nu}=\pd_{\m}A_{\n}-\pd_{\n}A_{\m}\;\;\;,\;\;\;H_{\mu\nu\rho}=
\pd_{\mu}B_{\nu\rho}+{\rm cyclic}\;,
\ee
\be
\hat G_{\m\n\rho\s}=G_{\mu\nu\rho\sigma}+(F_{\mu\nu}B_{\rho\sigma}+{\rm
5~~ permutations})\;.
\ee
This is the type-IIA effective action in the Einstein frame.
We can go to the string frame by $g_{\m\n}\to e^{-\s}g_{\m\n}$. The
ten-dimensional dilaton is   $\Phi=3\s$.
The action is
\bea
\tilde S_{10}&=&{1\over 2\kappa^2}\int
d^{10}x\sqrt{g}e^{-\Phi}\left[\left(R+(\nabla\Phi)^2-{1\over
12}H^2\right)-{1\over 2\cdot 4!}\hat G^2 -{1\over 4}F^2\right]+
\nn\\
&&+{1\over
2\kappa^2 (48)^2}\int B\wedge G\wedge G
\,.\label{6055}\eea
Note that the kinetic terms of the $R$-$R$ fields $A_{\mu}$ and
$C_{\m\n\rho}$
do not have dilaton dependence at the tree level, as advocated in
section
\ref{RRR}.

$\bullet$ \underline{Type-IIB supergravity}. It contains
the graviton ($g_{\mu\nu}$), two antisymmetric tensors
($B^i_{\mu\nu}$), two scalars $\phi^i$, a self-dual four-index
antisymmetric tensor $T^+$,  two Majorana-Weyl gravitini
and two Majorana-Weyl fermions
of the same chirality.
The theory is chiral but anomaly-free, as we will see further on.
The self-duality
condition
implies that the field strength $F$ of the four-form is equal to its
dual.
This equation cannot be obtained from a covariant action.
Consequently, for type-IIB supergravity, the best we can do is to write
down the
equations of motion \cite{IIB}.

There is an SL(2,$\R$) global invariance in this theory, which
transforms
the antisymmetric tensor and scalar doublets (the metric as well as
the four-form are invariant).
We will denote by $\phi$ the dilaton that  comes from the ($NS-NS$)
sector and by $\chi$ the scalar that comes from the $R$-$R$ sector.
Define the complex scalar
\be
S=\chi+ie^{-\phi/2}
\,.\label{303}\ee
Then, SL(2,$\R)$ acts by fractional transformations on $S$ and linearly
on $B^i$
\be
S \to {aS+b\over cS+d}\;\;\;,\;\;\; \left(\matrix{B^N_{\mu\nu}\cr
B^{R}_{\mu\nu}\cr}\right)\to \left(\matrix{d&-c\cr -b&a\cr}\right)
 \left(\matrix{B^N_{\mu\nu}\cr
B^{R}_{\mu\nu}\cr}\right)
\,,\label{304}\ee
where $a,b,c,d$ are real with $ad-bc=1$.
$B^N$ is the $NS$-$NS$ antisymmetric tensor while $B^R$ is the $R$-$R$
antisymmetric tensor.
When we set the four-form to zero, the rest of the equations of motion
can be obtained from the following action
\be
S^{IIB}={1\over 2\kappa^2}\int d^{10}x\sqrt{-\det g}\left[R-{1\over 2}
{\partial S\partial\bar S\over S_2^2}
-{1\over 12}{|H^R+ SH^N|^2\over S_2}\right]
\,,\label{IIB}\ee
where $H$ stands for the field strength of the antisymmetric tensors.
Obviously (\ref{IIB}) is SL(2,$\R$) invariant.

\renewcommand{\theequation}{\thesection.\arabic{equation}}
\section{Anomalies\label{anomal}}
\setcounter{equation}{0}

An anomaly is the breakdown of a classical symmetry in the quantum
theory.
Two types of symmetries can have anomalies, global or local (gauge)
symmetries.
In the following we will be interested in anomalies of local
symmetries.
If a local symmetry has an anomaly, this implies that
longitudinal
degrees of freedom no longer  decouple.
This signals problems
with unitarity.
In two dimensions, anomalies are not fatal.
The example of the chiral Schwinger model
(U(1) gauge theory coupled to a massless fermion)
indicates
that one can include the extra degrees of freedom and obtain a
consistent
theory.
However, we do not yet know how to implement this procedure in  more
than two dimensions.
Thus, we will impose an absence of anomalies.

Consider the physical effective action of a theory containing gauge
fields as well as a metric
$\Gamma^{\rm eff}[A_{\mu},g_{\mu\nu},\dots]$.
The gauge current and the energy-momentum tensor are
\be
J^{\mu}={\delta\Gamma^{\rm eff}\over {\delta
A_{\mu}}}\;\;\;,\;\;\;T^{\mu\nu}
={1\over \sqrt{-g}}{\delta\Gamma^{\rm eff}\over {\delta g_{\mu\nu}}}
\,.\label{305}\ee
The variation of the effective action under a gauge transformation
$\delta_{\Lambda} A=[a,\Lambda]$ is
\be
\delta_{\Lambda}\Gamma^{\rm eff}={\rm Tr}\int ~D_{\mu}\Lambda ~
{\delta\Gamma^{\rm eff}\over {\delta A_{\mu}}}=
{\rm Tr}\int ~\Lambda~D_{\mu}{\delta\Gamma^{\rm eff}\over {\delta
A_{\mu}}}
=\int ~Tr~[\Lambda~D_{\mu}J^{\mu}]
\,,\label{306}
\ee
where we have used integration by parts.
Consequently, iff $D_{\mu}J^{\mu}\not= 0$ there is an anomaly in the
gauge
symmetry.
Similar remarks apply to the invariance under diffeomorphisms:
\be
\delta_{diff}\Gamma^{\rm
eff}=\int~(\nabla^{\mu}\epsilon^{\nu}+\nabla^{\nu}\epsilon^{\mu})
{\delta\Gamma^{\rm eff}\over {\delta g_{\mu\nu}}}=\int
\epsilon^{\mu}\nabla_{\nu}T^{\mu\nu}
\,.\label{307}\ee
Thus, a gravitational anomaly implies the non-conservation of
the stress-tensor in the quantum theory.

\begin{figure}
\begin{center}
\leavevmode
\epsfxsize=13cm
\epsffile{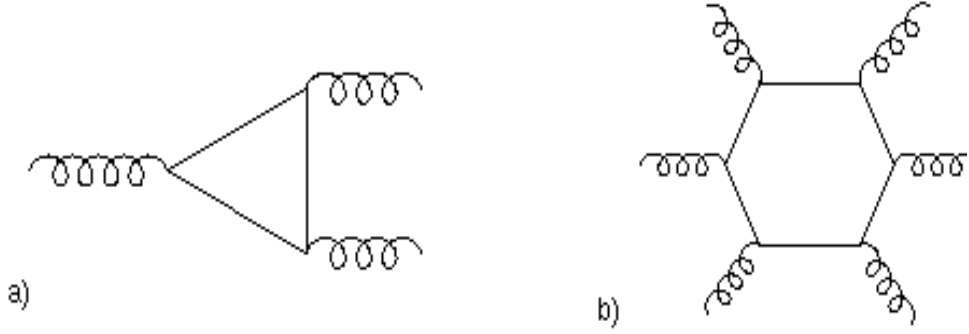}
\end{center}
\vspace{-.5cm}
\caption[]{\it a) The anomalous triangle diagram in four dimensions. b)
The anomalous hexagon diagram in ten dimensions.}
\label{f13}\end{figure}

Anomalies in field theory appear due to UV problems. Consider the
famous triangle graph in four dimensions (Fig. \ref{f13}a).
It is superficially linearly divergent, and gauge invariance reduces
this  to a logarithmic divergence.
If the fermions going around the loop are non-chiral, we can
regularize the diagram using Pauli-Villars regularization and we can
easily show
that the graph vanishes when one of the gauge field polarizations is
longitudinal. There is no anomaly in this case.
However, if the fermions are chiral,
Pauli-Villars (or any other
regularization) will break gauge invariance, which will not be
recovered
when the regulator mass is going to infinity.

\begin{figure}
\begin{center}
\leavevmode
\epsffile{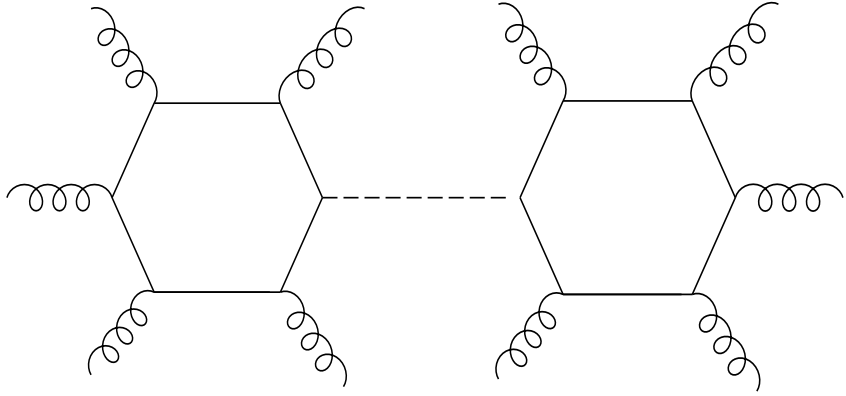}
\end{center}
\vspace{-.5cm}
\caption[]{\it Two-loop diagram with physical external legs in which
longitudinal modes propagate.}
\label{f14}\end{figure}

In ten dimensions, the leading graph that can give a contribution to
anomalies is the hexagon diagram depicted in Fig. \ref{f13}b.
The external lines can be either gauge bosons or gravitons.
It can be shown that only the completely symmetric part of the graph
gives a non-trivial contribution to the anomaly.
Non-symmetric contributions can be canceled by local counterterms.
If the diagram were non-zero when one of the external lines is
longitudinal, then this will imply that the unphysical polarizations
will propagate in the two-loop diagram in Fig. \ref{f14}.

We will consider the linearized approximation, which is relevant for
the leading hexagon diagram: $F=F_0+A^2$, $F_0=dA$ and $A\to
A+d\Lambda$.
Here, $\Lambda$ is the gauge parameter matrix (zero-form).
The anomaly due to the hexagon diagram with gauge fields in the
external lines
can have the following general form
\be
\left.\delta \Gamma\right|_{\rm gauge} \sim \int
d^{10}x~\left[c_1{\rm Tr}[\Lambda F_0^5]+c_2{\rm Tr}[\Lambda
F_0]{\rm Tr}[F_0^4]+c_3{\rm Tr}[\Lambda F_0](Tr[F_0^2])^2\right]
\,,\label{308}\ee
where powers of forms are understood as wedge products.
For comparison, the similar expression in four dimensions is
proportional to
${\rm Tr}[\Lambda F_0^2]$.
The three different coefficients $c_i$ correspond to the three group
invariants  ${\rm Tr}[T^6]$, ${\rm Tr}[T^4]Tr[T^2]$ and $({\rm
Tr}[T^2])^3$ of a given
group generator $T$ in a symmetric group trace.
There is a similar result for the gravitational anomaly, where the
role of $F$
is played by the O(D) two-form
$R^{ab}_{\mu\nu}=e^{a}_{\rho}e^{b}_{\s}
{R^{\rho\s}}_{\mu\nu}$.
The matrix valued two-form $R$ is obtained by multiplying $R^{ab}$ with
the O(D) adjoint matrices $T^{ab}$.
It can be written in terms of the spin connection one-form $\omega$
as $R=d\omega+\omega^2$.
Considering the anomaly diagram with graviton external lines we
obtain
\be
\left.\delta\Gamma\right|_{\rm grav} \sim \int
d^{10}x~\left[d_1Tr[\Theta R_0^5]+
d_2{\rm Tr}[\Theta R_0]{\rm Tr}[R_0^4]+d_3{\rm Tr}[\Theta R_0]({\rm
Tr}[R_0^2])^2\right]
\,.\label{314}\ee
Finally, by considering some of the external lines to be gauge bosons
and some to be gravitons we obtain the mixed anomaly
\bea
\left.\delta \Gamma\right|_{\rm mixed}&\sim& \int d^{10}x~\left[
e_1{\rm Tr}[\Lambda F_0]{\rm Tr}[R_0^4]+e_2{\rm Tr}[\Theta R_0]{\rm
Tr}[F_0^4]+\right.
\nn\\
&&\left.+e_3{\rm Tr}[\Theta R_0]
({\rm Tr}[F_0^2])^2+e_4{\rm Tr}[\Lambda F_0]({\rm Tr}[R_0])^2\right]\,.
\eea
There is also another potential term
${\rm Tr}[\Lambda F_0]{\rm Tr}[F_0^2]{\rm Tr}[R_0^2]$, but it can be
removed by a local
counterterm.

There is a geometric construction that provides the full anomaly
from the leading linearized piece (for a more complete discussion
see, \cite{GSW}, p. 343).
First, the anomaly satisfies the so-called Wess-Zumino consistency
condition,
which reflects the group structure of gauge transformations.
Let $G(\Lambda)=\delta \Gamma/\delta \lambda$. Then
\be
\delta_{\Lambda_{1}}G(\Lambda_{2})-\delta_{\Lambda_{2}}G(\Lambda_{1})
=G([\Lambda_{1},\Lambda_{2}])
\,.\label{309}\ee
The field strengths transform as follows under gauge transformations
and diffeomorphisms
\be
\delta F=[F,\Lambda]\;\;\;,\;\;\;\delta R=[R,\Theta]
\,.\label{310}\ee
It is straightforward to show that the traces ${\rm Tr}[R^m]$, ${\rm
Tr}[F^m]$
are gauge-invariant and closed:
\be
d~{\rm Tr}[R^m]=d~{\rm Tr}[F^m]=0
\,.\label{311}\ee
Also, the traces are non-zero for even $m$.
In order to construct the anomaly $D$-form in $D$-dimensions we start
with the most general gauge-invariant and closed $(D+2)$-form
$I^{D+2}(R,F)$,
which can be written as a linear combination of products of even
traces of $F,R$.
Since $I^{D+2}$ is closed, it can be written (locally) as
\be
I^{D+2}(R,F)=d \Omega^{D+1}(\omega,A)
\,,\label{312}\ee
where the $(D+1)$-form $\Omega^{D+1}$ is no longer  gauge-invariant,
but changes under gauge transformations as
\be
\delta_{\Lambda}\Omega^{D+1}(\omega,A)=d\Omega^{D}(\omega,A,\Lambda)
\,.\label{313}\ee
This is required by the fact that $I^{D+2}$ is gauge-invariant.
Finally, the $D$-dimensional anomaly is the piece of $\Omega^{D}$
linear in
$\Lambda$.

Except for the irreducible part of the gauge anomaly proportional
to ${\rm Tr}\Lambda F_0^5$ and ${\rm Tr}[\Theta R_0^5]$, the rest can
be canceled,
if it
appears in a suitable linear combination.
This is known as the Green-Schwarz mechanism.

Assume that the reducible part of the anomaly factorizes as follows
\bea
\left.\delta\Gamma\right|_{\rm reduc} &\sim&
 \int d^{10}x~\left({\rm Tr}[\Lambda F_0]+{\rm Tr}[\Theta
R_0]\right)\left(
a_1{\rm Tr}[F_0^4]+a_2{\rm Tr}[R_0^4]+\right.\nn\\
&&\left. + a_3({\rm Tr}[F_0^2])^2 +a_4({\rm Tr}[R_0^2])^2 + a_5Tr[F_0^2]{\rm Tr}[R_0^2]\right)\,.
\eea
We have seen that, in N=1 supergravity, the field strength of the
antisymmetric tensor is shifted by the gauge Chern-Simons form. We
can also add the gravitational Chern-Simons form (it will contribute
four derivative terms):
\be
\hat H=dB+\Omega^{CS}(A)+\Omega^{CS}(\omega)
\,.\label{317}\ee
We have also seen that this addition makes the $B$-form transform
under gauge
transformations and diffeomorphisms to keep $\hat H$  invariant.
Since
\be
\delta_{\Lambda}\Omega^{CS}(A)=d{\rm Tr}[\Lambda dA]\;\;\;,\;\;\;
\delta_{\Theta}\Omega^{CS}(\omega)=d{\rm Tr}[\Theta d\omega]
\label{318}\ee
the antisymmetric tensor must transform as
\be
\delta B=-{\rm Tr}[\Lambda F_0+\Theta R_0]
\,.\label{319}\ee
Thus, the counterterm
\be
\Gamma_{\rm counter}\sim \int d^{10}x~B~\left(
a_1{\rm Tr}[F_0^4]+a_2{\rm Tr}[R_0^4]+a_3({\rm Tr}[F_0^2])^2+\right.
\label{320}\ee
$$\left.+a_4({\rm Tr}[R_0^2])^2
+a_5{\rm Tr}[F_0^2]{\rm Tr}[R_0^2]\right)
$$
can cancel the reducible anomaly.
This mechanism can work also in other dimensions, provided there
exists an antisymmetric tensor in the theory.
There are also generalizations of this mechanism in theories with
more than
one antisymmetric tensor. Such theories can be obtained by
compactifying
superstring theories down to six dimensions.

What kind of fields can contribute to the anomalies?
First of all they have to be massless, secondly they must be chiral.
Chirality exists in even dimensions, and fields that can be chiral are
(spin
1/2) fermions, (spin 3/2) gravitini and (anti)self-dual antisymmetric
tensors
$B_{\mu_1\dots\mu_{D/2-1}}$.
Their field strength $F=dB$ is a $D/2$-form that is (anti)self-dual
\be
 F_{\mu_1\dots\mu_{D/2}}=\pm {i\over
(D/2)!}\epsilon_{\mu_1\dots\mu_{D}}F^{\mu_{D/2+1}\dots\mu_{D}}
\,.\label{321}\ee

For gravitational anomalies to appear, we must have chiral
representations
of the Lorentz group O(1,D-1). They exist in $D=4k+2$ dimensions.
For gauge anomalies, we must have chiral representations of the gauge
group
G. This can happen in even dimensions and when the gauge group admits
complex representations.

We will now give the contributions to the anomalies coming from the
various
chiral fields.
As we argued before, the anomaly is completely characterized by a
closed,
gauge-invariant (D+2)-form.
By an orthogonal transformation we can bring the $D\times D$
antisymmetric
matrix $R_0$ to the following block-diagonal form
\be
R_0=\left(\matrix{0&x_1& 0&0 & &\dots & \cr -x_1&0&0&0&&\dots&\cr
0&0&0&x_2&&\dots&\cr 0&0&-x_2&0&&\dots&\cr
&\dots&\dots&\dots&\dots&\dots&\cr &\dots&&&&0&x_{D/2}\cr
&\dots&&&&-x_{D/2}&0\cr}\right)
\,.\label{322}\ee
Then ${\rm Tr}[R_0^{2m}]=2(-1)^m\sum_{i}x_{i}^{2m}$.
The contribution to the gravitational
\linebreak[4]
anomaly of a spin
$\half$ fermion is given by \cite{AG}
\be
\hat I_{1/2}(R)=\prod_{i=1}^{D/2}\left({x_i/2\over
\sinh(x_i/2)}\right)
\,.\label{323}
\ee
In the previous formula, we have to expand it in a  series that
contains forms of various orders and
pick the piece that is a (D+2)-form.
Similarly, we have the following contributions for chiral gravitini:
\be
I_{3/2}(R)=\hat I_{1/2}(R)\left(-1+2\sum_{i=1}^{D/2}\cosh(x_i)\right)
\label{324}\ee
and self-dual tensors
\be
I_A(R)=-{1\over 8}\prod_{i=1}^{D/2}\left({x_i\over \tanh(x_i)}\right)
 \,.\label{325}\ee
The gravitini and self-dual tensors do not contribute to gauge or
mixed anomalies, since they cannot be charged under the gauge group.
However the spin 1/2 fermions can transform non-trivially and their
total contribution to anomalies is given by
\be
I_{1/2}(R,F)=Tr\left[e^{iF}\right]\hat I_{1/2}(R)
\,.\label{326}\ee
Assuming $D=10$ and expanding the formulae above we obtain
\bea
\left.I_{1/2}(R,F)\right|_{\rm 12-form}&=&-{{\rm Tr}[F^6]\over
720}+  {Tr[F^4]Tr[R^2]\over 24\cdot 48}+ \nn\\
&&-{Tr[F^2]\over 256}\left({Tr[R^4]\over 45}+{(Tr[R^2])^2\over
36}\right)+ \nn\\
&&
+{n\over 64}\left({Tr[R^6]\over 5670}+{Tr[R^2]Tr[R^4]\over 4320}
+{(Tr[R^2])^3\over 10368}\right)\,,\nn
\eea
where $n$ is the total number of spin 3/2 fermions.
\bea
\left.I_{3/2}(R)\right|_{\rm 12-form}&=&-{495\over
64}\left({Tr[R^6]\over 5670}+{Tr[R^2]Tr[R^4]\over 4320}
+{(Tr[R^2])^3\over 10368}\right)+\nn\\
&&+{Tr[R^2]\over 384}\left(Tr[R^4]+{(Tr[R^2])^2\over 4}\right)\,,\nn
\eea
\be
\left. I_A(R)\right|_{\rm 12-form}=\left.\hat I_{1/2}(R)\right|_{\rm
12-form}-\left.I_{3/2}(R)\right|_{\rm 12-form}
\,.\label{329}\ee
The anomaly contributions $I_{1/2}$ and $I_{3/2}$ given above
correspond to Weyl fermions.
Since in ten dimensions we also have Majorana-Weyl fermions,
their contribution to anomalies is half of the above.

We are now in a position to examine which ten-dimensional  theories are
free of anomalies.

The theory of N=1 supergravity without matter contains a Majorana-Weyl
gravitino and a
Majorana-Weyl
spin $\half$ fermion of opposite chirality.
It can easily be checked from the formulae above that this is
anomalous.

Type-IIA supergravity is non-chiral and thus trivially anomaly-free.
Type-IIB, however, is chiral and contains two Majorana-Weyl gravitini
contributing
$I_{3/2}$ to the anomaly, two Majorana-Weyl fermions of the opposite
chirality
contributing $-I_{1/2}$, and a self-dual tensor contributing $-I_{A}$.
The total anomaly can be seen to vanish from (\ref{329}).

We will  now consider N=1 supergravity coupled to vector multiplets.
The gaugini have the same chirality as the gravitino.
The total anomaly is
\be
2I^{N=1}=I_{3/2}(R)-I_{1/2}(R)+I_{1/2}(R,F)
\,.\label{330}\ee
We should at least require that the irreducible anomaly corresponding
to
the traces of $R^6$ and $F^6$ cancels.
The ${\rm Tr}[R^6]$ cannot be written as a product of lower traces,
since
the group O(10) has an independent Casimir of order 6.
Thus, the coefficient of the ${\rm Tr}[R^6]$ term in (\ref{330}) must
vanish.
This implies that $n=496$. Since the gaugini are in the adjoint
representation of
the gauge group, their number $n$ is the dimension of the gauge
group.
We obtain that a necessary (but not sufficient) condition for
anomaly
cancelation is $\mbox{dim} G=496$.
Inserting $n=496$ in (\ref{330}) we obtain
\be
96I^{\rm total}=-{{\rm Tr}[F^6]\over 15}+{{\rm Tr}[R^2]{\rm
Tr}[F^4]\over
24}+{{\rm Tr}[R^2]{\rm Tr}[R^4]\over 8}+{({\rm Tr}[R^2])^3\over 32}-
\label{331}\ee
$$-{{\rm Tr}[F^2]\over 960}\left(4{\rm Tr}[R^4]+5({\rm
Tr}[R^2])^2\right)\,.
$$
It is obvious from the above that the only hope for canceling the
leftover
anomaly is to
be able to use the Green-Schwarz mechanism.
It would work if we could factorize $I^{\rm total}$.
This will happen iff
\be
{\rm Tr}[F^6]={1\over 48}{\rm Tr}[F^2]{\rm Tr}[F^4]-{1\over 14400}({\rm
Tr}[F^2])^3
\,.\label{332}\ee
Then
\be
96I^{\rm total}=\left({\rm Tr}[R^2]-{1\over 30}{\rm Tr}[F^2]\right)~X_8
\,,\label{333}\ee
with
\be
X_8={{\rm Tr}[F^4]\over 24}-{({\rm Tr}[F^2])^2\over 720}-{{\rm
Tr}[F^2]{\rm Tr}[R^2]\over
240}
+{{\rm Tr}[R^4]\over 8}+{({\rm Tr}[R^2])^2\over 32}\;,
\label{334}\ee
and the rest of the anomaly can be canceled via the Green-Schwarz
mechanism.
The only non-trivial condition that remains is (\ref{332}).

Consider first the gauge group to be O(N).
Then the following formulae apply \cite{GSW}
\be
{\rm Tr}[F^6]=(N-32){\rm tr}[F^6]+15{\rm tr}[F^2]{\rm tr}[F^4]
\,,\label{335}\ee
\be
{\rm Tr}[F^4]=(N-8){\rm tr}[F^4]+3({\rm tr}[F^2])^2\;\;\;,\;\;\;{\rm
Tr}[F^2]=(N-2){\rm tr}[F^2]
\,,\label{336}\ee
where Tr stands for the trace in the adjoint and tr for the trace
in the fundamental of O(N).

\vskip .4cm
\noindent\hrulefill
\nopagebreak\vskip .2cm
{\large\bf Exercise}.  Show that the factorization condition
(\ref{332})
and $\mbox{dim} G=496$ are satisfied by G=O(32).
Thus, the type-I and heterotic string theories with G=O(32) are
anomaly free.
\nopagebreak\vskip .2cm
\noindent\hrulefill
\vskip .4cm

Consider now $\rm G=E_8\times E_8$, which has also dimension 496.
$\rm E_8$ has no independent Casimirs of order 4 and 6,
\be
{\rm Tr}[F^6]={1\over 7200}({\rm Tr}[F^2])^3\;\;\;,\;\;\;{\rm
Tr}[F^4]={1\over
100}({\rm Tr}[F^2])^2
\,.\label{337}\ee

\vskip .4cm
\noindent\hrulefill
\nopagebreak\vskip .2cm
{\large\bf Exercise}.  Verify that $\rm E_8\times E_8$ satisfies
(\ref{332}).
Thus, the E$_8\times$E$_8$ heterotic string is also anomaly-free.
Check also that the groups E$_{8}\times$U(1)$^{248}$ and U(1)$^{496}$
are
anomaly-free. No known ten-dimensional string theory corresponds to
these groups.
\nopagebreak\vskip .2cm
\noindent\hrulefill
\vskip .4cm

The presence of the Green-Schwarz counterterm (\ref{320}) necessary for
the
cancelation
of the reducible anomaly, was checked by a one-loop computation in
the heterotic string \cite{Le}.
Moreover a direct relation between modular invariance and the absence
of anomalies was obtained.

{}From (\ref{317}) we obtain that (G=O(32))
\be
d\hat H={\rm tr}[R^2]-{1\over 30}{\rm Tr}[F^2]
\,.\label{338}\ee
Integrating (\ref{330}) over any closed four-dimensional submanifold,
we obtain the important constraint to be satisfied by the background
fields:
\be
\int ~{\rm tr}[R^2]={1\over 30}\int ~{\rm Tr}[F^2]
\,.\label{339}\ee

\vskip .4cm
\noindent\hrulefill
\nopagebreak\vskip .2cm
{\large\bf Exercise}. Consider the non-supersymmetric $\rm O(16)\times
O(16)$
heterotic string in ten dimensions.
It is a chiral theory with fermionic content transforming as
$(V,V)$, $(\bar S,1)$ and $(1,\bar S)$ under the gauge group.
$S$ stands for the 128-dimensional spinor representation of O(16).
Use
\be
{\rm tr}_S[F^6]=16{\rm tr}[F^6]-15{\rm tr}[F^2]{\rm tr}[F^4]+{15\over
4}({\rm tr}[F^2])^3
\,,\label{340}\ee
\be
{\rm tr}_S[F^4]=-8{\rm tr}[F^4]+6({\rm tr}[F^2])^2\;\;\;,\;\;\;{\rm
tr}_S[F^2]=16{\rm tr}[F^2]
\,,\label{341}\ee
with tr$_S$ the trace in the spinor representation space and tr the
trace in the
fundamental representation space to show that
the theory is anomaly-free. What is the Green-Schwarz counterterm?
Are there  any other chiral, non-supersymmetric anomaly-free theories
in ten dimensions?
\vskip .6cm
{\bf\large Exercise}. Consider the  $\rm E_8\times E_8$
heterotic
string in ten dimensions. This theory has a symmetry ${\cal I}$ that
interchanges the two $\rm E_8$ factors. Consider the $Z_2$ orbifold of
this theory
with respect to the symmetry transformation $g=(-1)^{F+1}\cdot {\cal
I}$.
Construct the modular-invariant partition function (you will need the
duplication formulae for the $\th$-functions that you can find in
appendix A).
What is the gauge group and the massless spectrum?
Is this theory supersymmetric? Chiral? Anomaly-free?
\nopagebreak\vskip .2cm
\noindent\hrulefill
\vskip .4cm

\renewcommand{\theequation}{\thesection.\arabic{equation}}
\section{Compactification and supersymmetry breaking\label{comp}}
\setcounter{equation}{0}

So far, we have considered superstring theories in ten non-compact
dimensions. However, our direct physical interest is in
four-dimensional theories.
One way to obtain them is to make use of the Kaluza-Klein idea:
consider some of the dimensions to be curled-up into a compact
manifold, leaving only four non-compact dimensions.
As we have seen in the case of the bosonic strings, exact solutions
to equations of motion of a string theory correspond to a CFT.
In the case of type-II string theory, they would correspond to a (1,1)
superconformal FT, and to a (1,0)
superconformal FT in the heterotic case.
The generalization of the concept of compactification to four
dimensions,
for example, is to replace the original flat non-compact CFT with
another one, where four dimensions are still flat but the rest is
described by an arbitrary CFT with the appropriate central charge.
This type of description is more general than that of a geometrical
compactification, since there are CFTs with no geometrical
interpretation.
In the following, we will examine both the geometric point of view and
the CFT point of view, mainly via orbifold compactifications.

\renewcommand{\theequation}{\thesubsection.\arabic{equation}}
\subsection{Toroidal compactifications\label{torcomp}}
\setcounter{equation}{0}

The simplest possibility is that the ``internal compact" manifold be a
flat torus. This can be considered as a different background of the
ten-dimensional theory, where we have given constant expectation
values
to internal metric and other background fields.

Consider first the case of the heterotic string compactified
to $D<10$ dimensions.
It is rather straightforward to construct the partition function of
the compactified theory.
There are now $D-2$ transverse non-compact coordinates, each
contributing
$\sqrt{\tau_2}\eta\bar\eta$.
There is no change in the contribution of the left-moving world-sheet
fermions and   16 right-moving compact coordinates.
Finally the contribution of the $10-D$ compact coordinates is given
by
(\ref{159}).
Putting everything together we obtain
\be
Z^{\rm heterotic}_D={\Gamma_{10-D,10-D}(G,B)~\bar\Gamma_H\over
\tau_2^{D-2\over 2}\eta^8\bar\eta^8}~{1\over
2}\sum_{a,b=0}^1~(-1)^{a+b+ab}~{\th^4[^a_b]\over \eta^4}
\,,\label{342}\ee
where $\bar\Gamma_H$ stands for the lattice sum for either O(32) or
$\rm E_8\times E_8$;
$G_{\a\b}$, $B_{\a\b}$ are the constant expectation values of the
internal
$(10-D)$-dimensional metric and antisymmetric tensor.
It is not difficult to find the massless spectrum of the theory.
The original ten-dimensional metric gives rise to the $D$-dimensional
metric,
$(10-D)$ U(1) gauge fields and $\half (10-D)(11-D)$ scalars.
The antisymmetric tensor produces a $D$-dimensional antisymmetric
tensor,
$(10-D)$ $U(1)$ gauge fields and $\half (10-D)(9-D)$ scalars (the
internal
components
of the gauge fields).
The ten-dimensional dilaton gives rise to another scalar.
Finally the dim $H$ ten-dimensional gauge fields give rise to dim $H$
gauge fields and $(10-D)\cdot$ dim $H$ scalars.
Similar reduction works for the fermions.

We will consider in more detail the scalars $Y^a_\alpha$ coming from
the ten-dimensional
vectors, where $a$ is the adjoint index and $\alpha$ the
internal index taking values $1,2,\dots,10-D$.
The non-abelian field strength (\ref{294}) contains a term without
derivatives.
Upon dimensional reduction this gives rise to a potential term for the
(Higgs) scalars $Y^a_{\alpha}$:
\be
V_{\rm Higgs}\sim
{f^a}_{bc}{f^a}_{b'c'}~G^{\a\g}~G^{\b\d}~Y_{\a}^{b}~
Y_{\b}^c ~Y_{\g}^{b'}~Y^{c'}_{\d}
\,.\label{343}\ee
This potential has flat directions (continuous families of minima)
when
$Y^a_{\a}$ takes constant expectation values in the Cartan subalgebra
of the Lie algebra.
We will  label these values by $Y^I_{\a}$, $I=1,2,\dots,16$.
This is a normal Higgs phenomenon and it generates a mass matrix for
the gauge fields
\be
\left[m^2\right]^{ab}\sim
G^{\a\b}~{f^{ca}}_{d}~{f^{cb}}_d'~Y^{d}_{\a}~Y^{d'}_{\b}
\,.\label{344}\ee
This mass matrix has rank-H generic zero eigenvalues. The gauge
fields belonging to the Cartan remain massless while all the other
gauge fields
get a non-zero mass.
Consequently, the gauge group is broken to the Cartan
$\sim \rm U(1)^{{\rm rank}-H}$.
If we also turn on these expectation values, then the heterotic
compactified partition function becomes

\be
Z^{\rm heterotic}_D={\Gamma_{10-D,26-D}(G,B,Y)\over \tau_2^{D-2\over
2}\eta^8\bar\eta^8}~{1\over
2}\sum_{a,b=0}^1~(-1)^{a+b+ab}~{\th^4[^a_b]\over \eta^4}
\,,\label{345}\ee
where the derivation of the $\Gamma_{10-D,26-D}$ lattice sum is
described in
detail in Appendix B.

The $(10-D)(26-D)$ scalar fields $G,B,Y$ are called moduli since they
can have arbitrary expectation values.
Thus, the heterotic string compactified down to $D$ dimensions
is essentially a continuous family of $vacua$
parametrized by the expectation values of the moduli that describe
the geometry of the internal manifold $(G,B)$ and the (flat) gauge
bundle ($Y$).

Consider now the tree-level effective action for the bosonic
massless
modes in the toroidally compactified theory.
It can be obtained by direct dimensional reduction of the
ten-dimensional heterotic effective action, which in the $\s$-model
frame\footnote{This is also called the ``string frame".}
is given by
(\ref{255}) with the addition of the gauge fields
\be
\a'^{8}~S_{10-d}^{\rm heterotic}=  \int d^{10} x\sqrt{-{\rm
det}~G_{10}}e^{-\Phi}
\left[R+(\nabla\Phi)^2-{1\over 12}\hat H^2-{1\over 4}{\rm
Tr}[F^2]\right]
+{\cal O}(\alpha')
\,.\label{346}\ee
The massless fields in $D$ dimensions are obtained from those of the
ten-dimensional theory by assuming that the latter do not depend on
the internal coordinates $X^{\a}$.
Moreover we keep only the Cartan gauge fields since they are the only
ones that will remain massless for generic values of the Wilson lines
$Y^I_{\a}$, $I=1,2,\dots,16$.
So, the gauge kinetic terms abelianize  ${\rm Tr}[F^2]\to
\sum_{I=1}^{16}
F^I_{\mu\nu}
F^{I,\m\n}$ with
\be
F^I_{\m\n}=\pd_{\m}A^I_{\n}-\pd_{\n}A^I_{\m}
\,.\label{347}\ee
Also
\be
\hat H_{\m\n\rho}=\pd_{\m}B_{\n\rho}-{1\over 2}\sum_I
A^I_{\m}F^I_{\n\rho}
+{\rm cyclic}
\,,\label{349}\ee
where we have neglected the gravitational Chern-Simons contribution,
since it is of higher order in $\a'$.

There is a standard ansatz to define the $D$-dimensional fields,
such that the gauge invariances of the compactified theory
are simple.
This is given in Appendix C.
In this way we obtain
\bea
S^{\rm heterotic}_{D}=\int ~d^Dx~\sqrt{-{\rm
det}~G}e^{-\Phi}\left[R+\pd^{\m}\Phi\pd_{\m}\Phi
-{1\over 12} \hat H^{\m\n\rho}\hat H_{\m\n\rho}-\right.\nn\\
\left.-{1\over 4}(\hat M^{-1})_{ij}
F^{i}_{\m\n}F^{j\m\n}+{1\over 8}{\rm Tr}(\pd_{\m} \hat M\pd^{\m}
\hat M^{-1})\right]\,,\label{348}\quad\quad
\eea
where $i=1,2,\dots,36-2D$ and
\be
\hat H_{\m\n\rho}=\pd_{\m}B_{\n\rho}-{1\over 2}L_{ij}
A^i_{\m}F^j_{\n\rho}
+{\rm cyclic}
\,.\label{350}\ee
The moduli scalar matrix $\hat M$ is given in (\ref{B5}).
This action has a continuous O(10-D,26-D) symmetry.
If $\Lambda \in $ O(10-D,26-D) is a $(36-2D)\times (36-2D)$ matrix
then
\be
\hat M\to \Omega ~\hat M~\Omega^T\;\;\;,\;\;\;A_{\m}\to \Omega\cdot
A_{\m}
\label{351}\ee
leaves the effective action invariant.
However, we know from the exact string theory treatment that the
presence
of the massive states coming from the lattice break this symmetry to
the discrete infinite subgroup O(10-D,26-D,$\Z)$.
This is the group of T-duality symmetries.
The $(10-D)(26-D)$ scalar action in (\ref{348}) is the
$\rm O(10-D,26-D)/(O(10-D)
\times O(26-D)$ $\s$-model.

We can also go to the Einstein frame by (\ref{258}), in which the
action
becomes
\bea
S^{\rm heterotic}_D=\int ~d^Dx~\sqrt{-{\rm det}~G_E}\left[R-{1\over
D-2}\pd^{\m}\Phi\pd_{\m}\Phi
-{e^{-{4\Phi\over D-2}}\over 12}\hat H^{\m\n\rho}\hat
H_{\m\n\rho}-\right.
\nn\\
\left.-{e^{-{2\Phi\over D-2}}\over 4}(\hat M^{-1})_{ij}
F^{i}_{\m\n}F^{j\m\n}+{1\over 8}Tr(\pd_{\m} \hat M\pd^{\m}
\hat M^{-1})\right]\,.\quad\quad\quad\quad
\label{352}\eea

For $D=4$, the ten-dimensional gravitino gives rise to 4
four-dimensional Majorana gravitini. Consequently, the four-dimensional
compactified theory
has N=4 local SUSY.
The relevant massless N=4 supermultiplets are the supergravity
multiplet
and the vector multiplet.
The supergravity multiplet contains the metric, six vectors (the
graviphotons), a scalar and an antisymmetric tensor, as well as four
Majorana gravitini and four  Majorana spin $\half$ fermions.
The vector multiplet contains a vector, four Majorana spin $\half$
fermions
and six scalars.
In total we have, apart from the SUGRA multiplet, 22 vector multiplets.

In $D=4$ the antisymmetric tensor is equivalent (on-shell) via a
duality
transformation to a pseudoscalar $a$, the ``axion".
It is defined (in the Einstein frame) by
\be
e^{-2\phi}\hat H_{\m\n\rho}={{\e_{\m\n\rho}}^{\s}\over \sqrt{-{\rm
det}~
g_{E}}}\na_{\s}a
\,.\label{353}\ee
This definition is such that the $B_{\m\n}$ equations of motion
$\nabla^{\m}e^{-\Phi}\hat H_{\m\n\rho}=0$ are
automatically solved by substituting (\ref{353}).
However the Bianchi identity for $\hat H$ from (\ref{350})
\be
{\epsilon^{\m\n\rho\s}\over \sqrt{-{\rm
det}~g_E}}\pd_{\m}\hat H_{\n\rho\s}=-L_{ij}
F^{i}_{\m\n}\tilde F^{j,\m\nu}
\,,\label{ 354}\ee
where
\be
\tilde F^{\m\n}={1\over 2}{\e^{\m\n\rho\s}\over \sqrt{-{\rm
det}~g_{E}}}
F_{\rho\s}\;,
\label{355}\ee
becomes, after substituting (\ref{353}), an equation of motion for the
axion:
\be
\na^{\mu}~e^{2\phi}\na_{\m}a=-{1\over 4}F^{i}_{\m\n}\tilde F^{j,\m\n}
\,.\label{356}\ee
This equation can be obtained from the ``dual" action
\bea
\tilde S^{\rm heterotic}_{D=4}=\int
d^4x\sqrt{-{\rm det}~g_{E}}\;\left[R-{1\over
2}\pd^{\m}\phi\pd_{\m}\phi
-{1\over 2}e^{2\phi}\pd^{\m}a\pd_{\m}a + \right. \nn\\
\left. -{1\over 4}e^{-\phi}(
M^{-1})_{ij}
F^{i}_{\m\n}F^{j,\m\n} + {1\over 4}a\;L_{ij}F^{i}_{\m\n}\tilde
F^{j,\m\n}+ \right.\nn\\
\left. +{1\over 8}{\rm Tr}(\pd_{\m}M\pd^{\m}M^{-1})\right]\,.\quad\quad
\label{357}\eea
Finally defining the complex $S$ field
\be
S=a+i\; e^{-\phi}\;,
\label{358}\ee
we can write the action as
\bea
\tilde S^{\rm heterotic}_{D=4}=\int
d^4x\sqrt{-{\rm det}~g_{E}}\;\left[R-{1\over 2}{\pd^{\m}S\pd_{\m}\bar
S\over {\rm Im}S^2}-{1\over 4}{\rm Im}S(M^{-1})_{ij}
F^{i}_{\m\n}F^{j,\m\n}\right.
\nn\\
\left.+{1\over 4}{\rm Re}S\;L_{ij}F^{i}_{\m\n}\tilde
F^{j,\m\n}+{1\over 8}{\rm Tr}(\pd_{\m}M\pd^{\m}M^{-1})\right]\;.\quad\quad\quad
\label{359}\eea
{}From the definition (\ref{358}), $1/{\rm Im}S$ is the
string loop-expansion parameter (heterotic string coupling constant).
As we will see later on, the 4-d heterotic string has a
non-perturbative SL(2,$\Z)$ symmetry acting on $S$ by fractional
transformations
and as electric-magnetic duality on the abelian gauge fields.
The scalar field $S$ takes values in the upper-half plane,
SL(2,$\R$)/U(1).
The rest of the  scalars take values in the coset space
O(6,22)/O(6)$\times$O(22).

We will briefly describe here the toroidal compactification
of type-II string theory to four dimensions.
It can be shown that in closed string theory with a compact dimension
of radius $R$, a duality transformation $R\to 1/R$ is accompanied
by a reversal of the chirality of the left-moving spinor groundstate.
This is explained in more detail in the last section.
Thus, type-IIA theory with radius $R$ is equivalent to type-IIB theory
with radius $1/R$. Once we compactify on a torus, both theories are
non-chiral.
We need only examine the type-IIA theory reduction to $D=4$.
First the two Majorana-Weyl gravitini and fermions  give rise to eight
$D=4$
Majorana gravitini and 48 spin $\half$ Majorana fermions.
Thus, the $D=4$ theory has maximal N=8 supersymmetry.
The ten-dimensional metric produces the four-dimensional metric, 6 U(1)
vectors and 21
scalars.
The antisymmetric tensor produces (after four-dimensional dualization),
6 U(1)
vectors
and 16 scalars.
The dilaton gives an extra scalar.
The $R$-$R$ U(1) gauge field gives one gauge field and 6 scalars.
The $R$-$R$ three-form gives a three-form (no physical degrees of
freedom in
four dimensions)
15 vectors and 26 scalars.
All of the above degrees of freedom form the N=8
supergravity multiplet that contains the graviton, 28 vectors, 70
scalars,
8 gravitini and 48 fermions.

\vskip .4cm
\noindent\hrulefill
\nopagebreak\vskip .2cm
{\large\bf Exercise}. Start from the ten-dimensional type-IIA effective
action in (\ref{6055}) and by using toroidal dimensional reduction
(you will find relevant formulae in appendix C) derive the
four-dimensional
effective action. Dualize all two-forms.
\nopagebreak\vskip .2cm
\noindent\hrulefill
\vskip .4cm

\subsection{Compactification on non-trivial manifolds\label{compact}}
\setcounter{equation}{0}

The next step would be to attempt to compactify the ten-dimensional
theories on non-flat manifolds.
Such backgrounds, however, must satisfy the string equations of motion.
As we described in a previous section, this is equivalent to
conformal
invariance of the associated $\s$-model.
When the background fields are slowly varying, the $\a'$ expansion is
applicable
and to leading order the background must satisfy the low-energy
effective
field equations of motion.

We will be  interested in ground-states for which the four-dimensional
world
is flat.
In the  most general case, such a ground-state is given by the tensor
product of a four-dimensional non-compact flat CFT and an internal
conformal field
theory.
A CFT with appropriate central charge and symmetries
is an exact solution of the (tree-level) string equations of motion
to all orders in $\a'$.
In the heterotic case, this internal CFT must have left N=1
invariance
and $(c,\bar c)=(9,22)$.
In the type-II case it must have both left and right N=1
superconformal invariance and $(c,\bar c)=(9,9)$.
If the CFT has a ``large volume limit", then an $\a'$ expansion is
possible
and we can recover the leading $\s$-model (geometrical) results.

It is also of interest, for the compactified theory, to have some
left-over
supersymmetry at the compactification scale.
For phenomenological purposes we eventually need N=1 supersymmetry,
since it is the only case that admits chiral representations.
Although the very low energy world is not supersymmetric, we do need
some supersymmetry beyond Standard-Model energies for hierarchy
reasons.

In the effective field theory approach, we assume that some bosonic
fields
acquire expectation values that satisfy the equations of motion, while
the
expectation
values of the fermions are zero (to preserve $D=4$ Lorentz
invariance).
In the generic case, a background breaks all the supersymmetries of
flat
ten-dimensional space.
A supersymmetry will be preserved, if the associated variation of
the fermion fields vanish.
This gives a set of first order equations.
If they are satisfied for at least one supersymmetry, then the
full equations of motion will also be satisfied.
Another way to state this is by saying that every compact manifold that
preserves at least one SUSY, is a solution of the  equations of motion.

We will consider here the case of the heterotic string on a space
that is locally $M_4\times K$ with $M_4$ the four-dimensional
Minkowski space
and $K$ some six-dimensional compact manifold.
Splitting indices into Greek indices for $M_4$ and Latin indices for
$K$,
we have the following supersymmetry variations (in the Einstein
frame) of the ten-dimensional  heterotic action
\be
\d \psi_{\mu}=\na_{\m}\e+{\sqrt{2}\over
32}e^{2\Phi}\left(\gamma_{\mu}\gamma_5
\otimes H\right)\e
\,,\label{360}\ee
\be
\d \psi_{m}=\na_m \e+{\sqrt{2}\over 32}e^{2\Phi}\left(\gamma_{m}H-12
H_m\right)\e
\,,\label{361}\ee
\be
\d \l=\sqrt{2}(\gamma^m\na_m\Phi)\e+{1\over 8}e^{2\Phi}H\e
\,,\label{362}\ee
\be
\d \chi^a=-{1\over 4}e^{\Phi}F^a_{m,n}\gamma^{mn}\e
\,,\label{363}\ee
where $\psi$ is the gravitino, $\l$ is the dilatino and $\chi^a$ are
the
gaugini;
$\e$ is a spinor (the parameter of the supersymmetry transformation).
Furthermore we used
\be
H=H_{mnr}~\gamma^{mnr}\;\;\;,\;\;\;H_m=H_{mnr}~\gamma^{nr}
\,.\label{364}\ee
The ten-dimensional $\Gamma$-matrices can be constructed from the $D=4$
matrices $\gamma^{\mu}$, and the internal matrices $\gamma^m$, as
\be
\Gamma^{\m}=\gamma^{\mu}\otimes {\bf 1}_6\;\;\;,\;\;\;
\Gamma^{m}=\gamma^{5}\otimes \gamma^m
\,,\label{365}\ee
\be
\gamma^5={i\over
4!}\e_{\mu\nu\rho\s}\gamma^{\mu\nu\rho\s}\;\;\;,\;\;\;
\gamma={i\over 6!}\sqrt{\rm det g}\e_{mnrpqs}\gamma^{mnrpqs}\;.
\label{366}\ee
$\gamma$ is the analog of $\gamma^5$ for the internal space.

If, for some value of the background fields, the equations $\d ({\rm
fermions})
=0$ admit a solution, namely a non-trivial, globally defined spinor
$\e$,
then the background is N=1 supersymmetric.
If more than one solution exist, then we will have extended
supersymmetry.
This problem was considered in \cite{chsw} with the assumption that
$H_{mnr}=0$.
The conditions for the existence of N=1 supersymmetry in four
dimensions for
$H=0$
can be summarized as follows: the dilaton must be constant and the
manifold $K$ must admit a Killing spinor
$\xi$,
\be
\na_m~\xi=0
\,.\label{367}\ee
Moreover this condition implies that $K$ is a Ricci-flat ($R_{mn}=0$)
K\"ahler manifold.
Finally the background (internal) gauge fields must satisfy
\be
F^a_{mn}\gamma^{mn}~\xi=0
\label{368}\ee
and (\ref{338}) then becomes
\be
{R^{rs}}_{[mn}R_{pq]rs}={1\over 30}F^a_{[mn}F^a_{pq]}
\,.\label{369}\ee

In a generic six-dimensional manifold, the spin connection is in
$\rm O(6)\sim SU(4)$.
If the manifold is K\"ahler, then the spin connection is in
$\rm U(3)\subset SU(4)$.
Finally, if the Ricci tensor vanishes, the spin connection is an
SU(3) connection.
Such manifolds are known as Calabi-Yau (CY) manifolds.

A simple way to solve (\ref{368}) and (\ref{369}) is to embed the spin
connection
$\omega\in \rm SU(3)$ into the gauge connection $A\in \rm O(32)$ or
$\rm E_8\times E_8$.
The only embedding of SU(3) in O(32) that satisfies (\ref{369})
is the one in which $\rm O(32)\ni 32\to 3+\bar 3+{\rm singlets}\in
SU(3)$.
In this case O(32) is broken down to $\rm U(1)\times O(26)$ (this is
the
subgroup
that commutes with SU(3)).
The U(1) is ``anomalous", namely the sum of the U(1) charges
$\rho=\sum_i q^i$ of the massless states is not zero.
This anomaly is apparent, since we know that the string theory is not
anomalous.
What happens is that the Green-Schwarz mechanism implies here that
there is a
one-loop coupling of the form $\rho~B\wedge F$. This gives a mass to
the U(1) gauge
field and it cannot appear as a low-energy symmetry.
There is a more detailed discussion of this phenomenon in section
\ref{anomu}.
The leftover gauge group O(26) has only non-chiral representations.

More interesting is the case of $\rm E_8\times E_8$.
$\rm E_8$ has a maximal $\rm SU(3)\times E_6$ subgroup, under which the
adjoint of $\rm E_8$ decomposes as $\rm E_8\ni {\bf 248}\to ({\bf
8,1})\otimes
({\bf 3,27})\otimes ({\bf \bar 3,\overline{27}})\otimes ({\bf 1,78})\in
SU(3)\times
E_6$.
Embedding the spin connection in one of the $\rm E_8$ in this fashion
solves (\ref{369}).
The unbroken gauge group in this case is $\rm E_6\times E_8$.
Let $N_L$ be the number of massless left-handed Weyl fermions in four
dimensions
transforming in the $\bf 27$ of $\rm E_6$ and $N_R$ the same number for
the
$\bf \overline{27}$.
The number of net chirality (number of ``generations") is $|N_L-N_R|$;
it can be obtained by an index theorem on the CY manifold.
The $\bf 27$'s transform as the $\bf 3$ of SU(3) and the
$\bf\overline{27}$
transform in the $\bf\bar 3$ of SU(3).
Thus, the number of generations is the index of the Dirac operator on
$K$
for the fermion field $\psi_{\a A}$, where $\a$ is a spinor index and
$A$ is a $\bf 3$ index.
It can be shown \cite{chsw}, that the index of the Dirac operator, and
thus the number of generations, is equal to $|\chi(K)/2|$,
where $\chi(K)$ is the Euler number of the manifold $K$.

The above considerations are correct to leading order in $\alpha'$.
At higher orders we expect, generically, corrections and only some
statements
about the massless states survive these corrections.

As another example we will consider the compactification of type-II
theory
on the K3 manifold down to six dimensions.
K3 is a topological class of  four-dimensional compact,
Ricci-flat, K\"ahler manifolds without isometries.
Such manifolds have SU(2)$\subset$O(4) holonomy and are also
hyper-K\"ahler.
The hyper-K\"ahler condition is equivalent  to the existence of three
integrable complex
structures
that satisfy the SU(2) algebra.
It can be shown that a left-right symmetric N=1 supersymmetric
$\sigma$-model on such manifolds is exactly conformally
invariant and has
N=4 superconformal symmetry on both sides.
Moreover, K3 has a covariantly constant spinor, so that the
type-II
theory compactified on it has N=2 supersymmetry in six dimensions
(and N=4 if further compactified on a two-torus).
It would be useful for latter purposes to briefly describe the
cohomology
of K3.
There is a harmonic  zero-form that  is constant (since the manifold is
compact and connected). There are no harmonic one-forms.
There is one (2,0) and one (0,2) harmonic forms as well as 20 (1,1)
forms.
The (2,0), (0,2) and one of the (1,1) K\"ahler forms are self-dual, the
other
19 (1,1)
forms are anti-self-dual.
There are no harmonic three-forms and a unique four-form (the volume
form).
More details on the geometry and topology of K3 can be found in
\cite{asp}.

Consider first the type-IIA theory and derive the massless
bosonic
spectrum in six dimensions.
To find the massless states coming from the ten-dimensional
metric $G$, we make the following decomposition
\be
G_{MN}\sim h_{\m\n}(x)\otimes \phi(y)+A_{\mu}(x)\otimes f_{m}(y)+
\Phi(x)\otimes h_{mn}(y)
\,,\label{370}\ee
where $x$ denotes the six-dimensional non-compact flat coordinates and
$y$ are
the internal coordinates. Also $\mu=0,1,\dots,5$ and  $m=1,2,3,4$ is
a K3 index.
Applying the ten-dimensional equations of motion to the metric $G$ we
obtain
that
$h_{\mu\nu}$ (the six-dimensional graviton) is massless if
\be
\square_y~\phi(y)=0
\,.\label{371}\ee
The solutions to this equation are the harmonic zero-forms,
and there is only one of them. Thus, there is one massless graviton in
six dimensions.
$A_{\mu}(x)$ is massless if $f_m(y)$ is covariantly constant on
K3.
Thus, it must be a harmonic one-form and there are none on K3.
Consequently, there are no massless vectors coming from the metric.
$\Phi(x)$ is a massless scalar if $h_{mn}(y)$ satisfies the
Lichnerowicz
equation
\be
-\square
h_{mn}+2R_{mnrs}h^{rs}=0\;\;\;,\;\;\;\na^mh_{mn}=g^{mn}h_{mn}=0
\,.\label{372}\ee
The solutions of this equation can be constructed out of the three
self-dual
harmonic two-forms $S_{mn}$ and the 19 anti-self-dual two-forms
$A_{mn}$.
Being harmonic, they satisfy the following equations ($R_{mnrs}$ is
anti-self-dual)
\be
\square f_{mn}-R_{mnrs}f^{rs}=\square f_{mn}+2R_{mrsn}f^{rs}=0
\,,\label{373}\ee
\be
\na_mA_{np}+\na_pA_{mn}+\na_n A_{pm}=0\;\;\;,\;\;\;\na^m A_{mn}=0
\,.\label{374}\ee
Using these equations and the self-duality properties it can be
verified
that solutions to the Lichnerowicz equation are given
by
\be
h_{mn}=A^p_m S_{pm}+A^p_n S_{pm}
\,.\label{375}\ee
Thus, there are $3\cdot 19=57$ massless scalars.
There is an additional massless scalar (the volume of K3)
corresponding to constant rescalings of the K3 metric, that
obviously preserves the
Ricci-flatness condition.
We obtain in total $58$ scalars.
The ten-dimensional dilaton also gives an extra  massless scalar in
six dimensions.

There is a similar expansion for the 2-index antisymmetric tensor:
\be
B_{MN}\sim B_{\m\n}(x)\otimes \phi(y)+B_{\mu}(x)\otimes f_{m}(y)+
\Phi(x)\otimes B_{mn}(y)
\,.\label{376}\ee
The masslessness condition implies that the zero-, one- and two-forms
($\phi$,
$f_m$, $B_{mn}$ respectively) be harmonic.
We obtain  one massless two-index antisymmetric tensor and 22
scalars in six dimensions.

{}From the $R$-$R$ sector we have a one-form that, following the same
procedure,
gives a massless vector and a three-form that gives a massless
three-form, and 22 vectors in six dimensions.
A massless three-form in six dimensions is equivalent to a massless
vector
via a duality transformation.

In total we have  a graviton, an antisymmetric tensor, 24 vectors
and 81
scalars.
The two gravitini in ten dimensions give rise to two Weyl gravitini in
six dimensions.
Their internal wavefunctions are proportional to the covariantly
constant spinor that
exists on K3.
The gravitini
preserve their original chirality. They have therefore opposite
chirality.
The relevant representations of (non-chiral or (1,1)) N=2
supersymmetry in six dimensions are:

$\bullet$ \underline{The vector multiplet}. It contains a vector, two
Weyl spinors of opposite chirality and four scalars.

$\bullet$ \underline{The supergravity multiplet}. It contains the
graviton, two Weyl
gravitini of opposite chirality, 4 vectors, an antisymmetric tensor,
a scalar
and 4 Weyl fermions of opposite chirality.

We conclude that the six-dimensional massless content of type-IIA
theory
on K3 consists of the supergravity multiplet and 20 U(1) vector
multiplets.
N=(1,1) supersymmetry in six dimensions is sufficient
to fix the two-derivative
low-energy couplings of the massless fields.
The bosonic part is

\bea
S^{IIA}_{\rm K3}=\int
d^6x\sqrt{-{\rm det}~G_6}e^{-\Phi}\left[R+\na^{\m}\Phi\na_{\m}\Phi
-{1\over 12}H^{\m\n\rho}H_{\m\n\rho}+\right.\quad\quad\nn\\
\left.+{1\over 8}{\rm Tr}(\pd_{\m} \hat M\pd^{\m}\hat
M^{-1})\right]-{1\over
4}\int
d^6x\sqrt{-{\rm det}~G}(\hat M^{-1})_{IJ}
F^{I}_{\m\n}F^{J\m\n}+\nn\\
+ {1\over 16}\int d^6x
\e^{\m\n\rho\s\tau\upsilon}B_{\m\n}F^{I}_{\rho\s}\hat L_{IJ}
F^{J}_{\tau\upsilon}\,,\quad\quad
\label{377}\eea
where $I=1,2,\dots,24$.
Supersymmetry and the fact that there are 20 vector multiplets
restricts the $4\cdot 20$ scalars to live on the coset space
$\rm O(4,20)/O(4)\times O(20)$ and there will be a continuous O(4,20)
global symmetry. Thus, they were parametrized by the matrix $\hat M$
as in (\ref{B5})
with $p=4$, where $\hat L$ is the invariant O(4,20) metric.
Here $H_{\m\n\rho}$ does not contain the Chern-Simons term. Note
also
the absence of the dilaton-gauge field coupling.
This is due to the fact that the gauge fields come from the $R$-$R$
sector.

Observe that type-IIA theory on K3 gives exactly the same massless
spectrum
as the heterotic string theory compactified on $T^4$.
The low-energy actions (\ref{348}) and (\ref{377}) are different,
though.
As we will see later on, there is a non-trivial and interesting
relation between the two.

Now consider the type-IIB theory compactified on K3 down
to six dimensions.
The $NS$-$NS$ sector bosonic fields ($G,B,\Phi$) are the same as in the
type-IIA theory and we obtain again a graviton, an antisymmetric
tensor and 81 scalars.

{}From the $R$-$R$ sector we have another scalar, which gives a
massless
scalar in D=6, another two-index antisymmetric tensor, which gives, in
six dimensions, a two-index antisymmetric tensor and 22 scalars and a
self-dual
4-index antisymmetric tensor, which gives 3 self-dual 2-index
antisymmetric tensors and 19 anti-self-dual 2-index antisymmetric
tensors and scalar.
Since we can split a 2-index antisymmetric tensor into a self-dual
and an anti-self dual part we can summarize the bosonic spectrum in the
following way: a
graviton,
5 self-dual and 21 anti-self-dual antisymmetric tensors, and 105
scalars.

Here, unlike the type-IIA case we obtain two massless Weyl gravitini of
the same chirality.
They generate a chiral N=(2,0) supersymmetry in six dimensions.
The relevant massless representations are:

$\bullet$ (2,0) \underline{The SUGRA multiplet}. It contains the
graviton, 5 self-dual antisymmetric tensors, and two left-handed Weyl
gravitini.

$\bullet$ (2,0) \underline{The tensor multiplet}. It contains an
anti-self-dual antisymmetric tensor, 5 scalars and 2 Weyl fermions of
 chirality opposite to that of the  gravitini.

The total massless spectrum forms the supergravity multiplet and
21 tensor multiplets.
The theory is chiral but anomaly-free.
The scalars live on the coset space $\rm O(5,21)/O(5)\times O(21)$ and
there
is a global O(5,21) symmetry.
Since the theory involves self-dual tensors, there is no covariant
action
principle,
but we can write covariant equations of motion.

 \vskip .4cm
\noindent\hrulefill
\nopagebreak\vskip .2cm
{\large\bf Exercise}. Use the results on anomalies to show that the
O(5,21),
(2,0), six-dimensional supergravity is anomaly-free.
\vskip .8cm
{\large\bf Exercise}. Consider compactifications of type-IIA,B theories
to four dimensions. Greek indices describe the four-dimensional part,
Latin ones the six-dimensional internal part.
Repeat the analysis at the beginning of this section and find the
conditions
for the internal fields $g_{mn},B_{mn},\Phi$ as well as $A_m,C_{mnr}$
for
type-IIA
and $\chi,B^{RR}_{mn},F^+_{mnrst}$ for type-IIB  so that
the effective four-dimensional theory has N=1,2,4 supersymmetry in flat
space.
\nopagebreak\vskip .2cm
\noindent\hrulefill
\vskip .4cm

\renewcommand{\theequation}{\thesubsection.\arabic{equation}}
\subsection{World-sheet versus spacetime supersymmetry\label{bd}}
\setcounter{equation}{0}

There is an interesting relation between world-sheet and spacetime
supersymmetry.
We will again consider first the case of the heterotic string with
D=4
flat Minkowski space.
An N-extended supersymmetry algebra in four dimensions is generated by
N Weyl
supercharges $Q^I_a$ and their Hermitian conjugates $\bar
Q^I_{\dot\a}$
satisfying the algebra
\bea
\{Q_{\a}^I,Q_{\b}^J\}&=&\e_{\a\b}Z^{IJ}\;,\nn\\
\{\bar Q_{\dot\a}^I,Q_{\dot\b}^J\}&=&\e_{\dot\a\dot\b}\bar Z^{IJ}
\;,\nn\\ \{Q_{\a}^I,\bar
Q_{\dot\a}^J\}&=&\d^{IJ}~\s^{\mu}_{\a\dot\a}P_{\m}
\,,\label{399}\eea
where $Z^{IJ}$ is the antisymmetric central charge matrix.

\def\S{\Sigma}
\def\Sb{\bar \Sigma}

As we have seen in section \ref{svertex}, the spacetime supersymmetry
charges
can be constructed from the massless fermion vertex at zero momentum.
In our case we have
\be
Q^I_{\a}={1\over 2\pi i}\oint dz~ e^{-\phi/2}S_{\a}\S^I
\;\;\;,\;\;\;\bar Q^I_{\dot\a}={1\over 2\pi i}\oint dz~
e^{-\phi/2}C_{\dot\a}\Sb^I
\,,\label{419}\ee
where $S,C$ are the spinor and conjugate spinor of O(4) and
$\S^I,\Sb^I$ are
operators in the $R$ sector of the internal CFT with conformal weight
$\tfrac{3}{8}$.
We will also need
\be
:e^{q_1\phi(z)}::e^{q_2\phi(w)}:=(z-w)^{-q_1q_2}
:e^{(q_1+q_2)\phi(w)}:+\dots
\,,\label{420}\ee
\be
S_{\a}(z)C_{\dot\a}(w)=\s^{\mu}_{\a\dot\a}~\psi^{\mu}(w)+{\cal
O}(z-w)
\,,\label{421}\ee
\bea
S_{\a}(z)S_{\b}(w)&=&{\e_{\a\b}\over \sqrt{z-w}}+{\cal O}(\sqrt{z-w})
\;,\nn\\
C_{\dot\a}(z)C_{\dot\b}(w)&=&{\e_{\dot\a\dot\b}\over
\sqrt{z-w}}+{\cal O}(\sqrt{z-w})
\,,\label{422}\eea
Using the above and imposing the anticommutation relations
(\ref{399})
we find that the internal operators must satisfy the following OPEs:
\be
\S^I(z)\Sb^J(w)={\d^{IJ}\over
(z-w)^{3/4}}+(z-w)^{1/4}~J^{IJ}(w)+\dots
\,,\label{423}\ee
\bea
\S^I(z)\S^J(w)&=&(z-w)^{-1/4}\Psi^{IJ}(w)+\dots\;,\nn\\
 \Sb^I(z)\Sb^J(w)&=&(z-w)^{-1/4}\bar\Psi^{IJ}(w)+\dots
\,,\label{424}\eea
where $J^{IJ}$ are some internal theory operators with weight 1 and
$\Psi^{IJ},\bar \Psi^{IJ}$ have weight 1/2.
The central charges are given by $Z^{IJ}=\oint \Psi^{IJ}$.
The $R$ fields $\S,\Sb$ have square root branch cuts
 with respect to the internal  supercurrent
\be
G^{int}(z)\S^I(w)\sim (z-w)^{-1/2}\;\;\;,\;\;\;
G^{int}(z)\Sb^I(w)\sim (z-w)^{-1/2}\;\;\;.\;\;\;
\label{425}\ee
BRST invariance of the fermion vertex implies  that the OPE
$(e^{-\phi/2}
S_{\alpha}\S^I)(e^{\phi}G)$ does have a single pole term.
This in turn implies that there are  no more singular terms
in (\ref{425}).

Consider an extra scalar X with two-point function
\linebreak[4]\mbox{$\langle X(z)X(w)\rangle =-\log(z-w)$}.
Construct the dimension-$\half$ operators
\be
\l^I(z)=\S^I(z)e^{iX/2}\;\;\;,\;\;\;\bar\l(z)=\Sb^I(z)e^{-iX/2}
\,.\label{426}\ee
Using (\ref{423}) and (\ref{424}) we can verify the following OPEs
\bea
\l^I(z)\bar\l^J(w)&=&{\d^{IJ}\over z-w}+\hat J^{IJ}+{\cal O}(z-w)
\,,\label{427}\\
\l^I(z)\l^J(w)&=&e^{iX}\Psi^{IJ}+{\cal
O}(z-w)\;,\\
\bar\l^I(z)\bar\l^J(w)&=&e^{-iX}\bar\Psi^{IJ}+{\cal
O}(z-w)
\,,\label{428}\eea
where $\hat J^{IJ}=J^{IJ}+\tfrac{i}{2}\d^{IJ}\pd X$.
Thus, $\l^I,\bar\l^I$ are N complex free fermions and they generate
an O(2N)$_1$ current algebra.
Moreover, this immediately shows that $\Psi^{IJ}=-\Psi^{JI}$.
Thus the original fields belong to the coset $\rm O(2N)_1/U(1)$.
It is not difficult to show that O(2N)$_1\sim $U(1)$\times$SU(N)$_1$.
The U(1) is precisely the one generated by $\pd X$.
We can now compute the OPE of the Cartan currents $\hat J^{II}$
\be
\hat J^{II}(z)\hat J^{JJ}(w)={\d^{IJ}\over (z-w)^2}+{\rm regular}
\,,\label{429}\ee
from which we obtain
\be
J^{II}(z)J^{JJ}(w)={\d^{IJ}-3/4\over (z-w)^2}+{\rm regular}
\,.\label{430}\ee

$\bullet$ \underline{N=1 spacetime supersymmetry}. In this case, there
is a single
field $\S$ and a single current that we will call $J$
\be
J=2J^{11}\;\;\;,\;\;\;J(z)J(w)={3\over (z-w)^2}+{\rm regular}
\ee
and no $\Psi$ operator.
Computing the three-point function
\be
\langle J(z_1) \S(z_2)\Sb(z_3)\rangle={3\over 2}{z_{23}^{1/4}\over
z_{12}
z_{13}}
\label{431}\ee
we learn that $\S,\Sb$ are affine primaries with U(1) charges $3/2$ and
$-3/2$ respectively.
Bosonize the U(1) current and separate the charge degrees of freedom
\be
J=i{\sqrt{3}}\pd \Phi\;\;\;,\;\;\;\S=e^{{i\sqrt{3}}\Phi/2}W^+
\;\;\;,\;\;\;\Sb=e^{-i\sqrt{3}\Phi/2}W^-
\,,\label{432}\ee
where $W^{\pm}$ do not depend on $\Phi$.
If we write the internal Virasoro operator as $T^{int}=\hat
T+T_{\Phi}$
with $T_{\Phi}=-(\pd\Phi)^2/2$, then $\hat T$ and $T_{\Phi}$ commute.
The fact that the dimension of the $\S$ fields is equal to the U(1)
charge
squared over 2 implies that $W^{\pm}$ have dimension zero and thus
must be proportional to the identity.
Consequently $\S,\Sb$ are pure vertex operators of the field $\Phi$.

Now consider the internal supercurrent and expand it in U(1)
charge eigenoperators
\be
G^{int}=\sum_{q\geq 0}e^{iq\Phi}T^{(q)}+e^{-q\Phi}T^{(-q)}
\,,\label{433}\ee
where the operators $T^{(\pm q)}$ do not depend on $\Phi$.
Then, (\ref{425}) implies that $q$ in (\ref{433}) can only take the
value
$q=1/\sqrt{3}$.
We can write $G^{int}=G^++G^-$ with
\be
J(z)G^{\pm}(w)=\pm{G^{\pm}(w)\over (z-w)}+\dots
\,.\label{434}\ee
Finally the N=1 superconformal algebra satisfied by $G^{int}$ implies
that, separately, $G^{\pm}$ are Virasoro primaries with weight 3/2.
Moreover the fact that $G^{int}$ satisfies (\ref{203}) implies
that $J,G^{\pm},T^{int}$ satisfy the N=2 superconformal algebra
(\ref{211})-(\ref{215}) with $c=9$.
The reverse argument is obvious: if the internal CFT has N=2
invariance,
then one can use the (chiral) operators of charge $\pm 3/2$ to
construct the spacetime supersymmetry charges.
In section \ref{N=2superconformal} we have shown, using the spectral
flow,
that such Ramond operators are always in the spectrum since they are
the images of the $NS$ ground-state.

We will describe here how the massless spectrum emerges from the
general properties of the internal N=2 superconformal algebra.
As discussed in section  \ref{N=2superconformal}, in the $NS$ sector
of the internal N=2 CFT, there are two relevant ground-states, the
vacuum
$|0\rangle$ and the chiral ground-states $|h,q\rangle=|1/2,\pm
1\rangle$.
We have also the four-dimensional left-moving world-sheet fermion
oscillators $\psi^{\mu}_r$, the four-dimensional right-moving bosonic
oscillators
$\bar a_{n}$.
Also in the right-moving sector of the internal CFT we have, apart from
the vacuum state, a collection of $\bar h=1$ states.
Combining the internal ground-states, we obtain:
\be
|h,q;\bar h\rangle~~~:~~~
|0,0;0\rangle\;\;,\;\;|0,0;1\rangle^I\;\;,\;\;
|1/2,\pm 1;1\rangle^i\;\;,
\ee
where the indices $I=1,2,\cdots,M$, $i=1,2\cdots,\bar M$ count the
various such states.
The physical massless bosonic states are:
\begin{itemize}
\item $\psi^{\mu}_{-1/2}\bar a^{\nu}_{-1}|0,0;0\rangle$, which provide
the graviton, antisymmetric tensor and dilaton.

\item $\psi^{\mu}_{-1/2}|0,0;1\rangle^I$. They provide massless vectors
of gauge group with dimension $M$.

\item $|1/2,\pm 1;1\rangle^i$. They provide $\bar M$ complex scalars.

\end{itemize}

Taking into account also the fermions, from the $R$ sector, we can
organize the massless spectrum in multiplets of N=1 four-dimensional
supersymmetry.
Using the results of Appendix D, we obtain the N=1 supergravity
multiplet,
one tensor multiplet (equivalent under a duality transformation to a
chiral
multiplet),
$M$ vector multiplets and $\bar M$ chiral multiplets.

$\bullet$ \underline{N=2 spacetime supersymmetry}. In this case there
are
two fields $\S^{1,2}$ and four currents $J^{IJ}$.
Define $J^s=J^{11}+J^{22}$, $J^3=(J^{11}-J^{22})/2$ in order to
diagonalize (\ref{430}):
\bea
J^s(z)J^s(w)&=&{1\over (z-w)^2}+\dots\,,\;\;\nn\\
J^3(z)J^3(w)&=&{1/2\over (z-w)^2}+\dots\,,\;\;\nn\\
J^s(z)J^3(w)&=&\cdots
\,.\label{435}\eea
In a similar fashion we can show that under ($J^s,J^3$),
$\S^1$ has charges $(1/2,1/2)$, $\S_2$~$(1/2, -1/2)$,
$\Sb^1$~$(-1/2,-1/2)$
and $\Sb^2$~$(-1/2,1/2)$.
Moreover their charges saturate their conformal weights so that if we
bosonize the currents then the fields $\S,\Sb$ are pure vertex
operators
\be
J^s=i\pd\phi\;\;\;,\;\;\;J^3={i\over \sqrt{2}}\pd\chi
\,,\label{436}\ee
 \be
\S^1=\exp\left[{i\over 2}\phi+{i\over \sqrt{2}}\chi\right]
\;\;\;,\;\;\;\S^2=\exp\left[{i\over 2}\phi-{i\over
\sqrt{2}}\chi\right]
\,,\label{437}\ee
\be
\Sb^1=\exp\left[-{i\over 2}\phi-{i\over \sqrt{2}}\chi\right]
\;\;\;,\;\;\;\Sb^2=\exp\left[-{i\over 2}\phi+{i\over
\sqrt{2}}\chi\right]\;.
\ee
Using these in (\ref{423}) we obtain that
$J^{12}=\exp[i\sqrt{2}\chi]$
and $J^{21}=\exp[-i\sqrt{2}\chi]$.
Thus, $J^3,J^{12},J^{21}$ form the current algebra $\rm SU(2)_1$.
Moreover, $\Psi^{12}=\exp[i\phi]$, $\bar\Psi^{12}=\exp[-i\phi]$.

We again consider the internal supercurrent and expand it in charge
eigenstates.
Using (\ref{422}) we can verify that the charges that can appear
are $(\pm 1,0)$ and $(0,\pm 1/2)$.
We can split
\bea
&G^{int}=G_{(2)}+G_{(4)}\;\;\;,\;\;\;
G_{(2)}=G_{(2)}^++G_{(2)}^-
\;,&\nn\\
&G_{(4)}=G_{(4)}^++G_{(4)}^-
\,,&\label{438}\eea
\bea
J^s(z)G_{(2)}^{\pm}(w)&=&\pm{G_{(2)}^{\pm}(w)\over z-w}+\dots
\,,\;\;\;\nn\\
J^3(z)G_{(4)}^{\pm}(w)&=&\pm{1\over 2}{G_{(4)}^{\pm}(w)\over
z-w}+\dots
\,,\label{439}\eea
\be
J^s(z)G_{(4)}^{\pm}(w)={\rm finite}\;\;\;,\;\;\;
J^3(z)G_{(2)}^{\pm}(w)={\rm finite}
\ee
\be
G_{(2)}^{\pm}=e^{\pm i\phi}Z^{\pm}
\,.\label{440}\ee
$Z^{\pm}$ are dimension-1 operators. They can be written
in terms of scalars as $Z^{\pm}=i\pd X^{\pm}$.
The vertex operators $e^{\pm i\phi}$ are those of a complex free
fermion.
Thus, the part of the internal theory corresponding to $G^{(2)}$ is a
free
two-dimensional CFT with $c=3$.
Finally it can be shown that the SU(2) algebra acting on
$G_{(4)}^{\pm}$
supercurrents generates two more supercurrents that form
the N=4 superconformal algebra
(\ref{400})-(\ref{402}) with $c=6$.

Since there is a complex free fermion $\psi=e^{i\phi}$ in the $c=3$
internal CFT we can construct two massless vector boson states
$\psi_{-1/2}\bar a^{\m}_{-1}
|p\rangle$ and $\bar\psi_{-1/2}\bar a^{\m}_{-1}
|p\rangle$. One of them is the graviphoton belonging to the N=2
supergravity
multiplet while the other is the vector belonging to the vector-tensor
multiplet
(to which the dilaton and $B_{\m\n}$ also belong).
The vectors of massless vector multiplets correspond to states of the
form
$\psi^{\mu}_{-1/2}\bar J^a_{-1}|p\rangle$, where $\bar J^a$ is a
right-moving affine current.
The associated massless complex scalar of the vector multiplet
corresponds to the state
$\psi_{-1/2}\bar J^a_{-1}|p\rangle$.
Massless hypermultiplet bosons arise from the N=4 internal CFT.
As already described in section \ref{N=4superconformal}, an N=4
superconformal
CFT with $c=6$ always contains states with $\Delta =\half$ that
transform as two conjugate
doublets of the SU(2)$_1$ current  algebra.
Combining them with a right-moving operator with $\bar \Delta=1$ gives
the
four massless scalars of a hypermultiplet.

$\bullet$ \underline{N=4 spacetime supersymmetry}. In this case going
through
the same analysis we find that one  out of the four diagonal
currents, namely $J^{11}+J^{22}+J^{33}+J^{44}$, is null and thus
identically zero.

\vskip .4cm
\noindent\hrulefill
\nopagebreak\vskip .2cm
{\large\bf Exercise}. Bosonize the leftover three currents, write
the $\S,\Sb$ fields as vertex operators and show that in this case
the left-moving internal CFT has to be a toroidal one.
\nopagebreak\vskip .2cm
\noindent\hrulefill
\vskip .4cm

The six graviphotons participating in the N=4 supergravity multiplet
are states of the form $\bar a^{\m}_{-1}\psi^I_{-1/2}|p\rangle$
where $I=1,\dots,6$ and the $\psi^I$ are the fermionic partners of
the six left-moving currents of the toroidal CFT mentioned above.

In all the above, there are no constraints due to spacetime SUSY on
the right-moving side of the heterotic string.
We will use the notation ($p,q$) to denote $p$ left-moving
superconformal
symmetries and $q$ right-moving ones.
To summarize,  in the D=4 heterotic string the internal CFT has
at least (1,0) invariance. If it has (2,0) then we have N=1 spacetime
SUSY.
If we have $c=3$ $(2,0)\oplus c=6$ $(4,0)$ then we have N=2 in
spacetime.
Finally, if we have six free left-moving coordinates then we have N=4
in
four-dimensional spacetime.

In the type-II theory, the situation is similar, but here the
supersymmetries
can come from either the right-moving and/or the left-moving side.
For example, N=1 spacetime supersymmetry needs (2,1) world-sheet SUSY.
For N=2 spacetime supersymmetry there are two possibilities.
Either (2,2), in which one supersymmetry comes from the left and one
from the right, or $c=3$ $(2,0)\oplus c=6$ $(4,0)$ on one side only,
in which both spacetime supersymmetries come from this side.

More details on this can be found in \cite{BDFM,BD}.

\renewcommand{\theequation}{\thesubsection.\arabic{equation}}
\subsection{Heterotic orbifold compactifications with N=2 supersymmetry
\label{hn2}}
\setcounter{equation}{0}

In this section we will consider exact orbifold CFTs to provide
compactification spaces that reduce the maximal supersymmetry in
four dimensions.
We will focus for concreteness on the heterotic string.

We have already seen in section \ref{torcomp} that toroidal
compactification
of the heterotic string down to four dimensions, gives a theory with
N=4
supersymmetry.
What we would like to do is to consider orbifolds of this theory
that have N=1,2 spacetime supersymmetry.
We will have to find orbifold symmetries under which some of the four
four-dimensional
gravitini are not invariant. They will be projected out of the
spectrum and we will be left with a theory that has less
supersymmetry.
To find such symmetries we will have to look carefully at the
vertex
operators of the gravitini first.
We will work in the light-cone gauge and it will be convenient to
bosonize
the eight transverse left-moving fermions $\psi_i$ into four
left-moving
scalars.
Pick a complex basis for the fermions
\be
\psi^0={1\over \sqrt{2}}(\psi^3+i\psi^4)\;\;\;,\;\;\;\psi^1={1\over
\sqrt{2}}(\psi^5+i\psi^6)
\,,\label{3777}\ee
\be
\psi^2={1\over \sqrt{2}}(\psi^7+i\psi^8)\;\;\;,\;\;\;\psi^3={1\over
\sqrt{2}}(\psi^9+i\psi^{10})
\label{378}\ee
and similarly for $\bar \psi^I$.
They satisfy
\be
\langle \psi^I(z)\bar \psi^J(w)\rangle={\d^{IJ}\over
z-w}\;\;\;,\;\;\;
\langle \psi^I(z)\psi^J(w)\rangle=\langle \bar \psi^I(z)\bar
\psi^J(w)\rangle=0\,.\label{379}\ee
The four Cartan currents of the left-moving $\rm O(8)_1$ current
algebra
$J^I=\psi^I\bar\psi^I$ can be written in terms of four free bosons as
\be
J^I(z)=i\pd_z \phi^I(z)\;\;\;,\;\;\;\langle\phi^I(z)\phi^J(w)\rangle
=-\d^{IJ}\log(z-w)
\,.\label{3799}\ee
In terms of the bosons
\be
\psi^I=:e^{i\phi^I}:\;\;\;,\;\;\; \bar\psi^I=:e^{-i\phi^I}:
\label{380}\ee
The spinor primary states are given by
\be
V(\e_I)=:\exp\left[{i\over 2}\sum_{I=0}^3\e_I~\phi^I\right]:
\,,\label{381}\ee
with $\e_I=\pm 1$.
This operator has $2^4=16$ components and contains both the $S$ and
the $C$ $\rm O(8)$ spinor.

The fermionic system has an O(8) global symmetry (the zero mode part
of the
$\rm O(8)_1$ current algebra.
Its $\rm U(1)^4$ abelian subgroup acts as
\be
\psi^I\to e^{2\pi i \theta^I}\psi^I\;\;\;,\;\;\;\bar\psi^I\to
e^{-2\pi i \theta^I}\bar\psi^I
\,.\label{382}\ee
This acts equivalently on the bosons as
\be
\phi^I\to \phi^I+2\pi~\theta^I
\,.\label{383}\ee
A $Z_2$ subgroup of the $\rm U(1)^4$ symmetry, namely
$\theta^I=1/2$ for all $I$, is the $(-1)^F$ symmetry.
Under this transformation, the fermions are odd as they should be and
the spinor vertex
operator
transforms with a phase $\exp[i\pi(\sum_I~\e^I)/2]$.
Thus,

$\bullet$ $\sum_I~\e^I=4 k$, $k\in \Z$ corresponds to the spinor $S$.

$\bullet$ $\sum_I~\e^I=4 k+2$, $k\in \Z$ corresponds to the conjugate
spinor $C$.

The standard GSO projection picks one of the two spinors, let us say
the $S$.
Consider the massless physical  vertex operators  given by
\be
V^{\pm,\e}=\bar\pd X^{\pm}~V_S(\e)e^{ip\cdot
X}\;\;\;,\;\;\;X^{\pm}={1\over \sqrt{2}}
(X^3\pm iX^4)
\,.\label{384}\ee
The boson $\phi^0$ was constructed from the $D=4$ light-cone spacetime
fermions
and thus carries four-dimensional  helicity.
The $X^{\pm}$ bosons also carry four-dimensional  helicity $\pm 1$.
The subset of the vertex operators in (\ref{384}) that
corresponds to the gravitini are $\bar\pd X^+ V(\e^0=1)$, with
helicity $3/2$,
and $\bar\pd X^- V(\e^0=-1)$, with helicity $-3/2$.
Taking also into account the GSO projection we find four helicity
($\pm 3/2$)
states, as we expect in an N=4 theory.

Consider the maximal subgroup O(2)$\times $O(6)$\subset $O(8)
where the O(2) corresponds to the four-dimensional helicity.
The O(6) symmetry is an internal symmetry from the
four-dimensional point of
view.
It is the so-called $R$-symmetry of N=4 supersymmetry, since the
N=4
supercharges transform as the four-dimensional spinor of O(6), and
O(6)
is an automorphism of the N=4 supersymmetry algebra (with vanishing
 central charges) .
Since the supercharges are used to generate the states of an N=4
supermultiplet, the various states inside the multiplet have
well-defined
transformation
properties under the O(6) $R$-symmetry.
Here are some useful examples:

$\bullet$ \underline{The N=4 SUGRA multiplet}. It contains the
graviton (singlet of O(6))
four Majorana gravitini (spinor of O(6)), six graviphotons (vector
of O(6)),
four Majorana fermions (conjugate spinor of O(6)), and two scalars
(singlets).

$\bullet$ \underline{The massless spin 3/2 multiplet}. It contains a
gravitino (singlet),
four vectors (spinor), seven Majorana fermions (vector plus singlet)
and eight
scalars (spinor + conjugate spinor).

$\bullet$ \underline{The massless vector multiplet}. It contains a
vector (singlet), four Majorana fermions (spinor) and six scalars
(vector).

If we break the O(6) $R$-symmetry, then we will break the N=4
structure of supermultiplets. This will break N=4
supersymmetry.

We will  now search for symmetries of the CFT that will reduce,  after
orbifolding, the supersymmetry.
In order to preserve Lorentz invariance, the symmetry should not act
on the
four-dimensional supercoordinates $X^{\mu}$,
$\psi^{\mu}$.\footnote{There is an
exception to this statement, but I will not consider this further.}
The rest are symmetries acting on the internal left-moving fermions
and a simple class are  the discrete subgroups of the $\rm U(1)^3$
subgroup
of O(6)
acting on the fermions.
There are also symmetries acting on the bosonic $(6,22)$ compact CFT.
An important constraint on such symmetries is to leave the internal
supercurrent
\be
G^{\rm int}=\sum_{i=5}^{10}\psi^i~\pd~X^i
\label{385}\ee
invariant.
The reason is that $G^{\rm int}$ along with $G^{D=4}$ (which is
invariant
since we are not acting on the $D=4$ part) define the constraints
(equations
of motion)
responsible for the absence of ghosts. Messing them up can jeopardize
the unitarity of the orbifold theory.

We will start with a simple example of a $Z_2$ orbifold that will
produce N=2
supersymmetry in four dimensions.
Consider setting the Wilson lines to zero for the moment and pick
appropriately
the internal six-torus $G,B$ so that the $(6,22)$ lattice factorizes as
$(2,2)\otimes(4,4)\otimes (0,16)$.
This lattice has a symmetry that changes the sign of all the (4,4)
bosonic coordinates.
To keep the internal supercurrent invariant we must also change the
sign
of the fermions $\psi^i$, $i=7,8,9,10$.
This corresponds to shifting the associated bosons
\be
\phi^{2}\to\phi^2+\pi\;\;\;,\;\;\;\phi^{3}\to\phi^3-\pi
\,.\label{386}\ee
An immediate look at the four  gravitini vertex operators indicates
that two of them are invariant while the other two transform with a
minus sign.
We have exactly what we need. We are not yet done, though.

\vskip .4cm
\noindent\hrulefill
\nopagebreak\vskip .2cm
{\large\bf Exercise}. Compute the partition function of the above
orbifold. Show that it is not modular-invariant.
\nopagebreak\vskip .2cm
\noindent\hrulefill
\vskip .4cm

We must make a further action somewhere else.
What remains is the $(0,16)$ part.
Consider the case in which it corresponds to the $\rm E_8\times E_8$
lattice.
As we have mentioned already,
E$_8\ni {\bf [248]\to [120]\oplus[128]}\in$O(16).
Decomposing further with
respect to the $\rm SU(2)\times SU(2)\times O(12)$ subgroup of O(16),
we
obtain:
\bea
{\bf [120]\to
[3,1,1]\oplus[1,3,1]\oplus[1,1,66]\oplus[2,1,12]\oplus[1,2,12]}\nn\\
\in {\rm SU(2)\times SU(2)\times O(12)}\,,\quad\quad
\label{387}\eea
\be
{\bf [128]\to [2,1,32]\oplus[1,\bar 2,32]} \in {\rm SU(2)
\times SU(2)\times O(12)}
\,.\label{388}\ee
The action on $\rm E_8$ will be to take the spinors (the [$\bf 2$]'s)
of the
two SU(2) subgroups
to minus themselves, but keep the conjugate spinors (the [$\bf\bar
2$]'s)
invariant.
This projection keeps the $\bf [3,1,1],[1,3,1], [1,1,66], [1,\bar
2,32]$
representations that combine to form the group $\rm E_7\times SU(2)$.
This can be seen by decomposing the adjoint of $\rm E_{8}$ under its
$\rm SU(2)
\times E_7$ subgroup.
\be
{\rm E_8}\ni {\bf [248]\to [1,133]\oplus[3,1]\oplus [2,56]}\in {\rm
SU(2)\times
E_7}
\,,\label{389}\ee
where in this basis the above transformation corresponds to $\bf [3]\to
[3]$
and $\bf [2]\to -[2]$.
The reason why we considered  a more complicated way in terms of
orthogonal groups is that, in this language, the construction of the
orbifold blocks is straightforward.

We will  now construct the various orbifold blocks.
The left-moving fermions contribute
\be
{1\over 2}\sum_{a,b=0}^1~(-1)^{a+b+ab}~{\th^2[^a_b]\th[^{a+h}_{b+g}]
\th[^{a-h}_{b-g}]\over \eta^4}
\,.\label{390}\ee
The bosonic (4,4) blocks can be constructed in a fashion similar to
(\ref{232}).
We obtain
\be
Z_{(4,4)}[^0_0]={\Gamma_{4,4}\over
\eta^4\bar\eta^4}\;\;\;,\;\;\;Z_{(4,4)}[^h_g]=2^4{\eta^2\bar
\eta^2\over \th^2[^{1-h}_{1-g}]\bar\th^2[^{1-h}_{1-g}]}
\;\;\;,\;\;\;(h,g)\not=(0,0)
\,.\label{391}\ee
The blocks of the $\rm E_8$ factor in which our projection acts are
given
by
\be
{1\over
2}\sum_{\g,\d=0}^1~{\bar\th[^{\g+h}_{\d+g}]\bar\th[^{\g-h}_{\d-g}]
\bar\th^6[^{\g}_{\d}]\over \bar\eta^8}
\,.\label{392}\ee
Finally there is a (2,2) toroidal and an $\rm E_8$ part that are
not touched by  the projection.
Putting all things together we obtain the heterotic partition
function of the $Z_2$ orbifold
\be
Z^{\rm heterotic}_{N=2}={1\over
2}\sum_{h,g=0}^1~{\Gamma_{2,2}~\bar\Gamma_{E_8} Z_{(4,4)}[^h_g]\over
\t_2\eta^4\bar\eta^{12}}~{1\over
2}\sum_{\g,\d=0}^1~{\bar\th[^{\g+h}_{\d+g}]\bar\th[^{\g-h}_{\d-g}]
\bar\th^6[^{\g}_{\d}]\over \bar\eta^8}\times
 \label{393}\ee
$$\times {1\over
2}\sum_{a,b=0}^1~(-1)^{a+b+ab}~{\th^2[^a_b]\th[^{a+h}_{b+g}]
\th[^{a-h}_{b-g}]\over \eta^4}\,.
$$

\vskip .4cm
\noindent\hrulefill
\nopagebreak\vskip .2cm
{\large\bf Exercise}. Show that the above partition function is
modular-invariant.
Find the bosonic ($a=0$) massless spectrum. In particular show that,
from the untwisted sector ($h=0$) we obtain
the graviton, an antisymmetric tensor, vectors in the adjoint of
$\rm G=U(1)^4\times
SU(2)\times E_7\times E_8$, a complex scalar in the adjoint of the
gauge group G, 16 more neutral scalars and scalars transforming as
four copies of
the $\bf [56,2]$ representation of $\rm E_7\times SU(2)$.
{}From the twisted sector $(h=1)$, show that we obtain scalars only
transforming
as 32 copies of the $\bf [56,1]$ and 128 copies of the $\bf [1,2]$.
\nopagebreak\vskip .2cm
\noindent\hrulefill
\vskip .4cm

As mentioned before, this four-dimensional
 theory has N=2 local supersymmetry.
The associated $R$-symmetry is SU(2), which rotates the two
supercharges.
We will  describe the relevant massless representations and their
transformation properties under the $R$-symmetry.

$\bullet$ \underline{The SUGRA multiplet} contains the graviton
(singlet), two Majorana gravitini (doublet) and a vector (singlet).

$\bullet$ \underline{The vector multiplet} contains a vector
(singlet) two Majorana fermions (doublet), and a complex (two real)
scalars (singlets).

$\bullet$ \underline{The vector-tensor multiplet} contains a vector
(singlet), two Majorana fermions (doublet), a real scalar (singlet)
and an antisymmetric tensor (singlet).

$\bullet$ \underline{The hypermultiplet} contains two Majorana
fermions (singlets)
and four scalars (two doublets).

We can now arrange the massless states into N=2 multiplets.
We have the SUGRA multiplet, a vector-tensor multiplet (containing the
dilaton),
a vector multiplet in the adjoint of U(1)$^2\times $SU(2)$\times
$E$_7\times $E$_8$;
the rest are hypermultiplets transforming under SU(2)$\times $E$_7$ as
$\bf 4[1,1]+[2,56]+8[1,56]+32[2,1]$.

We will  also further investigate  the origin of the SU(2)
$R$-symmetry.
Consider the four real left-moving fermions $\psi^{7,\dots,10}$.
Although they transform with a minus sign under the orbifold action,
their O(4)$\sim $SU(2)$\times $SU(2) currents, being bilinear in the
fermions, are invariant.
Relabel the four real fermions as $\psi^0$ and $\psi^a$,
$a=1,2,3$.
Then,  the $\rm SU(2)_1\times SU(2)_1$ current algebra is generated
by
\be
J^a=-{i\over 2}\left[\psi^0\psi^a+{1\over
2}\e^{abc}\psi^b\psi^c\right]
\;\;\;,\;\;\;\tilde J^a=-{i\over 2}\left[\psi^0\psi^a-{1\over
2}\e^{abc}\psi^b\psi^c\right]
 \,.\label{394}\ee
Although both SU(2)'s are invariant in the untwisted sector, the
situation in the twisted sector is different.
The O(4) spinor ground-state decomposes as $\bf [4]\to [2,1]+[1,2]$
under $\rm SU(2)\times SU(2)$.
The orbifold projection acts trivially on the spinor of the first
SU(2)
and with a minus sign on the spinor of the second.
The orbifold projection breaks the second SU(2) invariance.
The remnant $\rm SU(2)_1$ invariance becomes the $R$-symmetry of the
N=2
theory.
Moreover, the only operators (relevant for massless states) that
transform
non-trivially under the SU(2) are the (quaternionic) linear
combinations
\be
V^{\pm}_{\a\b}=\pm i(\d_{\a\b}\psi^0\pm i\s^a_{\a\b}~\psi^a)\;,
\label{397}\ee
which  transform as the $\bf [2]$ and $\bf [\bar 2]$
respectively, as well as the $\bf [2]$ spinor in the $R$-sector.
We obtain
\be
V^+_{\a\g}(z)V^+_{\g\b}(w)=V^-_{\a\g}(z)V^-_{\g\b}(w)={\d_{\a\b}\over
z-w}-2\s^a_{\a\b}(J^a(w)-\tilde J^a(w))+{\cal O}(z-w)
\,,\label{395}\ee
\be
V^+_{\a\g}(z)V^-_{\g\b}(w)={3\d_{\a\b}\over z-w}+4\s^a_{\a\b}~\tilde
J^a(w)+{\cal O}(z-w)
\,,\label{396}\ee
\be
V^-_{\a\g}(z)V^+_{\g\b}(w)={3\d_{\a\b}\over z-w}-4\s^a_{\a\b}~
 J^a(w)+{\cal O}(z-w)
\,,\label{398}\ee
where a summation over $\g$ is implied.

This $\rm SU(2)_1$ current algebra combines with
four operators
of conformal weight $3/2$ to make the N=4 superconformal algebra in
any theory with N=2 spacetime supersymmetry agree with the general
discussion of section \ref{bd}.

In an N=2 theory, the complex scalars that are partners of the
gauge bosons belonging to the Cartan of the gauge group are moduli
(they have no potential). If they acquire generic expectation values,
they break the gauge group down to the Cartan.
All charged hypermultiplets also get masses.

A generalization of the above orbifold, where all
Higgs expectation values are turned on, corresponds to splitting
the original (6,22) lattice to (4,4)$\oplus$(2,18).
We perform a $Z_2$ reversal in the $(4,4)$, which will break $\rm
N=4\to
N=2$.
In the leftover lattice we can only perform a $Z_2$ translation
(otherwise the supersymmetry will be broken further).
We will perform a translation by $\e/2$, where $\e\in L_{2,18}$.
Then the partition function is
\be
Z^{\rm heterotic}_{N=2}={1\over
2}\sum_{h,g=0}^1~{\Gamma_{2,18}(\e)[^h_g]~Z_{(4,4)}[^h_g]\over
\t_2\eta^4\bar\eta^{20}}~ {1\over
2}\sum_{a,b=0}^1~(-1)^{a+b+ab}~{\th^2[^a_b]\th[^{a+h}_{b+g}]
\th[^{a-h}_{b-g}]\over \eta^4}
\,,\label{441}\ee
the shifted lattice sum $\Gamma_{2,18}(\e)[^h_g]$ is described in
Appendix B.

\vskip .4cm
\noindent\hrulefill
\nopagebreak\vskip .2cm
{\large\bf Exercise}. Show that (\ref{441}) is modular-invariant
if  $\e^2/2=$1~mod(4).
\nopagebreak\vskip .2cm
\noindent\hrulefill
\vskip .4cm

The theory depends on the $2\times 18$ moduli of
$\Gamma_{2,18}(\e)[^h_g]$
and the 16 moduli in $Z_{4,4}[^0_0]$.
There are, apart from the tensor multiplet, another 18 massless
vector multiplets.
The $2\times 18$ moduli are the scalars of these vector multiplets.
There are also 4 neutral hypermultiplets whose scalars are the
untwisted
(4,4)
orbifold moduli.
At special submanifolds of the vector multiplet moduli space, extra
massless vector
multiplets and/or hypermultiplets can appear.
We have seen such a symmetry enhancement already at the level of the
CFT.

The local structure of the vector moduli space is that of
$\rm O(2,18)/O(2) \times O(18)$.
{}From the real moduli, $G_{\a\b},B_{\a\b},Y^I_{\a}$ we can
construct
the 18 complex moduli $T=T_1+iT_2,U=U_1+iU_2,W^I=W^I_1+iW^I_2$ as
follows
\bea
G&=&{T_2-{W_2^IW_2^I\over 2U_2}\over U_2}\left(\matrix{1&U_1\cr
U_1&|U|^2\cr}\right)\,,\;\;\;\nn\\B&=&\left(T_1-{W_1^IW_2^I\over
2U_2}\right)
\left(\matrix{0&1\cr -1&0\cr}\right)
\label{454}\eea
and $W^I=-Y^I_2+UY_1^I$.
There is also one more complex scalar, the $S$ field with
${\rm Im}~ S=S_2=e^{-\phi}$,
whose real part is the axion $a$, which comes from dualizing the
antisymmetric tensor.
The tree-level prepotential and K\"ahler potential are
\be
f =S(TU-\half W^IW^I)\;\;\;,\;\;\;K=-\log(S_2)-\log\left[U_2T_2-\half
W_2^IW_2^I\right]
\,.\label{455}
\ee

The hypermultiplets belong to the quaternionic manifold
$\rm O(4,4)/O(4) \times O(4)$.
Since N=2 supersymmetry does not permit neutral couplings between
vector- and
hypermultiplets, and the dilaton belongs to a vector multiplet, the
hypermultiplet moduli space does not receive perturbative or
non-perturbative corrections.

In this class of N=2 ground-states, we will consider the helicity
supertrace
$B_2$ which traces
the presence of N=2 (short) BPS multiplets.\footnote{You will find the
definition of helicity supertraces and their relation to BPS
multiplicities in Appendix D.}
The computation is straightforward, using the results of Appendices E
and F.
We obtain
\bea
\tau_2~B_2&=&\tau_2~\langle \l^2\rangle
=\Gamma_{2,18}[^0_1]{\bar\th^2_3\bar\th_4^2\over
\bar\eta^{24}}-\Gamma_{2,18}[^1_0]{\bar\th^2_2\bar\th_3^2\over
\bar\eta^{24}}-\Gamma_{2,18}[^1_1]{\bar\th^2_2\bar\th_4^2\over
\bar\eta^{24}}
\nn\\
&=&{\Gamma_{2,18}[^0_0]+\Gamma_{2,18}[^0_1]\over 2}\bar
F_1-{\Gamma_{2,18}[^0_0]-\Gamma_{2,18}[^0_1]\over 2}\bar
F_1+\nn\\
&&-{\Gamma_{2,18}[^1_0]+\Gamma_{2,18}[^1_0]\over 2}\bar
F_+-{\Gamma_{2,18}[^1_0]-\Gamma_{2,18}[^1_0]\over 2}\bar F_-
\label{460}
\eea
with
\be
\bar F_1={\bar\th^2_3\bar\th_4^2\over \bar\eta^{24}}\;\;\;,\;\;\;
\bar F_{\pm}={\bar\th^2_2(\bar\th_3^2\pm\bar\th_4^2)\over
\bar\eta^{24}}
\,.\label{461}\ee
For all N=2 heterotic ground-states, $B_2$ transforms as
\be
\tau\to\tau+1\;\;:\;\;B_2\to B_2\;\;\;,\;\;\;\tau\to -{1\over
\tau}\;\;:\;\;
B_2\to \tau^2 ~B_2
\,.\label{462}\ee

All functions $\bar F_i$ have positive coefficients and have the
expansions
\be
F_{1}={1\over q}+\sum_{n=0}^{\infty}d_{1}(n)q^n={1\over q}+16+156
q+{\cal O}(q^2)
\,,\label{463}\ee
\be
F_{+}={8\over q^{3/4}}+q^{1/4}\sum_{n=0}^{\infty}d_{+}(n)q^n={8\over
q^{3/4}}+8q^{1/4}(30+481 q+{\cal O}(q^2))
\,,\label{464}\ee
\be
F_{-}={32\over q^{1/4}}+q^{3/4}\sum_{n=0}^{\infty}d_{-}(n)q^n=
{32\over q^{1/4}}+32q^{3/4}(26+375q+{\cal O}(q^2))
\,.\label{465}\ee
Also the lattice sums $\half
(\Gamma_{2,18}[^h_0]\pm\Gamma_{2,18}[^h_1])$
have
positive multiplicities.
Overall plus signs correspond to vector-like multiplets, while minus
signs correspond to hyper-like multiplets.
The contribution of the generic massless multiplets is given by the
constant coefficient of $F_1$; it agrees with what we expected:
$16=20-4$
since we have the supergravity multiplet and 19 vector multiplets
contributing 20 and 4 hypermultiplets contributing $-4$.

We will analyze the BPS mass-formulae associated with (\ref{460}).
We will use the notation for the  shift vector
$\e=(\vec\e_L;\vec\e_R,\vec\zeta)$, where $\e_L,\e_R$ are
two-dimensional integer vectors and $\zeta$ is a vector in the
$\rm O(32)/Z_2$
lattice.
We also have the modular-invariance constraint
$\e^2/2=\vec\e_L\cdot\vec\e_R-
\vec\zeta^2/2=1~(\mbox{mod}~4)$.

Using the results of Appendix B we can write the BPS mass-formulae
associated
to the lattice sums above.
For $h=0$
the mass-formula is
\be
M^2={|-m_1U+m_2+Tn_1+(TU-{1\over 2}
\vec W^2)n_2+\vec W\cdot \vec Q|^2\over 4 \;S_2
\left(T_2U_2-{1\over 2}{\rm
Im}\vec W^2\right)}
\,,\label{466}\ee
where $\vec W$ is the 16-dimensional complex vector of Wilson lines.
When the integer
\be
\rho=\vec m\cdot \vec \e_R+\vec n\cdot\e_L-\vec Q\cdot\vec\zeta
\label{467}\ee
is even, these states are vector-like multiplets with multiplicity
function
$d_1(s)$ of (\ref{463})
and
\be
s=\vec m\cdot\vec n-{1\over 2}\vec Q\cdot\vec Q\;;
\label{468}\ee
when $\rho$ is odd, these states are hyper-like multiplets with
multiplicities $d_1(s)$.
In the $h=1$ sector the mass-formula is
\bea
M^2&=&\left|(m_1+\half\e_L^1)U-(m_2+\half\e_L^2)-T(n_1+\half\e_R^1)+\right.\nn\\
&&-(TU-\half
\vec W^2)(n_2+\half\e_R^2) +  \nn\\
&&\left. -\vec W\cdot (\vec Q+\half\vec \zeta)\right|^2 /  4 \;S_2
\left(T_2U_2-\half{\rm
Im}\vec W^2\right)
\,.\label{469}\eea
The states with $\rho$ even are vector-multiplet-like with
multiplicities
$d_+(s')$, with
\be
s'=\left(\vec m+{\vec\e_L\over 2}\right) \cdot\left(\vec n
+{\vec \e_R\over 2}\right)-{1\over 2}\left(\vec Q+{\vec\zeta\over
2}\right)\cdot\left(\vec Q+{\vec\zeta\over 2}\right)
\,,\label{470}\ee
while the states with $\rho$ odd are hypermultiplet-like with
multiplicities
$d_-(s')$.

\renewcommand{\theequation}{\thesubsection.\arabic{equation}}
\subsection{Spontaneous supersymmetry breaking}
\setcounter{equation}{0}

We have seen in the previous section that we can break maximal
supersymmetry by
the orbifolding procedure.
The extra gravitini are projected out of the spectrum.
However, there is  a major difference between freely acting and
non-freely
acting orbifolds with respect to the restoration of the broken
supersymmetry.

To make the difference transparent, consider the $Z_2$ twist on $T^4$
described before, under which two of the gravitini transform with a
minus sign and are thus projected out.
Consider now doing at the same time a $Z_2$ shift in one direction
of the extra (2,2) torus. Take the two cycles to be orthogonal, with
radii $R,R'$, and do an $X\to X+\pi$ shift on the first cycle.
The oscillator modes are invariant but the vertex operator states
$|m,n\rangle
$ transform with a phase $(-1)^m$.
This is a freely-acting orbifold, since the action on the circle is
free.
Although the states of the two gravitini, $\bar
a^{\mu}_{-1}|S_a^I\rangle$
$I=1,2$ transform  with a minus sign under the twist, the states
$\bar a^{\mu}_{-1}|S_a^I\rangle\otimes |m=1,n\rangle$ are invariant!
They have the spacetime quantum numbers of two gravitini, but they are
not massless any more.
In fact, in the absence of the state $|m=1,n\rangle$
they would be massless,
but now we have an extra contribution to the mass coming  from that
state:
\be
m^2_{L}={1\over 4}\left({1\over
R}+nR\right)^2\;\;\;,\;\;\;m^2_{R}={1\over 4}\left({1\over
R}-nR\right)^2
\,.\label{471}\ee
The matching condition $m_L=m_R$ implies $n=0$, so that the mass of
these states is $m^2=1/4R^2$.
These are massive gravitini and in this theory, the N=4 supersymmetry
is broken spontaneously to N=2. In field theory language,
the effective
field theory is a gauged version of N=4 supergravity where the
supersymmetry
is spontaneously broken to N=2 at the minimum of the potential.

Although there seems to be little difference between such ground-states
and
the previously discussed ones, this is misleading.

We will note here some important differences between explicit and
spontaneous breaking of supersymmetry.

$\bullet$ In spontaneously broken supersymmetric theories, the
behavior at high energies is softer than the opposite case.
If supersymmetry is spontaneously broken, there are still leftover
broken Ward identities that govern the short distance properties of
the theory.
In such theories there is a characteristic energy scale, namely the
gravitino mass $m_{3/2}$ above which supersymmetry is effectively
restored.
A scattering experiment at energies $E>>m_{3/2}$ will reveal
supersymmetric physics. This has important implications for such
effects as the running
of low-energy couplings. We will come back to this later on.

$\bullet$ There is also a technical difference. As we already argued,
in the case of the freely-acting orbifolds, the states coming from
the twisted sector
have moduli-dependent masses that are generically non-zero (although
they can become zero at special values of the moduli space).
This is unlike non-freely acting orbifolds, where the twisted sector
masses are independent of the original moduli and one
obtains generically massless states from the twisted sector.

$\bullet$ In ground-states with spontaneously broken supersymmetry, the
super-symmetry-breaking scale $m_{3/2}$ is a freely-sliding scale since
it depends on moduli
with arbitrary expectation values.
In particular, there are corners of the moduli space where $m_{3/2}\to
0$,
and physics becomes supersymmetric at all scales.
These points are an infinite distance away using the natural metric
of the moduli scalars.
In our simple example from above $m_{3/2}\sim 1/R\to 0$ when $R\to
\infty$.
At this point, an extra dimension of spacetime becomes non-compact and
supersymmetry
is restored in five dimensions. This behavior is generic in all
ground-states
where the free action comes from translations.

Consider the class of N=2 orbifold ground-states we described in
(\ref{441}).
If the (2,18) translation vector $\e$ lies within the (0,16) part
of the lattice, then the breaking of $\rm N=4\to N=2$ is ``explicit".
When, however, $(\vec\e_{L},\vec\e_R)\not=(\vec 0,\vec 0)$ then the
breaking is spontaneous.

In the general case, there is no global identification of the
massive gravitini inside the moduli space due to surviving duality
symmetries.
Consider the following change in the previous simple example.
Instead of the $(-1)^m$ translation action, pick instead $(-1)^{m+n}$.
In this case there are two candidate states with the quantum numbers
of the gravitini: $\bar a^{\mu}_{-1}|S_a^I\rangle\otimes
|m=1,n=0\rangle$ with mass
$m_{3/2}\sim 1/R$, and $\bar a^{\mu}_{-1}|S_a^I\rangle\otimes
|m=0,n=1\rangle$
with mass $\tilde m_{3/2}\sim R$. In the region of large $R$ the
first set of states behaves like massive gravitini, while in the
region of small $R$
it is the second set that is light.

\renewcommand{\theequation}{\thesubsection.\arabic{equation}}
\subsection{Heterotic N=1 theories and chirality in four dimensions}
\setcounter{equation}{0}

So far, we have seen how, using orbifold techniques, we can get rid of
two gravitini and end up with N=2 supersymmetry.
We can carry this procedure one step further in order to reduce
the supersymmetry to N=1.

\vskip .4cm
\noindent\hrulefill
\nopagebreak\vskip .2cm
{\large\bf Exercise}. Consider splitting the (6,22) lattice in the
N=4 heterotic string as (6,22)=
$\oplus_{i=1}^3$(2,2)$_i\oplus$(0,16).
Label the coordinates of each two-torus as $X^{\pm}_i$, $i=1,2,3$.
Consider the following $Z_2\times Z_2$ orbifolding action:
The element $g_1$ of the first $Z_2$ acts with a minus sign on the
coordinates of the first and second two-torus, the element $g_2$ of the
second $Z_2$
acts with a minus sign on the coordinates of the first and third
torus, and
$g_{1}g_{2}$ acts with a minus sign on the coordinates of the second
and third torus.
Show that only one of the four gravitini survives this $Z_2\times
Z_2$ projection.
\nopagebreak\vskip .2cm
\noindent\hrulefill
\vskip .4cm

To ensure modular invariance we will have to act also on the gauge
sector.
We will assume to start from the $\rm E_8\times E_8$ string, with the
$\rm E_8$'s fermionically realized. We will split the 16 real fermions
realizing the first
$\rm E_8$ into groups of 10+2+2+2.
The $Z_2\times Z_2$ projection will act in a similar way in the three
groups
of two fermions each, while the other ten will be invariant.

The partition function for this $Z_{2}\times Z_{2}$ orbifold is
straightforward:
\bea
Z^{N=1}_{Z_2\times Z_2}&=&{1\over \t_2~\eta^2\bar\eta^2}~
{1\over 4}\sum_{h_1,g_1=0,h_2,g_2=0}^1
   {1\over 2}\sum_{\alpha,\beta=0}^{1}(-)^{\alpha+\beta+\alpha\beta}\times\nn\\
&&\times\frac{\vartheta[^{\alpha}_{\beta}]}{\eta}
\frac{\vartheta[^{\alpha+h_1}_{\beta+g_1}]}{\eta}
\frac{\vartheta[^{\alpha+h_2}_{\beta+g_2}]}{\eta}
\frac{\vartheta[^{\alpha-h_1-h_2}_{\beta-g_1-g_2}]}{\eta}~
{}~{\bar\Gamma_{8}\over \bar\eta^8}{}~
{Z^1_{2,2}[^{h_1}_{g_1}]Z^2_{2,2}[^{h_2}_{g_2}]
Z^3_{2,2}[^{h_1+h_2}_{g_1+g_2}]}\times\nn\\
&&\times{1\over 2}\sum_{\bar\a,\bar\b=0}^1~\frac{{\bar\vartheta}
[^{\bar\alpha}_{\bar\beta}]^5}{{\bar\eta}^5}
\frac{{\bar\vartheta}[^{{\bar\alpha}+h_1}_{{\bar\beta}+g_1}]}
{{\bar\eta}}
\frac{{\bar\vartheta}[^{{\bar\alpha}+h_2}_{{\bar\beta}+g_2}]}
{{\bar\eta}}
\frac{{\bar\vartheta}
[^{{\bar\alpha}-h_1-h_2}_{{\bar\beta}-g_1-g_2}]}
{{\bar\eta}}
\,.\label{472}
\eea

We will  find the massless spectrum, classified in
multiplets of
N=1 supersymmetry.
We have of course the N=1 supergravity multiplet.
Next we consider the gauge group of this ground-state.
It comes from the untwisted sector, so we will have to impose the extra
projection
on the gauge group of the N=2 ground-state.
The graviphoton, vector partner of the dilaton, and the two U(1)'s
coming from the $T^2$ are now projected out.
The E$_8$ survives.

\vskip .4cm
\noindent\hrulefill
\nopagebreak\vskip .2cm
{\large\bf Exercise}. Show that the extra $Z_2$ projection on
$E_7 \times SU(2)$
gives
$E_6 \times U(1) \times U(1)'$.
The adjoint of E$_6$ can be written as the adjoint of O(10) plus the
O(10) spinor plus a U(1).
\nopagebreak\vskip .2cm
\noindent\hrulefill
\vskip .4cm

Thus, the gauge group of this ground-state is
E$_8\times$E$_6\times$U(1)$\times$U(1)'
and we have the appropriate vector multiplets.
There is also the linear multiplet containing the antisymmetric
tensor and the dilaton.
Consider the rest of the states that form N=1 scalar
multiplets.
Notice first that there are no massless states charged under the
E$_8$.

\vskip .4cm
\noindent\hrulefill
\nopagebreak\vskip .2cm
{\large\bf Exercise}. Show that the charges of scalar multiplets
under  $E_6 \times U(1) \times U(1)'$ and their multiplicities are
those of
tables 1 and 2 below.
\nopagebreak\vskip .2cm
\noindent\hrulefill
\vskip .4cm

\bigskip
\centerline{
\begin{tabular}{|c|c|c|c|c|}\hline
${\rm E_6}$ & ${\rm U(1)}$ &{\rm U(1)'}& Sector & Multiplicity\\
\hline\hline
{\bf 27}&1/2&1/2&Untwisted &1\\\hline
{\bf 27}&-1/2&1/2&Untwisted&1\\\hline
{\bf 27}&0&-1&Untwisted&1\\\hline
1&-1/2&3/2&Untwisted&1\\\hline
1&1/2&3/2&Untwisted&1\\\hline
1&1&0&Untwisted&1\\\hline
1&1/2&0&Twisted&32\\\hline
1&1/4&3/4&Twisted&32\\\hline
1&1/4&-3/4&Twisted&32\\\hline
\end{tabular}}
\bigskip

\centerline{Table 1: Non-chiral massless content of the $Z_2\times
Z_2$
orbifold.}
\bigskip

 \centerline{
\begin{tabular}{|c|c|c|c|c|}\hline
${\rm E_6}$ & ${\rm U(1)}$ &{\rm U(1)'}& Sector & Multiplicity\\
\hline\hline
{\bf 27}&0&1/2&Twisted&16\\ \hline
{\bf 27}&1/4&-1/4&Twisted&16\\ \hline
{\bf 27}&-1/4&-1/4&Twisted&16\\ \hline
1&0&3/2&Twisted&16\\ \hline
1&3/4&-3/4&Twisted&16\\ \hline
1&-3/4&-3/4&Twisted&16\\ \hline
\end{tabular}}
\bigskip

\centerline{Table 2: Chiral massless content of the $Z_2\times Z_2$
orbifold.}
\bigskip\bigskip

As we can see, the spectrum of the theory is chiral. For example, the
number of
$\bf 27$'s minus the number of $\overline{\bf 27}$'s is 3$\times$16.
However, the theory is free of gauge anomalies.

More complicated orbifolds give rise to different gauge groups and
spectra, even with some phenomenological interpretations.
A way to construct such ground-states, which can be systematized, is
provided
by the fermionic construction \cite{ferm}.
We will not continue further in this
direction, but we refer the reader to \cite{dienes} which summarizes
known N=1 heterotic ground-states with a realistic spectrum.

\renewcommand{\theequation}{\thesubsection.\arabic{equation}}
\subsection{Orbifold compactifications of the type-II string}
\setcounter{equation}{0}

In section \ref{compact} we have considered the compactification of
the
ten-dimensional type-II string on the four-dimensional manifold
K3.
This provided a six-dimensional theory with N=2 supersymmetry.
Upon toroidal compactification on an extra $T^2$ we obtain a
four-dimensional theory with N=4 supersymmetry.

We will consider here a $Z_2$ orbifold compactification to six
dimensions
with N=2 supersymmetry and we will argue that it describes the
geometric
compactification on K3 that we considered before.

If we project out the orbifold transformation that acts on the $T^4$
by reversing
the sign of the coordinates (and similarly for the world-sheet
fermions
both on the left and the right), we will obtain a ground-state (in
six dimensions)
with half the supersymmetries, namely two.
The partition function is the following

\be
Z^{II-\l}_{6-d}={1\over 2}\sum_{h,g=0}^1~ {Z_{(4,4)}[^h_g]\over
\t_2^2\eta^4\bar\eta^{4}}~\times {1\over
2}\sum_{a,b=0}^1~(-1)^{a+b+ab}~{\th^2[^a_b]\th[^{a+h}_{b+g}]
\th[^{a-h}_{b-g}]\over \eta^4}\times
 \label{473}\ee
$$\times {1\over 2}\sum_{\bar a,\bar b=0}^1~(-1)^{\bar a+\bar b+\l
\bar a\bar b}~{\bar \th^2[^{\bar a}_{\bar b}]\bar \th[^{\bar
a+h}_{\bar b+g}]
\bar \th[^{\bar a-h}_{\bar b-g}]\over \bar \eta^4}\;,
$$
where $Z_{4,4}[^h_g]$ are the $T^4/Z_2$ orbifold blocks in
(\ref{391}) and $\l=0,1$ corresponds to type-IIB,A respectively.

We will  find the massless bosonic spectrum.
In the untwisted $NS$-$NS$ sector we obtain the graviton, antisymmetric
tensor
the dilaton and 16 scalars (the moduli of the $T^4/Z_2$).
In the $NS$-$NS$ twisted sector we obtain 4$\cdot$16 scalars.
The total number of scalars (apart from the dilaton) is 4$\cdot$20.
Thus, the massless spectrum of the $NS$-$NS$ sector is the same as that
of the K3 compactification in section \ref{compact}.

In the $R$-$R$ sector we will have to distinguish A from B.
In the type-IIA theory, we obtain 7 vectors and a three-form from the
$R$-$R$
untwisted sector and another
16 vectors from the $R$-$R$ twisted sector.
In type-IIB we obtain 4 two-index antisymmetric tensors and 8 scalars
from
the  $R$-$R$ untwisted sector and 16 anti-self-dual two-index
antisymmetric tensors and 16 scalars from the $R$-$R$ twisted sector.
Again this agrees with the K3 compactification.

To further motivate the fact that we are describing a CFT realization
of the
string moving on the K3 manifold, let us look more closely into the
cohomology
of $T^4/Z_2$.
We will use the two complex coordinates that describe the $T^4$,
$z_{1,2}$.
The $T^4$ has one zero-form, the constant, 2 (1,0) one-forms
($dz_1,dz_2$),
two (0,1) one-forms ($d\overline{z_1},d\overline{z_2}$), one (2,0) form
($dz_1\wedge dz_2$)
one (0,2) form ($d\overline{z_1}\wedge d\overline{z_2}$), and 4 (1,1)
forms ($dz_i\wedge d\overline{z_j}$).
Finally there are four three-forms and one four-form.
Under the orbifolding $Z_2$, the one- and three-forms are projected
out and we are left with a zero-form, a four-form, a (0,2), (2,0) and
4 (1,1) forms.
However the $Z_2$ action has  16 fixed-points on $T^4$, which
become singular in the orbifold.
To make a regular manifold we excise a small neighborhood around each
singular
point. The boundary is $S^3/Z_2$ and we can paste a
Ricci-flat manifold with the same boundary. The relevant manifold with
this property  is the zero-size limit of
the Eguchi-Hanson gravitational instanton.
This is the simplest of a class of four-dimensional non-compact
hyper-K\"ahler
manifolds known as Asymptotically Locally Euclidean (ALE) manifolds.
The three-dimensional manifold at infinity has the structure
$S^3/\Gamma$.
$\Gamma$ is one of the simple finite subgroups of SU(2).
The SU(2) action on $S^3$ is the usual group action (remember that
$S^3$ is the group manifold of SU(2)). This action induces an action of
the finite subgroup
$\Gamma$.
The finite simple SU(2) subgroups have an A-D-E classification.
The A series corresponds to the $Z_{N}$ subgroups. The Eguchi-Hanson
space
corresponds to $N=2$.
The D-series corresponds to the $D_N$ subgroups of SU(2), which are
$Z_N$
groups augmented by an extra $Z_2$ element.
Finally, the three exceptional cases correspond to the dihedral,
tetrahedral
and icosahedral groups.
The reader can find more information on the Eguchi-Hanson space
in \cite{eh}.
This space carries an anti-self dual (1,1) form. Thus, in total, we
will obtain
16 of them.
We have eventually obtained the cohomology of the K3 manifold, which is
of course at a singular limit.
We can also compute the Euler number. Suppose we have a manifold $M$
that  we divide by the action of an abelian group $G$ of order $g$;
we excise a set of fixed-points $F$ and we paste some regular
manifold $N$ back.
Then the Euler number is given by
\be
\chi={1\over g}[\chi(M)-\chi(F)]+\chi(N)
\,.\label{474}\ee
Here $\chi(T^4)=0$, $F$ is 16 fixed-points with $\chi=1$ each, while
$\chi=2$ for each Eguchi-Hanson instanton and we have sixteen of them
so that in total
$\chi(T^4/Z_2)=24$, which is the Euler number of K3.
The orbifold can be desingularized by moving away from zero instanton
size. This procedure is called  a ``blow-up" of the orbifold
singularities. In the orbifold CFT description, it corresponds to
marginal perturbations by the twist operators, or in string theory
language to changing
the expectation values of the scalars that are generated by the
sixteen orbifold twist fields.
Note that at the orbifold limit, although the K3 geometry is
singular, the associated string theory is not.
There are points in the moduli space though, where string theory
becomes singular. We will return later to the interpretation of such
singularities.

\renewcommand{\theequation}{\thesection.\arabic{equation}}
\section{Loop corrections to effective couplings in string theory
\label{loop}}
\setcounter{equation}{0}

So far, we have described ways of obtaining four-dimensional string
ground-states with or without supersymmetry and with various particle
contents.
Several ground-states have  the correct structure at tree level to
describe the
supersymmetric Standard Model particles and interactions.
However, to test further agreement with experimental data,
loop corrections should be incorporated.
In particular, we know that, at low energy, coupling constants run with
energy
due to loop contributions of charged particles.
So we need a computational framework to address similar issues
in the context of string theory.
We have mentioned before the relation between a ``fundamental theory"
(FT)
and its associated effective field theory (EFT), at least at
tree level.
Now we will have to take loop corrections into account.
In the EFT we will have to add the quantum corrections coming from
heavy particle loops.
Then we can calculate with the EFT where the quantum effects are
generated only by the light states.
In order  to derive the loop-corrected EFT, we will have, for every
given
amplitude of
light states, to do a computation in the FT where both
light and heavy states propagate in the loops; we will also have to
subtract the same
amplitude calculated in the EFT with only light states propagating in
the loops.
The difference (known as $threshold$ $correction$) is essentially the
contribution to a particular process of heavy states only.
We will have to incorporate this into the effective action.

We will  look in some more detail at the essential issues of such
computations.
Start from a given string theory calculation.
We will only deal here with one-loop corrections, although higher-loop
ones can be computed as well.
A one-loop amplitude in string theory will be some integrated
correlation function on the torus, which will be modular-invariant and
integrated in the fundamental domain.
As we have mentioned before, there are no UV divergences in string
theory
and, unlike a similar calculation in field theory, there is no need
for an UV cutoff.
Such calculations are done in the first quantized framework, which
means that we are forced to work on-shell.
This means that  there will be (physical) IR divergences, since
massless
particles on-shell propagate in the loop.
Formally, the amplitude will be infinite.
This IR divergence is physical, and the way we deal with it in field
theory is to allow the external momenta to be off-shell.
In any case, the IR divergence will cancel when we subtract the EFT
result from the ST result.

There are several methods to deal with the IR divergence in one-loop
calculations,
each with its merits and drawbacks.

$\bullet$ The original approach, due to Kaplunovsky \cite{Kap}, was to
compute appropriate two-point functions of gauge fields on the torus,
remove wave function factors that would make this amplitude vanish
(such a two-point function on-shell is required to vanish by
gauge invariance), and regularize  the IR divergence by inserting a
regularizing factor for the massless states.
This procedure gave the gauge-group-dependent corrections and the
first concrete calculation of the moduli-dependent threshold
corrections was done
in \cite{DKL}.
However, modular invariance is broken by such a regularization, and
the prescription of removing vanishing wavefunction factors does not
rest on a solid basis.

$\bullet$ Another approach, followed in \cite{ant}, is to calculate
derivatives of threshold corrections with respect to moduli.
Threshold corrections depend on moduli, since the masses of massive
string states do.
This procedure is free of IR divergences (the massless states drop
out)
and modular-invariant.
However, there are still vanishing wavefunction factors that need to be
removed by hand, and this approach cannot calculate moduli-independent
constant contributions to the thresholds.

$\bullet$ In \cite{kk} another approach was described which solved all
previous
problems. It provides the rigorous framework to calculate one-loop
thresholds.
The idea is to curve four-dimensional spacetime, which
provides a
physical IR cutoff on the spectrum.
This procedure is IR-finite, modular-invariant, free of ambiguities,
and allows the calculation of thresholds.
On the other hand, this IR regularization breaks maximal supersymmetry
(N=4
in heterotic and N=8 in type-II).
It preserves, however, any smaller fraction of supersymmetry.
It also becomes messy when applied to higher derivative operators.

The last method is the rigorous method of calculation.
We will describe it, in a subsequent section without going into all the
 details.
Since the result in several cases is not much different from that
obtained by the other methods, for simplicity, we will do some of the
calculations using the second method.

We will consider string ground-states that have N$\geq 1$
supersymmetry.
Although we know that supersymmetry is broken in the low-energy
world, for hierarchy reasons it should be broken at a low enough
scale $\sim $ 1 TeV.
If we assume that the superpartners have masses that are not far away
from the
supersymmetry breaking scale, their contribution to thresholds are
small.
Thus, without loss of generality, we will assume the presence of
unbroken N=1 supersymmetry.

\renewcommand{\theequation}{\thesubsection.\arabic{equation}}
\subsection{Calculation of gauge thresholds}
\setcounter{equation}{0}

We will first consider N=2 heterotic ground-states with an explicit
$T^2$.
This will provide a simple way to calculate derivatives of the
correction with respect to the moduli. Afterwards, we will derive a
general
formula for the corrections.
Such groundstates have a geometrical interpretation as a
compactification
on K3$\times T^2$ with a gauge bundle of instanton number 24.

In N=2 ground-states, the complex moduli that belong to vector
multiplets are the $S$
field
($1/S_2$ is the string coupling), the moduli $T,U$ of the two-torus
and several Wilson lines $W^I$, which we will keep to zero, so that we
have an unbroken non-abelian group.
Because of N=2 supersymmetry, the gauge couplings can depend only on
the vector moduli. We will focus here on the dependence  on $S,T,U$.
At tree level, the gauge coupling for the non-abelian factor $G_i$ is
given
by
\be
\left.{1\over g_i^2}\right|_{\rm tree}={k_i\over \gs^2}={k_i S_2}
\,,\label{475}\ee
where $k_i$ is the central element of the right-moving affine $G_i$
algebra, which generates the gauge group $G_i$.
The gauge boson vertex operators are
\be
V^{\m,a}_{G}\sim (\pd X^{\mu}+i(p\cdot \psi)\psi^{\m})~\bar
J^a~e^{ip\cdot X}\,.
\label{476}\ee
In the simplest ground-states, all non-abelian factors have $k=1$. We
will keep $k$ arbitrary.

The term in the effective action we would like to calculate is
\be
\int ~d^4 x {1\over g^2(T_i)}F^a_{\m\n}F^{a,\m\n}
\,,\label{477}\ee
where the coupling will depend in general on the vector moduli.
Therefore we must calculate a three-point amplitude on the torus, with
two
gauge fields and one modulus. We must also be in the even
spin-structures.
The odd spin-structure gives a contribution proportional to the
$\e$-tensor and is thus a contribution to the renormalization of the
$\theta$ angle.
The term that is  quadratic in momenta will give
us the derivative with respect to the appropriate modulus of the
correction to the gauge coupling.
The vertex operators for the torus moduli $T,U$ are given by
\be
V_{\rm modulus}^{IJ}= (\pd X^{I}+i(p\cdot \psi)\psi^{I})~\bar \pd
X^J~e^{ip\cdot X}
\,.\label{478}\ee
So we must calculate
\bea
I_{\rm 1-loop}&=&\int \langle V^{a,\mu}(p_1,z)V^{b,\n}(p_2,w)V_{\rm
modulus}^{IJ}(p_3,0)\rangle \nn\\
&\sim& \delta^{ab}(p_1\cdot
p_2\eta^{\m\n}-p_1^{\m}p_{2}^{\nu})F^{IJ}(T,U)+{\cal O}(p^4)
\,,\label{479}\eea
where $p_1+p_2+p_3=0$, $p_i^2=0$.
Because of supersymmetry, in order to get a non-zero result we will
have to contract the 4 $\psi^{\mu}$ fermions in (\ref{479}), which
gives us two powers of momenta. Therefore, to quadratic order, we can
set the
vertex operators $e^{ip\cdot X}$ to 1.
The only non-zero contribution to $F^{IJ}$ is
\be
F^{IJ}=\int_{\cal F}{d^2\tau\over \t_2^2}\int {d^2 z\over\t_2}
\int d^2 w ~\langle \psi(z)\psi(w)\rangle^2~\langle \bar J^a(\bar
z)\bar J^b(\bar w)\rangle
{}~\langle\pd X^I(0)\bar \pd X^J(0)\rangle
\,.\label{480}\ee
The normalized fermionic two-point function on the torus for an even
spin-structure is given by the Szeg\"o kernel
\be
S[^a_b](z)=
\left.\langle
\psi(z)\psi(0)\rangle\right|^a_b={\th[^a_b](z)\th_1'(0)\over
\th_1(z)\th[^a_b](0)}={1\over z}+\ldots
\,.\label{481}\ee
It satisfies the following identity:
\be
S^2[^a_b](z)={\cal P}(z)+4\pi i \pd_{\t}\log{\theta[^a_b](\tau)
\over \eta(\t)}
\,,\label{4813}\ee
so that all the spin-structure dependence is in the $z$-independent
second term.
We will have to weight this correlator with the partition function.
Since the first term is spin-structure independent it will not
contribute
for ground-states with N$\geq 1$ supersymmetry, where the partition
function vanishes.
For the N=2 ground-states described earlier, the spin-structure sum of
the
square of the fermion correlator can be evaluated directly:
\bea
\langle\langle\psi(z)\psi(0)\rangle\rangle&=&\half\sum_{(a,b)\not=
(1,1)}~(-1)^{a+b+ab}{\th^2[^a_b]\th[^{a+h}_{b+g}]
\th[^{a-h}_{b-g}]\over \eta^4}
S^2[^a_b](z)\nn\\
&=& 4\pi^2~\eta^2~\th[^{1+h}_{1+g}]\th[^{1-h}_{1-g}]
\,,\label{4811}\eea
where we have used (\ref{t11}) and the Jacobi identity (\ref{t15}).

The two-point function of the currents is also simple:
\bea
\langle \bar J^a(\bar z)\bar
J^b(0)\rangle&=&{k\delta^{ab}\over 4\pi^2}\bar\pd^2_{\bar
z}\log\bar\th_1(\bar
z)+{\rm Tr}[J_0^aJ_0^b]\nn\\
&=&
\d^{ab}\left({k\over 4\pi^2}\bar\pd^2_{\bar z}\log\bar\th_1(\bar
z)+{\rm Tr}[Q^2]
\right)
\,,\label{482}\eea
where Tr$[Q^2]$ stands for the conventionally normalized trace into the
whole string spectrum of the quadratic Casimir of the group $G$.
This can be easily computed by picking a single Cartan generator
squared and performing the trace. In terms of the affine characters
$\chi_R(v_i)$ this trace is
\linebreak[4]
\mbox{$\pd_{v_1}^2\chi_R(v_i)/(2\pi i)^2|_{v_i=0}$.}
This is the normalization for the quadratic Casimir standard in field
theory, which for a representation R is defined as
\linebreak[4]
\mbox{Tr$[T^aT^b]=I_2(R)\d^{ab}$}, where $T^a$ are the
matrices in the R representation. The field theory normalization
corresponds
to picking the squared length of the  highest root to be 1.
For the fundamental of SU(N) this implies the value 1 for the Casimir.
Also the spin~$j$ representation of SU(2) gives $2j(j+1)(2j+1)/3$.

Finally,
$\langle\pd X^I(0)\bar \pd X^J(0)\rangle$ gets contributions
from zero modes only, and it can be easily calculated, using the
results of
section \ref{cscalars}, to be
\bea
\langle\pd X^I(0)\bar \pd X^J(0)\rangle=
\frac{\sqrt{\mbox{det}G}}{(\sqrt{\t_{2}} \eta
\bar{\eta})^2}\sum_{\vec{m},\vec{n}}~(m^I+n^I\t)
(m^J+n^J\bar\t)\times
\nn\\
\times~ \exp\left[-\frac{\pi(G_{KL} + B_{KL})}{\t_2} (m_K + n_K\t)
(m_L + n_L\bar{\t})\right]\,.\label{483}
\eea
A convenient basis for the $T^2$ moduli is given by (\ref{178}).
In this basis we have
\be
V_{T_i}=v_{IJ}(T_i)\pd X^I\bar \pd X^J
\,,\label{484}\ee
with
\be
v(T)=-{i\over 2U_2}\left(\matrix{1&U\cr
\overline{U}&|U|^2\cr}\right)\;\;\;,\;\;\;v(U)={iT_2\over
U_2^2}\left(\matrix{
1&\overline{U}\cr \overline{U}&\overline{U}^2\cr}\right)
\,,\label{485}\ee
$v(\overline{T})=\overline{v(T)}$,
$v(\overline{U})=\overline{v(U)}$.
Then
\be
\langle V_{T_i}\rangle =-{\t_2\over 2\pi}\pd_{T_i}{\Gamma_{2,2}\over
\eta^2\bar \eta^2}
\,.\label{486}\ee

Using (\ref{At}) we
obtain for the one-loop correction to the gauge coupling in the N=2
ground-state
\be
{\partial\over \pd T_i}\left.{16\pi^2\over g^2_i}\right|_{\rm
1-loop}\sim {\partial\over \pd T_i}\int_{\cal F}{d^2\t\over
\t_2^2}{\t_2\Gamma_{2,2}\over
\bar\eta^4}~Tr^{int}_R\left[(-1)^F\left(Q_i^2-{k_i\over
4\pi\tau_2}\right)\right]+{\rm constant}
\,.\label{487}\ee

The internal theory consists of the (4,20) part of the original
theory,
which carries N=4 superconformal invariance on the left.
The derivative with respect to the moduli kills the IR divergence
due to the massless states.

For the remainder of this section, we will be cavalier about IR
divergences and vanishing wavefunctions. This will be dealt with
rigorously in the next section.
For a general string ground-state (with or without supersymmetry), we
can
parametrize its partition function as
in (\ref{F2}), where we have separated the bosonic and fermionic
contributions coming from the non-compact four-dimensional part.
In particular, the $\th$-function carries the helicity-dependent
contributions due to the fermions.
What we are now computing is the two-point amplitude of two gauge
bosons
at one loop.
When there is no supersymmetry, the $\pd X$ factors of the vertex
operators contribute $\langle \pd_z X(z)X(0)\rangle^2$, where we have
to use the torus
propagator for the non-compact bosons:
\be
\langle X(z,\bar z)X(0)\rangle=-\log|\th_1(z)|^2+2\pi{{\rm Im}z^2\over
\tau_2}
\,.\label{4815}\ee

In this case the $z$-integral we will have to perform is
\bea
\int{d^2 z\over \t_2}(S^2[^a_b](z)-\langle X\pd
X\rangle^2)~\left({k\over 4\pi^2}\bar\pd^2\log\bar\th_1(\bar z)+{\rm
Tr}[Q^2]\right)
=\nn\\
= 4\pi i\pd_{\t}\log{\th[^a_b]\over \eta}~\left({\rm Tr}[Q^2]-{k\over
4\pi\tau_2}
\right)\label{4812}\eea
where we have used (\ref{4813}), (\ref{intw}) and (\ref{intw2}).
The total threshold correction is, using (\ref{F2})
\be
Z_2^I=\left.{16\pi^2\over g_I^2}\right|_{\rm 1-loop}={1\over 4\pi^2}
\int_{\cal F}{d^2\t\over \t_2}
{1\over \eta^2\bar\eta^2}\sum_{\rm even}4\pi
i\pd_{\t}\left({\th[^a_b]\over
\eta}\right)~{\rm Tr}_{\rm int}\left[Q_I^2-{k_I\over 4\pi\tau_2}
\right][^a_b]
\,,\label{4814}\ee
where the trace is taken in the $[^a_b]$ sector of the internal CFT.
Note that the integrand is modular-invariant.
This result is general.
The measure  $\int_{\cal F}{d^2\t\over \t_2}$ will give an IR
divergence
as $\t_2\to\infty$ coming from constant parts of the integrand.
The constant part precisely corresponds to the contributions of the
massless states.
The derivative on the $\vartheta$-function gives a factor proportional
to
$s^2-1/12$, where $s$ is the helicity of a massless state.
The $k/\t_2$ factor accompanying the group trace gives an IR-finite
part.
Thus, the IR-divergent contribution to the one-loop result is
\be
\left.{16\pi^2\over g^2_I}\right|^{\rm IR}_{\rm 1-loop}=\int_{\cal
F}{d^2\t\over \t_2}~{\rm Str}~Q_I^2~\left({1\over 12}-s^2\right)
\,,\label{488}\ee
where Str stands for the supertrace.
\begin{figure}
\begin{center}
\leavevmode
\epsffile{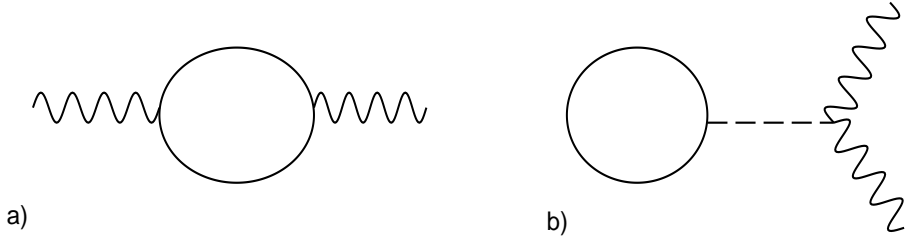}
\end{center}
\vspace{-.5cm}
\caption[]{\it a) One-loop gauge coupling correction due to charged
particles. b) Universal one-loop correction.}
\label{f15}\end{figure}

Inserting by hand a regularizing factor $e^{-\a'\mu^2\t_2}$ we obtain
\be
\left.{16\pi^2\over g^2_I}\right|^{\rm IR}_{\rm
1-loop}=b_I~\log(\mu^2\alpha')
+{\rm finite}
\,,\label{489}\ee
where $b_I$ is the conventional one-loop $\beta$-function coefficient
\be
\left.b_I={\rm Str}~Q^2_I~\left({1\over 12}-s^2\right)\right|_{\rm
massless}
\,.\label{490}\ee
We will try to better understand the origin of the various terms in
(\ref{4814}).
The term proportional to Tr$[Q^2]$ comes from conventional diagrams
where the two external gauge bosons are coupled to a loop of charged
 particles
(Fig.~\ref{f15}a).

\begin{figure}
\begin{center}
\leavevmode
\epsffile{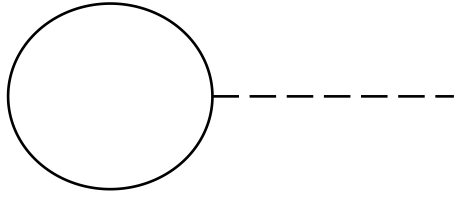}
\end{center}
\vspace{-.5cm}
\caption[]{\it One-loop dilaton tadpole}
\label{f16}\end{figure}

The second term proportional to $k$ seems bizarre, since all particles
contribute to it, charged or not.
This is a stringy correction to the gauge couplings due to the presence
of the gravitational sector. There is no analog of it in field theory.
Roughly speaking this term would arise from diagrams like the one
shown in Fig. \ref{f15}b.
Two external gauge bosons couple to the dilaton (remember that there is
a
tree-level universal coupling of the dilaton to all gauge bosons) and
then the dilaton couples to a loop of any string state (the dilaton
coupling is universal).
One may object that this diagram, being one-particle-reducible, should
not
be included as a correction to the coupling constants.
Moreover, it seems to imply that there is a non-zero dilaton tadpole
at one loop (Fig. \ref{f16}).
When at least N=1 supersymmetry is unbroken, we can show that the
dilaton tadpole in Fig. \ref{f16} is zero.
However, the diagram in Fig. \ref{f15}b still contributes owing to a
delicate
cancelation of the zero from the tadpole and the infinity coming from
the dilaton propagator on-shell.
This type of term is due to a modular-invariant regularization, of the
world-sheet short-distance singularity present when two vertex
operators collide.
We will see in the next section that these terms arise in a background
field calculation due to the gravitational back-reaction
to background gauge fields.
It is important to notice that such terms are truly universal, in the
sense that they are independent of the gauge group in question.
Their presence is essential for modular invariance.

There is an analogous diagram contributing to the one-loop
renormalization
of the $\theta$-angles. At tree-level, there is a (universal) coupling
of the antisymmetric tensor to two gauge fields due to the presence of
Chern-Simons terms. This gives rise to a parity-odd contribution like
the one in Fig.
\ref{f15}b, where now the intermediate state is the antisymmetric
tensor.

It was obvious from the previous calculation that the universal terms
came
as contact terms from the singular part of the correlator of affine
currents.
In open string theory, the gauge symmetry is not realized by a current
algebra on the world-sheet, but by charges (Chan-Paton) factors
attached to the end-points of the open string.
Thus, one would think that such universal contributions are absent.
However, even in the open string case, such terms appear in an indirect
way,
since the Planck scale has a non-trivial correction at one-loop for
$N\leq 2$ supersymmetry, unlike the heterotic case \cite{tI}.

We will show here that there are no corrections, at one loop, to the
Planck
mass in heterotic ground-states with $N\geq 1$ supersymmetry.
The vertex operator for the graviton is
\be
V_{\rm grav}=\e_{\mu\nu}(\pd X^{\mu}+ip\cdot \psi \psi^{\m})\bar \pd
X^{\nu}\,.
\ee
We have to calculate the two-point function on the torus and keep the
${\cal O}(p^2)$ piece.
On the supersymmetric side only the fermions contribute and produce a
$z$-independent contribution.
On the right, we obtain a correlator of scalars, which has to be
integrated over the torus.\footnote{Strictly speaking the amplitude is
zero on-shell but we can remove the wave-function factors. A rigorous
way to calculate it, is by
calculating the four-point amplitude of gravitons, and extract the
${\cal O}(p^2)$ piece.}
The result is proportional to
\be
\int {d^2z\over \t_2}\langle X\bar\pd_{\bar z}^2X\rangle =\int
{d^2z\over \t_2}\left(
\bar\pd_{\bar z}^2\log\bar \th_1(\bar z)+{\pi\over \tau_2}\right)=0
\,.\label{491}\ee

In the presence of at least N=1  supersymmetry there is no one-loop
renormalization of
the Planck mass in the heterotic string.
Similarly it can be shown that there are no wavefunction
renormalizations
for the other universal fields, namely the antisymmetric tensor and the
dilaton.

\renewcommand{\theequation}{\thesubsection.\arabic{equation}}
\subsection{On-shell infrared regularization}
\setcounter{equation}{0}

As mentioned in the previous section, the one-loop corrections to the
effective
coupling constants are calculated on-shell and are IR-divergent.
Also, for comparison with low-energy data, the moduli-independent
piece is also essential.
In this section we will provide a framework for this calculation.

Any four-dimensional heterotic string ground-state is described by  a
world-sheet
CFT, which is a product of a flat non-compact CFT describing
\linebreak[4]
Minkowski
space with $(c,\bar c)=(6,4)$ and an internal compact CFT with $(c,\bar
c)=(9,22)$.
Both must have N=1 superconformal invariance on the left, necessary for
the decoupling of ghosts.

To regulate the IR divergence on-shell, we will modify the
four-dimensional part.
We will consider the theory in a background with non-trivial
four-dimensional curvature
and other fields $B_{\m\n},\Phi$ so that the string spectrum acquires
a mass gap. Thus, all states are massive on-shell and there will be no
IR divergences.
The curved background must satisfy the exact string equations of
motion.
Consequently, it should correspond to an exact CFT.
We will require the following properties:

$\bullet$ The string spectrum must have a mass gap $\mu^2$.
          In particular, chiral fermions should be regularized
consistently.

$\bullet$ We should be able to take the limit $\mu^2\to 0$.

$\bullet$ It should have $(c,\bar c)=(6,4)$ so that we will not have to
modify the internal CFT.

$\bullet$ It should preserve as many spacetime supersymmetries of the
original theory as possible.

$\bullet$ We should be able to calculate the regularized quantities
relevant for the effective field theory.

$\bullet$ The theory should be modular-invariant (which guarantees the
absence
of anomalies).

$\bullet$ Such a regularization should be possible also at the
effective field theory level. In this way, calculations in the
fundamental theory can be matched
without any ambiguity to those of the effective field theory.

There are several CFTs with the properties required above.
It can be shown that the thresholds will not depend on which we choose.
We will pick a simple one, which corresponds to the $\rm SO(3)_{N}$ WZW
model times a free boson with background charge.
The background fields corresponding to this CFT are:

\be
ds^2=G_{\m\n}dx^{\m}dx^{\n}=(dX^0)^2+{N\over
4}\left(d\a^2+d\b^2+d\g^2+2\sin(\b/\sqrt{\a'})~d\a d\g\right)
\,,\label{492}\ee
where the Euler angles take values $\b\in[0,\sqrt{\a'}\pi]$,
$\a,\g\in[0,2\sqrt{\a'}\pi]$.
Thus, the three-space is almost a sphere of radius squared equal to
$N$.
For unitarity, N must be a positive even integer.
\be
B_{\m\n}dX^{\m}\wedge dX^{\n}={N\over 2}\cos(\b/\sqrt{\a'})~d\a\wedge
d\g
\;\;\;,\;\;\;\Phi={X^0\sqrt{\a'}\over \sqrt{N+2}}
\,.\label{493}\ee
The linear dilaton implies that the scalar $X^0$ has a background
charge
$Q^2=1/(N+2)$.
We will also work in Euclidean space.
The spectrum of operators in the $X^0$ part of the CFT with background
charge is
given by $\Delta=E^2+/4\a'(N+2)+$integers, where $E$ is a continuous
variable,
the ``energy". In the SO(3) theory the conformal weights are
$j(j+1)/\a'(N+2)+$ integers. The ratio $j(j+1)/\a'(N+2)$ plays the role
of $\vec p^2$ of flat space.
So
\be
L_0=-{1\over 2\a'}+E^2+{1\over 4\a'(N+2)}+{j(j+1)\over \a'(N+2)}+\ldots
\,.\label{495}\ee
All the states now have masses shifted by a mass gap $\m^2$
\be
\m^2={\ms^2\over 2(N+2)}\;\;\;,\;\;\;\ms={1\over\sqrt{\a'}}
\,.\label{494}\ee
Taking $N\to \infty$, $\m\to 0$,  we recover the flat space theory.
Moreover, this CFT preserves the original supersymmetries up to N=2.
The partition function of the new CFT is known, and after some
manipulations
we can write the partition function of the IR-regularized theory as
\be
Z(\mu)=\Gamma(\m/\ms)~Z(0)
\,,\label{496}\ee
where $Z(0)$ is the original partition function and
\bea
\Gamma(\m/\ms)&=&\left.4\sqrt{x}{\pd\over \pd
x}[\rho(x)-\rho(x/4)]\right|_{x=N+2}
\;,\nn\\
\rho(x)&=&\sqrt{x}\sum_{m,n\in Z}\exp\left[-{\pi x\over
\t_2}|m+n\t|^2\right]
\,.\label{497}\eea
Here,
$\Gamma(\m/\ms,\t)$ is modular-invariant and $\Gamma(0)=1$.

The background we have employed has another interesting interpretation.
It is a neutral heterotic five-brane of charge N and zero size
\cite{chs}.
The background fields are those of an axion-dilaton instanton.

We will now need to turn on background gauge fields and compute the
one-loop amplitude
as a function of these background fields. The quadratic part will
provide the one-loop correction to the gauge coupling constants.
The perturbation of the theory that turns on gauge fields is
\be
\delta I=\int d^2 z\left(A^a_{\m}(X)\pd
X^{\m}+F^a_{\m\n}\psi^{\m}\psi^{\nu}\right)\bar J^a
\,.\ee
In this background, there is such a class of perturbations, which are
an exact solution of the string equations of motion:
\be
\delta I=\int d^2 z~B^a\left(J^3+i\psi^1\psi^2\right)\bar J^a
\,,\label{498}\ee
where $J^3$ is the current belonging to the SO(3) current algebra of
the WZW model and $\psi^i$, $i=1,2,3$, are the associated free
fermions.
It turns out that for this choice, the one-loop free energy can be
computed exactly as a function of $B^a$.
I will spare you the details of the calculation, which can by found
in \cite{kk}.
Going through the procedure described above, we finally obtain the
expression (\ref{4814}), but with a factor of $\Gamma(\m/\ms)$ inserted
into the
modular integral, which renders this expression IR-finite.
We will denote the one-loop regularized result as $Z^I_{2}(\mu/\ms)$.
So, to one-loop order, the gauge coupling can be written as the sum
of the tree-level and one-loop result
\be
k_I{16\pi^2\over \gs^2}+Z_2^I(\m/\ms)
\,,\label{500}\ee
where $\gs$ is the string coupling.

In order to evaluate the thresholds, we must perform a similar
calculation
in the EFT and subtract the string from the EFT result.
The EFT result (with the same IR regulator) can be obtained from the
string result by the following operations:

$\bullet$ Do the trace on the massless sector only.

$\bullet$ Only the momentum modes contribute to the regularizing
function $\Gamma(\mu/\ms)$. We will denote this piece by $\Gamma_{\rm
EFT}(\m/\ms)$.

$\bullet$ The EFT result is UV-divergent. We will have to regularize
separately
this UV divergence. We will use dimensional regularization in the
$\overline{DR}$ scheme.
With these changes, the field theory result for the tree-level and
one-loop contributions reads
\be
{16\pi^2\over g_{I~\rm bare}^2}+b_I(4\pi)^{\epsilon}\int_{0}^{\infty}
{dt \over t^{1-\e}}\Gamma_{\rm EFT}\left({\m\over
\sqrt{\pi}\ms},t\right)
\,.\label{499}\ee
The extra factor of $\sqrt{\pi}$ comes in since $t=\pi \t_2$ and we
chose $\ms$ as the EFT renormalization scale.
In the $\overline{DR}$ scheme the relation between the bare and running
coupling constant is
\be
{16\pi^2\over g_{I~\rm bare}^2}={16\pi^2\over
g_{I}^2(\mu)}-b_I(4\pi)^{\epsilon}\int_{0}^{\infty}
{dt \over t^{1-\e}}e^{-t\m^2/M^2}
\,.\label{501}\ee
Putting (\ref{501}) into (\ref{499}) and identifying the result with
(\ref{500}),
we obtain
\be
\left.{16\pi^2\over g_{I}^2(\mu)}\right|_{\overline{DR}}=
k_I{16\pi^2\over \gs^2}+Z^I_2(\m/\ms)-b_I(2\gamma+2)
\,,\label{502}\ee
where $\g=0.577\ldots$ is the Euler-Mascheroni constant.
We can separate the IR piece from $Z_2^I$ using
\be
\int_{\cal F}{d^2\tau\over \t_2}\Gamma(\m/\ms)=\log{\ms^2\over \mu^2}
+\log{2e^{\g+3}\over \pi\sqrt{27}}+{\cal O}\left({\m\over \ms}\right)
\,,\label{503}
\ee
in order to rewrite the effective running coupling in the limit $\mu\to
0$
as
\be
\left.{16\pi^2\over g_{I}^2(\mu)}\right|_{\overline{DR}}
=k_I{16\pi^2\over \gs^2}+b_I\log{\ms^2\over \mu^2}
+b_I\log{2e^{1-\g}\over \pi\sqrt{27}}+\Delta_I
\,,\label{504}\ee
\be
\Delta_I=\int_{\cal F}{d^2\t\over \t_2}\left[
{1\over |\eta|^4}\sum_{\rm even}{i\over \pi}\pd_{\t}
\left({\th[^a_b]\over \eta}\right)~{\rm Tr}_{\rm int}
\left[Q_I^2-{k_I\over 4\pi\tau_2} \right][^a_b]-b_I\right]
\,.\label{505}\ee
This is the desired result, which produces the string corrections to
the EFT
running coupling in the $\overline{DR}$ scheme.
We will call $\Delta_I$ the string threshold correction to the
associated gauge coupling.
It is IR-finite for generic values of the moduli.
However, as we will see below, at special values of the moduli, extra
states
can become massless.
If such states are charged, then there will be an additional IR
divergence in the string threshold, which will modify the
$\beta$-function.

\renewcommand{\theequation}{\thesubsection.\arabic{equation}}
\subsection{Gravitational thresholds}
\setcounter{equation}{0}

We have seen above that the two-derivative terms in the effective
action concerning the universal sector ($G_{\m\n},B_{\m\n},\Phi$)
do not receive corrections in supersymmetric ground-states.
However, there are higher derivative terms that do.
A specific example is the $R^2$ term and its parity-odd counterpart
$R\wedge R$ (four derivatives) whose one-loop $\beta$-function is
the conformal anomaly in four dimensions.
In theories without supersymmetry, the corrections to these terms are
unrelated.
In theories with supersymmetry the two couplings are related by
supersymmetry.
The one-loop correction to $R^2$ can be obtained from the ${\cal
O}(p^4)$ part of the one-loop two-graviton amplitude, summed over the
even spin-structures.
The odd spin-structure will give the renormalization of $R\wedge R$.
Going through the same steps  as above we obtain (assuming $N\geq 1$
supersymmetry)
\be
\Delta_{\rm grav}=\int_{\cal F}{d^2\t\over \t_2}\left[
{1\over |\eta|^4}\sum_{\rm even}{i\over \pi}\pd_{\t}
\left({\th[^a_b]\over \eta}\right)~{\hat{\bar E}_2\over 12}~C^{\rm
int}[^a_b]
-b_{\rm grav}\right]
\,,\label{506}\ee
where the modular form $\hat{\bar E}_2$ is defined in (\ref{E9}).
The $R^2$ and $R\wedge R$ run logarithmically in four dimensions,  and
the coefficient of the logarithmic term $90b_{\rm grav}$ is the
conformal anomaly.
A scalar contributes 1 to the conformal anomaly, a Weyl fermion
${7\over 4}$,
a vector $-13$, a gravitino $-{233\over 4}$, an antisymmetric tensor
91,
and a graviton 212.
Again $\Delta_{\rm grav}$ is IR-finite, since we have subtracted the
contribution of the massless states $b_{\rm grav}$.

\renewcommand{\theequation}{\thesubsection.\arabic{equation}}
\subsection{Anomalous U(1)'s\label{anomu}}
\setcounter{equation}{0}

It turns out that some N=1 ground-states contain U(1) gauge fields that
are
``anomalous". We have seen this already in our earlier discussion on
compactifications of the heterotic string.
The term ``anomalous"  indicates that the sum of U(1) charges of all
massless states charged under the U(1) is not zero.\footnote{Also the
higher odd traces of the charge are non-zero.}
In a standard field theory, this would imply the existence of a mixed
(gauge-gravitational) anomaly
in the theory.
However, in string theory things work a bit differently.

In the presence of an ``anomalous" U(1), under a gauge transformation
the effective action is not invariant. There is a one-loop term (gauge
anomaly) proportional
to $(\sum_i q^i)F\wedge F$. For the theory to be gauge-invariant there
should be some other term in the effective action that cancels the
anomalous variation.
Such a term is $(\sum_i q^i)B\wedge F$.
You remember from the chapter on anomaly cancelation that $B$ has an
anomalous
transformation law under gauge transformations, $\delta B=\e F$.
This gives precisely the term we need to cancel the one-loop gauge
anomaly.
There is another way to argue on the existence of this term.
In 10-d there was an anomaly canceling term of the form $B\wedge F^4$.
Upon compactifying to four dimensions, this will give rise to a term
$B\wedge F$
with proportionality factor $\int~F\wedge F\wedge F$ computed in the
internal theory.
The coefficient of such a term at one loop can be computed directly.
The torus one-point function of the associated world-sheet current,
being proportional to the charge trace, is non-zero.
Moreover the coupling is parity-violating so it will come from the odd
spin-structure.
Consider the two-point function of an antisymmetric tensor and the
``anomalous" U(1)
gauge boson in the odd spin-structure of the torus.
In this case, one of the vertex operators must be put in the zero
picture and
an insertion of the zero mode of the supercurrent is needed:
\bea
\zeta_{U(1)}=\left.\e^1_{\m\n}\e^2_{\rho}\int{\d^2 z\over
\t_2}\left\langle (\pd x^{\m}+ip_1\cdot \psi \psi^{\m})\bar\pd
X^{\n}e^{ip_{1}\cdot X}\right|_{z}\times \right.\nn\\
\times \left.\left.\left.\psi^{\rho}\bar J e^{ip_{2}\cdot
X}\right|_{0}\oint dw(\psi^{\s}\pd X^{\s}+G^{\rm int})\right
|_{w}
\right\rangle
\,.\label{523}\eea
If N$\geq 2$ there are more than four fermion zero modes and the
amplitude
vanishes.
For N=1 there are exactly four zero modes, and we will have to use the
four fermions in order to obtain a non-zero answer.
This produces an $\e$-tensor. The leading (in momentum) non-zero piece
is
\be
\zeta_{U(1)}=\e^1_{\m\n}\e^2_{\rho}\e^{a\m\rho\s}p_1^a\langle \pd
X^{\s}\bar \pd X^{\n}(\bar z)\rangle \langle \bar J\rangle+{\cal
O}(p^2)
\,.\label{524}\ee
The integral over the scalar propagator produces a constant. Putting
everything together we obtain
\be
\zeta_{U(1)}\sim \int_{\cal F}{d^2\t\over \t_2^2}~{1\over \bar
\eta^2}{\rm Tr}[(-1)^F~Q]_{R}
\,,\label{525}\ee
where the trace is taken in the R sector of the internal CFT. The odd
spin-structure projects on the internal elliptic genus, which is
antiholomorphic. Thus, the full integrand is anti-holomorphic,
modular-invariant and
 has no pole at infinity (the tachyon is chargeless).
Such a function is a constant and it can be found by looking at the
limit
$\t_2\to \infty$, where only massless states contribute.
This constant is the sum of the U(1) charges of the massless states:
\be
\zeta_{U(1)}\sim \sum_{i, \rm massless} ~q^i
\,.\label{526}
\ee
It is easy to see that an ``anomalous" U(1) symmetry is spontaneously
broken.
Consider the relevant part of the EFT in the Einstein frame:
\be
S=\int\sqrt{\det G}\left[-{1\over
12}e^{-2\phi}H_{\m\n\rho}H^{\m\n\rho}+\zeta B\wedge F\right]
\,;\label{527}\ee
dualizing the $B$ to a pseudo-scalar axion field $a$ we obtain
\be
\tilde S= \int\sqrt{\det G}~e^{2\phi}(\pd_{\m} a+\zeta A_{\m})^2
\,.\label{528}\ee
Consequently, the gauge field acquires a mass $\sim \zeta \ms$.
There is also a Fayet-Iliopoulos D-term generated, that produces a
potential
for the dilaton and the scalars charged under the anomalous U(1).
The coefficient of this term can be calculated using the (auxiliary)
vertex
for a D-term \cite{D}, $J\bar J$, where $J$ is the internal U(1)
current of the N=2 superconformal algebra and $\bar J$ is the U(1)
world-sheet current of the anomalous U(1).

\vskip .4cm
\noindent\hrulefill
\nopagebreak\vskip .2cm
{\large\bf Exercise} Calculate the one-point function of $J\bar J$ on
the torus and show that the result is again given by (\ref{525}).
\nopagebreak\vskip .2cm
\noindent\hrulefill
\vskip .4cm

The generated potential is of the form
\be
V_D\sim \zeta e^{\phi}\left(e^{-\phi}+\sum_{i}q^ih_i|c_i|^2\right)^2
\,,\label{529}\ee
where $c_i$ are massless scalars with charge $q_i$  under the anomalous
U(1) with helicity $h_i$.

\renewcommand{\theequation}{\thesubsection.\arabic{equation}}
\subsection{N=1,2 examples of threshold corrections}
\setcounter{equation}{0}

We will examine here some sample evaluations of the one-loop threshold
corrections described in the previous sections.
Consider the N=2 heterotic ground-state described
in section \ref{hn2}.
The partition function was given in (\ref{393}).
The gauge group is E$_8\times$E$_7\times$SU(2)$\times$U(1)$^2$ (apart
from the graviphoton and the vector partner of the dilaton).
{}From (\ref{490}) we find that, up to the group trace, a vector
multiplet
contributes $-1$ and a hypermultiplet 1 to the $\beta$-function.

First we will compute the sum over the fermionic $\th$-functions
appearing in (\ref{505}).
\be
{i\over 2\pi}{1\over 2}\sum_{\rm even}~(-1)^{a+b+ab}~\pd_{\t}
\left({\th[^a_b]\over \eta}\right)
{\th[^a_b]\th[^{a+h}_{b+g}]\th[^{a-h}_{b-g}]\over \eta^3}~
{Z_{4,4}[^h_g]\over |\eta|^4}
=4{\eta^2\over \bar\th[^{1+h}_{1+g}]\bar\th[^{1-h}_{1-g}]}
\,,\label{509}\ee
for $(h,g)\not=(0,0)$ and gives zero for $(h,g)=(0,0)$.

We will also compute the group trace for E$_8$.
The level is $k=1$ and the E$_8$ affine character is
\be
\bar\chi^{\rm E_8}_{0}(v_i)={1\over 2}\sum_{a,b=0}^1{\prod_{i=1}^8\bar
\th[^a_b](v_i)\over \bar \eta^8}
\,.\label{507}\ee
Then
\be
\left[\left.{1\over (2\pi i)^2}\pd_{v_1}^2-{1\over
4\pi\t_2}\right]\bar\chi^{E_8}_{0}(v_i)\right|_{v_i=0}
={1\over 12}(\hat{\bar E}_2\bar E_4-\bar E_6)
\,,\label{508}\ee
which gives the correct value for the Casimir of the adjoint of E$_8$,
namely 60.
Using also
\be
{1\over 2}\sum_{(h,g)\not=(0,0)}\sum_{a,b=0}^1{\bar\th[^a_b]^6
\bar\th[^{a+h}_{b+g}]\bar\th[^{a-h}_{b-g}]\over
\bar\th[^{1+h}_{1+g}]\bar\th[^{1-h}_{1-g}]}=-{1\over 4}{\bar E_6\over
\bar\eta^6}
\,,\label{510}\ee
and putting everything together, we obtain $b_{\rm E_8}=-60$ and
\be
\Delta_{\rm E_8}=\int_{\cal F}{d^2\t\over \t_2}\left[-{1\over
12}\Gamma_{2,2}{
\hat{\bar E}_2\bar E_4\bar E_6-\bar E_6^2\over \bar\eta^{24}}+60\right]
\,.\label{5110}\ee

\vskip .4cm
\noindent\hrulefill
\nopagebreak\vskip .2cm
{\large\bf Exercise} Calculate the threshold for the $\rm E_7$ group.
The $\rm E_7$ group trace is given by
\be
\left[{\rm Tr} Q_{\rm E_7}^2-{1\over
4\pi\t_2}\right]=\left[\left.{1\over (2\pi i)^2}\pd_{v}^2-{1\over
4\pi\t_2}\right]{1\over 2}\sum_{a,b}{\bar\th[^a_b](v)\bar\th^5[^a_b]
\bar\th[^{a+h}_{b+g}]\bar\th[^{a-h}_{b-g}]\over \bar\eta^8}
\right|_{v=0}\;.
\ee
Show that the $\beta$-function coefficient is 84 ($I_2(133)=36,
I_2(56)=12$) and
\be
\Delta_{\rm E_7}=\int_{\cal F}{d^2\t\over \t_2}\left[-{1\over
12}\Gamma_{2,2}{
\hat{\bar E}_2\bar E_4\bar E_6-\bar E_4^3\over \bar\eta^{24}}-84\right]
\,.\label{5111}\ee
Show also that $b_{\rm SU(2)}$=84.
\nopagebreak\vskip .2cm
\noindent\hrulefill
\vskip .4cm
The difference between the two thresholds has a simpler form:
\be
\Delta_{\rm E_8}-\Delta_{\rm
E_7}=-144\Delta\;\;\;,\;\;\;\Delta=\int_{\cal F}{d^2\t\over
\t_2}(\Gamma_{2,2}-1)\;.
\label{diffe}\ee
The integral can be computed \cite{DKL} with the result
\be
\Delta=-\log\left[4\pi^2 T_2 U_2|\eta(T)\eta(u)|^4|\right]\;.
\label{diffe2}\ee
As we will show in the next section, (\ref{diffe}) written as
$\Delta_i-\Delta_j=(b_i-b_j)\Delta$ applies to all K3$\times T^2$
ground-states of the heterotic string.
Taking the large volume limit $T_2\to \infty$ in (\ref{diffe2}) we
obtain
\be
\lim_{T_2\to \infty} \Delta={\pi \over 3}T_2+{\cal O}(\log T_2)\;.
\ee
In the decompactification limit, the difference of gauge thresholds
behaves
as the volume of the two-torus.
A similar result applies to the individual thresholds.
This can be understood from the fact that  a six-dimensional gauge
coupling scales as [length].

Consider now the N=1 $Z_2\times Z_2$ orbifold ground-state with gauge
group \mbox{E$_8\times$E$_6\times$U(1)$\times$\\U(1)'}.
The partition function depends on the moduli ($T_i,U_i$) of the
3 two-tori (``planes").
In terms of the orbifold projection there are three types of sectors:

$\bullet$ N=4 sectors. They correspond to $(h_i,g_i)=(0,0)$ and have
$N=4$ supersymmetry structure. They give no correction to the gauge
couplings.

$\bullet$ N=2 sectors. They correspond to one plane being untwisted
while the other two are twisted. There are three of them and they have
an N=2 structure.
For this reason their contribution to the thresholds is similar to the
ones we described above.

$\bullet$ N=1 sectors. They correspond to all planes being twisted.
Such sectors do not depend on the untwisted moduli ($T_i,U_i$), but
they may depend on twisted moduli.

The structure above is generic in N=1 orbifold ground-states of the
heterotic
string.

In our example, the N=1 sectors do not contribute to the thresholds,
so we can directly write down the the E$_8$ and E$_6$ threshold
corrections
as
\be
\Delta_{\rm E_8}^{N=1}=\int_{\cal F}{d^2\t\over \t_2}\left[-{1\over
12}\sum_{i=1}^3\Gamma_{2,2}(T_i,U_i){
\hat{\bar E}_2\bar E_4\bar E_6-\bar E_6^2\over \bar\eta^{24}}+{3\over
2}60\right]
\,,\label{5112}\ee
\be
\Delta_{\rm E_6}^{N=1}=\int_{\cal F}{d^2\t\over \t_2}\left[-{1\over
12}\sum_{i=1}^3\Gamma_{2,2}(T_i,U_i){
\hat{\bar E}_2\bar E_4\bar E_6-\bar E_4^3\over \bar\eta^{24}}-{3\over
2}84\right]
\,.\label{5113}\ee
The extra factor $3/2$ in the $\beta$-function coefficient comes as
follows.
There is a $1/2$ because of the extra $Z_2$ orbifold projection
relative to the
$Z_2$ N=2 orbifold ground-state and a factor of 3 due to the three
planes contributing.
This is what we would expect  from the massless spectrum. Remember that
there are no scalar multiplets charged under the E$_8$.
So the E$_8$ $\beta$-function comes solely from the N=1 vector
multiplet
and using (\ref{490}) we can verify that it is 3/2 times that of an N=2
 vector multiplet.

The structure we have seen in the $Z_2\times Z_2$ orbifold ground-state
generalizes
to more complicated N=1 orbifolds.
It is always true that the untwisted moduli dependence of the threshold
corrections comes only from the N=2 sectors.

We will also analyze here thresholds in N=2 ground-states where N=4
supersymmetry is spontaneously broken to N=2.
We will pick a simple ground-state described in section \ref{hn2}.
It is the usual $Z_2$ orbifold acting on $T^4$ and one of the E$_8$
factors, but it is also accompanied by a $Z_2$ lattice shift $X^1\to
X^1+\pi$ in one of the coordinates of the left over two-torus. This is
a freely-acting orbifold
and we have two massive gravitini in the spectrum.
The geometrical interpretation is that of a compactification on a
manifold that is locally of the form K3$\times T^2$ but {\em not
globally}.
Its partition function is
\bea
Z_{N=4\to N=2}&=&{1\over
2}\sum_{h,g=0}^1~{1\over \t_2|\eta|^4}{\Gamma_{2,2}[^h_g]\over
|\eta|^4}~{\bar\Gamma_{\rm E_8}\over \bar\eta^8}
Z_{(4,4)}[^h_g]~{1\over
2}\sum_{\g,\d=0}^1~{\bar\th[^{\g+h}_{\d+g}]\bar\th[^{\g-h}_{\d-g}]
\bar\th^6[^{\g}_{\d}]\over \bar\eta^8}\times
 \nn\\
&&\times {1\over
2}\sum_{a,b=0}^1~(-1)^{a+b+ab}~{\th^2[^a_b]\th[^{a+h}_{b+g}]
\th[^{a-h}_{b-g}]\over \eta^4}\;,
\label{543}\eea
where $\Gamma_{2,2}[^h_g]$ are the translated torus blocks described in
Appendix B. In particular there is a $Z_2$ phase $(-1)^{g~m_1}$ in the
lattice sum and $n_1$ is shifted to $n_1+h/2$.

\vskip .4cm
\noindent\hrulefill
\nopagebreak\vskip .2cm
{\large\bf Exercise} Show that the gauge group of this ground-state
is the same as the usual $Z_2$ orbifold, namely
E$_8\times$E$_7\times$SU(2)
$\times$U(1)$^2$.
Show that there are also four neutral massless hypermultiplets and one
transforming as [2,56].
Confirm that there are no massless states coming from the twisted
sector.
Use (\ref{l22}) to show that the mass of the two massive gravitini is
given by
\be
m_{3/2}^2={|U|^2\over T_2U_2}
\,.\label{544}\ee
Show that the $\beta$-functions here are:
\be
b_{\rm E_8}=-60\;\;\;,\;\;\;b_{\rm E_7}=-12\;\;\;,\;\;\;b_{\rm
SU(2)}=52.
\ee
\nopagebreak\vskip .2cm
\noindent\hrulefill
\vskip .4cm

After a straightforward evaluation \cite{deco} we obtain that the
thresholds can be written as
\be
\Delta_I=b_I\Delta+\left({\tilde b_I\over 3}-b_I\right)\d-k_I Y
\,,\label{546}\ee
where $b_I$ are the $\beta$ functions of this ground-state, while
$\tilde b_I$ are those of the standard $Z_2$ orbifold (without the
torus translation).
Moreover
\be
\Delta=\int_{\cal F}{d^2\t\over
\t_2}\left[\sum'_{h,g}\Gamma_{2,2}[^h_g]-1\right]=
-\log\left[{\pi^2\over 4}
|\th_4(T)|^4|\th_2(U)|^4T_2U_2\right]
\,,\label{547}\ee
\be
\d= \int_{\cal F}{d^2\t\over
\t_2}~\sum'_{h,g}\Gamma_{2,2}[^h_g]\bar\s[^h_g]
\,,\label{548}\ee
\be
Y=\int_{\cal F}{d^2\t\over \t_2}~\sum'_{h,g}\Gamma_{2,2}\left[{1\over
12}{\hat{\bar E}_2\over
\bar\eta^{24}}\bar\Omega[^h_g]+\bar\rho[^h_g]+40\bar \s[^h_g]\right]
\,,\label{549}\ee
where
\be
\Omega[^0_1]={1\over 2}E_4\th_3^4\th_4^4(\th_3^4+\th_4^4)
\,,\label{5515}\ee
and its modular transforms
\be
\s[^h_g]=-{1\over 4}{\th^{12}[^h_g]\over \eta^{12}}
\,,\label{550}\ee
\be
\rho[^0_1]=f(1-x)\;\;\;,\;\;\;\rho[^1_0]=f(x)\;\;\;,
\;\;\;\rho[^1_1]=f(x/(x-1))
\,,\label{551}\ee
with $x=\th_2^4/\th_3^4$ and
\be
f(x)={4(8-49x+66x^2-49x^3+8x^4)\over 3x(1-x)^2}
\,.\label{552}\ee

We would be interested in the behavior of the above thresholds, in the
limit in which N=4 supersymmetry is restored: $m_{3/2}\to 0$ or $T_2\to
\infty$.
{}From (\ref{547}), $\Delta\to -\log[T_2]+\ldots$ while the other
contributions
vanish in this limit.
This is different from the large volume behaviour of the standard $Z_2$
thresholds (\ref{5110}), which we have shown to diverge linearly with
the volume $T_2$.
The difference of behavior can be traced to the enhanced supersymmetry
in the second example.
There are two parts of the spectrum of the second ground-state: states
with masses
below $m_{3/2}$, which have effective N=2 supersymmetry and contribute
logarithmically to the thresholds, and states with masses above
$m_{3/2}$,
which have effective N=4 supersymmetry and do not contribute.
When we lower $m_{3/2}$, if there are always charged states below it,
they will give a logarithmic divergence. This is precisely the case
here.
We could have turned on Wilson lines in such a way that there are no
charged states below $m_{3/2}$ as $m_{3/2}\to 0$. In such a case the
thresholds will vanish in the limit.

\renewcommand{\theequation}{\thesubsection.\arabic{equation}}
\subsection{N=2 universality of thresholds}
\setcounter{equation}{0}

For ground-states with N=2 supersymmetry the threshold corrections have
some universality properties.
We will demonstrate this in ground-states that come from N=1
six-dimensional theories compactified
further to four dimensions on $T^2$.
First we observe that the derivative of the helicity $\th$-function
that appears in the threshold formula essentially computes the
supertrace of the helicity squared.
Only short N=2 multiplets contribute to the supertrace and consequently
to
the thresholds (see Appendix D). This projects on the elliptic genus of
the internal CFT, which was defined in section \ref{N=2superconformal}.
Thus, the gauge and gravitational thresholds can be written as
\be
\Delta_I= \int_{\cal F}{d^2\t\over \t_2}\left[{\Gamma_{2,2}\over
\bar\eta^{24}}
\left({\rm Tr}[Q^2_I]-{k_I\over 4\pi\t_2}\right)\bar \Omega-b_I\right]
\,,\label{511}\ee
\be
\Delta_{\rm grav}= \int_{\cal F}{d^2\t\over
\t_2}\left[{\Gamma_{2,2}\over \bar\eta^{24}}{\hat{\bar E}_2\over
12}\bar \Omega-b_{\rm grav}\right]
\,.\label{512}\ee
The function $\bar \Omega$ is constrained by modular invariance to be
a weight-ten modular form, without singularities inside the fundamental
domain.
This is unique up to a constant
\be
\bar\Omega=\xi \bar E_4 \bar E_6
\,.\label{513}\ee
Consider further the integrand of $\Delta_I-k_I\Delta_{\rm grav}$
(without
the $b_I$ and $b_{\rm grav}$).
We find that it is antiholomorphic with at most a single pole at
$\t=i\infty$;
thus, it must be of the form $A_I\overline{j}+B_I$, where $A_I,B_I$ are
constants and $j$ is the modular-invariant function defined in
(\ref{E10}).
Consequently, the thresholds can be written as \cite{univ}
\be
\Delta_I= \int_{\cal F}{d^2\t\over \t_2}\left[\Gamma_{2,2}\left(
{\xi k_I\over 12}{\hat{\bar E}_2\bar E_4\bar E_6\over
\bar\eta^{24}}+A_I\overline{j}
+B_I\right)-b_I\right]
\,,\label{514}\ee
\be
\Delta_{\rm grav}= \xi\int_{\cal F}{d^2\t\over
\t_2}\left[\Gamma_{2,2}{\hat{\bar E}_2\bar E_4\bar E_6\over
12\bar\eta^{24}}-b_{\rm grav}\right]
\,.\label{515}\ee
We can now fix the constants as follows.
In the gauge threshold, there should be no $1/\bar q$ pole (the tachyon
is not charged), which gives
\be
A_I=-{\xi k_I\over 12}
\,.\label{516}\ee
Also the constant term is the $\beta$-function, which implies
\be
744\, A_I+B_I-b_{I}+ k_i\,  b_{\rm grav} = 0
\,,\label{517}\ee
with
$b_{\rm grav}=-22\xi$ from (\ref{515}).
Finally $b_{\rm grav}$ can be computed from the massless spectrum.
Using the results of the previous section we find that
\be
b_{\rm grav}={22-N_V+N_H\over 12}
\,,\label{518}\ee
where $N_V$ is the number of massless vector multiplets (excluding the
graviphoton and the vector partner of the dilaton) and $N_H$ is the
number of massless hypermultiplets.

\begin{centering}
\begin{figure}
\leavevmode
\epsfxsize=13cm
\epsffile[80 230 530 530]{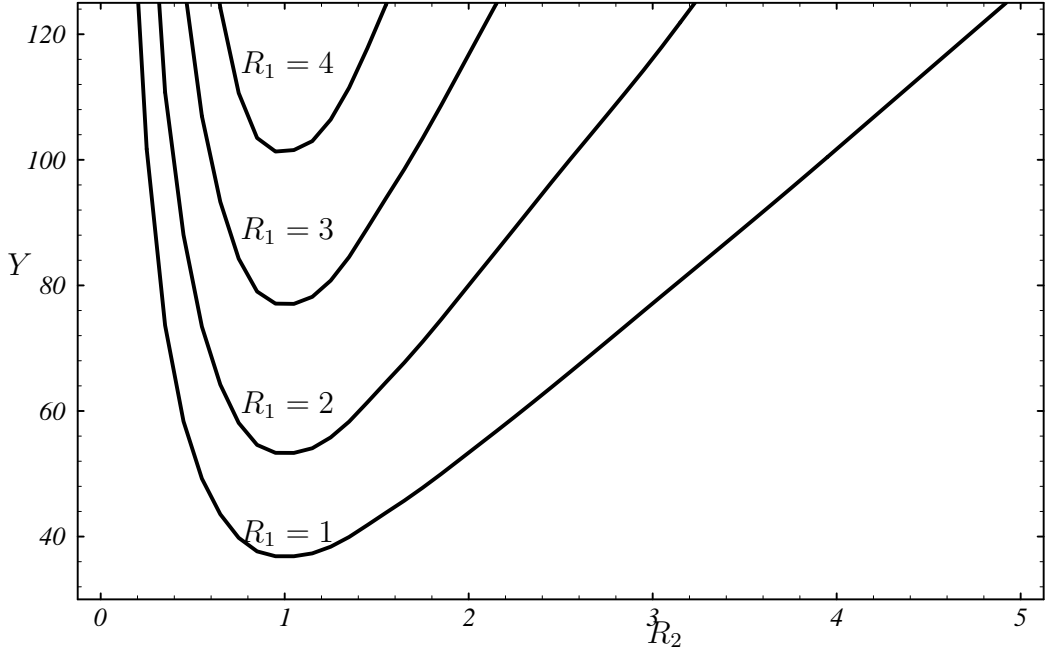}
\caption{\it Plots of the universal thresholds $Y(R_1,R_2)$ as a function
of $R_2$ for $R_1=1,2,3,4$.}
\vskip -2cm\hskip 8cm $R_2$
\vskip -5.5cm\hskip -.5cm $Y$
\vskip -3.3cm\hskip 2.6cm $R_1=4$
\vskip 1.6cm\hskip2.6cm$R_1=3$
\vskip 1.7cm\hskip2.6cm$R_1=2$
\vskip 1.1cm\hskip 2.6cm$R_1=1$
\vskip 3.5cm
\label{f17}\end{figure}
\end{centering}
\vskip .4cm

Moreover 6-d gravitational anomaly cancelation implies that
$N^{d=6}_H-N^{d=6}_V-29N_T^{d=6}=273$ where $N^{d=6}_{V,H}$ are the
number of six-dimensional vector and hypermultiplets while $N_T^{d=6}$
is the number of
six-dimensional tensor multiplets. For perturbative heterotic
ground-states
$N_T^{d=6}=1$ and we obtain $N^{d=6}_H-N^{d=6}_V=244$.
Upon toroidal compactification to four dimensions we obtain an extra 2
vector
multiplets (from the supergravity multiplet).
Thus, in four dimensions, $N_H-N_V=242$ and from (\ref{518}) we obtain
$b_{\rm grav}=22$, $\xi=-1$ for all such ground-states.
The thresholds
are now completely fixed in terms of the $\b$-functions of massless
states:
\be
\Delta_I=b_I~\Delta-k_I~Y
\,,\label{519}\ee
with
\bea
\Delta&=&  \int_{\cal F}{d^2\t\over \t_2}\left[\Gamma_{2,2}(T,U)-1\right]\nn\\
&=&-\log(4\pi^2|\eta(T)|^4|\eta(U)|^4~{\rm Im}T~{\rm Im}U)
\,,\label{520}\eea
\be
Y={1\over 12} \int_{\cal F}{d^2\t\over \t_2}~\Gamma_{2,2}(T,U)\left[
{\hat{\bar E}_2\bar E_4\bar E_6\over \bar \eta^{24}}-\overline{j}+
1008\right]
\,,\label{521}\ee
and
\be
\Delta_{\rm grav}= -\int_{\cal F}{d^2\t\over \t_2}\left[
\Gamma_{2,2}{\hat{\bar E}_2\bar E_4\bar E_6\over
12~\bar\eta^{24}}-22\right]
\,.\label{522}\ee

As can be seen from (\ref{504}), the universal term $Y$ can be absorbed
into a redefinition of the tree-level string coupling.
We can then write
\be
{16\, \pi^2\over g_{I}^2(\mu)}=k_{I}{16\, \pi^2\over g_{\rm
renorm}^2}+
b_{I}\log {M_{s}^2\over \mu^2}+\hat \Delta_I
\,,\label{540}
\ee
where we have defined a ``renormalized" string coupling by
\be
g_{\rm renorm}^2={\gs^2\over 1-{Y\over 16\, \pi^2}\,\gs^2 }
\,.\label{541}
\ee
Of course, such a coupling is meaningful, provided it appears as the
natural
expansion parameter in several amplitudes that are relevant for the
low-energy string physics. In general, this might not be the case as
a consequence
of some arbitrariness in the decomposition (\ref{519}), which is not
valid in general. Examples of
this
kind arise in N=1 ground-states as
well as in certain more general N=2 ground-states.
It is important to keep in mind that this ``renormalized"
string coupling is defined here in a {\it moduli-dependent} way.
This moduli dependence affects the string unification
\cite{kk}. Indeed, as we will see in the sequel, when proper
unification of the couplings appears, namely when $\hat\Delta_I$ can
be written
as $b_I\Delta$, their common value at the
unification scale is $g_{\rm renorm}$, which therefore plays the role
of a phenomenologically relevant parameter.
Moreover,
the unification scale
turns out to be proportional to $M_s$. The latter can
be
expressed in terms of
the ``low-energy" parameters
$g_{\rm renorm}$
and $M_P$, by using the fact that the Planck mass is not renormalized:
\be
\ms={M_{P}
\, g_{\rm renorm} \over \sqrt{1 + \frac{Y}{16\, \pi^2}\, g_{\rm
renorm}^2}}
\,.\label{542}
\ee
How much $Y$, which is moduli-dependent, can affect the running of the
gauge couplings can be seen from its numerical evaluation.
We take $T=iR_1R_2$ and $U=iR_1/R_2$, which corresponds to two
orthogonal circles of radii $R_{1,2}$. The values of $Y$ are plotted as
functions
of $R_{1,2}$ in Fig. \ref{f17}.

\renewcommand{\theequation}{\thesubsection.\arabic{equation}}
\subsection{Unification}
\setcounter{equation}{0}

Conventional unification of gauge interactions in a Grand Unified
Theory
(GUT) works by embedding the low-energy gauge group into a simple
unified
group G, which at tree level gives the following relation between the
unified gauge coupling $g_U$ of G and the low-energy gauge couplings
\be
{1\over g_I^2}={k_I\over g_U^2}
\,,\label{553}\ee
where $k_I$ are group theory coefficients that describe the embedding
of the low-energy gauge group into G.
Taking into account the one-loop running of couplings this relation
becomes,
in the $\overline{DR}$ scheme:
\be
{16\pi^2\over g_I^2(\mu)}=k_I{16\pi^2\over g_U^2}+b_I\log{M_U^2\over
\m^2}
\,.\label{554}\ee

In string theory, the high-energy gauge group need not be simple.
We have seen that (\ref{553}) is valid without this hypothesis, where
now the interpretation of $k_I$ is different. $k_I$ here are the levels
of the associated current algebras responsible for the gauge group.
Moreover, in string theory we have unification of gravitational and
Yukawa interactions as well.
We will  further study the string running of gauge couplings given
in (\ref{504}).
We would like to express it in terms of a measurable mass scale such as
the Planck mass, which is given in (\ref{542}).
We will assume for simplicity the  case of N=2 thresholds (\ref{519}).
We obtain a formula similar to (\ref{554}) with
\be
g_U=g_{\rm renorm}={\gs\over \sqrt{1-{\gs^2 Y\over 16\pi^2}}}
\,,\label{555}\ee
\be
M^2_U={2e^{1-\gamma}\over \pi\sqrt{27}}e^{\Delta}M_P^2\gs^2=
{2e^{1-\gamma}\over \pi\sqrt{27}}e^{\Delta}M_P^2{g_U\over
\sqrt{1+{g_U^2 Y\over 16\pi^2}}}
\,.\label{556}\ee
Both the ``unified" coupling and ``unification mass" are functions of
the moduli.
Moreover, they depend not only on the gauge-dependent threshold
$\Delta$
but also on the gauge independent-correction $Y$.

The analysis of the running of couplings in more realistic string
ground-states is summarized in \cite{dienes} where we refer the reader
for  a more detailed account.

\renewcommand{\theequation}{\thesection.\arabic{equation}}
\section{Non-perturbative string dualities: a foreword\label{nonpert}}
\setcounter{equation}{0}

In this chapter we will give a brief guide to some recent developments
towards understanding the non-perturbative aspects of string theories.
This was developed in parallel with similar progress in the context
of supersymmetric field theories \cite{s,sw}.
We will not discuss here the field theory case. The interested reader
may consult several comprehensive review articles \cite{n1,nn2}.
We would like to point out however that the field theory
non-perturbative
dynamics is naturally understood in the context of string theory
and there was important cross-fertilization between the two
disciplines.

We have seen that in ten dimensions there are five distinct, consistent
supersymmetric string theories, type-IIA,B, heterotic
(O(32),E$_8\times$E$_8$)
and the unoriented O(32) type-I theory that contains also open strings.
The two type-II theories have N=2 supersymmetry while the others have
only N=1.
An important question we would like to address is: Are these strings
theories
different or they are just different aspects of the same theory?

\begin{figure}[t]
\begin{center}
\leavevmode
\epsfxsize=13cm
\epsffile{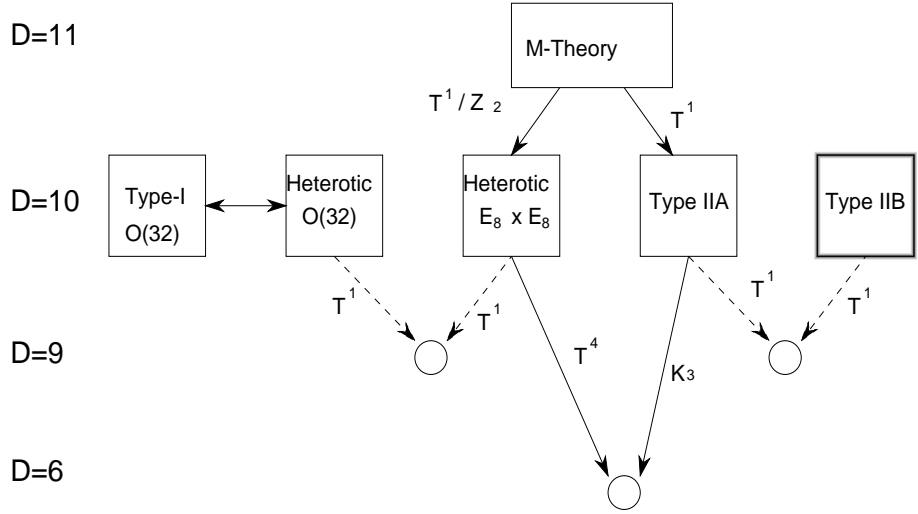}
\caption[]{\it The web of duality symmetries between string theories.
Broken lines
correspond to perturbative duality connections. Type-IIB in ten
dimensions  is supposed
to be self-dual under SL(2,$\Z$).}
\label{f19}
\end{center}\end{figure}

In fact, by compactifying one dimension on a circle we can show that we
can connect the two heterotic theories as well as the two type-II
theories.
This is schematically represented with the broken arrows in Fig.
\ref{f19}.

We will first show how the heterotic O(32) and E$_8\times$E$_8$
theories are connected in $D=9$.
Upon compactification on a circle of radius $R$ we can also turn on 16
Wilson lines according to our discussion in section \ref{torcomp}.
The partition function of the O(32) heterotic theory then can be
written as
\be
Z^{\rm O(32)}_{D=9}={1\over (\sqrt{\tau_2}\eta\bar\eta)^7}
{\Gamma_{1,17}(R,Y^I)\over
\eta\bar\eta^{17}}~{1\over
2}\sum_{a,b=0}^1~(-1)^{a+b+ab}~{\th^4[^a_b]\over \eta^4}
\,,\label{557}\ee
where the lattice sum $\Gamma_{1,17}$ was given explicitly in
(\ref{B4}).
We will focus on some special values for the Wilson lines $Y^I$, namely
we will take eight of them to be zero and the other eight to be 1/2.
Then the lattice sum (in Lagrangian representation (\ref{B1})) can be
rewritten as
$$
\Gamma_{1,17}(R)=R\sum_{m,n\in Z}\exp\left[-{\pi R^2\over \t_2}|m+\t
n|^2\right]{1\over 2}\sum_{a,b}~\bar\th^8[^a_b]~\bar\th^8[^{a+n}_{b+m}]
$$
\be={1\over 2}\sum_{h,g=0}^1 \Gamma_{1,1}(2R)[^h_g]~{1\over
2}\sum_{a,b}~\bar\th^8[^a_b]~\bar\th^8[^{a+h}_{b+g}]
\,,\label{558}\ee
where $\Gamma_{1,1}[^h_g]$ are the $Z_2$ translation blocks of the
circle partition function
\be
\Gamma_{1,1}(R)[^h_g]=R\sum_{m,n\in Z}\exp\left[-{\pi R^2\over \t_2}
\left|\left(m+{g\over 2}\right)+\tau\left(n+{h\over
2}\right)\right|^2\right]
\label{559}\ee
\be
={1\over R}\sum_{m,n\in Z}~(-1)^{mh+ng}~\exp\left[-{\pi\over \t_2
R^2}|m+\t n|^2\right]
\,.\label{560}\ee
In the $R\to \infty$ limit, (\ref{559}) implies that $(h,g)=(0,0)$
contributes in the sum in (\ref{558}) and we end up with the O(32)
heterotic string in ten dimensions.
In the $R\to 0$ limit the theory decompactifies again, but from
(\ref{560})
we deduce that all $(h,g)$ sectors contribute equally in the limit.
The sum on $(a,b)$ and $(h,g)$ factorizes and we end up with the
E$_8\times$ E$_8$ theory in ten dimensions.
Both theories are different limiting points (boundaries) in the moduli
space of toroidally compactified heterotic strings.

In the type-II case the situation is similar.
We compactify on a circle. Under an $R\to 1/R$ duality
\be
\pd X^9\to \pd X^9\;\;\;,\;\;\;\psi^9\to\psi^9\;\;\;,\;\;\;\bar\pd X^9
\to -\bar\pd X^9 \;\;\;,\;\;\; \bar\psi^9\to -\bar\psi^9
\,.\label{561}\ee
Due to the change of sign of $\bar\psi^9$ the projection in the $\bar
R$ sector is reversed.
Consequently the duality maps type-IIA to type-IIB and vice versa.
We can also phrase this in the following manner: the $R\to \infty$
limit of the toroidally compactified type-IIA string gives the type-IIA
theory in ten dimensions. The $R\to 0$ limit gives the type-IIB theory
in ten dimensions.

Apart from these perturbative connections, today we have evidence that
all supersymmetric string theories are connected.
Since they look very different in perturbation theory, the connections
necessarily involve strong coupling.

First, there is evidence that the type-IIB theory has an SL(2,$\Z$)
symmetry that, among other things, inverts the coupling constant
\cite{sch}.
Consequently, the strong coupling limit of type-IIB is isomorphic to
the perturbative type-IIB theory.
Upon  compactification this symmetry combines with the
perturbative $T$-duality symmetries to produce a large discrete duality
group known as the $U$-duality group, which is the discretization of
the non-compact continuous symmetries of the maximal effective
supergravity theory.
In table 3 below, the $U$-duality groups are given for various
dimensions.
They were conjectured to be exact symmetries in \cite{ht}.
A similar remark applies to non-trivial compactifications.

\vskip 1cm
\centerline{
\begin{tabular}{|c|c|c|c|}\hline
Dimension&SUGRA symmetry& T-duality&U-duality\\ \hline\hline
10A&SO(1,1,\R)/Z$_2$&{\bf 1}&{\bf 1}\\\hline
10B&SL(2,\R)&{\bf 1}&SL(2,\Z)\\\hline
9&SL(2,\R)$\times$O(1,1,\R)&Z$_2$&SL(2,\Z)$\times$Z$_2$\\\hline
8&SL(3,\R)$\times$SL(2,\R)&O(2,2,\Z)&SL(3,\Z)$\times$SL(2,\Z)\\\hline
7&SL(5,\R)&O(3,3,\Z)&SL(5,\Z)\\\hline
6&O(5,5,\R)&O(4,4,\Z)&O(5,5,\Z)\\\hline
5&E$_{6(6)}$&O(5,5,\Z)&E$_{6(6)}$(\Z)\\\hline
4&E$_{7(7)}$&O(6,6,\Z)&E$_{7(7)}$(\Z)\\\hline
3&E$_{8(8)}$&O(7,7,\Z)&E$_{8(8)}$(\Z)\\\hline
\end{tabular}}
\bigskip

\centerline{{ Table 3}: Duality symmetries for the compactified type-II
string.}
\vskip 1cm

Also, it can be argued that the strong coupling limit of type-IIA
theory
is described by an eleven-dimensional theory named ``M-theory"
\cite{va}.
Its low-energy limit is eleven-dimensional supergravity.
Compactification of M-theory on a circle of very small radius gives the
perturbative type-IIA theory.

If instead we compactify M-theory on the $Z_2$ orbifold of the circle
$T^1/Z_2$
then we obtain the heterotic E$_8\times$E$_8$ theory \cite{ee}.
When the circle is large the heterotic theory is strongly coupled,
while for small radius it is weakly coupled.

Finally, the strong coupling limit of the O(32) heterotic string theory
is the type-I O(32) theory and vice versa \cite{h-I}.

There is another non-trivial non-perturbative connection in six
dimensions:
the strong coupling limit of the six-dimensional toroidally
compactified heterotic string is given by the type-IIA theory
compactified on K3 and vice versa \cite{ht}.

\begin{figure}
\begin{center}
\leavevmode
\epsffile{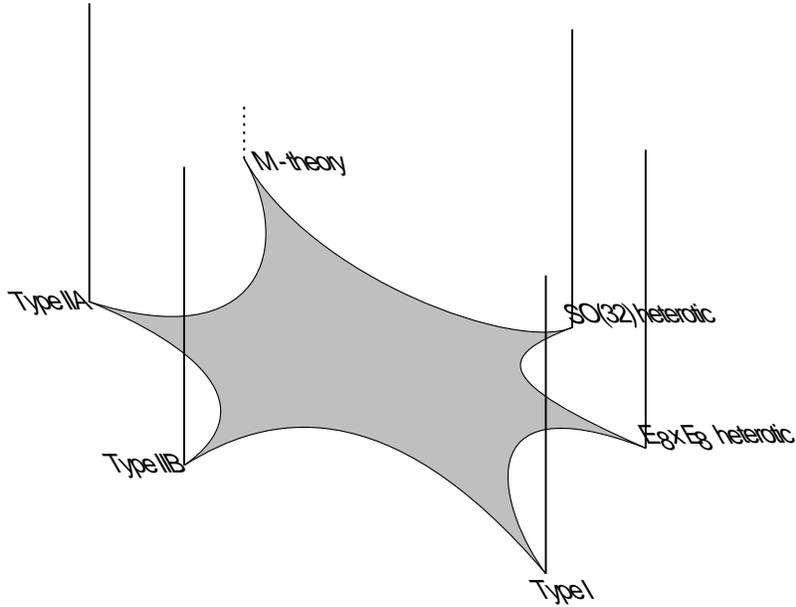}
\caption[]{\it A unique theory and its various limits.}
\label{f20}
\end{center}\end{figure}

Thus, we are led to suspect that there is an underlying ``universal"
theory
whose various limits in its ``moduli" space produce the weakly coupled
ten-dimensional supersymmetric string theories as depicted in Fig.
\ref{f20}
 (borrowed from \cite{P2}).
The correct description of this theory is unknown, although there is a
proposal
that it might have a matrix description \cite{matrix}, inspired from
D-branes \cite{Dp}, which reproduces
the perturbative IIA string in ten dimensions~\cite{m10}.

We will provide with a few more explanations and arguments
supporting the non-perturbative connections mentioned above.
But before we get there, we will need some ``non-perturbative tools",
namely the notion of BPS
states and p-branes, which I will briefly describe.

\renewcommand{\theequation}{\thesubsection.\arabic{equation}}
\subsection{Antisymmetric tensors and p-branes\label{antisym}}
\setcounter{equation}{0}

We have seen that the various string theories have massless
antisymmetric tensors in their spectrum.
We will use the language of forms and we will represent a rank-p
antisymmetric tensor $A_{\m_1\m_2\ldots\m_p}$ by the associated p-form
\be
A_p\equiv
A_{\m_1\m_2\ldots\m_p}dx^{\m_1}\wedge\ldots\wedge dx^{\m_p}
\,.\label{562}\ee
Such p-forms transform under generalized gauge transformations:
\be
A_p\to A_p+d~\Lambda_{p-1},
\,,\label{563}\ee
where $d$ is the exterior derivative ($d^2=0$) and $\Lambda_{p-1}$ is a
$(p-1)$-form that serves as the parameter of gauge transformations.
The familiar case of (abelian) gauge fields corresponds to p=1.
The gauge-invariant field strength is
\be
F_{p+1}=d~A_{p}
\,.\label{564}\ee
satisfying the free Maxwell equations
\be
d^{*}F_{p+1}=0
\label{5644}\ee

The natural objects, charged under a (p+1)-form $A_{p+1}$, are
$p$-branes.
A $p$-brane is an extended object with $p$ spatial dimensions.
Point particles correspond to p=0, strings to p=1.
The natural coupling of $A_{p+1}$ and a p-brane is given by
\be
\exp\left[iQ_p\int_{\rm world-volume} A_{p+1}\right]=
\exp\left[iQ_p\int A_{\m_0\ldots\m_p}dx^{\m_0}\wedge\ldots\wedge
dx^{\m_p}\right]
\,,\label{565}\ee
which generalizes the Wilson line coupling in the case of
electromagnetism.
The world-volume of $p$-brane is (p+1)-dimensional.
Note also that this is precisely the $\s$-model coupling of the usual
string
to the $NS$ antisymmetric tensor in (\ref{248}).
The charge $Q_p$ is the usual electric charge for p=0 and the string
tension for p=1.
For the p-branes we will be considering, the (electric) charges will be
related to their tensions (mass per unit volume).

In analogy with electromagnetism, we can also introduce magnetic
charges.
First, we must define the analog of the magnetic field: the magnetic
(dual) form.
This is done by first dualizing the field strength and then rewriting
it as the exterior derivative of another form\footnote{This is
guaranteed by (\ref{5644}).} :
\be
d\tilde A_{D-p-3}=\tilde F_{D-p-2}=^*F_{p+2}=^*dA_{p+1}
\,,\label{566}\ee
where D is the the dimension of spacetime.
Thus, the dual (magnetic) form couples to $(D-p-4)$-branes that play
the role of magnetic monopoles with ``magnetic charges" $\tilde
Q_{D-p-4}$.

There is a generalization of the Dirac quantization condition to
general
p-form charges discovered by Nepomechie and Teitelboim \cite{nt}.
The argument parallels that of Dirac. Consider an electric p-brane
with charge $Q_p$ and a magnetic $(D-p-4)$-brane with charge $\tilde
Q_{D-p-4}$.
Normalize the forms so that the kinetic term is ${1\over 2}\int
^*F_{p+2}F_{p+2}$.
Integrating the field strength $F_{p+2}$ on a (D-p-2)-sphere
surrounding
the p-brane we obtain the total flux $\Phi=Q_p$.
We can also write
\be
\Phi=\int_{S^{D-p-2}}~^*F_{p+2}=\int_{S^{D-p-3}}~\tilde A_{D-p-3}
\,,\label{567}\ee
where we have used (\ref{566}) and we have integrated around the
``Dirac string".
When the magnetic brane circles the Dirac string it picks up a phase
$e^{i\Phi\tilde Q_{D-p-4}}$, as can be seen from (\ref{565}).
Unobservability of the string implies the Dirac-Nepomechie-Teitelboim
quantization condition
\be
\Phi \tilde Q_{D-p-4}=Q_{p}\tilde Q_{D-p-4}=2\pi N\;\;\;,\;\;\;n\in Z
\,.\label{568}\ee

\renewcommand{\theequation}{\thesubsection.\arabic{equation}}
\subsection{BPS states and bounds\label{bps}}
\setcounter{equation}{0}

The notion of BPS states is of capital importance in discussions of
non-perturbative duality symmetries.
Massive BPS states appear in theories with extended supersymmetry.
It just so happens that supersymmetry representations are sometimes
shorter than usual. This is due to some of the supersymmetry operators
being ``null",
so that they cannot create new states.
The vanishing of some supercharges depends on the relation between the
mass of a multiplet and some central charges appearing in the
supersymmetry algebra.
These central charges depend on electric and magnetic charges of the
theory as well as on expectation values of scalars (moduli).
In a sector with given charges, the BPS states are the lowest lying
states
and they saturate the so-called BPS bound which, for point-like states,
is of the form
\be
M\geq ~{\rm maximal~~eigenvalue~~of }\;\; Z
\,,\label{569}\ee
where $Z$ is the central charge matrix.
This is shown in Appendix D where we discuss in detail the
representations of extended supersymmetry in four dimensions.

BPS states behave in a very special way:

$\bullet$ At generic points in moduli space  they are absolutely
stable.
The reason is the dependence of their mass on conserved charges.
Charge and energy conservation prohibits their decay.
Consider as an example, the BPS mass formula
\be
M^2_{m,n}={|m+n\tau|^2\over \tau_2}\;\;,
\label{du2}
\ee
where $m,n$ are integer-valued conserved charges, and $\tau$ is a
complex modulus. This BPS formula is relevant for N=4, SU(2) gauge
theory, in a subspace of its moduli space.
Consider a BPS state with charges $(m_0,n_0)$, at rest, decaying into N
states
with charges $(m_i,n_i)$ and masses $M_i$, $i=1,2,\cdots,N$.
Charge conservation implies that $m_0=\sum_{i=1}^N m_i$,
$n_0=\sum_{i=1}^N n_i$.
The four-momenta of the  produced particles are $(\sqrt{M_i^2+\vec
p_i^2},\vec p_i)$ with $\sum_{i=1}^N \vec p_i=\vec 0$.
Conservation of energy implies
\be
M_{m_0,n_0}=\sum_{i=1}^N\sqrt{M_i^2+\vec p^2_i}\geq \sum_{i=1}^N
M_i\;\;.
\label{du1}\ee
Also in a given charge sector (m,n) the BPS bound implies that any mass
$M\geq M_{m,n}$, with $M_{m,n}$ given in (\ref{du2}).
Thus, from (\ref{du1}) we obtain
\be
M_{m_0,n_0}\geq \sum_{i=1}^N M_{m_i,n_i}\;\;,
\label{du3}
\ee
and the equality will hold if all  particles are BPS and are produced
at rest ($\vec p_i=\vec 0$).
Consider now the two-dimensional vectors $v_i=m_i+\tau n_i$ on the
complex $\tau$-plane, with length $||v_i||^2=|m_i+n_i\tau|^2$.
They satisfy $v_0=\sum_{i=1}^N v_i$.
Repeated application of the triangle inequality implies
\be
||v_0||\leq \sum_{i=1}^N ||v_i||\;\;.
\label{du4}
\ee
This is incompatible with energy conservation (\ref{du3}) unless
all vectors $v_i$ are parallel. This will happen only if $\tau$ is
real.
For energy conservation it should also be a rational number.
On the other hand, due to the SL(2,$\Z$) invariance of (\ref{du2}), the
inequivalent choices for $\tau$ are in the SL(2,$\Z)$ fundamental
domain and
$\tau$ is never real there. In fact, real rational values of $\tau$ are
mapped by SL(2,$\Z)$ to $\tau_2=\infty$, and since $\tau_2$ is the
inverse of the coupling constant, this corresponds to the degenerate
case of zero coupling.
Consequently, for $\tau_2$ finite, in the fundamental domain, the BPS
states of this theory are absolutely stable. This is always true in
theories
with more than eight conserved supercharges (corresponding to N$>2$
supersymmetry
in four dimensions).
In cases corresponding to theories with 8 supercharges, there are
regions in the moduli space, where BPS states, stable at weak coupling,
can decay at strong coupling. However, there is always a large region
around weak coupling where they are stable.

$\bullet$ Their mass-formula is supposed to be exact if one uses
renormalized values for the charges and moduli.
The argument is that quantum corrections would spoil the relation of
mass and charges, if we assume unbroken SUSY at the quantum level.
This would give incompatibilities  with the dimension of their
representations.
Of course this argument seems to have a loophole: a specific set of BPS
multiplets can combine into a long one. In that case, the above
argument
does not prohibit corrections.
Thus, we have to count BPS states modulo long supermultiplets.
This is precisely what helicity supertrace formulae do for us.
They are reviewed in detail in Appendix E.
Even in the case of N=1 supersymmetry there is an analog of BPS states,
namely the massless states.

There are several amplitudes that in perturbation theory obtain
contributions
from BPS states only.
In the case of eight conserved supercharges (N=2 supersymmetry in four
dimensions), all two-derivative terms as well as
$R^2$ terms are of that kind.
In the case of sixteen conserved supercharges (N=4 supersymmetry in
four dimensions), except the above terms, also the four derivative
terms as well as $R^4$, $R^2 F^2$ terms are of a similar kind.
The normalization argument of the BPS mass-formula makes another
important
assumption: as the coupling grows, there is no phase transition during
which supersymmetry is (partially) broken.

The BPS states described above can be realized as point-like soliton
solutions
of the relevant effective supergravity theory.
The BPS condition is the statement that the soliton solution leaves
part of the supersymmetry unbroken.
The unbroken generators do not change the solution, while the broken
ones
generate the supermultiplet of the soliton, which is thus shorter than
the generic supermultiplet.

So far we discussed point-like BPS states. There are however BPS
versions for extended objects (BPS p-branes).
In the presence of extended objects the supersymmetry algebra can
acquire central charges that are not Lorentz scalars (as we assumed in
Appendix D).
Their general form can be obtained from group theory, in which case one
sees
that they must be antisymmetric tensors, $Z_{\m_1\ldots\m_p}$.
Such central charges have values proportional to the charges $Q_{p}$ of
 p-branes. Then, the BPS condition would relate these charges with the
energy densities (p-brane tensions) $\mu_p$ of the relevant p-branes.
Such p-branes can be viewed as extended soliton solutions of the
effective theory. The BPS condition is the statement that the soliton
solution leaves
some of the supersymmetries unbroken.

\renewcommand{\theequation}{\thesubsection.\arabic{equation}}
\subsection{Heterotic/type-I duality in ten dimensions.\label{hetI}}
\setcounter{equation}{0}

We will start our discussion by describing heterotic/type-I duality
in ten dimensions.
It can be shown \cite{sen} that heterotic/type-I duality, along with
T-duality can reproduce all known string dualities.

Consider first the O(32) heterotic string theory.
At tree-level (sphere) and up to two-derivative terms, the (bosonic)
effective
action in the $\s$-model frame is
\be
S^{\rm het}=\int d^{10}x\sqrt{G}e^{-\Phi}\left[
R+(\nabla\Phi)^2-{1\over 12}\hat H^2-{1\over 4}F^2\right]
\,.\label{570}\ee

On the other hand, for the O(32) type-I string the leading order
two-derivative effective action is
\be
S^{I}=\int d^{10}x\sqrt{G}\left[e^{-\Phi}\left(
R+(\nabla\Phi)^2\right)-{1\over 4}e^{-\Phi/2}F^2-{1\over 12}\hat
H^2\right]
\,.\label{571}\ee
The different dilaton dependence here comes as follows: the Einstein
and dilaton terms come from the closed sector on the sphere ($\chi=2$).
The gauge kinetic terms come from the disk ($\chi=1$). Since the
antisymmetric tensor comes from the $R$-$R$ sector of
the closed superstring it does not have any dilaton
dependence on the sphere.

We will now bring both actions to the Einstein frame,
$G_{\m\n}=e^{\Phi/4}g_{\m\n}$:
\be
S^{\rm het}_E=\int d^{10}x\sqrt{g}\left[
R-{1\over 8}(\nabla\Phi)^2-{1\over 4}e^{-\Phi/4}F^2-{1\over
12}e^{-\Phi/2}\hat H^2\right]
\,,\label{572}\ee
\be
S^{I}_E=\int d^{10}x\sqrt{g}\left[
R-{1\over 8}(\nabla\Phi)^2-{1\over 4}e^{\Phi/4}F^2-{1\over
12}e^{\Phi/2}\hat H^2\right]
\,.\label{573}\ee

We observe that the two actions are related by $\Phi\to -\Phi$ while
keeping the other fields invariant. This seems to suggest that the weak
coupling of one
is the strong coupling of the other and vice versa.
Of course, the fact that the two actions are related by a field
redefinition
is not a surprise. It is known that N=1 ten-dimensional supergravity is
completely fixed once the gauge group is chosen.
It is interesting though, that the field redefinition here is just an
inversion of the ten-dimensional coupling.
Moreover, the two theories have perturbative expansions that are very
different.

We would like to go further and check if there are non-trivial checks
of
what is suggested by the classical N=1 supergravity.
However, once we compactify one direction on a circle of radius $R$ we
seem
to have a problem.
In the heterotic case, we have a spectrum that depends both on momenta
$m$ in the ninth direction as well as on windings $n$.
The winding number is the charge that couples to the string
antisymmetric tensor. In particular, it is the electric charge of the
gauge boson obtained from
$B_{9\m}$.
On the other hand, in type-I theory, as we have shown earlier, we have
momenta $m$ but no windings.
One way to see this is that the open string Neumann boundary conditions
forbid the string to wind around the circle.
Another way is by noting that the $NS$-$NS$ antisymmetric tensor that
could couple to windings has been projected out by our orientifold
projection.

However, we do have the $R$-$R$ antisymmetric tensor, but as we argued
in section \ref{RRR},
no perturbative states are charged under it.
There may be, however, non-perturbative states that are charged under
this antisymmetric tensor.
According to our general discussion in section \ref{antisym} this
antisymmetric tensor would naturally couple to a string, but this is
certainly not  the perturbative string.
How can we construct this non-perturbative string?

An obvious guess is that this is a solitonic string excitation of the
low-energy type-I effective action. Indeed, such a  solitonic solution
was constructed \cite{dab} and shown to have the correct zero mode
structure.

We can give a more complete description of this non-perturbative
string.
The hint is given from $T$-duality on the heterotic side, which
interchanges windings and momenta.
When it acts on derivatives of $X$ it interchanges
$\pd_{\s}X\leftrightarrow \pd_{\tau}X$.
Consequently, Neumann boundary conditions are interchanged with
Dirichlet ones.
To construct such a non-perturbative string we would have to use also
Dirichlet boundary conditions.
Such boundary conditions imply that the open string boundary is fixed
in spacetime. In terms of waves traveling on the string, it implies
that a
wave arriving at the boundary is reflected with a minus sign.
The interpretation of fixing the open string boundary in some
(submanifold)
of spacetime has the following interpretation: there is a solitonic
(extended)
object there whose fluctuations are described by open strings attached
to it.
Such objects are known today as D-branes.

Thus, we would like to describe our non-perturbative string as a
D1-brane.
We will localize it to the hyperplane $X^2=X^3=\ldots=X^9=0$. Its
world-sheet extends in the $X^0,X^1$ directions.
Such an object is schematically shown in Fig. \ref{f21}.
Its fluctuations can be described by two kinds of open strings:

\begin{figure}[t]
\begin{center}
\leavevmode
\epsffile{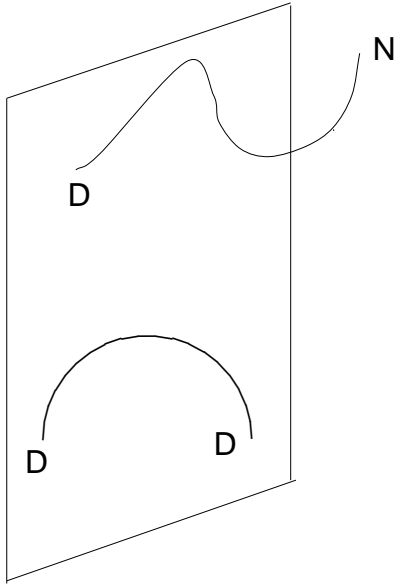}
\caption[]{\it Open string fluctuations of a D1-brane}
\label{f21}
\end{center}\end{figure}

$\bullet$ DD strings that have D-boundary conditions on both end-points
and are forced to move on the D1-brane.

$\bullet$ DN strings that have a D-boundary condition on one end, which
is stuck on the D1-brane, and N-boundary conditions on the other end,
which is free.

As we will see, this solitonic configuration breaks half of N=2
spacetime supersymmetry possible in ten dimensions.
It also breaks $SO(9,1) \to SO(8) \times SO(1,1)$. Moreover, we can put
it anywhere
in the transverse eight-dimensional space, so we expect eight bosonic
zero-modes
around it associated with the broken translational symmetry.
We will try to understand in more detail the modes describing the
world-sheet theory of the D1 string.
We can obtain them by looking at the massless spectrum of the open
string fluctuations around it.

Start with the DD strings.
Here $X^I,\psi^I,\bar\psi^I$, $I=2,\ldots,9$ have DD boundary
conditions while $X^{\m},\psi^{\m},\bar\psi^{\m}$, $\m=0,1$ have NN
boundary conditions.

For the world-sheet fermions NN boundary conditions imply
\be
{\rm NN~~NS~~sector}~~~~~~~~~\left.\psi+\bar\psi\right|_{\s=0}=\left.
\psi-\bar\psi\right|_{\s=\pi}=0
\,,\label{576}\ee
\be
{\rm NN~~R~~sector}~~~~~~~~~\left.\psi-\bar\psi\right|_{\s=0}=
\left.\psi-\bar\psi\right|_{\s=\pi}=0
\,.\label{577}\ee
The DD boundary condition is essentially the same with $\bar\psi\to-
\bar\psi$:
\be
{\rm DD~~NS~~sector}~~~~~~~~~\left.\psi-\bar\psi\right|_{\s=0}=\left.
\psi+\bar\psi\right|_{\s=\pi}=0
\,,\label{580}\ee
\be
{\rm DD~~R~~sector}~~~~~~~~~\left.\psi+\bar\psi\right|_{\s=0}=
\left.\psi+\bar\psi\right|_{\s=\pi}=0
\,,\label{581}\ee
and a certain action on the Ramond ground-state, which we will describe
below.

\vskip .4cm
\noindent\hrulefill
\nopagebreak\vskip .2cm
{\large\bf Exercise} Show that we have the following mode expansions
\be
X^I(\s,\t)=x^I+w^I\s+2\sum_{n\not= 0}{a^I_{n}\over n}e^{in\tau}\sin
(n\s)
\,,\label{574}\ee
\be
X^{\m}(\s,\t)=x^{\m}+p^{\m}\t-2i\sum_{n\not= 0}{a^{\m}_{n}\over
n}e^{in\tau}\cos (n\s)
\,.\label{575}\ee
In the $NS$ sector
\be
\psi^I(\s,\t)=\sum_{n\in Z}b^I_{n+1/2}e^{i(n+1/2)(\s+\t)}\;\;\;,\;\;\;
\psi^{\m}(\s,\t)=\sum_{n\in Z}b^{\m}_{n+1/2}e^{i(n+1/2)(\s+\t)}
\,,\label{578}\ee
while in the $R$ sector
\be
\psi^I(\s,\t)=\sum_{n\in Z}b^I_{n}e^{in(\s+\t)}\;\;\;,\;\;\;
\psi^{\m}(\s,\t)=\sum_{n\in Z}b^{\m}_{n}e^{in(\s+\t)}
\,.\label{579}\ee
Also
\be
\bar b^I_{n+1/2}=b^I_{n+1/2}\;\;\;,\;\;\;\bar b^I_{n}=-b^I_{n}
\,,\label{584}\ee
\be
\bar b^{\m}_{n+1/2}=-b^{\m}_{n+1/2}\;\;\;,\;\;\;\bar b^{\m}_{n}=
b^{\m}_{n}
\,.\label{585}\ee
\nopagebreak\vskip .2cm
\noindent\hrulefill
\vskip .4cm

The $x^I$ in (\ref{574}) are the position of the D-string in transverse
space.
There is no momentum in (\ref{574}), which implies that the state
wavefunctions would depend only on the $X^{0,1}$ coordinates, since
there is a continuous momentum in (\ref{575}). Thus, the states of this
theory ``live" on the world-sheet of the D1-string.
The usual bosonic massless spectrum would consist of a vector
$A_{\m}(x^0,x^1)$ corresponding to the state
$\psi^{\m}_{-1/2}|0\rangle$ and eight bosons
$\phi^I(x^0,X^1)$ corresponding to the states
$\psi^{I}_{-1/2}|0\rangle$\footnote{The GSO projection is always
present.}.
We will  now consider the action of the orientation reversal $\Omega$:
$\s\to -\s$, $\psi\leftrightarrow \bar \psi$.
Using (\ref{576})-(\ref{581}),
\be
\Omega~b^{\m}_{-1/2}|0\rangle= \bar b^{\m}_{-1/2}|0\rangle=-
b^{\m}_{-1/2}|0\rangle
\,,\label{582}\ee
\be
\Omega~b^{I}_{-1/2}|0\rangle= \bar b^{I}_{-1/2}|0\rangle=
b^{I}_{-1/2}|0\rangle
\,.\label{583}\ee
The vector is projected out, while the eight bosons survive the
projection.

We will now analyze the Ramond sector, where fermionic degrees of
freedom would come from.
The massless ground-state $|R\rangle$ is an SO(9,1) spinor satisfying
the usual
GSO projection
\be
\Gamma^{11}|R\rangle=|R\rangle
\,.\label{586}\ee
Consider now the $\Omega$ projection on that spinor.
In the usual $NN$ case $\Omega$ can be taken to commute with $(-1)^F$
and acts on the spinor ground-state as $-1$.
In the DD case the action of $\Omega$ on the transverse DD fermionic
coordinates is reversed compared to the NN case.
On the spinor this action is
\be
\Omega |R\rangle =-\Gamma^2\ldots\Gamma^9|R\rangle =|R\rangle
\,.\label{587}\ee
{}From (\ref{586}), (\ref{587}) we also obtain
\be
\Gamma^0\Gamma^1|R\rangle=-|R\rangle
\,.\label{588}\ee
If we decompose the spinor under SO(8)$\times$SO(1,1) the surviving
piece transforms as $8_-$ where $-$ refers to the SO(1,1) chirality
(\ref{588}).
As for the bosons, these fermions are functions of $X^{0,1}$ only.

To recapitulate, in the DD sector we have found the following massless
fluctuations moving on the world-sheet of the D1-string: 8 bosons and 8
chirality minus fermions.

Consider now the DN fluctuations.
In this case Chan-Paton factors are allowed in the free string end, and
the usual tadpole cancelation argument implies that there are 32 of
them.
In this case, the boundary conditions for the transverse bosons and
fermions become
\be
\left.\pd_{\t}X^I\right|_{\s=0}=0\;\;\;,\;\;\;
\left.\pd_{\s}X^I\right|_{\s=\pi}=0
\,,\label{589}\ee
\be
{\rm DN~~NS~~sector}~~~~~~~~~\left.\psi+\bar\psi\right|_{\s=0}=\left.
\psi+\bar\psi\right|_{\s=\pi}=0
\,,\label{590}\ee
\be
{\rm DN~~R~~sector}~~~~~~~~~\left.\psi-\bar\psi\right|_{\s=0}=
\left.\psi+\bar\psi\right|_{\s=\pi}=0
\,,\label{591}\ee
while they are NN in the longitudinal directions.

We observe that here, the bosonic oscillators are half-integrally
modded as in the twisted sector of $Z_2$ orbifolds. Thus, the
ground-state conformal weight is 8/16=1/2.
Also the modding for the fermions has been reversed between the $NS$
and $R$ sectors.
In the $NS$ sector the fermionic ground-state is also a spinor with
ground-state conformal weight 1/2. The total ground-state has
conformal weight
1 and only massive excitations are obtained in this sector.

In the $R$ sector there are massless states coming from the bosonic
ground-state combined with the O(1,1) spinor ground-state from the
longitudinal
Ramond fermions.
The usual GSO projection here is $\Gamma^0\Gamma^1=1$.
Thus, the massless modes in the DN sector are 32 chirality plus
fermions.

In total, the world-sheet theory of the D-string contains exactly what
we would
expect from the heterotic string in the physical gauge!
This is a non-trivial argument in favor of heterotic/type-I duality.

\vskip .4cm
\noindent\hrulefill
\nopagebreak\vskip .2cm
{\large\bf Exercise}. We have considered so far a D1-brane in type-I
theory.
Consider the general case of Dp-branes along similar lines.
Show that non-trivial configurations exist (compatible with GSO and
$\Omega$ projections) preserving half of the supersymmetry,  for
p=1,5,9.
The case p=9 corresponds to the usual open strings moving in 10-d
space.
\nopagebreak\vskip .2cm
\noindent\hrulefill
\vskip .4cm

The $R$-$R$ two-form couples to a one-brane (electric) and a five-brane
(magnetic). As we saw above, both can be constructed as D-branes.

We will describe now in some more detail the D5-brane, since it
involves
some novel features.
To construct a five-brane, we will have to impose Dirichlet boundary
conditions
in four transverse directions.
We will again have DD and NN sectors, as in the D1 case.
The massless fluctuations will have continuous momentum in the six
longitudinal
directions, and will describe fields living on the six-dimensional
world-volume of the five-brane.
Since we are breaking half of the original supersymmetry, we expect
that the
world-volume theory will have N=1 six-dimensional supersymmetry, and
the massless fluctuations will form multiplets of this supersymmetry.
The relevant multiplets are the vector multiplet, containing a vector
and a
gaugino, as well as the hypermultiplet, containing four real scalars
and
a fermion.
Supersymmetry implies that the manifold of the hypermultiplet scalars
is a hyper-K\"ahler manifold. When the hypermultiplets are charged
under the
gauge group, the gauge transformations are isometries of the
hyper-K\"ahler
manifold, of a special type: they are compatible with the
hyper-K\"ahler structure.

It will be important for our latter purposes to describe the Higgs
effect in this case.
When a gauge theory is in the Higgs phase, the gauge bosons become
massive
by combining with some of the massless Higgs modes.
The low-energy theory (for energies well below the gauge boson mass)
is described by the scalars that have not been  devoured by the gauge
bosons.
In our case, each (six-dimensional) gauge boson that becomes massive,
will
eat-up four scalars (a hypermultiplet). The left over low-energy theory
of the scalars will be described by a smaller hyper-K\"ahler manifold
(since supersymmetry is not broken during the Higgs phase transition).
This manifold is constructed by a mathematical procedure known as the
hyper-K\"ahler quotient.
The procedure ``factors out" the isometries of a hyper-K\"ahler
manifold
to produce a lower-dimensional manifold which is still hyper-K\"ahler.
Thus, the  hyper-K\"ahler quotient construction is describing the
ordinary
Higgs effect in six-dimensional N=1 gauge theory.

The D5-brane we are about to construct is mapped via heterotic/type-I
duality
to the NS5-brane of the heterotic theory. The NS5-brane has been
constructed
\cite{chs} as a soliton of the effective low-energy heterotic action.
The non-trivial fields, in the transverse space, are essentially
configurations
of axion-dilaton instantons, together with four-dimensional instantons
embedded in the O(32) gauge group. Such instantons have a size that
determines
the ``thickness" of the NS5-brane.
The massless fluctuations are essentially the moduli of the instantons.
There is a mathematical construction of this moduli space, as a
hyper-K\"ahler
quotient. This leads us to suspect \cite{si} that the interpretation of
this construction  is a Higgs
effect in the six-dimensional world-volume theory.
In particular, the mathematical construction implies that for N
coincident
NS5-branes, the hyper-K\"ahler quotient construction implies that an
Sp(N) gauge group is completely Higgsed. For a single five-brane, the
gauge group
is $\rm Sp(1)\sim SU(2)$.
Indeed, if the size of the instanton is not zero, the massless
fluctuations
of the NS5-brane form hypermultiplets only.
When the size becomes zero, the moduli space has a singularity, which
can be interpreted as the restoration of the gauge symmetry: at this
point
the gauge bosons become massless again.
All of this indicates that the world-volume theory of a single
five-brane
should contain an SU(2) gauge group, while in the case of N five-branes
the gauge group is enhanced to Sp(N), \cite{si}.

We will now return to our description of the massless fluctuations of
the D5-brane.
The situation parallels the D1 case that we have described in detail.
In particular, from the DN sectors we will obtain hypermultiplets only.
{}From the DD sector we can in principle obtain massless vectors.
However, as we have seen above, the unique vector that can appear is
projected
out by the orientifold projection.
To remedy this situation we are forced to introduce a Chan-Paton factor
for the Dirichlet end-points of the open string  fluctuations.
For a single D5-brane, this factor takes two values, $i=1,2$.
Thus, the massless bosonic states in the DD sector are of the form
\be
b_{-1/2}^{\mu}|p;i,j\rangle\;\;\;\;\;\;,
\;\;\;\;\;\;b_{-1/2}^{I}|p;i,j\rangle\;.
\ee
We have also seen that the orientifold projection $\Omega$ changes the
sign of $b_{-1/2}^{\mu}$ and leaves $b_{-1/2}^{I}$ invariant.
The action of $\Omega$ on the ground-state is $\Omega
|p;i,j\rangle=\e|p;j,i\rangle$. It interchanges the Chan-Paton factors
and can have a sign $\e=\pm 1$.
The number of vectors that survive the $\Omega$ projection depends on
this sign.
For $\e=1$, only one vector survives and the gauge group is O(2).
If $\e=-1$, three vectors survive and the gauge group is $\rm Sp(1)\sim
SU(2)$.
Taking into account our previous discussion, we must take $\e=-1$.
Thus, we have an Sp(1) vector multiplet.
The scalar states on the other hand will be forced to be
antisymmetrized in
the Chan-Paton indices. This will provide  a single hypermultiplet,
whose four scalars
describe the position of the D5-brane in the four-dimensional
transverse space.
Finally, the DN sector has an $i=1,2$ Chan-Paton factor on the D-end
and an $\a=1,2,\cdots,32$ factor on the Neumann end-point.
Consequently, we will obtain
a hypermultiplet transforming as $\bf (2,32)$ under $\rm Sp(1)\times
O(32)$
where Sp(1) is the world-volume gauge group and O(32) is the original
(spacetime) gauge group of the type-I theory.

In order  to describe N parallel coinciding D5-branes, the only
difference is that the Dirichlet Chan-Paton factor now takes 2N values.
Going through the same procedure as above we find in the DD sector,
Sp(N) vector multiplets,
and hypermultiplets transforming as a singlet (the center-of-mass
position coordinates) as well as the traceless symmetric tensor
representation of Sp(N)
of dimension $2N^2-N-1$.
In the DN sector we find a hypermultiplet transforming as $\bf (2N,32)$
under
$\rm Sp(N)\times O(32)$.

\vskip .4cm
\noindent\hrulefill
\nopagebreak\vskip .2cm
{\large\bf Exercise}: Consider N parallel coincident D1-branes in the
type-I theory.
Show that the massless excitations are a two-dimensional vector in the
adjoint representation of SO(N), eight scalars in the symmetric
representation of SO(N),
eight left-moving fermions in the adjoint of SO(N) and right-moving
fermions
transforming as $({\bf N,32})$ of SO(N)$\times$SO(32).
This is in agreement with matrix theory compactified on $S^1/Z_2$
\cite{rey}.
\nopagebreak\vskip .2cm
\noindent\hrulefill
\vskip .4cm

There are further checks of heterotic/type-I duality in ten dimensions.
BPS-saturated terms in the effective action match appropriately between
the two theories \cite{ts}.
You can find  a more detailed exposition of similar matters in
\cite{P2}.

The comparison becomes more involved and non-trivial  upon toroidal
compactification.
First, the spectrum of BPS states is richer and different in
perturbation theory in the two theories. Secondly, by adjusting moduli
both theories can be compared in the weak coupling limit.
The terms in the effective action that can be most easily compared are
the $F^4$,
$F^2R^2$ and $R^4$ terms. These are BPS-saturated and anomaly-related.
In the heterotic string, they obtain perturbative corrections at one
loop only.
Also, their non-perturbative corrections are due to instantons that
preserve half of the supersymmetry. Corrections due to generic
instantons, that break
more than 1/2 supersymmetry, vanish because of zero modes.
In the heterotic string the only relevant non-perturbative
configuration
is the NS5-brane. Taking its world-volume to be Euclidean and wrapping
it
supersymmetrically around a compact manifold (so that the classical
action is finite), it provides the relevant instanton configurations.
Since we need at least a six-dimensional compact manifold to wrap it,
we can immediately deduce that the BPS-saturated terms do not have
non-perturbative corrections for toroidal compactifications with more
than four non-compact directions.
Thus, for $D>4$ the full heterotic result is tree-level and one-loop.

In the type-I string the situation is slightly different.
Here we have both the D1-brane and the D5-brane that can provide
instanton configurations. Again, the D5-brane will contribute in four
dimensions.
However, the D1-brane has a two-dimensional world-sheet and can
contribute already in eight dimensions.
We conclude that, in nine dimensions, the two theories can be compared
in perturbation theory.
This has been done \cite{bk1}. They do agree at one loop. On the type-I
side, however, duality also implies contact contributions for
the factorizable terms $({\rm tr }R^2)^2$, tr$F^2{\rm tr} R^2$ and
$({\rm tr} F^2)^2$ coming
from surfaces with Euler number $\chi=-1,-2$.

In eight dimensions, the perturbative heterotic result is mapped via
duality to perturbative as well as non-perturbative type-I
contributions coming from the D1-instanton.
These have been computed and duality has been verified \cite{bk2}.

\renewcommand{\theequation}{\thesubsection.\arabic{equation}}
\subsection{Type-IIA versus M-theory.}
\setcounter{equation}{0}

We have mentioned in section \ref{SEA}, that the effective type-IIA
supergravity is the dimensional reduction of eleven-dimensional, N=1
supergravity.
We will see here that this is not just an accident \cite{ht,va}.

We will first review the spectrum of forms in type-IIA theory in ten
dimensions.

$\bullet$ $NS$-$NS$ two-form B. Couples to a string (electrically) and
a five-brane (magnetically). The string is the perturbative type-IIA
string.

$\bullet$ $R$-$R$ U(1) gauge field A$_{\mu}$. Can couple electrically
to particles
(zero-branes) and magnetically to six-branes. Since it comes from the
$R$-$R$ sector
no perturbative state is charged under it.

$\bullet$ $R$-$R$ three-form C$_{\m\n\rho}$. Can couple electrically to
membranes (p=2) and magnetically to four-branes.

$\bullet$ There is also the non-propagating zero-form field strength
and
ten-form field strength that would couple to eight-branes (see section
\ref{RRR}).

As stated in section  \ref{SEA}, the lowest-order type-IIA Lagrangian
is
$$
\tilde S^{IIA}={1\over 2\kappa^2_{10}}\left[\int
d^{10}x\sqrt{g}e^{-\Phi}\left[\left(R+(\nabla\Phi)^2-{1\over
12}H^2\right)-{1\over 2\cdot 4!}\hat G^2 -{1\over 4}F^2\right]+
\right.$$
\be+\left.
{1\over
(48)^2}\int B\wedge G\wedge G\right].\label{596}\ee
We are in the string frame.
Note that the $R$-$R$ kinetic terms do not couple to the dilaton as
already argued in section \ref{RRR}.

In the type-IIA supersymmetry algebra there is a central charge
proportional to the U(1) charge of the gauge field A:
\be
\{Q^1_{\a},Q^2_{\dot \a}\}=\d_{\a\dot\a} W
\,.\label{595}\ee
This can be understood, since this supersymmetry algebra is coming from
D=11
where instead of $W$ there is the momentum operator of the eleventh
dimension. Since the U(1) gauge field is the $G_{11,\m}$ component of
the metric, the momentum operator becomes the U(1) charge in the
type-IIA theory.
There is an associated BPS bound
\be
M\geq {c_0\over \lambda}|W|
\,,\label{597}\ee
where $\lambda=e^{\Phi/2}$ is the ten-dimensional string coupling and
$c_0$ some constant.
States that satisfy this equality are BPS-saturated and form smaller
supermultiplets.
As mentioned above all perturbative string states have $W=0$.
However, there is a soliton solution (black hole) of type-IIA
supergravity with the required properties.
In fact, the BPS saturation implies that it is an extremal black hole.
We would expect that quantization of this solution would provide a
(non-perturbative) particle state.
Moreover, it is reasonable to expect that the U(1) charge is quantized
in some units.
Then the spectrum of these BPS states looks like
\be
M ={c\over \lambda}|n|\;\;\;,\;\;\;n\in Z
\,.\label{598}\ee
At weak coupling these states are very heavy (but not as heavy as
standard solitons whose masses scale with the coupling as $1/\l^2$).
However, being BPS states, their mass can be reliably followed at
strong coupling, where they become light, piling up at zero mass as the
coupling becomes infinite.
This is precisely the behavior of Kaluza-Klein (momentum) modes as a
function of the radius. Since also the effective type-IIA field theory
is a dimensional reduction of the eleven-dimensional supergravity, with
$G_{11,11}$ becoming the string coupling,
we can take this seriously \cite{va} and claim that as $\l\to \infty$
type-IIA theory becomes some
eleven-dimensional theory whose low-energy limit is eleven-dimensional
supergravity.
We can calculate the relation between the radius of the eleventh
dimension
and the string coupling.
This was done essentially in section \ref{SEA}, where we described the
dimensional reduction of eleven-dimensional N=1 supergravity to ten
dimensions.
The radius of the eleventh dimension $R$ can be obtained from
(\ref{302})
to be $R=e^{\s}$.
The ten-dimensional type-IIA dilaton was found there to be $\Phi=3\s$.
Thus,
\be
R=\l^{2/3}
\,.\label{604}\ee

At strong type-IIA coupling,  $R\to\infty$ and the theory
decompactifies to eleven dimensions, while in the perturbative regime
the radius is small.

The eleven-dimensional theory (which has been named M-theory) contains
the three-form that can couple to a membrane and a five-brane.
Upon toroidal compactification to ten dimensions,
the membrane, wrapped around the circle, becomes the perturbative
type-IIA string that couples to $B_{\m\n}$.
When it is not winding around the circle, then it is the type-IIA
membrane coupling to the type-IIA three-form.
The M-theory five-brane descends to the type-IIA five-brane or, wound
around the circle, to the type-IIA four-brane.

\renewcommand{\theequation}{\thesubsection.\arabic{equation}}
\subsection{M-theory and the E$_8\times$E$_8$ heterotic string}
\setcounter{equation}{0}

M-theory has $Z_2$ symmetry under which the three-form changes sign.
We might consider an orbifold of M-theory compactified on a circle of
radius R, where the orbifolding symmetry
is $x^{11}\to -x^{11}$ as well as the $Z_2$ symmetry mentioned above
\cite{ee}.

The untwisted sector can be obtained by keeping the fields invariant
under the projection.
It is not difficult to see that the ten-dimensional metric and dilaton
survive the projection, while the gauge boson is projected out.
Also the three-form is projected out, while the two-form survives.
Half of the fermions survive, a Majorana-Weyl gravitino and a
Mayorana-Weyl fermion of opposite chirality.
Thus, in the massless spectrum, we are left with the N=1 supergravity
multiplet.
We do know by now that this theory is anomalous in ten dimensions.
We must have some ``twisted sector" that should arrange itself to
cancel the anomalies.
As we discussed in the section on orbifolds, $S^1/Z_2$ is a line
segment, with the fixed-points $0,\pi$ at the boundary. The
fixed-planes are two copies of
ten-dimensional flat space.
States coming from the twisted sector must be localized on these
planes.
We also have a symmetry exchanging the fixed planes, so we expect
isomorphic massless content coming from the two fixed planes.
It can also be shown that half of the anomalous variation is localized
at one fixed plane and the other half at the other.
The only N=1 multiplets that can cancel the anomaly symmetrically are
vector multiplets, and we must have 248 of them at each fixed plane.
The possible anomaly-free groups satisfying this constraint are $\rm
E_8\times E_8$
and U(1)$^{496}$.
Since there is no known string theory associated with the second
possibility,
it is natural to assume that we have obtained the E$_8\times$E$_8$
heterotic string theory.
A similar argument to that of the previous section shows that there is
a relation similar to (\ref{604}) between the radius of the orbifold
and the heterotic coupling.
In the perturbative heterotic string, the two ten-dimensional planes
are on top of each other and they move further apart as the coupling
grows.

The M-theory membrane survives in the orbifold only if one of its
dimensions is wound around the $S^1/Z_2$. It provides the perturbative
heterotic string.
On the other hand, the five-brane survives, and it cannot wind around
the orbifold direction. It provides the heterotic NS5-brane.
This is in accord with what we would expect from the heterotic string.
Upon compactification to four dimensions, the NS5-brane will give rise
to magnetically charged point-like states (monopoles).

\renewcommand{\theequation}{\thesubsection.\arabic{equation}}
\subsection{Self-duality of the type-IIB string}
\setcounter{equation}{0}

In section \ref{SEA} we have seen that the low-energy effective action
of the type-IIB theory in ten dimensions has an SL(2,$\R)$ global
symmetry. Its SL(2,$\Z$) subgroup was conjectured \cite{sch,ht} to be
an exact non-perturbative symmetry.

As described in section \ref{RRR}, the type-IIB theory in ten
dimensions
contains the following forms:

$\bullet$ The $NS$-$NS$ two-form $B^1$. It couples electrically to the
perturbative
type-IIB string (which we will call for later convenience the (1,0)
string) and magnetically to a five-brane.

$\bullet$ The $R$-$R$ scalar. It is a zero-form (there is a
Peccei-Quinn
symmetry associated with it) and couples electrically to a
($-1$)-brane. Strictly speaking this is an instanton whose
``world-volume" is a point in spacetime.
It also couples magnetically to a seven-brane.

$\bullet$ The $R$-$R$ two-form $B^2$. It couples electrically to a
(0,1) string
(distinct from the perturbative type-II string) and magnetically to
another (0,1) five-brane.

$\bullet$ The self-dual four-form. It couples to a self-dual
three-brane.

As we have mentioned before, the low-energy effective theory is
invariant under a continuous SL(2,$\R)$ symmetry, which acts by
fractional transformations
on the complex scalar $S$ defined in (\ref{303}) and linearly on the
vector of two-forms $(B^1,B^2)$, the four-form being invariant.
The part of SL(2,$\R$) that translates the scalar is a symmetry of the
full perturbative theory.

There is a (charge-one)  BPS instanton solution in type-IIB theory
given by the following configuration \cite{Din}
\be
e^{\phi/2}=\l+{c\over r^8}\;\;\;,\;\;\;\chi=\chi_0+i{c\over \l
(\l r^8+c)} \,,\label{607}\ee
where $r=|x-x_0|$, $x_0^{\m}$ being the position of the instanton, $\l$
is the string coupling far away from the instanton, $c=\pi\sqrt{\pi}$
is fixed by the requirement that the solution has minimal instanton
number and the other expectation values are trivial.

There is also a fundamental string solution, which is charged under
$B^1$ (the (1,0) string) found in \cite{fstring}.
It has a singularity at the core, which is interpreted as a source for
the fundamental type-IIB string.
Acting with $S\to -1/S$ transformation on this solution we obtain
\cite{sch}
a solitonic string solution (the (0,1) string) that is charged under
the $R$-$R$ antisymmetric tensor $B^2$.
It is  given by the following configuration \cite{sch}:
\be
ds^2=A(r)^{-3/4}[-(dx^0)^2+(dx^1)^2]+A(r)^{1/4}dy\cdot dy \,,
\ee
\be
S=\chi_0+i{e^{-\phi_0/2}\over \sqrt{A(r)}}
\,,\label{608}
\ee
\be
B^1=0\;\;\;,\;\;\;B^2_{01}={1\over \sqrt{\Delta} A(r)}
\,,\label{609}\ee
where
\be
A(r)=1+{Q\sqrt{\Delta}\over 3r^6}\;\;\;,\;\;\;Q={3\kappa^2 T\over
\pi^4}\;\;\;,\;\;\;
\Delta=e^{\phi_0/2}\left[\chi_0^2+e^{-\phi_0}\right]
\,.\label{610}\ee
Here, $\kappa$ is Newton's constant and $T=1/(2\pi\a')$ is the tension
of the perturbative
type-IIB string.
The tension of the (0,1) string can be calculated to be
\be
\tilde T=T\sqrt{\Delta}
\,.\label{611}\ee
In the perturbative regime, $e^{\phi_0}\to 0$, $\tilde T\sim
Te^{-\phi_0/4}$ is large, and the (0,1) string is very stiff.
Its vibrating modes cannot be seen in perturbation theory.
However, at strong coupling, its fluctuations become the relevant
low-energy
modes.
Acting further by SL(2,$\Z)$ transformations we can generate a
multiplet
of (p,q) strings with p,q relatively prime.
If such solitons are added to the perturbative theory, the continuous
SL(2,$\R$) symmetry is broken to SL(2,$\Z$).
All the (p,q) strings have a common massless spectrum given by the
type-IIB supergravity content. Their massive excitations are distinct.
Their string tension is given by
\be
T_{p,q}=T{|p+qS|\over \sqrt{S_2}}\;\;.
\label{pq}\ee

By compactifying the type-IIB theory on a circle of radius $R_B$, it
becomes equivalent to the IIA theory compactified on a circle.
On the other hand, the nine-dimensional type-IIA theory is M-theory
compactified on a two-torus.

{}From the type-IIB point of view, wrapping (p,q) strings around the
tenth dimension provides a spectrum of particles in nine dimensions
with masses
\be
M^2_B={m^2\over R_B^2}+(2\pi R_BnT_{p,q})^2+4\pi T_{p,q}(N_L+N_R)\;\;,
\label{pq1}\ee
where $m$ is the Kaluza-Klein  momentum integer, $n$ the winding number
and $N_{L,R}$ the string oscillator numbers.
The matching condition is $N_L-N_R=mn$, and BPS states are obtained for
$N_{L}=0$ or $N_R=0$.
We thus obtain the following BPS spectrum
\be
\left. M^2_B\right|_{\rm BPS}=\left({m\over R_B}+2\pi
R_BnT_{p,q}\right)^2\;\;.
\label{pq2}\ee
Since an arbitrary pair of integers $(n_1,n_2)$ can be written as
$n(p,q)$,
where $n$ is the greatest common divisor and $p,q$ are relatively prime,
we can rewrite the BPS mass formula above as
\be
\left. M^2_B\right|_{BPS}=\left({m\over R_B}+2\pi
R_BT{|n_1+n_2S|\over \sqrt{S_2}}\right)^2\;\;.
\label{pq3}\ee

In M-theory, compactified on a two-torus with area $A_{11}$ and modulus
$\tau$,
we have two types of (point-like) BPS states in nine dimensions:
KK states with mass $(2\pi)^2|n_1+n_2\tau|^2/(\tau_2A_{11})$ as well as
states
that are obtained by wrapping the M-theory membrane $m$ times around
the
two-torus, with mass $(m A_{11} T_{11})^2 $, where $T_{11}$ is the
tension of the membrane.
We can also write $R_{11}$ that becomes the IIA coupling as
$R_{11}^2=A_{11}/(4\pi^2\tau_2)$.
Thus, the BPS spectrum is
\be
M_{11}^2=(m(2\pi R_{11})^2\tau_2 T_{11})^2+{|n_1+n_2\tau|^2\over
R_{11}^2\tau_2^2}+\cdots\;\;\;,
\label{pq4}\ee
where the dots are mixing terms that we cannot calculate.
The two BPS mass spectra should be related by $M_{11}^2=\b M_B^2$, where
$\b\not =1$
since the masses are measured in different units in the two theories.
Comparing, we obtain
\be
S=\tau\;\;\;,\;\;\;{1\over R_B^2}=T T_{11}A_{11}^{3/2}\;\;\;,\;\;\;
\beta=2\pi R_{11}{\sqrt{\tau_2}T_{11}\over T}\;\;.
\label{pq5}\ee
An outcome of this is the calculation of the M-theory membrane tension
$T_{11}$ in terms of string data.

There are several more tests of the consistency of assuming the
SL(2,$\Z)$ symmetry in the IIB string, upon compactification. These
include instanton calculations in ten or lower dimensions
\cite{iibinst} as well as the existence of the ``F-theory" structure
\cite{f} describing non-perturbative vacua of the IIB string.

\renewcommand{\theequation}{\thesubsection.\arabic{equation}}
\subsection{D-branes are the type-II $R$-$R$ charged states}
\setcounter{equation}{0}

We have seen in section \ref{hetI} that D-branes defined by imposing
Dirichlet boundary conditions on some of the string coordinates
provided non-perturbative extended solitons required by
heterotic/type-I string duality.

Similar D-branes can also be constructed in type-II string theory,
the only difference being that, here, there is no orientifold
projection.
Also, open string fluctuations around them cannot have Neumann (free)
end-points.
As we will see, such D-branes will provide all $R$-$R$ charged states
required by the non-perturbative dualities of type-II string theory.

In the type-IIA theory we have seen that there are  (in principle)
allowed
 $R$-$R$ charged p-branes, with  p$=0,2,4,6,8$, while in the type-IIB
p$=-1,1,3,5,7$.
D-branes can be constructed with a number of coordinates having
D-boundary conditions being $9-$p$=1,2,\ldots,10$, which precisely
matches the full allowed p-brane spectrum of type-II theories.
The important question is: Are such D-branes charged under $R$-$R$
forms?

\begin{figure}
\begin{center}
\leavevmode
\epsffile{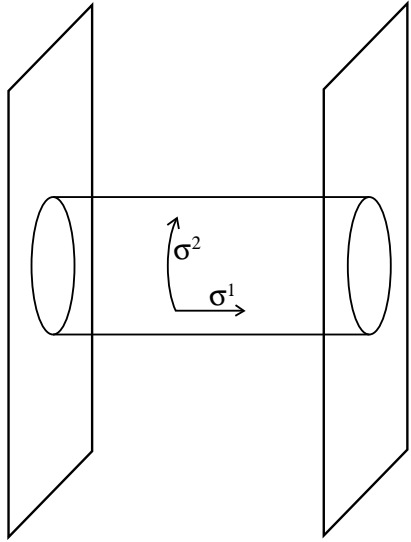}
\caption[]{\it D-branes interacting via the tree-level exchange of a closed
string.}
\label{f22}
\end{center}\end{figure}

To answer this question, we will have to study the tree-level
interaction
of two parallel Dp-branes via the exchange of a closed string
\cite{Dp},
depicted schematically in Fig. \ref{f22}.
For this interpretation time runs horizontally.
However, if we take time to run vertically, then, the same diagram can
be interpreted as a (one-loop) vacuum fluctuation of open strings with
their
end-points attached to the D-branes.
In this second picture we can calculate this  diagram  to be
\be
{\cal A}=2V_{p+1}\int {d^{p+1}k\over
(2\pi)^{p+1}}\int_{0}^{\infty}{dt\over 2t}
e^{-2\pi\a' tk^2-t{|Y|^2\over 2\pi\a'}}{1\over \eta^{12}(it)}{1\over 2}
\sum_{a,b}(-1)^{a+b+ab}\th^4[^a_b](it)
\label{612}\ee
$$=2V_{p+1}\int_{0}^{\infty}{dt\over 2t}(8\pi^2\a't)^{-{p+1\over 2}}
e^{-t{|Y|^2\over 2\pi\a'}}{1\over \eta^{12}(it)}{1\over 2}
\sum_{a,b=0}^1(-1)^{a+b+ab}\th^4[^a_b](it)\,.$$

$V_{p+1}$ is the world-volume of the p-brane, the factor of 2 is due to
the two end-points, $|Y|^2$ is the distance between the D-branes.
Of course the total result is zero, because of the $\th$-identity
(\ref{t14}).
This reflects the fact that the D-branes are BPS states and exert no
static force on each other.
However, our purpose is to disentangle the contributions of the various
intermediate massless states in the closed string channel.
This can be obtained by taking the leading $t\to 0$ behavior of the
integrand.
In order to do this, we have to perform a modular transformation $t\to
1/t$
in the $\th$- and $\eta$-functions. We obtain
\be
\left.{\cal A}\right|^{\rm closed~~string}_{\rm massless}=8(1-1)V_{p+1}
\int_{0}^{\infty}{dt\over t}(8\pi^2\a' t)^{-{p+1\over
2}}~t^4~e^{-{t|Y|^2
\over 2\pi\a'}}
\label{613}\ee
$$=2\pi(1-1)V_{p+1}(4\pi^2\a')^{3-p}G_{9-p}(|Y|)\;,
$$
where
\be
G_d(|Y|)={1\over 4\pi^{d/2}}\int_0^{\infty}{dt\over
t^{(4-d)/2}}e^{-t|Y|^2}
\label{614}\ee
is the massless scalar propagator in d dimensions.
The $(1-1)$ comes from the $NS$-$NS$ and $R$-$R$ sectors respectively.
Now consider the $R$-$R$ forms coupled to p-branes with action
\be
S={\a_p\over 2}\int F_{p+2}\;^*F_{p+2}+iT_{p}\int_{\rm branes}A_{p+1}
\,,\label{615}\ee
with $F_{p+2}=dA_{p+1}$.
Using this action, the same amplitude for an exchange of $A_{p+1}$
between two D-branes at distance $|Y|$ in the transverse space of
dimension $10-(p+1)=9-p$ is given by
\be
\left.{\cal A}\right|_{\rm field~~theory}={(iT_{p})^2\over
\a_{p}}V_{p+1}G_{9-p}(|Y|)
\,,\label{616}\ee
where the factor of volume is present since the $R$-$R$ field can be
absorbed or emitted at any point in the world-volume of the D-brane.
Matching with the string calculation we obtain
\be
{T^2_p\over \a_p}=2\pi(4\pi^2\a')^{3-p}
\,.\label{617}\ee
We will  now look at the DNT quantization condition which, with our
normalization
of the $R$-$R$ forms and $D=10$, becomes
\be
{T_pT_{6-p}\over \a_p}=2\pi n
\,.\label{618}\ee
{}From (\ref{617}) we can verify directly that D-branes satisfy this
quantization condition for the minimum quantum $n=1$!

Thus, we are led to accept that D-branes, with a nice (open) CFT
description of their fluctuations, describe non-perturbative extended
BPS states of the
type-II string carrying non-trivial $R$-$R$ charge.

We will now describe a uniform normalization of the D-brane
tensions.
Our starting point is the type-IIA ten-dimensional effective action
(\ref{596}).
The gravitational coupling $\kappa_{10}$ is given in terms of $\a'$ as
\be
2\kappa_{10}^{2}=(2\pi)^7\a'^4\;\;.
\label{nor1}\ee
We will also normalize all forms so that their kinetic terms
are $(1/4 \kappa_{10}^2)$ $\int d^{10}x F\otimes\tilde F$.
This corresponds to $\a_{p}=1/(2\kappa_{10}^2)$.
We will also define the tensions of various p-branes via their
world-volume action of the form
\be
S_{p}=-T_p\int_{W_{p+1}}d^{p+1}\xi ~e^{-\Phi/2}\sqrt{{\rm det} \hat
G}-iT_p\int
{}~A_{p+1}\;\;\;,
\label{nor2}\ee
where $\hat G$ is the metric induced on the world-volume
\be
\hat G_{\a\b}=G_{\m\n}{\pd X^{\mu}\over \pd \xi^{\a}}{\pd X^{\n}\over
\pd \xi^{\b}}
\label{nor3}\ee
and
\be
\int A_{p+1}={1\over (p+1)!}\int d^{p+1}\xi
{}~A_{\m_{1}\cdots\m_{p+1}}{\pd X^{\m_1}\over \pd \xi^{\a_1}}\cdots
{\pd X^{\m_{p+1}}\over \pd \xi^{\a_{p+1}}}\e^{\a_1\cdots\a_{p+1}}\;\;.
\label{nor4}\ee
The dilaton dependence will be explained in the next section.
The DNT quantization condition in (\ref{618}) becomes
\be
2\kappa_{10}^2 T_{p}T_{6-p}=2\pi n\;\;,
\ee
while (\ref{617}) and (\ref{nor1}) give
\be
T_p={1\over (2\pi )^p (\a')^{(p+1)/2}}\;\;.
\label{nor5}\ee
We have obtained the IIA theory from the reduction of
eleven-dimensional supergravity on a circle of volume $2\pi R_{11}=2\pi
\sqrt{a'}e^{\Phi/3}$.
Consequently, the M-theory gravitational constant is
\be
2\kappa_{11}^2=(2\pi)^{8}(\a')^{9/2}\;\;.
\label{nor6}\ee
The M-theory membrane, upon compactification of M-theory on a circle,
becomes the type-IIA D2-brane.
Thus, its
tension $T^M_2$ should be equal to the D2-brane tension:
\be
T^M_2=T_2={1\over (2\pi)^2(\a')^{3/2}}\;\;.
\label{nor7}\ee
Consider now the M-theory five-brane. It has a tension $T^M_5$ that can
be computed from the DNT quantization condition
\be
2\kappa_{11}^2 T_2^MT^M_5=2\pi\;\;\;\to\;\;\;\;T^M_5={1\over
(2\pi)^5(\a')^3}\;.
\label{nor8}\ee
On the other hand, wrapping one of the coordinates of the M5-brane
around the circle should produce the D4-brane and we can confirm that
\be
2\pi\sqrt{\a'}T^M_5=T_4\;.
\label{nor9}\ee

\renewcommand{\theequation}{\thesubsection.\arabic{equation}}
\subsection{D-brane actions}
\setcounter{equation}{0}

We will now derive the massless fluctuations of a single Dp-brane.
This parallels our detailed discussion of the type-I D1-brane.
The  difference here is that the open string fluctuations cannot have
free
ends.\footnote{Free end-points are interpreted as 9-branes and there
are none
in type-II string theory.}
Thus, only the DD sector is relevant.
Also there is no orientifold projection.
In the $NS$ sector, the massless bosonic states are a (p+1)-vector,
$A_{\m}$
corresponding to the state $b^{\mu}_{-1/2}|p\rangle$ and 9-p scalars,
$X^I$ corresponding to the states $b^I_{-1/2}|p\rangle$.
The $X^I$ represent the position coordinates of the Dp-brane in
transverse space.
These are the states we would obtain by reducing a ten-dimensional
vector
to p+1 dimensions.
Similarly, from the Ramond sector we obtain world-volume fermions that
are the reduction of a ten-dimensional gaugino to (p+1) dimensions.
In total we obtain the reduction of a ten-dimensional U(1) vector
multiplet
to p+1 dimensions.
The world-volume supersymmetry has 16 conserved supercharges. Thus, the
Dp-brane breaks half of the original supersymmetry as expected.

In order to calculate the world-volume action, we would have to
calculate the scattering of the massless states of the world-volume
theory.
The leading contribution comes from the disk diagram and is thus
weighted
with a factor of $e^{-\Phi/2}$.
The calculation is similar with the calculation of the effective action
in the ten-dimensional open oriented string theory.
The result there is the Born-Infeld action for the gauge field
\cite{bi}
\be
S_{BI}=\int d^{10}x~ e^{-\Phi/2}\sqrt{\det(\delta_{\m\n}+2\pi \a'
F_{\m\n})}\;\;.
\label{nor10}\ee
Dimensionally reducing this action, we obtain the relevant Dp-brane
action
from the disk.
There is a coupling to the spacetime background metric, which gives the
induced metric, (\ref{nor3}).
There is also a coupling to the spacetime $NS$ antisymmetric tensor.
This
can be seen as follows.
The closed string coupling to $B_{\m\n}$ and the vector $A_{\mu}$ can
be summarized in
\be
S_{B}={i\over 2\pi\a'}\int_{M_2}
d^2\xi ~\e^{\a\b}B_{\m\n}\pd_{a}x^{\mu}\pd_{\b}x^{\nu}-{i\over
2}\int_{B_1}ds ~A_{\mu}\pd_s x^{\m}\;\;,
\label{nor11}\ee
where $M_2$ is the two-dimensional world-sheet with one-dimensional
boundary
$B_1$.
Under a gauge transformation $\delta
B_{\m\n}=\pd_{\mu}\Lambda_{\n}-\pd_{\n}\Lambda_{\m}$, the above action
changes
by a boundary term,
\be
\delta S_{B}={i\over \pi\a'}\int_{B_1} ds ~\Lambda_{\m}\pd_s
x^{\m}\;\;.
\label{nor12}\ee
To reinstate gauge invariance, the vector $A_{\m}$ has to transform as
$\delta A_{\mu}={1\over 2\pi\a'}\Lambda_{\mu}$.
Thus, the gauge-invariant combination is
\be
{\cal F}_{\mu\nu}=2\pi \a' F_{\m\n}-B_{\mu\nu}=2\pi
\a'(\pd_{\mu}A_{\n}-\pd_{\n}A_{\m})-B_{\mu\nu}\;\;.
\label{nor13}\ee
We can now summarize the leading order Dp-brane action as
\be
S_{p}=-T_p\int_{W_{p+1}}d^{p+1}\xi ~e^{-\Phi/2}\sqrt{{\rm det}( \hat
G+{\cal F})}-iT_p\int ~A_{p+1}\;\;.
\label{nor14}\ee
As we saw in the previous section, the CP-odd term in the action
comes from the next diagram, the annulus.
There are however more CP-odd couplings coming from the annulus that
involve q-forms with q$<$p.
Their appearance is due to cancelation of anomalies, and we refer the
reader to \cite{inflow} for a detailed discussion.
We will present the result here.
It involves the roof-genus $\hat I_{1/2}(R)$ in (\ref{323}) and the
Chern character. Thus, (\ref{nor14}) is extended to
\bea
S_{p}=-T_p\int_{W_{p+1}}d^{p+1}\xi ~e^{-\Phi/2}\sqrt{{\rm det}( \hat
G+{\cal F})}+ \nn\\
-iT_p\int ~A\wedge {\rm Tr}[e^{i{\cal F}/2\pi}]\sqrt{\hat
I_{1/2}(R)}\;,
\label{nor15}\eea
where $A$ stands for a formal sum of all $R$-$R$ forms, and the
integration
picks up the (p+1)-form in the sum.

As an example we will consider the action of the  D1-string of type-IIB
theory.
The relevant forms that couple here is the $R$-$R$ two-form
$B^R_{\m\n}$
as well as the $R$-$R$ scalar (zero-form) $S_1$.
The action is
\be
S_{1}=-{1\over 2\pi \a'}\left[\int d^{2}\xi ~{|S|\over
\sqrt{S_2}}~\sqrt{{\rm det}( \hat G+{\cal F})}+i\int
\left(B^N+{iS_1\over 2\pi}{\cal F}\right)\right]\;\;,
\label{nor16}\ee
where $e^{-\Phi/2}=S_2$.
Note that ${|S|\over \sqrt{S_2}}=e^{-\Phi/2}$ when $S_1=0$.

We will now consider the effect of T-duality transformations on the
Dp-branes.
Consider the type-II theory with $x^9$ compactified on a circle of
radius R.
As we have mentioned earlier, the effect of a T-duality transformation
on open strings is to interchange N and D boundary conditions.
Consider first a Dp-brane not wrapping around the circle.
This implies that one of its transverse coordinates (Dirichlet) is in
the compact direction.
Doing a T-duality transformation $R\to \a'/R$, would change the
boundary conditions along $X^9$ to Neumann and would produce a
D(p+1)-brane wrapping
around the circle of radius $\a'/R$.
Thus, the Dp-brane has been transformed into a D(p+1)-brane.
The original Dp-brane action contains $T_p\int d^{p+1}\xi~e^{-\Phi/2}$.
The dilaton transforms under duality as
\be
e^{-\Phi/2}\to {\sqrt{\a'}\over R}e^{-\Phi/2}\;\;.
\ee
Consequently, $T_p\sqrt{\a'}/R=T_{p+1}(2\pi \a'/R)$ and we obtain
\be
T_{p+1}={T_p\over 2\pi \sqrt{\a'}}\;\;,
\label{nor17}\ee
which is in agreement with (\ref{nor5}).

On the other hand, if the Dp-brane was wrapped around the compact
direction, T-duality transforms it into a D(p-1)-brane.
This action of T-duality on the various D-branes is a powerful tool
for investigating non-perturbative physics.

So far, we have discussed a single Dp-brane, interacting with the
background type-II fields. An obvious question is: What happens when we
have more
than one parallel Dp-branes?
Consider first the case where we have N Dp-branes at the same point in
transverse space. Then, the only difference
with the previous analysis, is that we now include a Chan-Paton factor
$i=1,2,\cdots ,N$
at the open string end-points.
We now have $N^2$ massless vector states,
$b^{\mu}_{-1/2}|p;i,j\rangle$.
Going through the same procedure as before, we will find that the
massless fluctuations are described by the dimensional reduction of the
ten-dimensional
N=1 U(N) Yang-Mills multiplet on the world-volume of the brane (we have
oriented open strings here).
The U(1) factor of U(N) describes the overall center of mass of the
system.
If we take one of the Dp-branes and we separate it from the rest, the
open strings stretching between it and the rest N-1 of the branes
acquire a mass-gap
(non-trivial tension), and the massless vectors have a gauge group
which is
$\rm U(N-1)\times U(1)$.
In terms of the world-sheet theory, this is an ordinary Higgs effect.
For generic positions of the Dp-branes, the gauge group is $\rm
U(1)^N$.
The scalars that described the individual positions now become U(N)
matrices.
The world-volume action has a non-abelian generalization. In
particular,
to lowest order, it is the dimensional reduction of U(N)
ten-dimensional Yang-Mills:
\be
S^N_{p}=-T_p {\rm Str}\int_{W_{p+1}}d^{p+1}\xi ~e^{-\Phi/2}
(F_{\m\n}^2+2F_{\m I}^2+F_{IJ}^2)\;\;,
\label{nor18}\ee
where
\be
F_{\m\n}=\pd_{\m}A_{\n}-\pd_{\n}A_{\m}+[A_{\mu},A_{\n}]\;\;,
\label{nor19}\ee
\be
F_{\mu I}=\pd_{\m}X^I+[A_{\m},X^I]\;\;\;,\;\;\;F_{IJ}=[X^I,X^J]\;.
\label{nor20}\ee
Both $A_{\mu}$ and $X^I$ are U(N) matrices.
At the minimum of the potential, the matrices $X^I$ are commuting, and
can be simultaneously diagonalized. Their eigenvalues can be
interpreted
as the coordinates of the N Dp-branes.
Further information on D-branes can be found in \cite{P2}.

The dynamics of D-branes turns out to be very interesting. In
particular,
they behave differently from fundamental strings in sub-Planckian
energies
\cite{dbscat}.
They are interesting probes that can reach regimes not accessible by
strings.

One very interesting application of D-branes is the following.
Wrapped around compact manifolds, D-branes produce point-like $R$-$R$
charged particles in lower dimensions.
Such particles have an effective description as microscopic black
holes.
Using D-brane techniques, their multiplicity can be computed for fixed
charge and mass. It can be shown that this multiplicity agrees to
leading order
with the Bekenstein-Hawking entropy formula for classical black holes
\cite{vs}. The interested reader may consult \cite{mal} for a review.

\renewcommand{\theequation}{\thesubsection.\arabic{equation}}
\subsection{Heterotic/type-II duality in six and four dimensions}
\setcounter{equation}{0}

There is another non-trivial duality relation that we are going to
discuss
in some detail: that of the heterotic string compactified to six
dimensions on $T^4$ and the type-IIA string compactified on K3.
Both theories have N=2 supersymmetry in six dimensions.
Both theories have the same massless spectrum, containing the N=2
supergravity multiplet and twenty vector multiplets, as shown in
sections \ref{torcomp} and \ref{compact}.

The six-dimensional tree-level heterotic effective action in the
$\s$-model frame was given in (\ref{348}).
Going to the Einstein frame by $G_{\m\n}\to e^{\Phi/2}G_{\m\n}$, we
obtain
\bea
S^{\rm het}_{D=6}&=&\int ~d^6x~\sqrt{-G}\left[R-{1\over
4}\pd^{\m}\Phi\pd_{\m}\Phi
-{e^{-\Phi}\over 12}\hat H^{\m\n\rho}\hat H_{\m\n\rho}+\right.
\nn\\
&&\left.-{e^{-{\Phi\over 2}}\over 4}(\hat M^{-1})_{ij}
F^{i}_{\m\n}F^{j\m\n}+{1\over 8}{\rm Tr}(\pd_{\m} \hat M\pd^{\m}
\hat M^{-1})\right]\,.
\label{6188}
\eea
The tree-level type-IIA effective action in the $\s$-model frame was
also given
in (\ref{377}). Going again to the Einstein frame we obtain
\bea
S^{IIA}_{D=6}&=&\int d^6x\sqrt{-G}\left[R-{1\over
4}\pd^{\m}\Phi\pd_{\m}\Phi
-{1\over 12}e^{-\Phi}H^{\m\n\rho}H_{\m\n\rho}+\right.
\nn\\
&&\left.-{1\over 4}e^{\Phi/2}(\hat M^{-1})_{ij}
F^{i}_{\m\n}F^{j\m\n}+{1\over
8}{\rm Tr}(\pd_{\m}\hat M\pd^{\m}\hat M^{-1})\right]+\nn\\
&&+ {1\over 16}\int
d^6x
\e^{\m\n\rho\s\tau\varepsilon}B_{\m\n}F^{i}_{\rho\s}\hat
L_{ij}F^{j}_{\tau\varepsilon}\;,
\label{619}\eea
where $\hat L$ is the O(4,20) invariant metric.
Notice the following differences: the heterotic $\hat H_{\m\n\rho}$
contains the Chern-Simons term (\ref{350}), while the type-IIA one does
not.
The type-IIA action instead contains a parity-odd term coupling the
gauge
fields and $B_{\m\n}$.
Both effective actions have a continuous O(4,20,$\R$) symmetry, which
is broken in the string theory to the T-duality group O(4,20,$\Z$).

We will denote by a prime the fields of the type-IIA theory (Einstein
frame)
and without a prime those of the heterotic theory.

\vskip .4cm
\noindent\hrulefill
\nopagebreak\vskip .2cm
{\large\bf Exercise}. Derive the equations of motion stemming from the
actions (\ref{6188}) and (\ref{619}). Show that the two sets of
equations of motion  are equivalent via the following (duality)
transformations
\be
\Phi'=-\Phi\;\;\;,\;\;\;G'_{\m\n}=G_{\m\n}\;\;\;,\;\;\;\hat M'=\hat
M\;\;\;,\;\;\;
A'^i_{\mu}=A^{i}_{\mu}
\,,\label{620}\ee
\be
e^{-\Phi}\hat H_{\m\n\rho}={1\over
6}{{\e_{\m\n\rho}}^{\s\tau\varepsilon}
\over\sqrt{-G}}H'_{\s\tau\varepsilon}
\,,\label{621}\ee
where the data on the right-hand side are evaluated in the type-IIA
theory.
\nopagebreak\vskip .2cm
\noindent\hrulefill
\vskip .4cm

There is a way to see some indication of this duality by considering
the compactification of M-theory on $S^1\times$K3, which is equivalent
to type-IIA on K3.
As we have seen in a previous section, all vectors descend from the
$R$-$R$ one-
and three-forms of the ten-dimensional type-IIA theory, and these
descend
from the three-form of M-theory to which the membrane and five-brane
couple.
The membrane wrapped around $S^1$ would give a string in six
dimensions.
As in ten dimensions, this is the perturbative type-IIA string.
There is another string however, obtained by wrapping the five-brane
around the whole K3. This is the heterotic string \cite{ks}.

There is further evidence for this duality.
The effective action of type-IIA theory on K3 has a string solution,
singular at the core.
The zero mode structure of the string is similar to the perturbative
type-IIA string.
There is also a string solution that is regular at the core. This is a
solitonic string and analysis of its zero modes indicates that it has
the same (chiral) word-sheet structure as the heterotic
string.\footnote{We have already seen a similar phenomenon in the case
of the D1-string of type-I string theory.}
The string-string duality map (\ref{620})-(\ref{621}) exchanges the
roles of the two strings.
The type-IIA string now becomes regular (solitonic), while the
heterotic
string solution becomes singular.

We will now further compactify both theories on a two-torus down to
four dimensions and examine the consequences of the duality.
In both cases we use the standard Kaluza-Klein ansatz described in
Appendix C.
The four-dimensional dilaton becomes, as usual,
\be
\phi=\Phi-{1\over 2}\log[\det G_{\a\b}]
\,,\label{622}\ee
where $G_{\a\b}$ is the metric of $T^2$ and $B_{\a\b}=\e_{\a\b} B$ is
the antisymmetric tensor.
We obtain
\be
S_{D=4}^{\rm het}=\int d^4 x
\sqrt{-g}e^{-\phi}\left[R+L_{B}+L_{\rm gauge}+L_{\rm
scalar}\right]
\,,\label{623}\ee
where
\be
{\cal L}_{g+\phi}=R+\pd^{\m}\phi\pd_{\m}\phi
\,,\label{624}\ee
\be
{\cal L}_{B}=-{1\over 12}H^{\m\n\rho}H_{\m\n\rho}
\,,\label{625}\ee
with
\be
H_{\m\n\rho}=\pd_{\m}B_{\n\rho}-{1\over
2}\left[B_{\m\a}F^{A,\a}_{\n\rho}+A^{\a}_{\m}
F^{B}_{a,\n\rho}+\hat L_{ij}A^{i}_{\mu}F^{j}_{\n\rho}\right]+{\rm
cyclic}
\label{626}
\ee
$$\equiv \pd_{\m}B_{\n\rho}-{1\over 2}
L_{IJ}A^{I}_{\m}F^{J}_{\n\rho}+{\rm cyclic}
\,.$$
The matrix
\be
L=\left(\matrix{ 0&0&1&0& {\vec 0}\cr
0&0&0&1&\vec 0\cr
1&0&0&0&\vec 0\cr
0&1&0&0&\vec 0\cr
\vec 0&\vec 0&\vec 0&\vec 0&\hat L\cr}\right)
\label{627}\ee
is the O(6,22) invariant metric.
Also
\be
C_{\a\b}=\e_{\a\b}B-{1\over 2}\hat L_{ij}Y^i_{\a}Y^j_{\b}
\,,\label{6666}\ee
so that
\bea
{\cal L}_{\rm gauge}&=&-{1\over
4}\left\{\left[(\hat M^{-1})_{ij}+\hat L_{ki}\hat
L_{lj}Y^{k}_{\a}G^{\a\b}
Y^{l}_{\b}\right]
\;\;F^{i}_{\m\n}F^{j,\m\n}+G^{\a\b}\;\;F^{B}_{\a,\m\n}
F_{B,\b}^{\m\n}+\right.
\nn\\
&&
+\left[G_{\a\b}+C_{\g\a}G^{\g\d}C_{\d\b}+Y^{i}_{\a}(\hat
M^{-1})_{ij}
Y^{j}_{\b}\right] F^{A,a}_{\m\n}F_{A}^{\b,\m\n} +
\nn\\
&&
-2G^{\a\g}C_{\g\b} F^{B}_{\a,\m\n}F^{A,\b,\m\n}-
2\hat L_{ij}Y^{i}_{\a}G^{\a\b} F^{j}_{\m\n}F^{B,\m\n}_{\b}+
\nn\\
&&
\left.+2(Y^{i}_{\a}(\hat M^{-1})_{ij}+C_{\g\a}G^{\g\b}\hat
L_{ij}Y^{i}_{\b})\;\;
F^{a,A}_{\m\n}
F^{j,\m\n} \right\}
\nn\\
&\equiv&
 -{1\over 4}(M^{-1})_{IJ}F^{I}_{\m\n}F^{J,\m\n}\,,\label{628}
\eea
where the index I takes 28 values. For the scalars
\bea
{\cal L}_{\rm scalar}&=&\pd_{\m}\phi\pd^{\m}\phi
+ {1\over 8}{\rm Tr}[\pd_{\m}\hat M\pd^{\m}\hat M^{-1}] +
\nn\\
&&{1\over 2}G^{\a\b}(\hat M^{-1})_{ij}
\pd_{\m}Y^{i}_{\a}\pd^{\m}Y^{j}_{\b}
+{1\over 4}\pd_{\m}G_{\a\b}\pd^{\m}G^{\a\b} +
\nn\\
&&
+ \frac{1}{2 {\rm det} G}
\left[\pd_{\m}B+\e^{\a\b}\hat
L_{ij}Y^{i}_{\a}\pd_{\mu}Y^{j}_{\b}\right]
\left[\pd^{\m}B+\e^{\a\b}\hat
L_{ij}Y^{i}_{\a}\pd^{\mu}Y^{j}_{\b}\right]
\nn\\
&=&\pd_{\m}\phi\pd^{\m}\phi+{1\over 8}{\rm Tr}[\pd_{\m}M\pd^{\m}
M^{-1}]\,.
\label{629}\eea

We will now go to the standard axion basis in terms of the usual
duality transformation in four dimensions.
First we will go to the Einstein frame by
\be
g_{\m\n}\to e^{-\phi}g_{\m\n}
\,,\label{630}\ee
so that the action becomes
\bea
S^{\rm het,E}_{D=4}&=&\int
d^4x\sqrt{-g}\;\left[R-{1\over
2}\pd^{\m}\phi\pd_{\m}\phi
-{1\over 12}e^{-2\phi}H^{\m\n\rho}H_{\m\n\rho} + \right.
\nn\\
&&\left.-{1\over 4}e^{-\phi}(M^{-1})_{IJ}
F^{I}_{\m\n}F^{J,\m\n}+{1\over 8}{\rm Tr}(\pd_{\m}M\pd^{\m}
M^{-1})\right]
\,.\label{631}
\eea

The axion is introduced as usual:
\be
e^{-2\phi}H_{\m\n\rho}={{\e_{\m\n\rho}}^{\s}\over \sqrt{-g}}\pd_{\s}a
\,.\label{632}\ee
The transformed equations come from the
following
action:
\bea
\tilde S^{\rm het}_{D=4}&=&\int
d^4x\sqrt{-g}\;\left[R-{1\over
2}\pd^{\m}\phi\pd_{\m}\phi
-{1\over 2}e^{2\phi}\pd^{\m}a\pd_{\m}a+ \right.
\nn\\
&&-{1\over 4}e^{-\phi}(
M^{-1})_{IJ}
F^{I}_{\m\n}F^{J,\m\n}
+{1\over 4}a\;L_{IJ}F^{I}_{\m\n}\tilde
F^{J,\m\n}+
\nn\\
&&\left. + {1\over 8}{\rm Tr}(\pd_{\m}M\pd^{\m}M^{-1})\right]\;,
\label{633}
\eea
where
\be
\tilde F^{\m\n}={1\over 2}{\e^{\m\n\rho\s}\over \sqrt{-g}}
F_{\rho\s}
\,.\label{634}\ee

Finally, defining the complex S field
\be
S=a+i\; e^{-\phi}
\,,\label{635}\ee
we obtain
\bea
\tilde S^{\rm het}_{D=4}&=&\int
d^4x\sqrt{-g}\;\left[R-{1\over 2}{\pd^{\m}S\pd_{\m}\bar
S\over
{\rm Im}S^2}-{1\over 4}{\rm Im}S(M^{-1})_{IJ}
F^{I}_{\m\n}F^{J,\m\n} + \right.
\nn\\
&&
\left.+{1\over 4}{\rm Re}S\;L_{IJ}F^{I}_{\m\n}\tilde
F^{J,\m\n}+{1\over 8}{\rm Tr}(\pd_{\m}M\pd^{\m}M^{-1})\right]
\,.\label{637}
\eea

Now consider the type-IIA action (\ref{377}).
Going through the same procedure  and introducing the axion through
\be
e^{-2\phi}H_{\m\n\rho}={{\e_{\m\n\rho}}^{\s}\over \sqrt{-g}}
\left[\pd_{\s}a+{1\over
2}\hat L_{ij}Y^{i}_{\a}\d_{\s}Y^{j}_{\b}\e^{\a\b}\right]
\,,\label{641}\ee
we obtain the following four-dimensional  action in the Einstein frame:
\be
\tilde S_{D=4}^{IIA}=\int d^4 x
\sqrt{-g}\left[R+{\cal L}^{\rm even}_{\rm gauge}+{\cal L}^{\rm odd}_{\rm
gauge}+{\cal L}_{\rm scalar}\right]
\,,\label{638}\ee
with
\bea
{\cal L}^{\rm even}_{\rm gauge}&=&-{1\over 4}\int d^4x\sqrt{-g}\left[
e^{-\phi}G^{\a\b}
\left(F^{B}_{\a,\m\n}-B_{\a\g}F^{A,\g}_{\m\n}\right)
\left(F^{B,\m\n}_{\b}-B_{\a\d}F^{\d,\m\n}_{A}\right)+\right.
\nn\\
&&
+e^{-\phi}G_{\a\b}F^{A,\a}_{\m\n}F^{\b,\m\n}_{A}+
\nn\\
&&
\left.
\sqrt{{\rm det}G_{\a\b}}(\hat M^{-1})_{ij}\left(F^{i}_{\m\n}+Y^{i}_{\a}
F^{A,\a}_{\m\n}\right)
\left(F^{j,\m\n}+Y^{j}_{\b}F^{\b,\m\n}_{A}\right)\right]\,,
\eea
\bea
{\cal L}^{\rm odd}_{\rm gauge}&=&{1\over 2}\int d^4
x\e^{\m\n\rho\s}\left[{1\over
4}a
F^{B}_{\a,\m\n}F^{A,\a}_{\rho\s}+{1\over
2}\e^{\a\b}\hat
L_{ij}Y^{i}_{\b}F^{B}_{\a,\m\n}\left(F^{j}_{\rho\s}+{1\over
2}Y^{j}_{\g}F^{A,\g}_{\rho\s}\right)+
\right.
\nn\\
&&\left.
-{1\over 8} \e^{\a\b}\hat L_{ij}B_{\a\b}
\left(F^{i}_{\m\n}+Y^{i}_{\g}F^{A,\g}_{\m\n}\right)\left(
F^{j}_{\rho\s}+Y^{j}_{\d}F^{A,\d}_{\rho\s}\right)\right]\;,
\eea
\bea
{\cal L}_{\rm scalar}&=&-{1\over 2}(\pd\phi)^2+{1\over
4}\pd^{\m}G_{\a\b}\pd_{\m}
G^{\a\b}
-{1\over 2{\rm det} G}\pd_{\m} B\pd^{\m} B +
\nn\\
&&
{1\over 8}{\rm Tr}[\pd_{\m}\hat M\pd^{\m} \hat M^{-1}]
-{1\over 2}e^{2\phi}\left(\pd_{\m}a
 + {1\over 2}\hat
L_{ij}\e^{\a\b}Y^{i}_{\a}\pd^{\m}
Y^{j}_{\b}\right)^2 +
\nn\\
&&
-{1\over 2}e^{\phi}\sqrt{{\rm det}G_{\a\b}}(\hat M^{-1})_{ij}G^{\a\b}
\pd_{\m} Y^{i}_{\a}\pd^{\m} Y^{j}_{\b}
\,.
\eea

Now we will use unprimed fields to refer to the heterotic side and
primed ones for the type-II side.
We will now work out the implications of the six-dimensional
duality relations
(\ref{620}), (\ref{621}) in four dimensions.
{}From (\ref{620}), we obtain
\be
e^{-\phi}=\sqrt{{\rm det}G'_{\a\b}}\;\;\;,\;\;\;
e^{-\phi'}=\sqrt{{\rm det}G_{\a\b}}
\,,\label{643}\ee
\be
{G_{\a\b}\over \sqrt{{\rm det}G_{\a\b}}}={G'_{\a\b}\over \sqrt{{\rm
det}G'_{\a\b}}}\;\;\;,\;\;\;A'^{\a}_{\m}=A^{\a}_{\m}
\,,\label{644}\ee
\be
g_{\m\n}=g'_{\m\n}\;\;\;\;\;{\rm Einstein~~frame}
\,,\label{645}\ee
\be
\hat M'=\hat M\;\;\;,\;\;\;A^{i}_{\m}=A'^{i}_{\m}\;\;\;,\;\;\;
Y^{i}_{\a}=Y'^{i}_{\a}
\,.\label{646}\ee

Finally, the relation (\ref{621}) implies
\be
A=B'\;\;\;,\;\;\;A'=B\label{647}
\ee
and
\be
{1\over 2}{{\e_{\m\n}}^{\rho\s}\over
\sqrt{-g}}\e^{\a\b}F^{B'}_{\b,\rho\s}=
e^{-\phi}G^{\a\b}\left[F^{B}_{\b,\m\n}-C_{\b\g}F^{A,\g}_{\m\n}-\hat
L_{ij}
Y^{i}_{\b}F^{j}_{\m\n}\right]-{1\over 2}a{{\e_{\m\n}}^{\rho\s}\over
\sqrt{-g}}F^{A,\a}_{\rho\s}\,,\label{648}\ee
which is an electric-magnetic duality transformation on the $B_{\a,\m}$
gauge fields (see appendix H).
It is easy to check that this duality maps the scalar heterotic terms
to the type-IIA ones and vice versa.

In the following, we will keep the four moduli of the two-torus and the
sixteen
Wilson lines $Y^i_{\a}$.
In the heterotic case we will define the $T,U$ moduli of the torus
and the complex Wilson lines as
\be
W^i=W^i_1+iW_2^i=-Y^i_2+UY^i_1
\,,\label{649}\ee
\be
G_{\a\b}={T_2-{\sum_i(W^i_2)^2\over 2 U_2}\over
U_2}\left(\matrix{1&U_1\cr
U_1&|U|^2\cr}\right)\;\;\;,\;\;\;B=T_1-{\sum_i W_1^iW_2^i\over 2U_2}
\,.\label{650}\ee
Altogether we have the complex field $S \in SU(1,1)/U(1)$ (\ref{635})
and the $T$, $U$, $W^i$ moduli $\in$ ${{\rm O(2,18)}\over {\rm O(2)\times
O(18)}}$.
Then the relevant scalar kinetic terms can be written as
\be
{\cal L}^{\rm het}_{\rm scalar}=-{1\over 2}\pd_{z^i}\pd_{\bar z^j}K(z_k,\bar
z_k)~
\pd_{\m}z^i\pd^{\m}\bar z^j
\,,\label{651}\ee
where the K\"ahler potential is
\be
K=\log\left[S_2\left(T_2U_2-{1\over 2}\sum_i(W_2^i)^2\right)\right]
  \,.\label{652}\ee

In the type-IIA case the complex structure is different:
(\ref{649}) remains the same but
\be
G_{\a\b}={T_2\over U_2}\left(\matrix{1&U_1\cr
U_1&|U|^2\cr}\right)\;\;\;,\;\;\;B=T_1
\,.\label{653}\ee
Also
\be
S=a-{\sum_i W_1^iW_2^i\over
2U_2}+i\left(e^{-\phi}-{\sum_i(W^i_2)^2\over 2 U_2}\right)
\,.\label{654}\ee
Here $T\in $SU(1,1)/U(1) and $S,U,W^i\in {{\rm O(2,18)\over O(2)\times
O(18)}}$.
In this language the duality transformations become
\be
S'=T\;\;\;,\;\;\;T'=S\;\;\;,\;\;\;U=U'\;\;\;,\;\;\;W^i=W'^i
\,.\label{655}\ee

In the type-IIA string, there is an SL(2,$\Z)$ $T$-duality symmetry
acting on T
by fractional transformations.
This is a good symmetry in perturbation theory.
We also expect it to be a good symmetry non-perturbatively since, as we
argued
in section \ref{tduality}, it is a discrete remnant of a gauge symmetry
and is not expected to be broken by non-perturbative effects.
Then heterotic/type-II duality implies that there is an SL(2,$\Z)_S$
symmetry that acts on the coupling constant and the axion.
This is a non-perturbative symmetry from the point of view of the
heterotic string.
It acts as an electric-magnetic duality on all the 28 gauge fields.
In the field theory limit it implies an S-duality symmetry for N=4
super Yang-Mills theory in four dimensions.

We will finally see how heterotic/type-II duality acts on the 28
electric and
28 magnetic charges.
Label the electric charges by a vector ($m_1,m_2,n_1,n_2,q^i$),
where $m_i$ are the momenta of the two-torus, $n_i$ are the respective
winding numbers, and $q^i$ are the rest of the 24 charges.
For the magnetic charges we write the vector  ($\tilde m_1,\tilde
m_2,\tilde n_1,\tilde n_2,\tilde q^i$).
Because of (\ref{648}) we have the following duality map.
\be
\left(\matrix{m_1\cr m_2\cr n_1 \cr
n_2\cr q^i\cr}\right)
\to\left(\matrix{m_1\cr m_2\cr \tilde n_2 \cr-\tilde
n_1\cr q^i\cr}\right)\;\;,\;\;
\left(\matrix{\tilde m_1\cr \tilde m_2\cr \tilde n_1 \cr \tilde
n_2\cr\tilde q^i\cr}\right)
\to\left(\matrix{\tilde m_1 \cr \tilde m_2\cr -n_2\cr n_1\cr\tilde
q^i\cr}\right)
\,.\label{656}\ee
One can compute the spectrum of both short and intermediate BPS
multiplets.
The results of Appendix F are useful in this respect.
\vskip .4cm
\noindent\hrulefill
\nopagebreak\vskip .2cm
{\large\bf Exercise}. Find the BPS multiplicities on the heterotic
and type-IIA side in four dimensions.
\nopagebreak\vskip .2cm
\noindent\hrulefill
\vskip .4cm

There are indirect quantitative tests of this duality.
Compactifying the heterotic string to four dimensions with N=2
supersymmetry
can be dual to the type-IIA string compactified on a CY manifold
of a special kind (K3 fibration over $P^1$) \cite{kv,fhsv,re}.
In the heterotic theory, the dilaton is in a vector multiplet.
Consequently,  the vector multiplet moduli space has perturbative and
non-perturbative corrections,
while the hypermultiplet moduli space is exact.
In the dual type-II theory, the dilaton is in a hypermultiplet.
Consequently, the vector moduli space geometry has no corrections and
can be computed at tree level.
The duality map should reproduce all quantum corrections on
the heterotic side.
This has been done in some examples.  In this way, the one-loop
heterotic correction was obtained, which agreed with the heterotic
computation.
Moreover, all instanton effects were obtained this way.
Taking the field theory limit and decoupling gravity, the
Seiberg-Witten solution was verified for N=2 gauge theory.
This procedure gives also a geometric interpretation of the
Seiberg-Witten solution.
A review of these developments can be found in \cite{lere}.

There are also calculations in the type-II N=4 theory that translate
into non-perturbative effects on the heterotic side.
Such an example is the threshold correction to the $R^2$ term in the
effective action of the four dimensional theory, \cite{r2}.
On the type-II side, it can be argued that such a threshold comes from
one
loop only. In the heterotic language, it reproduces the tree level
result as well as non-perturbative corrections due to the Euclidean
heterotic five-brane.

\renewcommand{\theequation}{\thesection.\arabic{equation}}
\section{Outlook\label{out}}
\setcounter{equation}{0}

I hope I managed to  provide a certain flavor of what string theory is.
There is a lot of new structure appearing when compared to standard
field
theory.

Despite the many miraculous characteristics of string theory,
there are some major unresolved problems.
The most important in my opinion is to make contact with the real world
and more concretely to pin down the mechanism of supersymmetry breaking
and stability of the vacuum in that case.
Recent advances in our non-perturbative understanding of the theory
could help in this direction.

Also, the recent non-perturbative advances seem to require other
extended objects apart from strings.
This makes the following question resurface: What is string theory?
A complete formulation, which would include the required extended
objects is still lacking.

I think this  is an exciting period, because we seem to be at the verge
of understanding some of the mysteries of string theory and plausibly
of the high-energy real world.

\newpage
\addcontentsline{toc}{section}{Acknowledgments}
\section*{Acknowledgments}

I would like to thank A. Van Proeyen and W. Troost for giving me the
opportunity to present these lectures at the University of Leuven
and for their warm hospitality and support.
I am  indebted to Chris Van Den Broeck and Martijn Derix for
keeping notes and preparing part of the \LaTeX -file,
and Pieter Jan Desmet for additional corrections.
I would also like to thank F. Zwirner and the theory group at the
University of Padova, where part of these lectures were presented, for
their hospitality.
I would like to thank L. Alvarez-Gaum\'e, N. Obers, B. Pioline
and especially W. Troost for a critical reading of the manuscript.
Last but not least I would like to thank all colleagues who enlightened
me in the course of discussions.

\newpage
\addcontentsline{toc}{subsection}{Appendix A: Theta functions}
\section*{Appendix A: Theta functions}
\renewcommand{\theequation}{A.\arabic{equation}}
\setcounter{equation}{0}

\centerline{\bf Definition}

\be
\th[^a_b](v|\t)=\sum_{n\in Z}q^{{1\over 2}\left(n-{a\over
2}\right)^2}
e^{2\pi i\left(v-{b\over 2}\right)\left(n-{a\over 2}\right)}
\,,\label{t1}\ee
where $a,b$ are real and $q=e^{2\pi i\t}$.

\vskip .5cm
\centerline{\bf Periodicity properties}

\be
\th[^{a+2}_{\phantom{+}b}](v|\t)=\th[^a_b](v|\t)\;\;\;,\;\;\;
\th[^{\phantom{+}a}_{b+2}](v|\t)=e^{i\pi a}\th[^a_b](v|\t)
\,,\label{t2}\ee
\be
\th[^{-a}_{-b}](v|\t)=\th[^{a}_{b}](-v|\t)\;\;\;,\;\;\;
\th[^a_b](-v|\t)=
e^{i\pi ab}\th[^a_b](v|\t) ~~~(a,b\in Z)
\,.\label{t3}\ee

In the usual Jacobi/Erderlyi  notation we have $\th_1=\th[^1_1]$,
$\th_2=
\th[^1_0]$, $\th_3=\th[^0_0]$, $\th_4=\th[^0_1]$.

\vskip .5cm
\centerline{\bf Behaviour under modular transformations}

\be
\th[^a_b](v|\t+1)=e^{-{i\pi\over
4}a(a-2)}~\th[^{\phantom{a+}a}_{a+b-1}](v|\t)
\,,\label{t4}\ee
\be
\th[^a_b]\left({v\over \t}|-{1\over \t}\right)=\sqrt{-i\t}~
e^{{i\pi\over 2}ab+i\pi{v^2\over \t}}~\th[^{\phantom{*}b}_{-a}](v|\t)
\,.\label{t5}\ee

\vskip .5cm
\centerline{\bf Product formulae}

\be
\th_1(v|\t)=2q^{1\over 8}\sin[\pi v]\prod_{n=1}^{\infty}
(1-q^n)(1-q^ne^{2\pi iv})(1-q^ne^{-2\pi iv})
\,,\label{t6}\ee
\be
\th_2(v|\t)=2q^{1\over 8}\cos[\pi v]\prod_{n=1}^{\infty}
(1-q^n)(1+q^ne^{2\pi iv})(1+q^ne^{-2\pi iv})
\,,\label{t7}\ee\be
\th_3(v|\t)=\prod_{n=1}^{\infty}
(1-q^n)(1+q^{n-1/2}e^{2\pi iv})(1+q^{n-1/2}e^{-2\pi iv})
\,,\label{t8}\ee
\be
\th_4(v|\t)=\prod_{n=1}^{\infty}
(1-q^n)(1-q^{n-1/2}e^{2\pi iv})(1-q^{n-1/2}e^{-2\pi iv})
\,.\label{t9}\ee

Define also the Dedekind $\eta$-function:
\be
\eta(\t)=q^{1\over 24}\prod_{n=1}^{\infty}(1-q^n)
\,.\label{t10}\ee

It is related to the $v$ derivative of $\th_1$:

\be
{\partial\over \partial v}\th_1(v)|_{v=0}\equiv \th_1'=2\pi
{}~\eta^3(\t)
\label{t11}\ee
and satisfies

\be
\eta\left(-{1\over \t}\right)=~\sqrt{-i\t}~\eta(\t)
\,.\label{tt1}\ee

\vskip .5cm
\centerline{\bf v-periodicity formula}

\be
\th[^a_b]\left(v+{\epsilon_1\over 2}\t+{\epsilon_2\over 2}|\t\right)=
e^{-{i\pi\t\over 4}\epsilon_1^2-{i\pi\epsilon_1 \over 2}(2v-b)-
{i\pi \over 2}\epsilon_1\epsilon_2}
{}~\th[^{a-\epsilon_1}_{b-\epsilon_2}](v|\t)
\,.\label{t12}\ee

\vskip .5cm
\centerline{\bf Useful identities}

\be
\th_2(0|\t)\th_3(0|\t)\th_4(0|\t)=2~\eta^3
\,,\label{t13}\ee

\be
\th_2^4(v|\t)-\th_1^4(v|\t)=\th_3^4(v|\t)-\th_4^4(v|\t)
\,,\label{t14}\ee

\vskip .5cm
\centerline{\bf Duplication formulae}

\be
\th_2(2\tau)={1\over \sqrt{2}}\sqrt{\th^2_3(\tau)-\th^2_4(\tau)}
\;\;\;,\;\;\;
\th_3(2\tau)={1\over \sqrt{2}}\sqrt{\th^2_3(\tau)+\th^2_4(\tau)}
\,,\label{u2}\ee
\be
\th_4(2\tau)=\sqrt{\th_3(\tau)\th_4(\tau)}
\;\;\;,\;\;\;
\eta(2\tau)=\sqrt{{\th_2(\tau)~\eta(\tau)\over 2}}
\,.\label{tu14}\ee

\vskip .5cm
\centerline{\bf Jacobi identity}

\be
{1\over
2}\sum_{a,b=0}^1~(-1)^{a+b+ab}~\prod_{i=1}^4\th[^a_b](v_i)
=-\prod_{i=1}^4
{}~\th_1(v_i')
\,,\label{tt14}\ee
where
\be
v_1'={1\over 2}(-v_1+v_2+v_3+v_4)\;\;\;,\;\;\;v_2'={1\over
2}(v_1-v_2+v_3+v_4) \,,
\ee
\be
v_3'={1\over 2}(v_1+v_2-v_3+v_4)\;\;\;,\;\;\;v_4'={1\over
2}(v_1+v_2+v_3-v_4)\,.
\ee
Using (\ref{tt14}) and (\ref{t12}) we can show that
\be
{1\over 2}\sum_{a,b=0}^1(-1)^{a+b+ab}\prod_{i=1}^4
\th[^{a+h_i}_{b+g_i}](v_i)=-\prod_{i=1}^4
\th[^{1-h_i}_{1-g_i}](v_i')
\,.\label{t15}\ee
The Jacobi identity (\ref{t15}) is valid only when $\sum_i h_i=\sum_i
g_i=0$.
There is also a similar (IIA) identity
\be
{1\over
2}\sum_{a,b=0}^1~(-1)^{a+b}~\prod_{i=1}^4\th[^a_b](v_i)=
-\prod_{i=1}^4
{}~\th_1(v_i')+\prod_{i=1}^4\th_1(v_i)
\label{ttt14}\ee
and
\be
{1\over 2}\sum_{a,b=0}^1(-1)^{a+b}\prod_{i=1}^4
\th[^{a+h_i}_{b+g_i}](v_i)=-\prod_{i=1}^4
\th[^{1-h_i}_{1-g_i}](v_i')+\prod_{i=1}^4\th[^{1+h_i}_{1+g_i}](v_i)
\,.\label{tt15}\ee

The $\th$-functions satisfy the following heat equation

\be
\left[{1\over (2\pi i)^2}{\partial^2\over \partial v^2}-{1\over
i\pi}{
\partial\over \partial\t}\right]\th[^a_b](v|\t)=0
\label{t16}\ee
as well as
\be
{1\over 4\pi i}{\th_2''\over
\th_2}=\partial_{\tau}\log\th_2={i\pi\over 12}
\left(E_2+\th_3^4+\th_4^4\right)
\,,\ee
\be
{1\over 4\pi i}{\th_3''\over
\th_3}=\partial_{\tau}\log\th_3={i\pi\over 12}
\left(E_2+\th_2^4-\th_4^4\right)
\,,\ee
\be
{1\over 4\pi i}{\th_4''\over
\th_4}=\partial_{\tau}\log\th_4={i\pi\over 12}
\left(E_2-\th_2^4-\th_3^4\right)
\,,\ee
where the function $E_{2}$ is defined in
(\ref{E2}).

\vskip .5cm

\centerline{\bf The Weierstrass function}

\be
{\cal P}(z)=4\pi i\pd_{\t}\log~\eta(\tau)-\pd^2_z\log\th_1(z)={1\over
z^2}+{\cal O}(z^2)
\label{ww}\ee
is even and is the unique analytic function on the torus with a double
pole at zero.
\be
{\cal P}(-z)={\cal P}(z)\;\;\;,\;\;\;{\cal P}(z+1)={\cal P}(z+\t)={\cal
P}(z)
\,,\ee
\be
{\cal P}(z,\t+1)={\cal P}(z,\t)\;\;\;,\;\;\;{\cal P}\left({z\over \t},
-{1\over \t}\right)=\t^2~{\cal P}(z,\t)
\,.\ee
We will need the following torus integrals
\be
\int {d^2 z\over \t_2}~{\cal P}(z,\tau)=4\pi
i\pd_{\t}\log(\sqrt{\t_2}\eta)
\,,
\ee
\be
\int  {d^2 z\over \t_2}~|{\cal P}(z,\tau)|^2=|4\pi
i\pd_{\t}\log(\sqrt{\t_2}\eta)|^2
\,,\label{intw}\ee
\be
\int {d^2 z\over \t_2}~\overline{{\cal P}}(\bar z,\bar
\tau)\left[\pd_z\log\th_1(z)+2\pi i
{Imz\over \t_2}\right]^2=4\pi i \pd_{\t}\log(\eta\sqrt{\t_2})
\,,\label{intw2}\ee
\be
\int{d^2 z\over \t_2}~\pd^2_z~\log\th_1(z)=-{\pi\over \tau_2}
\,.\label{At}\ee
\vskip .8cm
\centerline{\bf Poisson Resumation}

Consider a function $f(x)$ and its Fourier transform $\tilde f$
defined as
\be
\tilde f(k)\equiv {1\over 2\pi}\int_{-\infty}^{+\infty}
f(x)e^{i k x}dx
\,.\label{t17}\ee
Then:
\be
\sum_{n\in Z}~f(2\pi n)=\sum_{n\in Z}~\tilde f(n)
\,.\label{t18}\ee

Choosing as $f$ an appropriate Gaussian function
we obtain:
\be
\sum_{n\in Z}e^{-\pi a n^2+\pi b n}={1\over \sqrt{a}}\sum_{n\in Z}
e^{-{\pi\over a}\left(n+i{b\over 2}\right)^2}
\,,\label{t19}\ee
\be
\sum_{n\in Z}~n~e^{-\pi a n^2+\pi b n}=-{i\over \sqrt{a}}\sum_{n\in
Z}
{}~{\left(n+i{b \over 2}\right)\over a}~e^{-{\pi\over
a}\left(n+i{b\over 2}\right)^2}
\,,\label{tt19}\ee
\be
\sum_{n\in Z}~n^2~e^{-\pi a n^2+\pi b n}={1\over \sqrt{a}}\sum_{n\in
Z}
{}~\left[{1\over 2\pi a}-{\left(n+i{b \over 2}\right)^2\over
a^2}\right]~e^{-{\pi\over a}\left(n+i{b\over 2}\right)^2}
\,.\label{tt20}\ee

The multidimensional generalization is (repeated indices are summed
over):

\be
\sum_{m_i\in Z}e^{-\pi m_im_jA_{ij}+\pi B_im_i}=
({\rm det}~A)^{-{1\over 2}}\sum_{m_i\in Z}e^{-\pi(m_k+iB_k/2)
(A^{-1})_{kl}(m_l+iB_l/2)}
\,.\label{t20}\ee

\addcontentsline{toc}{subsection}{Appendix B: Toroidal lattice sums}
\section*{Appendix B: Toroidal lattice sums}
\renewcommand{\theequation}{B.\arabic{equation}}
\setcounter{equation}{0}

We will consider here asymmetric lattice sums corresponding to $p$
left-moving bosons and $q$ right-moving ones.
To have good modular properties $p-q$ should be a multiple of eight.
We will consider here the case $q-p=16$ relevant for the heterotic
string. Other cases can be easily worked out using the same methods.

We will write the genus-one action using p bosons and 16 complex
right-moving fermions, $\psi^I(\bar z),\bar\psi^I(\bar z)$:
\be
S_{p,q}={1\over 4\pi}\int d^2\s\sqrt{{\rm det}~g}g^{ab}G_{\a\b}\pd_a
X^{\a}
\pd_b X^{\b}
+{1\over 4\pi}\int d^2\s \epsilon^{ab}B_{\a\b}\pd_a X^{\a} \pd_b
X^{\b}
+
\label{B1}\ee
$$+{1\over 4\pi}\int
d^2\s\sqrt{{\rm det}~g}~\sum_I~\psi^{I}(\bar\nabla+Y^I_{\a}(\bar\nabla
X^{\a})
 \bar\psi^{I}
\,,$$
where the torus metric is given in (\ref{117}).
We will take the fermions to be all periodic or antiperiodic.
A direct evaluation of the torus path integral along the line of
section
\ref{cscalars} gives
\bea
Z_{p,p+16}(G,B,Y)&=&{\sqrt{{\rm det~G}}\over
\t_2^{p/2}\eta^p\bar\eta^{p+16}} \times \nn\\
&&\times
\sum_{m^{\a},n^{\a}\in Z}\exp\left[-{\pi\over \t_2}(G+B)_{\a\b}
(m^{\a}+\t n^{\a})(m^{\b}+\bar\t n^{\b})\right]\times
\nn\\
&&\times {1\over
2}\sum_{a,b=0}^1~\prod_{I=1}^{16}~e^{i\pi(m^{\a}Y^I_{\a}
Y^I_{\b}n^{\b}-b~n^{\a}Y^I_{\a})}~\bar\th\left[^{a-2n^{\a}Y^I_{\a}}_
{b-2m^{\b}Y^I_{\b}}\right]
\nn\eea
\bea
={\sqrt{{\rm det~G}}\over
\t_2^{p/2}\eta^p\bar\eta^{p+16}}
\sum_{m^{\a},n^{\a}\in Z}\exp\left[-{\pi\over \t_2}(G+B)_{\a\b}
(m^{\a}+\t n^{\a})(m^{\b}+\bar\t n^{\b})\right]\times
\nn\eea
$$
\times \exp\left[-i\pi\sum_I n^{\a}\left(m^{\b}+\bar\t
n^{\b}\right)Y^I_{\a}Y^I_{\b}\right]~{1\over
2}\sum_{a,b=0}^1\prod_{I=1}^{16}
\bar\th[^a_b](Y^I_{\g}(m^{\g}+\bar\t n^{\g})|\bar\t) \,.
$$
Under modular transformations
\be
\t\to\t+1\;\;\;,\;\;\;Z_{p,p+16}\to e^{4\pi i/3}~Z_{p,p+16}
\,,\label{B3}\ee
while it is invariant under $\t\to -1/\t$.

Performing a Poisson resummation in $m^{\a}$ we can cast it in
Hamiltonian
form $Z_{p,p+16}=\Gamma_{p,p+16}/\eta^p\bar\eta^{p+16}$ with
\be
\Gamma_{p,p+16}(G,B,Y)=\sum_{m_{\a},n_{\a},Q_I}~q^{P_L^2/2}~\bar
q^{P_R^2/2}
\,,\label{B4}\ee
where $m_{\a},n^{\a}$ take arbitrary integer values, while $Q_I$ take
values in the even self-dual lattice $\rm O(32)/Z_2$.
To be concrete, the numbers $Q_I$ are either all integer or all
half-integer
satisfying in both cases the constraint $\sum_I~Q_I=$~even.
We will introduce the $(2p+16)\times (2p+16)$ symmetric matrix
\be
M=\left(\matrix{G^{-1}& G^{-1}C &G^{-1}Y^{t}\cr
C^{t}G^{-1}&G+C^{t}G^{-1}C+Y^{t}Y&C^{t}G^{-1}Y^{t}+Y^{t}\cr
YG^{-1}&YG^{-1}C+Y&{\bf 1}_{16}+YG^{-1}Y^{t}\cr}\right)
\,,\label{B5}\ee
where ${\bf 1}_{16}$ is the sixteen-dimensional unit matrix and
\be
C_{\a\b}=B_{\a\b}-{1\over 2}Y^{I}_{\a}Y^{I}_{\b}
\,.\label{B6}\ee
Introduce the O(p,p+16) invariant metric
\be
L=\left(\matrix{0&{\bf 1}_{p}&0\cr
{\bf 1}_{p}&0&0\cr
0&0&{\bf 1}_{16}\cr}\right)
\,.\label{B66}\ee
Then the matrix M satisfies:
\be
M^{T}LM=MLM=L\;\;\;,\;\;\;M^{-1}=LML
\,.\label{B7}\ee
Thus, $M\in $O(p,p+16).
In terms of $M$ the conformal weights are given by
\be
{1\over 2}P_L^2={1\over 4}(m^{\a},n_{\a},Q_{I})\cdot (M-L)\cdot\left(
\matrix{m^{\a}\cr n_{\a}\cr Q_{I}\cr}\right)
\,,\label{B8}\ee
\be
{1\over 2}P_R^2={1\over 4}(m^{\a},n_{\a},Q_{I})\cdot (M+L)\cdot\left(
\matrix{m^{\a}\cr n_{\a}\cr Q_{I}\cr}\right)
\,.\label{B99}\ee
The spin
\be
{1\over 2}P_R^2-{1\over 2}P_L^2=m^{\a}n_{\a}-{1\over 2}Q_IQ_I
\label{B9}\ee
is an integer.
When $Y=0$ the lattice sum factorizes
\be
\Gamma_{p,p+16}(G,B,Y=0)=\Gamma_{p,p}(G,B)~\bar \Gamma_{\rm O(32)/Z_2}
\,.\label{B10}\ee
It can be shown \cite{g} that for some special (non-zero) values
$\tilde Y^I_{\a}$ the lattice sum factorizes into the $(p,p)$ toroidal
sum and the lattice sum of $\rm E_8\times E_8$
\be\Gamma_{p,p+16}(G,B,Y=\tilde Y)=\Gamma_{p,p}(G',B')~\bar
\Gamma_{E_8\times E_8}
\,.\label{B11}\ee
Thus, we can continuously interpolate in $\Gamma_{p,p+16}$ between the
O(32) and E$_8\times $E$_8$ symmetric points.

Finally, the duality
group here
is O(p,p+16,$\Z)$.
An element of O(p,p+16,$\Z)$ is an integer-valued O(p,p+16) matrix.
Consider such a matrix  $\Omega$.
It satisfies $\Omega^T~L~\Omega=L$.
The lattice sum is invariant under the $T$-duality transformation
\be
\left(
\matrix{m^{\a}\cr n_{\a}\cr Q_{I}\cr}\right)\to
\Omega\cdot\left(\matrix{m^{\a}\cr n_{\a}\cr Q_{I}\cr}\right)
\;\;\;,\;\;\;M\to \Omega~M~\Omega^T
\,.\label{B12}\ee

In what follows, we will describe translation orbifold blocks for
toroidal CFTs.
Start from the $(d,d+16)$ lattice.
We will use the notation $\l=(m^{\a},n_{\a},Q_I)$ for a lattice
vector
with its O(p,p+16) inner product, which gives the invariant square
$\l^2=2m^{\a}n_{\a}-Q_{I}Q_{I}\in 2\Z$.
Perform a $Z_N$ translation by $\e/N\notin L$, where $\e$ is a
lattice vector.
The generalization of one-dimensional orbifold blocks (\ref{238}) is
straightforward:
\be
Z^N_{d,d+16}(\e)[^h_g]={\Gamma_{p,p+16}(\e)[^h_g]\over
\eta^p\bar\eta^{p+16}}={\sum_{\l\in L+\e{h\over N}}~e^{2\pi
ig~\e\cdot \l\over N}~q^{p^2_L/2}~\bar q^{p_{R}^2/2}\over
\eta^p\bar\eta^{p+16}}
\,\label{B13}\ee
where $h,g=0,1,\ldots,N-1$.
It has the following properties
\be
Z^N(-\e)[^h_g]=Z^N(\e)[^h_g]\;\;\;,\;\;\;Z^N(\e)[^{-h}_{-g}]=
Z^N(\e)[^h_g]
\,,\label{B14}\ee
\be
Z^N(\e)[^{h+1}_g]=\exp\left[-{i\pi g\e^2\over N}\right]Z^N(\e)[^h_g]
\;\;\;,\;\;\;Z^N(\e)[^h_{g+1}]=Z^N(\e)[^h_g]
\,,\label{B15}\ee
\be
Z^N(\e+N\e')[^h_g]=\exp\left[{2\pi i~ gh~\e\cdot\e'\over N}\right]
Z^N(\e)[^h_g]
\,.\label{B16}\ee
Under modular transformations
\be
\tau\to\tau+1\;\;\;:\;\;\;Z^N(\e)[^h_g]\to\exp\left[{4\pi i\over
3}+{i\pi h^2~\e^2\over N^2}\right]Z^N(\e)[^h_{h+g}]
\,,\label{B17}\ee
\be
\tau\to -{1\over \t}\;\;\;:\;\;\;Z^N(\e)[^h_g]\to \exp\left[-{2\pi
i~hg~\e^2\over N^2}\right]Z^N(\e)[^g_{-h}]
\,.\label{B18}\ee
Under O(p,p+16,$\Z$) duality transformations it transforms as
\be
Z^N(\e,\Omega M \Omega^T)[^h_g]=Z^N(\Omega\cdot \e,M)[^h_g]
\,,\label{B19}\ee
where $\Omega\in $O(p,p+16,$\Z$) and $M$ is the moduli matrix
(\ref{B5}).
The unbroken duality group consists of the subgroup of
O(p,p+16,$\Z$) transformations that preserve $\e$ modulo $N^2$ times
a
lattice vector.

\addcontentsline{toc}{subsection}{Appendix C: Toroidal Kaluza-Klein
reduction}
\section*{Appendix C: Toroidal Kaluza-Klein reduction}
\renewcommand{\theequation}{C.\arabic{equation}}
\setcounter{equation}{0}

In this appendix we will describe the Kaluza-Klein ansatz for
toroidal
dimensional reduction from 10 to $D<10$ dimensions.
A more detailed discussion can be found in \cite{ms}.
Hatted fields will denote the $(10-D)$-dimensional  fields and
similarly for the
indices.
Greek indices from the beginning of the alphabet will denote the
$10-D$
internal (compact) dimensions. Unhatted Greek indices from the middle
of
the alphabet will denote the $D$ non-compact dimensions.

The  standard form for the  10-bein is
\be
\hat e^{\hat r}_{\mh}=\left(\matrix{e^{r}_{\m}&
A^{\b}_{\m}E^{a}_{\b}\cr
0& E^{a}_{\a}\cr}\right)\;\;\;,\;\;\;\hat e^{\mh}_{\hat
r}=\left(\matrix{
e^{\m}_{r}& -e^{\n}_{r}A^{\a}_{\n}\cr
0&E^{\a}_{a}\cr}\right)
\,.\label{C1}\ee
For the metric we have
\be
\hat G_{\mh\nh}=\left(\matrix{g_{\m\n}+A_{\m}^{\a}G_{\a\b}A^{\b}_{\n}
& G_{\a\b}A^{\b}_{\m}\cr
G_{\a\b}A^{\b}_{\n}& G_{\a\b}\cr}\right)\;\;,\;\;\hat
G^{\mh\nh}=\left(\matrix{g^{\m\n}&-A^{\m\a}
\cr -A^{\n\a}&G^{\a\b}+A^{\a}_{\rho}A^{\b,\rho}\cr}\right)
\,.\label{C2}\ee
Then the part of the action containing the Hilbert term as well as
the dilaton
becomes
\bea
\a'^{D-2}S_{D}^{\rm heterotic}=\int d^D x
\sqrt{-{\rm det}~g}~e^{-\phi}\left[R+\pd_{\m}\phi\pd^{\m}\phi
+{1\over 4}\pd_{\m}G_{\a\b}\pd^{\m}G^{\a\b}+ \right.
\nn\\
\left.- {1\over 4}G_{\a\b}{F^{A}_{\m\n}}^{\a}
F_A^{\b,\mu\nu}\right]
\,,\label{C3}\eea
where
\be
\phi=\hat \Phi-{1\over 2}\log ({\rm det} G_{\a\b})
\,,\label{C4}\ee
\be
{F^A_{\m\n}}^{\a}=\pd_{\m}A^{\a}_{\n}-\pd_{\n}A^{\a}_{\m}
\,.\label{C5}\ee
We will now turn to the antisymmetric tensor part of the action:
\[
-{1\over 12}\int d^{10} x\sqrt{-{\rm det}~ \hat G}e^{-\hat \Phi}\hat
H^{\mh\nh\rh}\hat H_{\mh\nh\rh} =
\]
\be = -\int d^D x\sqrt{-{\rm det}~ g}~
e^{-\phi}\left[{1\over 4}H_{\m\a\b}H^{\m\a\b}
\label{C6}
+ {1\over 4}H_{\m\n\a}H^{\m\n\a}+{1\over
12}H_{\m\n\rho}H^{\m\n\rho}\right]\;,
\ee
where we have used $H_{\a\b\g}=0$ and
\be
H_{\m\a\b}=e^{r}_{\m}\hat e^{\mh}_{\hat r}\hat H_{\mh\a\b}=\hat
H_{\m\a\b}
\,,\label{C7}\ee
\be
H_{\m\n\a}=e^{r}_{\m}e^{s}_{\n}\hat e^{\mh}_{r}\hat e^{\nh}_{s}\hat
H_{\mh\nh\a}=\hat H_{\m\n\a}-A_{\m}^{\b}\hat
H_{\n\a\b}+A_{\n}^{\b}\hat H_{\m\a\b}
\,,\label{C8}\ee
\be
H_{\m\n\rho}= e^{r}_{\m}e^{s}_{\n}e^{t}_{\rho}\hat e^{\mh}_{r}\hat
e^{\nh}_{s}
\hat e^{\rh}_{t}\hat H_{\mh\nh\rh}=
\hat H_{\m\n\rho}+\left[-A_{\m}^{\a}\hat
H_{\a\n\rho}+A_{\m}^{\a}A_{\n}^{\b}\hat H_{\a\b\rho}+{\rm
cyclic}\right]
\,.\label{C9}\ee

Similarly,
\bea
\int d^{10} x\sqrt{-{\rm det}~ \hat G}~e^{-\hat
\Phi}\sum_{I=1}^{16}\hat
F^{I}_{\mh\nh}F^{I,\mh\nh}= \quad\quad\quad\quad\quad\quad\quad\quad\quad\quad\quad\quad\nn\\
= \int d^D x\sqrt{-{\rm det}~ g}~
e^{-\phi}\sum_{I=1}^{16}\left[\tilde F^{I}_{\m\n}\tilde
F^{I,\m\n}+2\tilde F^{I}_{\m\a}\tilde F^{I,\m\a}
\right]
\,,\label{C10}\eea
with
\be
Y^{I}_{\a}=\hat A^{I}_{\a}\;\;\;,\;\;\;A^{I}_{\m}=\hat
A^{I}_{\m}-Y^{I}_{\a}A^{a}_{\m}
\;\;\;,\;\;\;
\tilde F^{I}_{\m\n}=F^{I}_{\m\n}+Y^{I}_{\a}F^{A,\a}_{\m\n}
\label{C12}\ee
\be
\tilde F^{I}_{\m\a}=\pd_{\m}Y^{I}_{\a}\;\;\;,\;\;\;
F_{\m\n}^{I}=\pd_{\m}A^{I}_{\n}-\pd_{\n}A^{I}_{\m}
\,.\label{C14}\ee

We can now evaluate the D-dimensional antisymmetric tensor pieces
using (\ref{C7})-(\ref{C9}):

\be
\hat H_{\m\a\b}=\pd_{\m}\hat B_{\a\b}+{1\over
2}\sum_I\left[Y^{I}_{\a}\pd_{\m}
Y^{I}_{\b}-Y^{I}_{\b}\pd_{\m}Y^{I}_{\a}\right]
\,.\label{C15}\ee
Introducing
\be
C_{\a\b}\equiv\hat B_{\a\b}-{1\over 2}\sum_I Y^{I}_{\a}Y^{J}_{\b}
\,,\label{C16}\ee
we obtain from (\ref{C6})
\be
H_{\m\a\b}=\pd_{\m}C_{\a\b}+\sum_I Y^{I}_{\a}\pd_{\m}Y^{I}_{\b}
\,.\label{C166}\ee

Also
\be
\hat H_{\m\n\a}=
\pd_{\m}\hat B_{\n\a}-\pd_{\n}\hat B_{\m\a}+{1\over 2}\sum_I \left[
\hat A_{\n}^{I}\pd_{\m}Y_{\a}^{I}-\hat
A_{\m}^{I}\pd_{\n}Y^{I}_{\a}-Y^{I}_{\a}
\hat F^{I}_{\m\n}\right]
\,.\label{C17}\ee
Define
\be
B_{\m,\a}\equiv\hat B_{\m\a}+B_{\a\b}A_{\m}^{\b}+{1\over
2}\sum_I Y^{I}_{\a}A^{I}_{\m}
\,,\label{C18}\ee
\be
F^{B}_{\a,\m\n}=\pd_{\m}B_{\a,\n}-\pd_{\n}B_{\a,\m}
\,,\label{C19}\ee
we obtain from (\ref{C7})
\be
H_{\m\n\a}=F_{\a\m\n}^{B}-C_{\a\b}F^{A,\b}_{\m\n}-\sum_I Y^{I}_{\a}
F_{\m\n}^{I}
\,.\label{C20}\ee
Finally,
\be
B_{\m\n}=\hat B_{\m\n}+{1\over
2}\left[A^{\a}_{\m}B_{\n\a}+\sum_I A^{I}_{\m}
A^{\a}_{\n}Y^{I}_{\a}-(\m\leftrightarrow
\n)\right]-A^{\a}_{\m}A^{\b}_{\n}B_{\a\b}
\label{C21}\ee
and
\be
H_{\m\n\rho}=\pd_{\m}B_{\n\rho}-{1\over
2}\left[B_{\m\a}F^{A,\a}_{\n\rho}+A^{\a}_{\m}
F^{B}_{a,\n\rho}+\sum_I A^{I}_{\mu}F^{I}_{\n\rho}\right]+{\rm cyclic}
\label{C22}
\ee
$$\equiv \pd_{\m}B_{\n\rho}-{1\over 2}
L_{ij}A^{i}_{\m}F^{j}_{\n\rho}+{\rm cyclic}
$$
where we combined the $36-2D$ gauge fields
$A^{\a}_{\m},B_{\a,\mu},A^I_{\mu}$ into the uniform notation
$A^i_{\m}$, $i=1,2,\ldots,36-2D$ and
$L_{ij}$ is the O(10-D,26-D)-invariant metric (\ref{B66}).
We can combine the scalars $G_{\a\b},B_{\a\b},Y^I_{\a}$ into the
matrix $M$ given in (\ref{B5}).
Putting everything together,  the D-dimensional action becomes
\be
S^{\rm heterotic}_{D}=\int ~d^Dx~\sqrt{-{\rm
det}~g}e^{-\phi}\left[R+\pd^{\m}\phi\pd_{\m}\phi
-{1\over 12}\tilde H^{\m\n\rho}\tilde H_{\m\n\rho}-\right.
\label{C23}\ee
$$\left.-{1\over 4}(M^{-1})_{ij}
F^{i}_{\m\n}F^{j\m\n}+{1\over 8}{\rm Tr}(\pd_{\m} M\pd^{\m}
M^{-1})\right]\;.
$$

We will also consider here the KK reduction of a three-index
antisymmetric tensor $C_{\m\n\rho}$. Such a tensor appears in type-II
string theory and eleven-dimensional supergravity.
The action for such a tensor is
\be
S_{C}=-{1\over 2\cdot 4!}\int d^d x\sqrt{-G}~\hat F^2\;\;,
\label{C24}\ee
where
\be
\hat F_{\m\n\rho\s}=\partial_{\mu}\hat C_{\n\rho\s}-\partial_{\s}\hat
C_{\m\n\rho}+
\partial_{\rho}
\hat C_{\s\m\n}-\partial_{\n}\hat C_{\rho\s\m}\;\;.
\label{C25}\ee

We define the lower-dimensional components as
\be
C_{\a\b\g}=\hat C_{\a\b\g}\;\;\;,\;\;\;C_{\m\a\b}=\hat
C_{\m\a\b}-C_{\a\b\g}A^{\g}_{\m}\;\;,
\label{C26}\ee
\be
C_{\m\n\a}=\hat C_{\m\n\a}+\hat C_{\m\a\b}A^{\b}_{\n}-\hat
C_{\n\a\b}A^{\b}_{\m}+C_{\a\b\g}A^{\b}_{\mu}A^{\g}_{\n}\;\;,
\label{C27}\ee
\be
C_{\m\n\rho}=\hat C_{\m\n\rho}+\left(-\hat
C_{\n\rho\a}A^{\a}_{\mu}+\hat C_{\a\b\rho}A^{\a}_{\m}A^{\b}_{\n}+{\rm
cyclic}\right)-C_{\a\b\g}
A^{\a}_{\m}A^{\b}_{\n}A^{\g}_{\rho}\;\;.
\label{C28}\ee
Then,
$$
S_{C}=-{1\over 2\cdot 4!}\int d^D x\sqrt{-g}\sqrt{{\rm
det}G_{\a\b}}\left[
F_{\m\n\rho\s}F^{\m\n\rho\s}+4F_{\m\n\rho\a}F^{\m\n\rho\a}+
\right.$$\be
\left.+6
F_{\m\n\a\b}F^{\m\n\a\b}+4F_{\m\a\b\g}F^{\m\a\b\g}\right]\;,
\label{C29}\ee
where
\be
F_{\m\a\b\g}=\partial_{\m}C_{\a\b\g}\;\;\;,\;\;\;F_{\m\n\a\b}=
\partial_{\mu}C_{\n\a\b}-\partial_{\n}C_{\m\a\b}+
C_{\a\b\g}F^{\g}_{\m\n}\;,
\label{C30}\ee
\be
F_{\m\n\rho\a}=\partial_{\m}C_{\n\rho\a}+C_{\m\a\b}F^{\b}_{\n\rho}+
{\rm cyclic}\;\;,
\label{C31}\ee
\be
F_{\m\n\rho\s}=(\partial_{\m}C_{\n\rho\s}+{\rm
3~~perm.})+(C_{\rho\s\a}F^{\a}_{\m\n}+{\rm 5~~perm.})\;\;.
\label{C32}\ee

\addcontentsline{toc}{subsection}{Appendix D: N=1,2,4, D=4 supergravity
coupled to matter}
\section*{Appendix D: N=1,2,4, D=4 supergravity coupled to matter}
\renewcommand{\theequation}{D.\arabic{equation}}
\setcounter{equation}{0}

We will review here some facts about four-dimensional  supergravity
theories coupled to matter.

$\bullet$ \underline{N=1 supergravity}. Apart from the supergravity
multiplet,
we can have vector multiplets containing the vectors and their
Majorana gaugini, and chiral multiplets containing a complex scalar
and a Weyl spinor.
There is also the linear multiplet containing an antisymmetric
tensor, a scalar
and a Weyl fermion. However, this can be dualized into a chiral
multiplet,
but with an accompanying Peccei-Quinn symmetry.
\footnote{A Peccei-Quinn symmetry
is a translational symmetry of a scalar field, $\phi\to
\phi+$constant.}
The bosonic Lagrangian can be written as follows
\be
{\cal L}_{N=1}=-{1\over 2\kappa^2}R+G_{i\bar j}D_{\m}\phi^i
D^{\m}\bar \phi^{\bar j}+V(\phi,\bar \phi)+\sum_a{1\over
4g^2_{a}}[F_{\m\n}F^{\m\n}]_a
+{\theta_a\over 4}[F_{\m\n}\tilde F^{\m\n}]_a
\,.\label{445}\ee
The gauge group $G=\prod_a G_a$ is a product of simple or U(1)
factors;
$\phi^i$ are the complex scalars of the chiral multiplets, which in
general transform in some representation of the gauge group; $D_{\m}$
are the associated
covariant derivatives.

Supersymmetry requires the manifold of scalars to be K\"ahlerian,
\be
G_{i\bar j}=\pd _i \pd_{\bar j}K(\phi,\bar \phi)
\,.\label{446}\ee
The gauge couplings and $\theta$-angles must depend on the moduli
via a holomorphic function
\be
{1\over g_a^2}={\rm Re}~f_a(\phi)\;\;\;,\;\;\;\theta_a=-{\rm
Im}~f_a(\phi)
\,.\label{447}\ee
The holomorphic function $f_a$ must be gauge-invariant.
The scalar potential $V$ is also determined by a holomorphic
function.\footnote{
We will ignore D-terms.} the superpotential $W(\phi)$:
\be
V(\phi,\bar\phi)=e^{\kappa^2~K}\left( D_iWG^{i\bar i}\bar D_{\bar
i}\bar W
-3\kappa^2 |W|^2\right)
\,,\label{448}\ee
where
\be
D_i~W={\pd W\over \pd \phi^i}+\kappa^2{\pd K\over \pd\phi^i}W
\,.\label{449}\ee
Note that the potential of N=1 supergravity is not positive-definite.

There is an overall redundancy in the data ($K,f_a,W$).
The action is invariant under K\"ahler transformations
\be
K\to K+\Lambda(\phi)+\bar \Lambda(\bar \phi)\;\;\;,\;\;\;W\to
W~e^{-\Lambda}
\;\;\;,\;\;\;f_a\to f_a
\,.\label{450}\ee
It seems that this redundancy allows one to get rid of the
superpotential.
This is true if it has no singularities. Otherwise one obtains
a singular metric for the scalars. Further information can be found
in \cite{n=1}.

$\bullet$ \underline{N=2 supergravity}.
Apart from the supergravity multiplet we will have a number $N_V$ of
abelian vector multiplets and a number $N_H$ of hypermultiplets.
There is also an extra gauge boson, the graviphoton residing in the
supergravity multiplet.
Picking the gauge to be abelian is without loss of generality since
any non-abelian gauge group can be broken to the maximal abelian
subgroup
by
giving expectation values to the scalar partners of the abelian gauge
bosons.
Denote the graviphoton by $A^0_{\m}$, the rest of the gauge
bosons
by $A^i_{\mu}$, $i=1,2,\dots,N_V$, and the scalar partners of
$A^i_{\m}$
as $T^i,\bar T^i$.
Although the graviphoton does does not have a scalar partner, it is
convenient
to introduce one.
The theory  has a scaling symmetry, which allows us to set this scalar
equal to 1.
We will introduce the complex coordinates $Z^I$,
$I=0,1,2,\dots,N_V$,
which will parametrize the vector moduli space (VMS),~${\cal M}_V$.
The  $4 N_H$ scalars of the hypermultiplets parametrize the
hypermultiplet moduli space ${\cal M}_H$ and supersymmetry requires
this to be a quaternionic manifold.
The geometry of the full scalar manifold is that of a product, ${\cal
M}_V\times {\cal M}_H$.

N=2 supersymmetry implies that the VMS is not just a K\"ahler manifold,
but that it satisfies what is known as special geometry.
Special geometry eventually leads to the property that the full
action
of N=2 supergravity (we exclude hypermultiplets for the moment) can
be written
in terms of one function, which is holomorphic in the VMS coordinates.
This function, which we will denote by $F(Z)$, is called the
prepotential.
It must be a homogeneous function of the coordinates of degree 2:
$Z^I~F_I=2$, where $F_I={\pd F\over \pd Z^I}$.
For example, the K\"ahler potential is
\be
K=-\log\left[i(\bar Z^I~F_I-Z^{I}\bar F_I)\right]
\,,\label{442}\ee
which determines the metric $G_{I\bar J}=\pd_I\pd_{\bar J} K$ of
the kinetic terms of the scalars.
We can  fix the scaling freedom by setting $Z^0=1$, and then
$Z^I=T^I$
are the physical moduli.
The K\"ahler potential becomes
\be
K=-\log\left[2\left(f(T^i)+\bar f(\bar T^i)\right)-(T^i-\bar
T^i)(f_{i}-\bar f_i)\right]
\,,\label{443}\ee
where $f(T^i)=-iF(Z^0=1,Z^i=T^i)$.
The K\"ahler metric $G_{i\bar j}$ has the following property
\be
R_{i\bar j k\bar l}=G_{i\bar j}G_{k\bar l}+G_{i\bar l}G_{k\bar j}
-e^{-2K}W_{ikm}G^{m\bar m}\bar W_{\bar m\bar j\bar l}
\,,\label{444}\ee
where $W_{ijk}=\pd_i\pd_j\pd_k f$.
Since there is no potential, the only part of the bosonic
action left
to be specified is the kinetic terms for the vectors:
\be
{\cal L}^{\rm vectors}=-{1\over
4}\Xi_{IJ}F^I_{\m\n}F^{J,\m\n}-{\theta_{IJ}\over 4}F^I_{\mu\n}\tilde
F^{J,\m\n}
\,,\label{451} \ee
 where
\be
\Xi_{IJ}={i\over 4}[N_{IJ}-\bar
N_{IJ}]\;\;\;,\;\;\;\theta_{IJ}={1\over 4}
[N_{IJ}+\bar N_{IJ}]
\,,\label{452}\ee
\be
N_{IJ}=\bar F_{IJ}+2i{{\rm Im} ~F_{IK}~{\rm Im}~ F_{JL}Z^KZ^L\over
{\rm Im}~F_{MN}Z^MZ^N}
\,.\label{453}\ee
Here we see that the gauge couplings, unlike the N=1 case, are not
harmonic functions of the moduli.

N=2 BPS states\footnote{You will find
definitions and properties in Appendix E.} have masses of the form
\be
M^2_{BPS}={|e_I~Z^I+q^I~F_I|^2\over {\rm Im} (Z^I\bar F_I)}
\,,\label{bp}\ee
where $e_I,q^I$ are the electric and magnetic charges of the state.
Further reading can be found in \cite{n=2}.

$\bullet$ \underline{N=4 supergravity}. As we mentioned previously, in
the supergravity multiplet there is a complex scalar and
six graviphotons. In general we can also have  $N_{V}$ vector
multiplets containing six scalars and a vector each.
The local geometry of the scalar manifold is completely fixed to be
SL(2)/U(1)
$\rm \otimes O(6,6+N_{V})/O(6)\times O(6+N_{V})$.
The first factor is associated with the supergravity complex scalar
$S$,
while the second, with the vector multiplet scalars.
The bosonic action was given in (\ref{359}).
The BPS mass-formula is
\be
M^2_{BPS}={1\over 4 {\rm Im}~S} (\a^t+S\b^t)  M_{+} (\a+\bar
S\b)+ {1\over 2}\;\sqrt{(\a^t M_+\a)(\b^t M_+\b)-(\a^t M_+\b)^2}
\,,\label{457}\ee
where $\a,\b$ are integer-valued, $(12+N_V)$-dimensional vectors of
electric
and magnetic charges, $M$ is the moduli matrix in (\ref{B5})
and $M_+=M+L_{6,6+N_V}$.

$\bullet$ \underline{$4<N\leq 8$ supergravity}. There are no massless
matter
multiplets, and the Lagrangian is completely fixed by
supersymmetry.
We will not discuss any further detail, however.

\addcontentsline{toc}{subsection}{Appendix E: BPS multiplets and
helicity supertrace formulae}
\section*{Appendix E: BPS multiplets and\\ helicity supertrace
formulae\label{BPS}}
\renewcommand{\theequation}{E.\arabic{equation}}
\setcounter{equation}{0}

BPS states are important probes of non-perturbative physics in
theories with extended ($N\geq 2$) supersymmetry.

BPS states are special for the following reasons:

$\bullet$ Due to their relation with central charges, and although
they are massive, they form multiplets under extended SUSY which are
shorter than the generic massive multiplet.
Their mass is given in terms of their charges and moduli expectation
values.

$\bullet$ At generic points in moduli space  they are stable
because of energy and charge conservation.

$\bullet$ Their mass-formula is supposed to be exact if one uses
renormalized values for the charges and moduli.\footnote{In
theories with $N\geq 4$ supersymmetry there are no
renormalizations.}
The argument is that quantum corrections would spoil the relation of
mass and charges, and if we assume unbroken SUSY at the quantum level
there would be incompatibilities  with the dimension of their
representations.

In order to present the concept of BPS states we will briefly review
the representation theory of $N$-extended supersymmetry.
A more complete treatment can be found in \cite{BW}.
A general discussion of central charges in various dimensions can be
found
in \cite{central}.
We will concentrate here to four dimensions.
The anticommutation relations are
\be
\{Q_{\a}^I,Q_{\b}^J\}=\e_{\a\b}Z^{IJ}\;\;\;,\;\;\;
\{\bar Q_{\dot\a}^I,Q_{\dot\b}^J\}=\e_{\dot\a\dot\b}\bar Z^{IJ}
\;\;\;,\;\;\;\{Q_{\a}^I,\bar
Q_{\dot\a}^J\}=\d^{IJ}~2\s^{\mu}_{\a\dot\a}P_{\m}
\,,\label{3999}\ee
where $Z^{IJ}$ is the antisymmetric central charge matrix.

The algebra is invariant under the U(N) $R$-symmetry that rotates
$Q,\bar Q$.
We begin with a description of the representations of the algebra.
We will first assume that the central charges are zero.

$\bullet$ \underline{Massive representations}. We can go to the rest
frame
$P\sim(-M,\vec 0)$. The relations become
\be
\{Q_{\a}^I,\bar Q^J_{\dot\a}\}=2M\d_{\a\dot \a}\d^{IJ}
\;\;\;,\;\;\;\{Q^I_{\a},Q^J_{\b}\}=\{\bar Q^I_{\dot\a},\bar
Q^J_{\dot\b}\}
=0\,.\label{D1}\ee
Define the 2N fermionic harmonic creation and annihilation operators
\be
A^I_{\a}={1\over \sqrt{2M}}Q^I_{\a}\;\;\;,\;\;\;A^{\dagger
I}_{\a}={1\over \sqrt{2M}}\bar Q^I_{\dot\a}
\,.\label{D2}\ee
Building the representation is now easy. We start with Clifford
vacuum
$|\Omega\rangle$, which is annihilated by the $A^I_{\a}$ and we
generate
the representation by acting with the creation operators.
There are ${2N}\choose{n}$ states at the $n$-th oscillator level.
The total number of states is $\sum_{n=0}^{2N}$${2N}\choose{n}$, half
of them being bosonic
and half of them fermionic. The spin comes from symmetrization over
the spinorial indices. The maximal spin is the spin of the
ground-states plus $N$.

{\bf Example}. Suppose N=1 and the ground-state transforms into the
$[j]$ representation of SO(3).
Here we have two creation operators.
Then, the content of the massive representation is
$[j]\otimes([1/2]+2[0])
=[j\pm 1/2]+2[j]$.
The two spin-zero states correspond to the ground-state itself and to
the
state with two oscillators.

$\bullet$ \underline{Massless representations}. In this case we can
go to the frame $P\sim (-E,0,0,E)$.
The anticommutation relations now become
\be
\{Q^I_{\a},\bar Q^{J}_{\dot\a}\}=2\left(\matrix{2E&0\cr
0&0\cr}\right)
\d^{IJ}
\,,\label{D3}\ee
the rest being zero. Since $Q_2^I,\bar Q_{\dot 2}^I$ totally
anticommute,
they are represented by zero in a unitary theory.
We have $N$ non-trivial creation and annihilation operators
$A^I=Q_1^I/2\sqrt{E}$,$A^{\dagger~I}=\bar Q_1^I/2\sqrt{E}$,
and the representation is $2^N$-dimensional.
It is much shorter than the massive one.

$\bullet$ \underline{Non-zero central charges}. In this case the
representations are massive. The central charge matrix
can be brought by a U(N) transformation to block diagonal form as in
(\ref{322}), and we will label the real positive eigenvalues by $Z_m$.
We assume that
$N$ is even so that $m=1,2,\ldots,N/2$.
We will split the index $I\to (a,m)$: $a=1,2$ labels positions inside
the $2\times 2$ blocks while $m$ labels the blocks.
Then
\be
\{Q_{\a}^{am},\bar Q_{\dot\a}^{bn}\}=2M\d^{\a\dot\a}\d^{ab}\d^{mn}
\;\;\;,\;\;\;\{Q_{\a}^{am},Q_{\b}^{bn}\}=Z_n\e^{\a\b}\e^{ab}\d^{mn}
\,.\label{D4}\ee
Define the following fermionic oscillators
\be
A^{m}_{\a}={1\over
\sqrt{2}}[Q^{1m}_{\a}+\e_{\a\b}Q^{2m}_{\b}]\;\;\;,\;\;\;
B^{m}_{\a}={1\over \sqrt{2}}[Q^{1m}_{\a}-\e_{\a\b}Q^{2m}_{\b}]
\,,\label{D5}\ee
and similarly for the conjugate operators.
The anticommutators become
\be
\{A^m_{\a},A^n_{\b}\}=\{A^m_{\a},B^n_{\b}\}=\{B^m_{\a},B^n_{\b}\}=0
\,,\label{D6}\ee
\be
\{A^{m}_{\a},A^{\dagger
n}_{\b}\}=\d_{\a\b}\d^{mn}(2M+Z_n)\;\;\;,\;\;\;
\{B^{m}_{\a},B^{\dagger n}_{\b}\}=\d_{\a\b}\d^{mn}(2M-Z_n)
\,.\label{D7}\ee
Unitarity requires that the right-hand sides in (\ref{D7}) be
non-negative.
This in turn implies the Bogomolnyi bound
\be
M\geq {\rm max}\left[ {Z_{n}\over 2}\right]
\,.\label{D8}\ee
Consider $0\leq r\leq N/2$ of the $Z_n$'s to be equal to $2M$.
Then $2r$ of the $B$-oscillators vanish identically and we are left
with $2N-2r$ creation and annihilation operators.
The representation has $2^{2N-2r}$ states.
The maximal case $r=N/2$ gives rise to the short BPS multiplet whose
number of states are the same as in the massless multiplet.
The other multiplets with $0<r<N/2$ are known as intermediate BPS
multiplets.

Another ingredient that makes supersymmetry special is specific
properties of supertraces of powers of the helicity.
Such supertraces appear in loop amplitudes and they will be quite
useful.\footnote{The relation of loop corrections to supertraces
was first observed in \cite{str1}. General supertraces were computed
in \cite{str2}. The relationaship between $B_2$ and short multiplets
of the N=2 algebra in four dimensions was observed in \cite{str3}.
It was generalized to different amounts of supersymmetry in
\cite{bk1}.}
They can also be used to distinguish BPS states \cite{bk1,lec}.
We will define the helicity supertrace on a supersymmetry
representation $R$
as\footnote{In higher dimensions, traces over various Casimirs of the
little group have to be considered.}
\be
B_{2n}(R)={\rm Tr}_R[(-1)^{2\l}\l^{2n}]
\,.\label{D9}\ee
It is useful to introduce the ``helicity-generating function" of a
given supermultiplet R
\be
Z_{R}(y) = {\rm str}\  y^{2 \lambda}
\,.\label{D10}\ee
For a particle of spin $j$ we have
\be
Z_{[j]} = \cases{& $(-)^{2j} \Bigl( { y^{2j+1}-y^{-2j-1}
\over y- 1/y } \Bigr)$ \ \  {\rm massive} \cr
 &\ \cr
 & $(-)^{2j} ( y^{2j}+y^{-2j})$ \ \ \ \
  {\rm massless} \cr}
\,.\label{D11}
\ee
When tensoring representations the generating functionals get
multiplied,
\be
Z_{r\otimes {\tilde r}} = Z_r Z_{\tilde r}
\,.\label{D12}\ee
The supertrace of the $n$-th power of helicity can be extracted
from the generating functional through
\be
B_n(R) =   (y^2{d\over dy^2})^n \ Z_R(y)\vert_{y=1}
\,.\label{D13}\ee

For a supersymmetry representation constructed from a spin $[j]$
ground-state
by acting with $2m$ oscillators we obtain
\be
Z_{m}(y)=Z_{[j]}(y)(1-y)^m(1-1/y)^m
\,.\label{D14}\ee

We will now analyse in more detail N=2,4 supersymmetric
representations

$\bullet$ \underline{N=2 supersymmetry}. There is only one central
charge eigenvalue $Z$.
The long massive representations have the following content:
\be
L_j\;\;\;:\;\;\;[j]\otimes([1]+4[1/2]+5[0])
\,.\label{D33}\ee

When $M=Z/2$ we obtain the short (BPS) massive multiplet
\be
S_j\;\;\;:\;\;\;[j]\otimes(2[1/2]+4[0])
\,.\label{D34}\ee

Finally the massless multiplets have the following content
\be
M^0_{\l}\;\;\;:\;\;\;\pm (\l+1/2)+2(\pm\l)+\pm(\l-1/2)
\,.\label{D35}\ee
$\l=0$ corresponds to the hypermultiplet, $\l=1/2$ to the vector
multiplet
and $\l=3/2$ to the supergravity multiplet.

We have the following helicity supertraces
\be
B_{0}({\rm any~~rep})=0
\,,\label{D36}\ee
\be
B_2(M^0_{\l})=(-1)^{2\l+1}\;\;\;,\;\;\;B_2(S_j)=
(-1)^{2j+1}~D_j\;\;\;,\;\;\;
B_2(L_j)=0
\,.\label{D37}\ee

$\bullet$ \underline{N=4 supersymmetry}.  Here we have two eigenvalues
for the central charge matrix $Z_1\geq Z_2\geq 0$.
For the  generic massive multiplet, $M>Z_1$,
and all eight  raising operators
act non-trivially.  The representation is long,
 containing 128 bosonic
and 128 fermionic states.
The generic, long, massive multiplet can be generated by tensoring the
representation $[j]$ of its ground-state with the long fermionic
oscillator
representation of the N=4 algebra:

\be
{\rm L_j}\;\;:\;\; [j]\otimes
\left(42[0]+48[1/2]+27[1]+8[3/2]+[2]\right)
\,.\label{D15}\ee
It contains $128D_j$ bosonic degrees of freedom and $128D_j$
fermionic
ones ($D_j=2j+1$).
The  minimum-spin
massive long (ML) multiplet has $j=0$ and maximum spin 2  with the
following content:
\be
s=2\;\;{\rm  massive}\;\;{\rm
long}\;:\;\;42[0]+48[1/2]+27[1]+8[3/2]+[2]
\,.\label{a1}\ee

The generic  representation saturating the mass bound,
 $M=Z_{1}>Z_{2}$,  leaves  one unbroken  supersymmetry
 and is referred to as  massive intermediate BPS multiplet.
It can be obtained as
\be
I_j\;\;:\;\;[j]\otimes(14[0]+14[1/2]+6[1]+[3/2])
\label{D16}\ee
and contains 32$D_j$ bosonic and $32D_j$ fermionic states.
The minimum spin multiplet (j=0) has maximum spin 3/2 and content

\be
I_{3/2}\;\;:\;\;14[0]+14[1/2]+6[1]+[3/2]
\,.\label{D17}\ee

Finally, when $M=|Z_1|=|Z_2|$ the representation is a short
BPS representation.
It breaks half of the supersymmetries. For massive such
representations
we have the content
\be
S_j\;\;:\;\;[j]\otimes(5[0]+4[1/2]+[1])
\,,\label{D18}\ee
with $8D_j$ bosonic and $8D_j$ fermionic states.
The representation with minimum greatest spin is the one with $j=0$,
and
maximum spin 1:
\be
S_1\;\;:\;\;5[0]+4[1/2]+[1]
\,.\label{D19}\ee

Massless multiplets, which arise only when both  central charges
vanish, are thus
always short.
They have the following O(2) helicity content:
\be
M^0_{\l}\;\;:\;\;[\pm (\l+1)]+4[\pm(\l+1/2)]+6[\pm (\l)]+
4[\pm(\l-1/2)]+[\pm(\l-1)]
\,,\label{D20}\ee
with 16 bosonic and 16 fermionic states.
There is also the CPT self-conjugate vector representation ($V^0$)
(corresponding to $\l=0$)
with content $6[0]+4[\pm 1/2]+[\pm 1]$ and 8 bosonic and 8 fermionic
states.
For $\l=1$ we obtain  the spin-two massless supergravity
multiplet, which has the  helicity content
\be
M^0_1\;\;\;:\;\;[\pm 2]+4[\pm 3/2]+6[\pm 1]+4[\pm 1/2]+2[0]
\,.\label{D21}\ee
Long representations can be decomposed
into intermediate representations as
\be
L_{j}\to 2\;I_{j}+I_{j+1/2}+I_{j-1/2}
\,.\label{D22}\ee
When further, by varying the moduli, we can arrange that
$M=|Z_1|=|Z_2|$,
then the massive intermediate representations can break into massive
short
representations as
\be
I_j\to 2S_j+S_{j+1/2}+S_{j-1/2}
\,.\label{D23}\ee
Finally, when a short representation becomes massless, it decomposes
as follows into massless representations:
\be
S_j\to \sum_{\lambda=0}^{j}
M^0_{\lambda}
\;\;\;,\;\;  j-\l \in \Z
\,.\label{D24}\ee

By direct calculation we obtain the following helicity supertrace
formulae:

\be
B_n({\rm any\;\;rep})=0\;\;\;{\rm for}\;\;\;n=0,2
\,.\label{D25}\ee
The non-renormalization of the two derivative effective actions in
N=4 supersymmetry is based on (\ref{D25}).

\be
B_4(L_j)=B_4(I_j)=0\;\;\;,\;\;\;B_4(S_j)=(-1)^{2j}{3\over
2}D_j\label{D26}\ee
\be
B_4(M^0_{\lambda})=(-1)^{2\lambda}\;3\;
\;\;,\;\;\;B_4(V^0)={3\over 2}
\,.\label{D27}\ee
These imply that only short multiplets contribute in the
renormalization
of some terms in the four-derivative effective action in the presence
of N=4
supersymmetry.
It also strongly suggests that such corrections come only from one
order (usually one loop) in perturbation theory.

The following helicity sums will be useful when counting intermediate
multiplets in string theory:
\be
B_6(L_j)=0\;\;\;,\;\;\;B_6(I_j)=(-1)^{2j+1}{45\over
4}D_j\;\;\;,\;\;\;B_6(S_j)=(-1)^{2j}{15\over 8}D^3_j
\,,\label{D28}\ee
\be
B_6(M^0_{\lambda})=(-1)^{2\lambda}{15\over 4}(1+12\lambda^2)\;
\;\;,\;\;\;B_6(V^0)={15\over 8}
\,.\label{D29}\ee

Finally,
\be
B_8(L_j)=(-1)^{2j}{315\over
4}D_j\;\;\;,\;\;\;B_8(I_j)=(-1)^{2j+1}{105\over 16}D_j(1+D^2_j)
\,,\label{D30}\ee
\be
B_8(S_j)=(-1)^{2j}{21\over 64}D_j(1+2\;D^4_j)
\,,\label{D31}\ee
\be
B_8(M^0_{\lambda})=(-1)^{2\lambda}{21\over
16}(1+80\lambda^2+160\lambda^4)\;
\;\;,\;\;\;B_8(V^0)={63\over 32}
\,.\label{D32}\ee
The massive long N=4 representation is the same as the short massive
N=8 representation, which explains the result in (\ref{D30}).

Observe that the trace formulae above are in accord with the
decompositions
(\ref{D22})-(\ref{D24}).

\vskip .5cm

$\bullet$\underline{N=8 supersymmetry}.
The highest possible supersymmetry in four dimensions is N=8.
Massless representations ($T_0^{\lambda}$) have the following
helicity content
\be
(\l\pm 2)+8\left(\l\pm {3\over 2}\right)+28(\l\pm 1)+56\left(\l\pm
{1\over
2}\right)
+70(\l)\;\;.
\label{e1}\ee
Physical (CPT-invariant) representations are given by
$M_0^{\l}=T_0^{\l}+T_0^{-\l}$
and contain $2^8$ bosonic states and an equal number of fermionic ones
with the exception of the supergravity representation $M_0^0=T_0^0$
which is CPT-self-conjugate:
\be
(\pm 2)+8\left(\pm {3\over 2}\right)+28(\pm 1)+56\left(\pm {1\over
2}\right)+70(0)\;\;,
\label{e2}
\ee
and contains $2^7$ bosonic states.

Massive short representations ($S^j$), are labeled by the SU(2) spin j
of the ground-state and have the following content
\be
[j]\otimes \left([2]+8[3/2]+27[1]+48[1/2]+42[0]\right)\;\;.
\label{e3}
\ee
They break four (half) of the supersymmetries and contain $2^7\cdot
D_j$ bosonic states.
$S^j$ decomposes to massless representations as
\be
S^j\to \sum_{\l=0}^j\;M_0^{\l}\;\;,
\label{e222}\ee
where the sum runs on integer values of $\l$ if $j$ is integer and on
half-integer values if $j$ is half-integer.

There are three types of intermediate multiplets, which we list below:
\be
I_1^j\;:\;[j]\otimes
\left([5/2]+10[2]+44[3/2]+110[1]+165[1/2]+132[0]\right)\,,
\label{e4}
\ee
\bea
I_2^j\;:\;[j]\otimes
\left([3]+12[5/2]+65[2]+208[3/2]+429[1]+ \quad\quad\quad \right.
\nn\\
\left. + 572[1/2]+429[0]\right)\,,
\label{e5}
\eea
\bea
I_3^j\;:\;[j]\otimes
\left([7/2]+14[3]+90[5/2]+350[2]+910[3/2]+ \quad\quad\quad \right.
\nn\\
\left. + 1638[1] + 2002[1/2]+1430[0]\right)\,.
\label{e6}
\eea
They break respectively 5,6,7 supersymmetries.
They contain $2^9\cdot D_j$ ($I_1^j$), $2^{11}\cdot D_j$ ($I_2^j$)
and
$2^{13}\cdot D_j$ ($I_3^j$) bosonic states.

Finally, the long representations ($L^j$) (which  break all
supersymmetries )
are given by
$$
[j]\otimes
\left([4]+16[7/2]+119[3]+544[5/2]+1700[2]+
\right.$$\be
\left.+3808[3/2]+
6188[1]+7072[1/2]+4862
[0]\right).
\label{e7}
\ee
$L^j$ contains $2^{15}\cdot D_j$ bosonic states.

We also have the following recursive decomposition formulae:
\be
L^j\to I_3^{j+{1\over 2}}+2 I_3^j+I_3^{j-{1\over 2}}\;\;,
\label{e8}
\ee
\be
I_3^j\to I_2^{j+{1\over 2}}+2 I_2^j+I_2^{j-{1\over 2}}\;\;,
\label{e9}
\ee
\be
I_2^j\to I_1^{j+{1\over 2}}+2 I_1^j+I_1^{j-{1\over 2}}\;\;,
\label{e10}
\ee
\be
I_1^j\to S^{j+{1\over 2}}+2 S^j+S^{j-{1\over 2}}\;\;.
\label{e11}
\ee

All even helicity supertraces up to order six vanish for N=8
representations.
For the rest we obtain:
\be
B_8(M_0^{\l})=(-1)^{2\l}\;315\;\;,
\label{ee12}\ee
\be
B_{10}(M_0^{\l})=(-1)^{2\l}\;{4725\over 2}(6\l^2+1)\;\;,
\label{ee13}\ee
\be
B_{12}(M_0^{\l})=(-1)^{2\l}\;{10395\over 16}(240\l^4+240\l^2+19)\;\;,
\label{ee14}\ee
\be
B_{14}(M_0^{\l})=(-1)^{2\l}\;{45045\over
16}(336\l^6+840\l^4+399\l^2+20)\;\;,
\label{ee15}\ee
\be
B_{16}(M_0^{\l})=(-1)^{2\l}\;{135135\over 256}(7680\l^8+35840\l^6
+42560\l^4+12800\l^2+457)\;\;.
\label{ee16}\ee

The supertraces of the massless supergravity representation $M_0^0$
can be obtained from the above by setting $\l=0$ and dividing by a
factor of 2 to account for the smaller dimension of the
representation.

\be
B_8(S^j)=(-1)^{2j}\cdot {315\over 2}D_j\;\;,
\label{ee17}\ee
\be
B_{10}(S^j)=(-1)^{2j}\cdot {4725\over 8}D_j(D_j^2+1)\;\;,
\label{ee18}\ee
\be
B_{12}(S^j)=(-1)^{2j}\cdot {10395\over 32}D_j(3D_j^4+10D_j^2+6)\;\;,
\label{ee19}\ee
\be
B_{14}(S^j)=(-1)^{2j}\cdot {45045\over
128}D_j(3D_j^6+21D_j^4+42D_j^2+14)\;\;,
\label{ee20}\ee
\be
B_{16}(S^j)=(-1)^{2j}\cdot {45045\over
512}D_j(10D_j^{8}+120D_j^6+504D_j^4+560D_j^2+177)\;\;,
\label{ee21}\ee

\be
B_8(I_1^j)=0\;\;,
\label{ee22}\ee
\be
B_{10}(I_1^j)=(-1)^{2j+1}\cdot {14175\over 4}D_j\;\;,
\label{ee23}\ee
\be
B_{12}(I_1^j)=(-1)^{2j+1}\cdot {155925\over 16}D_j(2D_j^2+3)\;\;,
\label{ee24}\ee
\be
B_{14}(I_1^j)=(-1)^{2j+1}\cdot  {2837835\over
64}D_j(D_j^2+1)(D_j^2+4)\;\;,
\label{ee25}\ee
\be
B_{16}(I_1^j)=(-1)^{2j+1}\cdot {2027025\over
128}D_j(4D_j^6+42D_j^4+112D_j^2+57)\;\;,
\label{ee26}\ee

\be
B_8(I_2^j)=B_{10}(I_2^j)=0\;\;,
\label{ee27}\ee
\be
B_{12}(I_2^j)=(-1)^{2j}\cdot {467775\over 4}D_j\;\;,
\label{ee28}\ee
\be
B_{14}(I_2^j)=(-1)^{2j}\cdot {14189175\over 16}D_j(D_j^2+2)\;\;,
\label{ee29}\ee
\be
B_{16}(I_2^j)=(-1)^{2j}\cdot {14189175\over
32}D_j(6D_j^4+40D_j^2+41)\;\;,
\label{ee30}\ee

\be
B_8(I_3^j)=B_{10}(I_3^j)=B_{12}(I_3^j)=0\;\;,
\label{ee31}\ee
\be
B_{14}(I_3^j)=(-1)^{2j+1}\cdot {42567525\over 8}D_j\;\;,
\label{ee32}\ee
\be
B_{16}(I_3^j)=(-1)^{2j+1}\cdot {212837625\over 8}D_j(2D_j^2+5)\;\;,
\label{ee33}\ee

\be
B_8(L^j)=B_{10}(L^j)=B_{12}(L^j)=B_{14}(L^j)=0\;\;,
\label{ee34}\ee
\be
B_{16}(L^j)=(-1)^{2j}\cdot {638512875\over 2}D_j\;\;.
\label{ee35}\ee

A further check of the above formulae is provided by the fact that
they
respect the decomposition formulae of the various representations
 (\ref{e222}) and (\ref{e8})-\ref{e11}).

\addcontentsline{toc}{subsection}{Appendix F: Modular forms}
\section*{Appendix F: Modular forms}
\renewcommand{\theequation}{F.\arabic{equation}}
\setcounter{equation}{0}

In this appendix we collect some formulae
for modular forms, which are useful for analysing the spectrum of BPS
states and
BPS-generated one-loop corrections to the effective supergravity
theories.
A (holomorphic) modular form $F_{d}(\tau)$ of weight $d$ behaves as
follows under modular transformations:
\be
F_d(-1/\t)=\t^{d}F_d(\t)\;\;\;F_d(\t+1)=F_d(\t)
\,.\label{E1}\ee

We first list the  Eisenstein series:
\be
E_2={12\over i\pi}\partial_{\tau}\log\eta=1-24\sum_{n=1}^{\infty}
{nq^n\over 1-q^n}
\,,\label{E2}\ee
\be
E_4={1\over 2}\left(\vartheta_2^8+\vartheta_3^8+\vartheta_4^8\right)
=1+240\sum_{n=1}^{\infty}{n^3q^n\over 1-q^n}
\,,\label{E3}\ee
\be
E_6={1\over 2}\left(\vartheta_2^4+\vartheta_3^4\right)
\left(\vartheta_3^4+\vartheta_4^4\right)
\left(\vartheta_4^4-\vartheta_2^4\right)=
1-504\sum_{n=1}^{\infty}{n^5q^n\over 1-q^n}
\,.\label{E4}\ee
In counting  BPS states in string theory the following combinations
arise
\be
H_2\equiv {1-E_2\over 24}=\sum_{n=1}^{\infty}{nq^n\over 1-q^n}\equiv
\sum_{n=1}^{\infty}d_2(n)q^n
\,,\label{E5}\ee
\be
H_4\equiv {E_4-1\over 240}=\sum_{n=1}^{\infty}{n^3q^n\over
1-q^n}\equiv
\sum_{n=1}^{\infty}d_4(n)q^n
\,,\label{E6}\ee
\be
H_6\equiv {1-E_6\over 504}=\sum_{n=1}^{\infty}{n^5q^n\over
1-q^n}\equiv
\sum_{n=1}^{\infty}d_6(n)q^n
\,.\label{E7}\ee
We have the following arithmetic formulae for $d_{2k}$:
\be
d_{2k}(N)=\sum_{n|N}\;n^{2k-1}\;\;\;,\;\;\;k=1,2,3
\,.\label{E8}\ee

The $E_4$ and $E_6$ modular forms have  weight four and six,
respectively.
They generate the ring of modular forms.
However, $E_2$ is not exactly a modular form, but
\be
\hat E_2=E_2-{3\over \pi\t_2}
\label{E9}\ee
is a modular form of weight 2 but is not holomorphic any more.
The (modular-invariant) $j$ function and $\eta^{24}$ can be written
as
\be
j={E_4^3\over \eta^{24}}={1\over q}+744+\ldots\;\;\;,\;\;\;
\eta^{24}={1\over 2^6\cdot 3^3}\left[E_4^3-E_6^2\right]
\,.\label{E10}\ee

We will also introduce the covariant derivative on modular forms:
\be
F_{d+2}=\left({i\over \pi}\partial_{\t}+{d/2\over
\pi\t_2}\right)F_{d}
\equiv D_d\;F_d\;.
\ee
$F_{d+2}$ is a modular form of weight $d+2$ if $F_d$ has weight $d$.
The covariant derivative introduced above has the following
distributive
property:
\be
D_{d_1+d_2}~(F_{d_1}~F_{d_2})=F_{d_{2}}(D_{d_1}F_{d_1})+
F_{d_{1}}(D_{d_2}F_{d_2})
\,.\label{E11}\ee
The following relations and (\ref{E11}) allow the computation of any
covariant derivative
\be
D_2\;\hat E_2={1\over 6}E_4-{1\over 6}\hat E_2^2
\;\;\;,\;\;\;
D_4\;E_4={2\over 3}E_6-{2\over 3}\hat E_2\;E_4
\;\;\;,\;\;\;
D_{6}\;E_6=E_4^2-\hat E_2\;E_6
\,.\label{E14}\ee

Here we will give some identities between derivatives of
$\th$-functions and
modular forms. They are useful for trace computations in
string theory:
 \be
{\vartheta_{1}'''\over \vartheta_1'}=-\pi^2\;E_2
\;\;\;,\;\;\;
{\vartheta^{(5)}_1\over \vartheta_1'}=-\pi^2\;E_2\left(4\pi
i\partial_{\tau}
\log E_2-\pi^2 E_2\right)
\,,\label{E16}\ee
\be
-3{\vartheta^{(5)}_1\over \vartheta_1'}+5\left({\vartheta_1'''\over
\vartheta_1'}\right)^2=2\pi^4 E_4
\,,\label{E17}\ee
\be
-15{\vartheta^{(7)}_1\over \vartheta_1'}-{350\over
3}\left({\vartheta_1'''\over
\vartheta_1'}\right)^3+105{\vartheta_1^{(5)}\vartheta_1'''\over
\vartheta_1'^2}={80\pi^6\over 3}E_6
\,,\label{E18}\ee
\be
{1\over 2}\sum_{i=2}^4~{\th_i''~\th_i^7\over (2\pi i)^2}={1\over 12}
(E_2E_4-E_6)
\,.\label{E188}\ee

The function $\xi(v)$ that appears in string helicity-generating
partition functions is defined as
\be
\xi(v)=\prod_{n=1}^{\infty}{(1-q^n)^2\over (1-q^ne^{2\pi iv})
(1-q^ne^{-2\pi iv})}={\sin\pi v\over \pi}{\vartheta_1'\over
\vartheta_1(v)}\;\;\;\,\;\;\;\xi(v)=\xi(-v)
\,.\label{E19}\ee
It satisfies
\be
\xi(0)=1
\;\;\;,\;\;\;
\xi^{(2)}(0)=-{1\over 3}\left(\pi^2+{\vartheta_1'''\over
\vartheta_1'}\right)=-{\pi^2\over 3}(1-E_2)
\,,\label{E21}\ee
\be
\xi^{(4)}(0)={\pi^4\over 5}+{2\pi^2\over 3}{\vartheta_1'''\over
\vartheta_1'}
+{2\over 3}\left({\vartheta_1'''\over \vartheta_1'}\right)^2-{1\over
5}{\vartheta^{(5)}_1\over \vartheta_1'}={\pi^4\over
15}(3-10E_2+2E_4+5E_2^2)
\,,\label{E22}\ee
\bea
\xi^{(6)}(0)&=&-{\pi^6\over 7}-\pi^4{\vartheta_1'''\over \vartheta_1'}
-{10\pi^2\over 3}\left({\vartheta_1'''\over
\vartheta_1'}\right)^2+\pi^2
{\vartheta^{(5)}_1\over \vartheta_1'}+
\nn\\
&&-{10\over 3}
\left({\vartheta_1'''\over
\vartheta_1'}\right)^3+2{\vartheta_1^{(5)}\vartheta_1'''\over
\vartheta_1'^2}
-{1\over 7}{\vartheta^{(7)}_1\over \vartheta_1'}
\nn\\
&=&{\pi^6\over 63}(-9+63E_2-105E_2^2-42E_4+16E_6+
\nn\\
&& + 42E_2E_4+35E_2^3)\;,
\eea
where $\xi^{(n)}(0)$ stands for taking the $n$-th derivative with
respect
to $v$ and then setting $v=0$.

\addcontentsline{toc}{subsection}{Appendix G: Helicity string
partition functions}
\section*{Appendix G: Helicity string partition functions}
\renewcommand{\theequation}{G.\arabic{equation}}
\setcounter{equation}{0}

We have seen in section \ref{bps} that BPS states are important
ingredients in non-perturbative dualities. The reason is that their
special properties,
most of the time, guarantee that such states survive at strong
coupling.
In this section we would like to analyze ways of counting BPS states in
string perturbation theory.

An important point that should be stressed from the beginning is the
following:
a generic BPS state $is$ $not$ protected from quantum corrections.
The reason is that sometimes groups of short BPS multiplets can combine
into
long multiplets of supersymmetry. Such long multiplets  are not
protected from non-renormalization theorems.
We would thus like to count BPS multiplicities in such a way that only
``unpaired" multiplets contribute.
As explained in Appendix E, this can be done with the help of helicity
supertrace formulae. These have precisely the properties we need in
order to count BPS multiplicities that are protected from
non-renormalization theorems.
Moreover, multiplicities counted via helicity supertraces are
insensitive to moduli.
They are the generalizations of the elliptic genus, which is the
stringy generalization of the Dirac index.
In this sense, they are indices, insensitive to the details of the
physics.
We will show here how we can compute helicity supertraces in
perturbative string ground-states, and we will work out some
interesting examples.

We will introduce the helicity-generating partition
functions for D=4 string theories with N$\geq 1$ spacetime
supersymmetry.
The physical helicity in closed string theory $\l$ is a sum of the
left helicity $\l_L$ coming from the left-movers and the right
helicity $\l_R$
coming from the right-movers.
Then, we can consider the following helicity-generating partition
function
\be
Z(v,\bar v)=Str[q^{L_0}\;\bar q^{\bar L_0}e^{2\pi
iv\l_R-2\pi i\bar v\l_L}]
\,.\label{F1}\ee

We will first examine the heterotic string.
Four-dimensional vacua with at least N=1 spacetime supersymmetry
have
the following partition function
\be
Z^{\rm heterotic}_{D=4}={1\over \t_2\eta^2\bar \eta^2}\sum_{a,b=0}^1
{}~(-1)^{a+b+ab}~{\th[^a_b]\over \eta}~C^{\rm int}[^a_b]
\,,\label{F2}\ee
where we have separated the (light-cone) bosonic and fermionic
contributions of the four-dimensional  part.
$C[^a_b]$ is the partition function of the internal CFT with $(c,\bar
c)=(9,22)$ and at least (2,0) superconformal symmetry.
$a=0$ corresponds to the $NS$ sector, $a=1$ to the R
sector
and $b=0,1$ indicates the presence of the projection $(-1)^{F_L}$,
where
$F_L$ is the zero mode of the N=2, U(1) current.

The oscillators that would contribute to the left helicity are the
left-moving
light-cone bosons $\pd X^{\pm}=\pd X^3\pm i\pd X^4$ contributing
helicity $\pm 1$ respectively, and the light-cone fermions
$\psi^{\pm}$ contributing again
$\pm 1$ to the left helicity.
Only $\bar\pd X^{\pm}$ contribute to the right-moving helicity.
Calculating (\ref{F1}) is straightforward, with the result
\be
 Z^{\rm heterotic}_{D=4}(v,\bar v)={\xi(v)\bar\xi(\bar v)\over
\t_2\eta^2\bar \eta^2}\sum_{a,b=0}^1
{}~(-1)^{a+b+ab}~{\th[^a_b](v)\over \eta}~C^{\rm int}[^a_b]
\,,\label{F350}\ee
where $\xi(v)$ is given in (\ref{E19}).
This can be simplified using spacetime supersymmetry to
\be
 Z^{\rm heterotic}_{D=4}(v,\bar v)={\xi(v)\bar\xi(\bar v)\over
\t_2\eta^2\bar \eta^2}{\th[^1_1](v/2)\over \eta}~C^{\rm int}[^1_1](v/2)
\,,\label{F351}\ee
with
\be
C^{\rm int}[^1_1](v)={\rm Tr}_R[(-1)^{F^{\rm int}}~e^{2\pi i
v~J_0}~q^{L_0^{\rm int}-{3/8}}
{}~\bar q^{\bar L_0^{\rm int}-11/12}]
\,,\label{F4}\ee
where the trace is in the Ramond sector, and $J_0$ is the zero mode
of the U(1) current of the N=2 superconformal algebra;
$C^{\rm int}[^1_1](v)$ is the elliptic genus of the internal
(2,0)
theory and is antiholomorphic.
The leading term of $C^{\rm int}[^1_1](0)$ coincides with the Euler
number
in CY compactifications.

If we define
\be
Q={1\over 2\pi i}{\partial\over \partial v}\;\;\;,\;\;\;\bar
Q=-{1\over 2\pi i}{\partial\over \partial \bar v}
\,,\label{F5}\ee
then the helicity supertraces can be written as
\be
B_{2n}\equiv {\rm Str}[\l^{2n}]=\left.(Q+\bar Q)^{2n}\;Z^{\rm
heterotic}_{D=4}(v,\bar
v)\right|_{v=\bar v=0}
\,.\label{F6}\ee

Consider as an example the heterotic string on $T^6$ with
N=4, D=4 spacetime supersymmetry.
Its helicity partition function is
\be
Z^{\rm heterotic}_{N=4}(v,\bar v)={\vartheta^4_1(v/2)\over
\eta^{12}\bar \eta^{24}}\xi(v)\bar\xi(\bar v){\Gamma_{6,22}\over
\t_2}
\,.\label{F7}\ee

It is obvious that we need at least four powers of $Q$ in order to
get a non-vanishing contribution, implying $B_0=B_2=0$, in
agreement
with the N=4 supertrace formulae derived in Appendix E.
We will calculate $B_4$ which, according
to
(\ref{D26}),(\ref{D27}) is sensitive to short multiplets only:
\be
B_4=\langle (Q+\bar Q)^4\rangle=\langle
Q^4\rangle={3\over
2}{1\over \bar \eta^{24}}
\,.\label{F8}\ee
For the massless states the result agrees
with (\ref{D27}), as it should.
Moreover, from (\ref{D26}) we observe that massive short multiplets
with a bosonic ground-state give an opposite contribution from
multiplets with a fermionic ground-state. We learn that all such
short massive multiplets  in
the heterotic spectrum
are ``bosonic", with multiplicities given by the coefficients
of the~$\eta^{-24}$.

Consider further
\be
B_6=\langle(Q+\bar Q)^6\rangle= \langle
Q^6+15Q^4\bar Q^2\rangle
={15\over 8}{2-\bar E_2\over \bar\eta^{24}}
\,.\label{F9}\ee
Since there can be no intermediate multiplets in the perturbative
heterotic spectrum we get only contributions from the short
multiplets. An explicit analysis at low levels confirms the agreement
between (\ref{D26}) and
(\ref{F9}).

For type-II vacua, there are fermionic contributions to the helicity
from both the left-moving and the right-moving world-sheet fermions.
We will consider as a first  example the type-II string, compactified
on $T^6$
to four dimensions with maximal N=8 supersymmetry.

The light-cone helicity-generating partition function is
\bea
Z^{II}_{N=8}(v,\bar v)&=&{\rm Str}[q^{L_0}\;\bar q^{\bar L_0}e^{2\pi
iv\l_R-2\pi i\bar v\l_L}]=
\nn\\
&=&{1\over 4}\sum_{\a,\b=0}^{1}\sum_{\bar\a,\bar\b=0}^{1}\;
(-1)^{\a+\b+\a\b}{\vartheta[^{\a}_{\b}](v)
\vartheta^3[^{\a}_{\b}](0)\over \eta^4}\times
\nn\\
&&\times
(-1)^{\bar\a+\bar\b+\bar\a\bar
\b}{\bar\vartheta[^{\bar\a}_{\bar\b}]\bar\vartheta^3
[^{\bar\a}_{\bar\b}](0)\over \bar\eta^4}\;\;
{\xi(v)\bar\xi(\bar v)\over {\rm Im}\tau
|\eta|^4}\;\;{\Gamma_{6,6}\over
|\eta|^{12}}=
\nn\\
&=&{\Gamma_{6,6}\over {\rm Im}\tau }\;\;{\vartheta_1^4(v/2)\over
\eta^{12}}{\bar\vartheta_1^4(\bar v/2)\over
\bar\eta^{12}}\;\;\xi(v)\bar\xi(\bar v)\,.
\eea

It is obvious that in order to obtain a non-zero result, we need at
least a $Q^4$ on the left and a $\bar Q^4$ on the right.
This is in agreement with our statement in appendix E:
$B_0=B_2=B_4=B_6=0$ for an $N=8$ theory.
The first non-trivial case is $B_8$ and by straightforward
computation we obtain
\be
B_8=\langle(Q+\bar Q)^8\rangle =70\langle Q^4\bar
Q^4\rangle={315\over 2}{\Gamma_{6,6}\over {\rm Im}\tau }\;\;.
\label{cc6}\ee

At the massless level, the only N=8 representation is the supergravity
representation, which contributes $315/2$, in accordance with
(\ref{ee16}).
At the massive levels we have seen in appendix E that only short
representations
$S^j$ can contribute, each contributing $315/2\;(2j+1)$.
We learn from (\ref{cc6}) that all short massive multiplets
have $j=0$ and they are left and right ground-states of the type-II
CFT, thus breaking N=8 supersymmetry to N=4.
Since the mass for these states is
\be
M^2={1\over 4}p_L^2\;\;\;,\;\;\;\vec m\cdot\vec n=0\;\;,
\label{cc7}\ee
such multiplets exist for any (6,6) lattice vector satisfying the
matching condition.
The multiplicity coming from the rest of the theory is 1.

We will now compute the next non-trivial supertrace\footnote{We use
formulae from appendix F here.}
\bea
B_{10}&=&\langle(Q+\bar Q)^{10}\rangle =210\langle Q^6\bar Q^4+Q^4\bar
Q^6\rangle=
\nn\\
&=& -{4725\over 8\pi^2}{\Gamma_{6,6}\over {\rm Im}\tau }\left(
{\vartheta_1'''\over \vartheta_1'}+3\xi''+\;cc\right)={4725\over
4}{\Gamma_{6,6}\over {\rm Im}\tau }\;.
\label{cc8}\eea

In this  trace, $I_1$ intermediate representations can also in
principle contribute. Comparing (\ref{cc8}) with
(\ref{ee13}), (\ref{ee23}) we learn that
there are no $I_1$ representations in the perturbative string
spectrum.

Moving further:
\bea
B_{12}&=&\langle 495(Q^4\bar Q^8+Q^8\bar Q^4)+924Q^6\bar Q^6\rangle=
\nn\\
&=& \left[{10395\over 2}+{31185\over 64}(E_4+\bar E_4)\right]
{\Gamma_{6,6}\over {\rm Im}\tau } =
\nn\\
&=&\left[{10395\cdot 19\over 32}+{10395\cdot 45\over
4}\left({E_4-1\over 240}+cc\right)\right]{\Gamma_{6,6}\over {\rm
Im}\tau }\;.\label{cc9}
\eea
Comparison with (\ref{ee19}) indicates that the first term in the
above formula contains the contribution of the short multiplets.
Here however, $I_2$ multiplets can also contribute and the second term
in (\ref{cc9}) precisely describes their contribution.
These are string states that are ground-states either on the left or
on the right and comparing with (\ref{ee28}) we learn that their
multiplicities are given by $(E_4-1)/240$.
More precisely, for a given mass level with $p_L^2-p_R^2=4N >0$
the multiplicity of these representations
at that mass level is given  by the sum of cubes of all divisors of
N, $d_4(N)$ (see Appendix F):
\be
I_2^j\;\;:\;\; \sum_j (-1)^{2j}D_j =d_4(N)\;\;.
\label{cc10}\ee
They break N=8 supersymmetry to N=2.

The last trace to which  long multiplets do not contribute is
\bea
B_{14}&=&\langle (Q+\bar Q)^{14}\rangle= \label{cc11}
\\
&=& \left[{45045\over
32}20+{14189175\over
16} \left(2{E_4-1\over 240}+{1-E_6\over 504}+cc\right)
\right]{\Gamma_{6,6}\over {\rm Im}\tau}\,.
\nn\eea
Although in this trace $I_3$ representations can contribute, there are
no such representations in the perturbative string spectrum.
The first term in (\ref{cc11}) comes from short representations,
the second
from $I_2$ representations.
Taking into account (\ref{ee29}) we can derive the following sum rule
\be
I_2^j\;\;:\;\; \sum_j (-1)^{2j}D_j^3 =d_6(N)\;\;.
\label{cc12}\ee

The final example we will consider is also instructive because it shows
that
although a string ground-state can contain many BPS multiplets, most of
them are not protected from renormalization.
The relevant vacuum is the type-II string compactified on K3$\times
T^2$ down to four dimensions.

We will first start from the  $Z_{2}$ special point of the $K_3$
moduli space.
This is given by a  $Z_2$ orbifold of the four-torus. We can
write the one-loop vacuum amplitude as
\be
Z^{II}={1\over
8}\sum_{g,h=0}^{1}\sum_{\a,\b=0}^{1}\sum_{\bar\a,\bar\b=0}^{1}\;
(-1)^{\a+\b+\a\b}{\vartheta^2[^{\a}_{\b}]\over \eta^2}
{\vartheta[^{\a+h}_{\b+g}]\over \eta}
{\vartheta[^{\a-h}_{\b-g}]\over \eta}\times
\label{c1}\ee
$$\times(-1)^{\bar\a+\bar\b+\bar\a\bar \b}{\bar
\vartheta^2[^{\bar\a}_{\bar\b}]\over \bar\eta^2}
{\bar\vartheta[^{\bar\a+h}_{\bar\b+g}]\over \bar\eta}
{\bar\vartheta[^{\bar\a-h}_{\bar\b-g}]\over \bar\eta}\;\;
{1\over {\rm Im}\tau |\eta|^4}\;\;{\Gamma_{2,2}\over
|\eta|^4}\;\;Z_{4,4}[^h_g]\;,$$
where
\be
Z_{4,4}[^0_0]={\Gamma_{4,4}\over
|\eta|^8}\;\;\;,\;\;\;Z_{4,4}[^0_1]=16{|\eta|^4\over
|\vartheta_2|^4}={|\vartheta_3\vartheta_4|^4\over |\eta|^8}\;,
\label{c2a}\ee
\be
Z_{4,4}[^1_0]=16{|\eta|^4\over
|\vartheta_4|^4}={|\vartheta_2\vartheta_3|^4\over
|\eta|^8}\;\;\;,\;\;\;Z_{4,4}[^1_1]=16{|\eta|^4\over
|\vartheta_3|^4}={|\vartheta_2\vartheta_4|^4\over |\eta|^8}\;.
\label{c2b}\ee

We have N=4 supersymmetry in four dimensions. The mass formula of BPS
states
depends only on the two-torus moduli.
Moreover states that are ground-states both on the left and the right
will give
short BPS multiplets that break half of the supersymmetry.
On the other hand, states that are ground-states on the left but
otherwise
arbitrary on the right (and vice versa) will provide BPS states that
are intermediate multiplets breaking 3/4 of the supersymmetry.
Obviously there are many such states in the spectrum. Thus, we naively
expect
many perturbative intermediate multiplets.

We will now evaluate the helicity supertrace formulae.
We will first write the helicity-generating function,
\bea
Z^{II}(v,\bar v)&=&{1\over
4}\sum_{\a\b\bar\a\bar\b}(-1)^{\a+\b+\a\b+\bar\a+\bar\b+\bar\a\bar\b}
{\vartheta[^{\a}_{\b}](v)\vartheta[^{\a}_{\b}](0) \over \eta^6} \times
\nn\\
&&
\times {\bar\vartheta[^{\bar\a}_{\bar\b}](\bar v)\bar
\vartheta[^{\bar\a}_{\bar\b}](0)\over \bar\eta^6}\xi(v)\bar\xi(\bar
v)
C[^{\a\;\;\bar\a}_{\b\;\;\bar\b}]{\Gamma_{2,2}\over \tau_2}
\nn\\
&=&{\vartheta_1^2(v/2)\bar\vartheta^2_1(\bar v/2)\over
\eta^6\;\bar\eta^6}\xi(v)
\bar\xi(\bar v)C[^{1\;\;1}_{1\;\;1}](v/2,\bar v/2){\Gamma_{2,2}\over
\tau_2}
\;,
\label{c33}\eea
where we have used the Jacobi
identity in the second line;
$C[^{\a\;\;\bar\a}_{\b\;\;\bar\b}]$ is the partition function of the
internal (4,4) superconformal field theory in the various sectors.
Moreover
$C[^{1\;\;1}_{1\;\;1}](v/2,\bar v/2)$ is an even function of $v,\bar
v$ due to the SU(2) symmetry and
\be
C[^{1\;\;1}_{1\;\;1}](v,0)=8\sum_{i=2}^4\;{\vartheta_i^2(v)\over
\vartheta_i^2(0)}
\label{c34}\ee
is the elliptic genus of the (4,4) internal theory on K3.
Although we calculated the elliptic genus in the $Z_2$ orbifold limit,
the calculation is valid on the whole of K3 since the
elliptic genus does not depend on the moduli.

Let us first compute the trace of the fourth power of the helicity:
\be
\langle\l^4\rangle=\rangle (Q+\bar Q)^4\rangle =6\langle Q^2\bar Q^2
+Q^2\bar Q^4\rangle=36{\Gamma_{2,2}\over \tau_2}\;.
\label{c355}\ee
As expected, we obtain contributions from the the ground-states only,
but with arbitrary  momentum and winding on the (2,2) lattice.
At the massless level, we have the N=4 supergravity multiplet
contributing 3
and 22 vector multiplets contributing 3/2 each, making a total of 36,
in agreement with (\ref{c355}).
There is a tower of massive short multiplets at each mass level, with
mass
$M^2=p_L^2$, where $p_L$ is the (2,2) momentum. The matching condition
implies, $\vec m\cdot \vec n=0$.

We will further  compute the trace of the sixth power of the helicity,
to investigate the presence of intermediate multiplets:
\be
\langle\l^6\rangle=\rangle (Q+\bar Q)^6\rangle =15\langle Q^4\bar Q^2
+Q^2\bar Q^4\rangle=90{\Gamma_{2,2}\over \tau_2}\;,
\label{c35}\ee
where we have used
\be
\partial_v^2C[^{1\;\;1}_{1\;\;1}](v,0)|_{v=0}=
-16\pi^2\;E_2\;.
\label{c36}\ee

The only contribution again comes from the short multiplets, as
evidenced by
(\ref{D29}), since $22\cdot 15/8+13\cdot 15/4=90$.
We conclude that there are no contributions from intermediate
multiplets
in (\ref{c36}), although there are many such states in the spectrum.
The reason is that such intermediate multiplets pair up into long
multiplets.

We will finally comment on a problem where counting BPS multiplicities
is important.
This is the problem of counting black-hole microscopic states in the
case of maximal supersymmetry in type-II string theory.
For an introduction we refer the reader to \cite{mal}.
The essential ingredient is that, states can be constructed at weak
coupling,  using various D-branes. At strong coupling, these states
have the interpretation
of charged macroscopic black holes.
The number of states for given charges can be computed at weak
coupling.
These are BPS states. Their multiplicity can then be extrapolated to
strong coupling, and gives an entropy that scales as the classical area
of the black hole as postulated by Bekenstein and Hawking.
In view of our previous discussion, such an extrapolation is naive.
It is the number of unpaired multiplets that can be extrapolated at
strong coupling.
Here, however, the relevant states are the lowest spin vector
multiplets, which
as shown in appendix E always have positive supertrace.
Thus, the total supertrace is proportional to the overall number of
multiplets
and justifies the naive extrapolation  to strong coupling.

\addcontentsline{toc}{subsection}{Appendix H: Electric-Magnetic duality
in D=4}
\section*{Appendix H: Electric-magnetic duality in D=4}
\renewcommand{\theequation}{H.\arabic{equation}}
\setcounter{equation}{0}

In this appendix we will describe electric-magnetic duality
transformations
for free gauge fields.
We consider here a collection of abelian gauge fields in $D=4$.
In the presence of supersymmetry  we can write terms quadratic in the
gauge
fields as
\be
L_{\rm gauge}=-{1\over 8}{\rm Im}\int d^4 x\sqrt{-{\rm det}g}\; {\bf
F}^{i}_{\m\n}N_{ij}{\bf F}^{j,\m\n}
\,,\label{H1}\ee
where
\be
{\bf F}_{\m\n}=F_{\m\n}+i {}^\star F_{\m\n}\;\;\;,\;\;\;
{}^\star F_{\m\n}={1\over 2}{{\e_{\m\n}}^{\rho\s}\over \sqrt{-
g}}F_{\rho\s}
\,,\label{H2}\ee
with the property (in Minkowski space) that ${}^\star{}^\star F=-F$ and
\linebreak[4]
\mbox{${}^\star F_{\m\n} {}^\star F^{\m\n}=-F_{\m\n}F^{\m\n}$.}
In components,  the Lagrangian (\ref{H1}) becomes
\be
L_{\rm gauge}=-{1\over 4}\int d^4x
\left[\sqrt{-g}\;F^{i}_{\m\n}N^{ij}_2\;F^{j,\m\n}+F^{i}_{\m\n}
N_1^{ij}\;{}^\star F^{j,\m\n}\right]
\,.\label{H3}\ee

Define now the tensor that gives the equations of motion
\be
{\bf G}^{i}_{\m\n}=N_{ij}{\bf F}^{j}_{\m\n}=N_1\;F-N_2 \;{}^\star F+i
(N_2\;F+N_1\;{}^\star F)
\,,\label{H4}\ee
with $N=N_1+iN_2$.
The equations of motion can be written in the form
${\rm Im}\na^{\m}{\bf G}^{i}_{\m\n}=0$,
while the Bianchi identity is
${\rm Im}\na^{\m}{\bf F}^{i}_{\m\n}=0$,
or
\be
{\rm Im}\na^{\m}\left(\matrix{{\bf G}^{i}_{\m\n}\cr {\bf
F}^{i}_{\m\n}\cr}\right)=\left(\matrix{0\cr 0\cr}\right)
\,.\label{H5}\ee
Obviously any Sp(2r,R) transformation of the form
\be
\left(\matrix{{\bf G'}_{\m\n}\cr {\bf F'}_{\m\n}\cr}\right)=
\left(\matrix{A&B\cr C&D\cr}\right)\left(\matrix{{\bf G}_{\m\n}\cr
{\bf F}_{\m\n}\cr}\right)\,,\label{H6}\ee
where $A,B,C,D$ are $r\times r$ matrices ($CA^{t}-AC^{t}=0$,
$B^{t}D-D^{t}B=0$, $A^{t}D-C^{t}B={\bf 1}$),
preserves the collection of equations of motion and Bianchi
identities.
At the same time
\be
N'=(AN+B)(CN+D)^{-1}
\,.\label{H7}\ee

The duality transformations are
\be
F'=C(N_1\;F-N_2
\;{}^\star F)+D\;F\;\;\;,\;\;\;{}^\star F'=C(N_2\;F+N_1\;{}^\star F)+D\;{}^\star F
\,.\label{H8}\ee
In the simple case $A=D={\bf 0}$, $-B=C={\bf 1}$ they become
\be
F'=N_1\;F-N_2
\;{}^\star F\;\;\;,\;\;\;{}^\star F'=N_2\;F+N_1\;{}^\star F\;\;\;,\;\;\;N'=-{1\over N}
\,.\label{H9}\ee
When we perform duality with respect to one of the gauge fields (we
will call its component $0$) we have
\be
\left(\matrix{A&B\cr C&D\cr}\right)=\left(\matrix{{\bf 1}-e&-e\cr
e&{\bf 1}-e\cr}\right)\;\;\;,\;\;\;e=\left(\matrix{1&0&...\cr
0&0&...\cr
.&.&\cr}\right)\;,
\,.\label{H10}\ee
\be
N_{00}'=-{1\over N_{00}}\;\;,\;\;N'_{0i}={N_{0i}\over
N_{00}}\;\;,\;\;
N_{i0}'={N_{i0}\over
N_{00}}\;\;,\;\;N_{ij}'=N_{ij}-{N_{i0}N_{0j}\over N_{00}}
\,.\label{H11}\ee

Finally consider the duality generated by
\be
\left(\matrix{A&B\cr C&D\cr}\right)=\left(\matrix{{\bf 1}-e_1&e_2\cr
-e_2&{\bf
1}-e_1\cr}\right)\;,
\ee
\be
e_1=\left(\matrix{1&0&0&...\cr
0&1&0&...\cr
0&0&0&.\cr .&.&.&.\cr}\right)\;\;,\;\;e_2=\left(\matrix{0&1&0&...\cr
-1&0&0&...\cr
0&0&0&.\cr .&.&.&.\cr}\right)
\,.\label{H13}\ee
We will denote the indices in the two-dimensional  subsector where the
duality
acts by $\a,\b,\g, ...$
Then
\be
N'_{\a\b}=-{N_{\a\b}\over {\rm det}N_{\a\b}}\;\;\;N'_{\a
i}=-{N_{\a\b}\e^{\b\g}N_{\g i}\over {\rm det}N_{\a\b}}\;\;\;,\;\;\;
N'_{i\a}={N_{i\b}\e^{\b\g}N_{\a\g}\over {\rm det}N_{\a\b}}
\,,\label{H14a}\ee
\be
N'_{ij}=N_{ij}+{N_{i\a}\e^{\a\b}N_{\b\g}\e^{\g\d}N_{\d j}\over {\rm
det}N_{\a\b}}
\,.\label{H14b}\ee

Consider now the N=4 heterotic string in D=4.
The appropriate matrix N is
\be
N=S_{1}L+iS_{2}M^{-1}\;\;\;,\;\;\;S=S_1+iS_2
\,.\label{H15}\ee
Performing an overall duality as in (\ref{H9}) we obtain
\be
N'=-N^{-1}=-{S_{1}\over |S|^2}L+i{S_{2}\over |S|^2}M=-{S_{1}\over
|S|^2}L+i{S_{2}\over |S|^2}LM^{-1}L
\,.\label{H16}\ee
Thus, we observe that apart from an $S\to -1/S$ transformation on the
S field it also affects an O(6,22,$\Z)$ transformation by the matrix
$L$,
which interchanges windings and momenta of the six-torus.

The duality transformation that  acts only on S is given by
$A=D=0$, $-B=C=L$ under which
\be
N'=-LN^{-1}L=-{S_{1}\over |S|^2}L+i{S_{2}\over |S|^2}M^{-1}
\,.\label{H17}\ee
The full SL(2,$\Z$) group acting on $S$ is generated by
\be
\left(\matrix{A&B\cr
C&D\cr}\right)=\left(\matrix{a\;{\bf 1}_{28}&b\; L\cr
c\;L&d\;{\bf 1}_{28}\cr}\right)\;\;\;,\;\;\;ad-bc=1
\,.\label{H18}\ee

Finally the duality transformation, which acts as an O(6,22,$\Z$)
transformation,
is given by $A=\Omega$, $D^{-1}=\Omega^{t}$, $B=C=0$.

\end{document}